\documentstyle[12pt,twoside,graphicx]{book}

\newcommand{\be}{\begin{equation}}
\newcommand{\ee}{\end{equation}}
\newcommand{\ba}{\begin{array}}
\newcommand{\ea}{\end{array}}
\newcommand{\bea}{\begin{eqnarray}}
\newcommand{\eea}{\end{eqnarray}}
\newcommand{\bi}{\begin{itemize}}
\newcommand{\ei}{\end{itemize}}

\def\vec#1{\mathchoice{\mbox{\boldmath$\displaystyle\bf#1$}}
{\mbox{\boldmath$\textstyle\bf#1$}}
{\mbox{\boldmath$\scriptstyle\bf#1$}}
{\mbox{\boldmath$\scriptscriptstyle\bf#1$}}}

\def\bbbz{{\mathchoice {\hbox{$\sf\textstyle Z\kern-0.4em Z$}}
{\hbox{$\sf\textstyle Z\kern-0.4em Z$}}
{\hbox{$\sf\scriptstyle Z\kern-0.3em Z$}}
{\hbox{$\sf\scriptscriptstyle Z\kern-0.2em Z$}}}}

\def\i{\hbox{i}}

\def\e{\hbox{e}}

\chardef\ii="10

\def\bbbc{{\mathchoice {\setbox0=\hbox{\rm C}\hbox{\hbox
to0pt{\kern0.4\wd0\vrule height0.9\ht0\hss}\box0}}
{\setbox0=\hbox{$\textstyle\hbox{\rm C}$}\hbox{\hbox
to0pt{\kern0.4\wd0\vrule height0.9\ht0\hss}\box0}}
{\setbox0=\hbox{$\scriptstyle\hbox{\rm C}$}\hbox{\hbox
to0pt{\kern0.4\wd0\vrule height0.9\ht0\hss}\box0}}
{\setbox0=\hbox{$\scriptscriptstyle\hbox{\rm C}$}\hbox{\hbox
to0pt{\kern0.4\wd0\vrule height0.9\ht0\hss}\box0}}}}

\newcommand{\fcaption}[1]{
        \refstepcounter{figure}
        \setbox\@tempboxa = \hbox{\footnotesize Fig.~\thefigure. #1}
        \ifdim \wd\@tempboxa > 5in
           {\begin{center}
        \parbox{5in}{\footnotesize\smalllineskip Fig.~\thefigure. #1}
            \end{center}}
        \else
             {\begin{center}
             {\footnotesize Fig.~\thefigure. #1}
              \end{center}}
        \fi}
\newcommand{\tcaption}[1]{
        \refstepcounter{table}
        \setbox\@tempboxa = \hbox{\footnotesize Table~\thetable. #1}
        \ifdim \wd\@tempboxa > 5in
           {\begin{center}
        \parbox{5in}{\footnotesize\smalllineskip Table~\thetable. #1}
            \end{center}}
        \else
             {\begin{center}
             {\footnotesize Table~\thetable. #1}
              \end{center}}
        \fi}
\begin{document}
\begin{titlepage}
\begin{center}
\vspace*{1.5cm}
{\Large\bf Levon G. Mardoyan}
\footnote{Joint Institute for Nuclear Research, Dubna, Russia}$^,$\footnote{Yerevan State University, Yerevan, Armenia \\ e-mail: mardoyan@theor.jinr.ru, mardoyan@ysu.am}

\vspace*{1.0cm}

\vspace{1.5cm}

\vspace{0.5cm}
{\LARGE{\it\bf Superintegrability

\vspace{0.2cm}

and

\vspace{0.5cm}

Coulomb-Oscillator Duality}}

\vspace{1.5cm}

\vspace{1.0cm}

\vfill {\bf 2024}
\end{center}

\end{titlepage}

\vspace{0.3cm}

\tableofcontents

\newpage

\addcontentsline{toc}{chapter}{Introduction}

\chapter*{Introduction}
\pagestyle{myheadings}
\markboth{INTRODUCTION}{INTRODUCTION}

This review is devoted to the problem of Coulomb (dyon)--oscillator duality in non-relativistic quantum mechanics,
which is based on the so--called non--bijective quadratic transformations, i.e. Levi--Civita transformation \cite{LC},
Kustaanheimo--Stiefel (KS--transformation) \cite{KS} and Hurwitz transformation \cite{HUR}.

In classical mechanics, the Kepler-Coulomb and harmonic oscillator potentials differ from other potentials in that
during the finite motion of a particle, according to Bertrand's theorem \cite{Bertrand}, only in these fields the particle trajectories
are closed curves. This fact (completeness of the trajectory) in quantum mechanics results in random degeneration of energy
spectra with respect to the orbital quantum number $l$. This property is associated with the presence of a hidden symmetry group,
i.e. symmetry higher than the group of geometric symmetry of the system, which is directly connected with the symmetry of physical space.
An example of the latter is the group of three--dimensional rotations for the Schr\"{o}dinger equation with any centrally symmetric potential.

We can consider that the development of the theory of systems with hidden symmetry begins with the work of Laplace \cite{Laplace},
in which it was shown that in the Kepler problem, along with angular momentum, another vector quantity is conserved,
modulo equal to eccentricity, lying in the orbital plane and directed along the large ellipse axis. It is this additional integral of
motion that is the reason for the completeness of the trajectory in the Kepler problem. The Laplace additive integral of motion was
rediscovered with a significant delay by Runge \cite{Runge} and after him then by Lenz \cite{Lenz}.
A generalization of the Laplace-Runge-Lenz integral of motion to the case of quantum mechanics was used by Pauli \cite{Pauli}
to calculate the energy spectrum of the hydrogen atom \cite{LL}. An explanation of the random degeneracy of the energy spectrum of the
hydrogen atom led Fock to the concept of hidden symmetry of the hydrogen atom.
Fock showed \cite{FOCK1,FOCK2} that the Schr\"{o}dinger equation of the hydrogen atom, when moving to the momentum representation and stereographic
projections, turns into an integral equation that describes the free movement of a particle on a three-dimensional sphere and is invariant
under the group of four--dimensional rotations $O(4)$. In the input Schr\"{o}dinger equation this symmetry is not visible and
therefore was called hidden or dynamic symmetry. This $O(4)$ symmetry, which contains the group of three--dimensional rotations $O(3)$
as a subgroup, explains why in a Coulomb field, along with the orbital momentum, there should be another vector integral of motion.
The connection between the Pauli and Fock approaches was established by Bargman \cite{BARGMAN}.
He showed that the properly normalized Laplace--Runge--Lenz--Pauli vector $A_{i}$ and the orbital momentum operator $l_{i}$ together obey the commutation
relations for the algebra $o(4)$. For a continuous spectrum, instead of a three-dimensional sphere, we have to deal with a three-dimensional
hyperboloid, and instead of a finite-dimensional representation of the group $O(4)$, we have to deal with infinite-dimensional unitary
representations of the Lorentz group $O(3,1)$ \cite{PER-POP,PER-POP1,BANDER}.
The hidden symmetry of the multidimensional $(n>3)$ Kepler-Coulomb problem was considered in \cite{ALLILUEV,revai}, and the dynamic symmetry
group of the one-dimensional problem was first found by our group in \cite{DPST1,LMPST1}. For an isotropic harmonic oscillator, in addition
to the angular momentum $\vec{l}$, the tensor quantity $T_{ik}=p_{i}p_{k}+ x_{i}x_{k}$ is preserved \cite{DEMKOV1}. The accidental
degeneracy of the energy spectrum of an isotropic oscillator in terms of the hidden symmetry group $U(n)$ was also explained in
\cite{JAUCH1,JAUCH2,BAKER}.

It is important that the connection between the hidden symmetry groups of the Kepler-Coulomb problem $O(n+1)$ and the isotropic oscillator $U(n)$
leads in quantum mechanics to the concept of dyon-oscillator duality \cite{TERANT}. Let us recall that the dyon is a hypothetical particle introduced by
Schwinger \cite{schwinger}, which, unlike the Dirac monopole \cite{DIRAC}, is endowed with not only a magnetic but also an electric charge.
The dyon-oscillator duality in the Schr\"{o}dinger equations is as follows.

The Schr\"{o}dinger equation for an isotropic oscillator has two parameters – $E$ and cyclic frequency $\omega$. Quantization leads to the limitation
$E=\hbar \omega (N+D/2)$, where $N=0,1,2,…$ and $D$ is the dimension of the oscillator configuration space. If $\omega$ is fixed, then energy is
quantized, and this is the standard situation. Let us now imagine that the energy $E$ is fixed. Then $\omega$ is necessarily quantized and we are in a
non-standard situation. The question is that the non-standard situation corresponds to some kind of physics, i.e. is it possible to find a
transformation that translates an isotropic  oscillator into a physical system with a coupling constant $\alpha$ as a function of $E$, and with an
energy $\varepsilon$ depending on $\omega$?

If such a transformation exists, we can state that the "non--standard oscillator" is identical with this physical system, and the original oscillator
and the final system are dual to each other. Both the "standard" and "non-standard" modes are mutually exclusive, the original oscillator and the final
"charge--dyon" system are dual to each other, and this explains the significance of the term "dyon--oscillator duality".
Note also that the energy spectrum of the original system is only discrete, i.e. the particle undergoes a finite motion (for such cases they usually
say that we are a model is with confinement). Generally speaking, a finite system has both a discrete and a continuous spectrum, i.e. there is no
confinement in this model. However, unlike the first model, in the second model we have a monopole. There is some analogy between the dyone--oscillator duality
and the Seiberg--Witten duality, according to which gauge theories with strong coupling are equivalent to theories in which, on the one hand,
there is a weak coupling, and on the other, there are topologically non-trivial objects in the form of monopoles and dyons \cite{SW}.

All the systems we consider in the review are a subclass of integrable systems and they are usually called superintegrable \cite{Wojciechowski}.
It is known \cite{Arnold,PERELOMOV} that a classical system with $N$ degrees of freedom is completely integrable in the sense of Liouville
(or integrable in quadratures) if $N$ functionally independent integrals of motion in involution are known, i.e. for which the Poisson
brackets are equal to zero. In quantum mechanics, a system is called integrable if $N-1$ algebraically independent operators
are known that commute both with the Hamiltonian of the system and with each other.
For superintegrable systems, the number of independent integrals of motion can be equal to either $2N-1$ \cite{LL-1} or $2N-2$
\cite{EVANS-1,EVANS-2,EVANS-3}. In the first case, the system is called a maximally integrable system, and in the second, a minimally
superintegrable system. It is known that the Kepler--Coulomb problem is solved by the method of separation of variables in four
(spherical, parabolic, prolate spheroidal and spheroconical); and the isotropic oscillator, in eight (Cartesian, cylindrical, spherical,
prolate spheroidal, oblate spheroidal, elliptical cylindrical and ellipsoidal) coordinate systems.
From a physical point of view, the importance of separating variables in multiple coordinate systems is obvious. In the spectroscopic
problem of hydrogen-like systems, a spherical coordinate system is used, in the Stark effect, a parabolic coordinate system
is used, and in the two-center problem, an elongated spheroidal coordinate system is used. Thus, the choice of a specific basis is dictated
by considerations of convenience and there is often a need to move from one basis to another. The first interbasis transition can be
considered the well-known Rayleigh expansion, i.e. expansion of a plane wave into spherical waves.
In the theory of the hydrogen atom, the worst interbasis expansion is the result of Stone \cite{STONE}, who obtained the expansion of the parabolic
basis into the spherical basis in the momentum representation. Further, Park \cite{PARK}, within the framework of the group-theoretical method,
and Tarter \cite{TARTER}, using a purely analytical method, obtained Stone's result in the configuration space. By representing the spheroidal
basis of the hydrogen atom in the form of expansion over a spherical basis, Coulson and Joseph \cite{COJO} reduced the expansion problem to
solving an algebraic system of homogeneous equations.
The expansion of the parabolic Coulomb problem in terms of a spherical basis in a continuous spectrum was considered by Mayundar and Basu \cite{BESU}.
Interbasis expansions in an isotropic oscillator were first discussed in \cite{TOLAR,CHACON}. Using examples of quantum systems with Coulomb and
oscillator interactions, the works of our group \cite{M-1} -- \cite{DPST-2} developed effective methods for calculating the coefficients of
interbasis expansions (asymptotic method and orthogonality of radial wave functions with respect to orbital momentum).
It should be noted that all possible interbasis expansions for the two- and three-dimensional hydrogen atom, as well as for the circular
oscillator, are given in our monograph \cite{MONO}.

This review considers the following superintegrable systems:

1. Maximally superintegrable systems: four--dimensional and eight--dimensional isotropic oscillators, five--dimensional Kepler--Coulomb
problem and $SU(2)$--Yang--Coulomb monopole and three-dimensional MIC--Kepler problem.

2. Minimally superintegrable systems: generalized Kepler-Coulomb problem, generalized oscillator system, generalized MIC--Kepler problem
and four--dimensional double--singular oscillator.

The issues discussed in the review are distributed among the chapters as follows.

The first chapter, written on the basis of the works \cite{DMPST} -- \cite{M-8}, is devoted to the Hurwitz transformation and coulomb-oscillator duality. It is shown
that the nonbijective quadratic transformation (Hurwitz transformation) generated by the Keley matrix converts the problem of an eight--dimensional
isotropic oscillator into a five--dimensional Kepler--Coulomb problem. A parametrization is proposed in which the Hurwitz transformation is equivalent
to the Euler addition of angles. It is shown that in this parametrization the reduction of state vectors is realized by the addition theorem for
the Wigner $D$--function. The problem of an eight-dimensional oscillator in Euler coordinates is analyzed. The orthogonality of the radial wave functions
of centrally symmetric systems with respect to the orbital quantum number is proven and it is shown that this orthogonality condition is a
consequence of the accidental degeneracy of the energy spectrum.
In particular, it is shown that for the hydrogen atom this occurs not only in flat space but also in curved space of constant curvature.
The spherical and cylindrical bases of the eight--dimensional isotropic oscillator are constructed, and two representations are found for
the coefficients of the interbasis expansions: "sphere--cylinder" and "cylinder--sphere". Three-term recurrence relations are established
that generate a spheroidal basis of the eight-dimensional isotropic oscillator.
An analysis of the five-dimensional Coulomb problem in Euler coordinates is given. The spherical and parabolic bases of this system are calculated,
and Park and Tarter representations for the coefficients of sphero--parabolic and parabola--spherical interbasis expansions are derived.
Three-term recurrence relations are found that determine the form of the Coulson--Joseph amplitudes. The quantum mechanical problem of
scattering in a five-dimensional Coulomb field is solved and the hyperspherical and parabolic wave functions of the Coulomb problem in the
continuous spectrum are calculated.

The second chapter examines the dyon-oscillator duality. It is written based on the works \cite{M-9} -- \cite{M-15}. Within the analytical approach,
using a generalized version of the Hurwitz transformation, a five-dimensional system composed of a Yang monopole and an isospin particle,
coupled to each other by $SU(2)$ and Coulomb interaction ($SU(2)$--Yang--Coulomb monopole), is constructed from an eight-dimensional isotropic
oscillator. The Clebsch--Gordan expansion due to the space-calibration coupling is calculated.
It is shown that the variables in the Schr\"{o}dinger equation for the coupled "charge--$SU(2)$--monopole" system are separated in five-dimensional
hyperspherical, parabolic and spheroidal coordinates. The hyperspherical and parabolic wave functions and the energy spectrum of this system
are found, and the multiplicity of degeneracy of the energy spectrum is calculated. It is shown that the coefficients of the expansion of a
parabolic basis in a hyperspherical one are expressed through the Clebsch--Gordan coefficients of the $SU(2)$ group.
Three-term recurrence relations are derived that govern the expansion coefficients of the spheroidal basis of the $SU(2)$--Yang--Coulomb monopole
in hyperspherical and parabolic terms, respectively. An analogue of the Runge-Lenz vector for the $SU(2)$--Yang-Coulomb monopole is found and the
hidden symmetry group $SO(6)$ is established. It is also shown that the hidden symmetry group allows one to calculate the energy spectrum of a
system purely algebraically. Both hyperspherical and parabolic wave functions of the $SU(2)$--Yang--Coulomb monopole in the continuous spectrum were found,
and the quantum mechanical problem of scattering of charged particles in the field of the $SU(2)$--Yang--Coulomb monopole was solved.

The third chapter is written on the basis of a series of works \cite{M-16} -- \cite{M-20}. This chapter discusses the problem of separating variables into
the Schr\"{o}dinger equation for the MIC-Kepler problem describing the motion of a charged particle in the field of a Dirac dyon and establishes
the dyon(Coulomb)--oscillator correspondence. A linear transformation is found that connects the fundamental, spherical and parabolic bases
of the MIC-Kepler problem. Using this transformation, three-term recurrence relations are derived that generate the spheroidal basis of
this system. The linear Stark effect is considered, and it is shown that a constant electric field completely removes
the degeneracy of energy levels in terms of the azimuthal quantum number. The spherical and parabolic wave functions of the charge-dyon system in the
continuous spectrum are calculated. It is shown that the "parabola-sphere" and "sphere-parabola" expansion
coefficients  are proportional to the hypergeometric function $_{3}F_{2}(...|1)$. The quantum mechanical problem
of scattering of charged particles in the field of the Dirac dyon has been solved.  It is shown that the MIC--Kepler problem is dual to the four-dimensional
isotropic oscillator and that the duality transformation is a generalized version of the Kustaanheimo--Stiefel \cite{KS} transformation (KS--transformation).
The wave functions of the four-dimensional isotropic oscillator in Cartesian, Eulerian, double polar, spherical coordinates and the coefficients of their
mutual expansions are found. Three-term recurrence relations are derived for the expansion coefficients of the spheroidal basis in the spherical and
parabolic bases for the MIC--Kepler problem and in the Eulerian and double polar basis for the four-dimensional isotropic oscillator. Spheroidal corrections
to the spherical and parabolic bases of the MIC--Kepler problem, as well as to the Eulerian and double polar bases of the four-dimensional isotropic oscillator,
are calculated. A correspondence has been established between the wave functions of the four-dimensional isotropic oscillator and the MIC--Kepler problem.

The fourth chapter is written based on the works \cite{M-21} -- \cite{M-25}. The problems discussed in this chapter relate to the quantum mechanics of ring-shaped potentials.
An explicit expression for the ring-shaped matrix connecting ring-shaped functions related to different values of the axiality parameter is obtained.
A connection between this matrix and the $6j$-Wigner symbols is found. Special attention is paid to the expansion of a hoop-shaped function into
spherical ones. The motion of a quantum particle in a ring-shaped model with a zero "bare" potential is studied. The bases of this model
factorized in spherical and cylindrical coordinates are obtained.
A formula is derived that generalizes the Rayleigh expansion of a plane wave into spherical waves in a ring-shaped model. A dynamic system generalizing
the Kepler--Coulomb and Hartmann systems is considered. It is shown that the variables in the Schr\"{o}dinger equation for this generalized Kepler--Coulomb
system can be separated in a prolate spheroidal coordinate system. The coefficients of interbasis expansions between three bases (spherical, parabolic
and spheroidal) are studied in detail. It is found that the coefficients of the expansion of a parabolic basis into a spherical one and the inverse
expansion can be expressed in terms of the Clebsch--Gordan coefficients of the $SU(2)$ group, analytically extended to real values of the argument.
It is shown that the expansion coefficients of the spheroidal basis in terms of the spherical and parabolic bases satisfy the three-term recurrence relations.
A study was carried out of a non-relativistic quantum mechanical dynamic system that generalizes a three-dimensional isotropic harmonic oscillator. It is
shown that the variables in the Schr\"{o}dinger equation for a generalized oscillatory system are separated in spherical, cylindrical and spheroidal (prolate
and oblate) coordinate systems. The problem of interbasis expansions of the wave functions of a generalized oscillator system is completely solved.
It is shown that the coefficients of the direct and inverse expansions of a cylindrical basis on a spherical basis are expressed in terms of the Clebsch--Gordan
coefficients of the $SU(2)$ group and, as in the case of the generalized Kepler--Coulomb system are analytically extended to the region of arbitrary real values
of the moment. Three--term recurrence relations are derived that generate spheroidal (prolate and oblate) bases of the generalized oscillatory system and
spheroidal corrections to the spherical and cylindrical bases.

In the fifth chapter, we propose a new model of a minimal superintegrable system, which we called the generalized MIC--Kepler problem \cite{M-26}.
It is shown that the variables in the Schr\"{o}dinger equation for the generalized MIC--Kepler problem in spherical, parabolic and prolate
spheroidal coordinate systems are separated \cite{M-27}. Integrals of motion are found that are responsible for the separation of variables in the coordinate systems
indicated above. Orthonormal spherical and parabolic wave functions are obtained. The orthogonality of the radial wave function with respect to the
orbital momentum is proven and the problem of interbasis expansions is solved. Three-term recurrence relations are established for the coefficients
of expansions of the spheroidal basis in the spherical and parabolic bases of the generalized MIC-Kepler problem. It is shown that the four-dimensional
double singular oscillator is dual to the generalized MIC--Kepler problem and the duality transformation is again a generalized version of the KS--transformation \cite{KS}.
The integrals of motion of the four-dimensional double singular oscillator responsible for the separation of variables in the Schr\"{o}dinger equation in Euler,
double-polar and spheroidal coordinates are given \cite{M-28}. Spherical and parabolic wave functions were calculated for the generalized MIC–-Kepler system in
the continuous spectrum \cite{M-29}. It is shown that the parabola–sphere and sphere–parabola expansion coefficients are expressed through the generalized hypergeometric
function $_{3}F_{2}(... | 1)$. The quantum mechanical problem of scattering in the generalized MIC–-Kepler system is solved \cite{M-29}.

In conclusion, I am grateful to my colleagues and co-authors S. Bellucci, A.A. Gusev, L.S. Davtyan, M. Kibler, I.V. Lutsenko, A.P. Nersessian,
L.S. Petrosyan, M.G. Petrosyan, G.S. Pogosyan, H.A. Sarkisyan, A.N. Sissakian, V.M. Ter-Antonyan, S.I. Vinitsky, V. Yeghikyan and
A. Yeranyan for the long-term and fruitful scientific cooperation.

\newpage
\chapter{Coulomb-oscillator analogy}
\markboth{CHAPTER 1. COULOMB-OSCILLATOR ANALOGY}{}

\section{Coulomb-oscillator analogy}
\markboth{CHAPTER 1. COULOMB-OSCILLATOR ANALOGY}{1.1. Coulomb-oscillator analogy}

The purpose of this section is to illustrate one property of the Schr\"{o}dinger equation,
which we call the Coulomb-oscillator duality.

Consider the radial equation for a D-dimensional isotropic oscillator
\bea \
\label{1.1.1}
\frac{d^{2}R}{du^{2}}+\frac{D-1}{u}\frac{dR}{du}-\frac{L(L+D-2)}{u^{2}}R
+\frac{2\mu_{0}}{\hbar^{2}}\left(E-\frac{\mu_{0}\omega^{2}u^{2}}{2}\right)R=0.
\eea
Here the $R$-radial wave function of a $D$-dimensional isotropic oscillator $(D>2)$,
$L=0,1,2,...$ - are the eigenvalues of the global moment operator.

After substituting $r=u^{2}$, equation (\ref{1.1.1}) becomes the equation
\bea \frac{d^2 R}{d r^2}+\frac{d-1}{r}\frac{d R}{d
r}-\frac{l(l+d-2)}{r^2}R +\frac{2\mu_0}{\hbar^2}
\left(\varepsilon+\frac{\alpha}{r}\right)R=0, \label{1.1.2} \eea
where $d=D/2+1$, $l=L/2$,
\bea
\varepsilon=-\mu_0\omega^2/8, \qquad \alpha=E/4.
\label{1.1.3}
\eea
This is a very unexpected result. If $D=4,6,8,10...$, then $d=3,4,5,6,...$, and equation (\ref{1.1.2})
is similar in form to the radial equation for the $d$-dimensional Kepler-Coulomb problem. For odd $D>2$,
the quantity $d$ runs through half-integer values and therefore cannot have the meaning of the dimension
of space in the generally accepted sense. Further, $l$ takes not only integer, but also integer values,
which means it has the meaning of a total moment and the question arises of where the fermionic degree
of freedom comes from. We will answer this question later. Finally, as noted above, equations (\ref{1.1.1})
and (\ref{1.1.2}) are dual to each other, and the duality transformation is the transformation $r=u^{2}$.

The condition $r=u^{2}$ in Cartesian coordinates has the form
\bea 
x_0^2+x_1^2 +\dots +x_{d-1}^2=\left(u_0^2+u_1^2+\dots
+u_{D-1}^2\right)^2, \label{1.1.4} 
\eea
which is called the Euler identity. According to the Hurwitz theorem
\cite{HUR}, if $x_i$ $(i=0,1,\dots,d-1)$ is a bilinear combination $u_\mu$
$(\mu=0,1,\dots,D-1)$, then identity (\ref{1.1.4}) is valid only for the
following pairs of numbers:
\begin{eqnarray*}
(D,\,d) = (1,\,1),\,(2,\,2),\,(4,\,3),\,(8,\,5).
\end{eqnarray*}
However, in [78,79] it is shown that Euler's identity also holds for the pair (D,\,d) = (16,\,9).
The "magic" numbers $D=1,2,4,8,16$ are directly related to the fact of the existence of four fundamental
algebraic structures: real numbers, complex numbers, quaternions and octanions. Moreover, transformation
(\ref{1.1.4}) establishes a connection between two fundamental problems of mechanics: the oscillator and
the Kepler-Coulomb problems.

The transformation $(D,d)=(1,1)$ connects the problem of a linear oscillator with the problem of a
one-dimensional Coulomb anion \cite{TERANT}.

The transformation $(D,d)=(2,2)$ is the Levi-Cevita transformation known from celestial mechanics \cite{LC}.
This transformation is a duality transformation, which transforms the problem of a circular oscillator
into the problem of a two-dimensional anion \cite{NTAT,ARTUR,NTA}.

Further, the transformation corresponding to a pair of numbers $(D,d)=(4,3)$ in celestial mechanics is called the Kustaanheimo-Stiefel
transformation (KS-transformation) \cite{KS}. The KS-transformation reduces the problem of a four-dimensional isotropic oscillator into
the well-known MIC-Kepler system \cite{Iwai-Uwano} - \cite{PRIS-2}. The superintegrable MIC-Kepler system was constructed by Zwanziger
\cite{ZWANZIG} and rediscovered by McIntosh and Cisneros \cite{MIC} and is a generalization of the Coulomb problem in the presence of a
Dirac monopole \cite{DIRAC}. There are generalizations of the MIC-Kepler system to a three-dimensional sphere \cite{GRITKUROT}
and a hyperboloid \cite{NERPOG}. The MIC-Kepler system was considered from different points of view in \cite{MLADENOV-1} - \cite{M-33}.

The Hurwitz transformation, which corresponds to the case $(D,d)=(8,5)$ reduces transforms the problem of an eight-dimensional isotropic
oscillator into a five-dimensional Coulomb problem \cite{DMPST,Lambert}. Supplementing the Hurwitz transformation with three angles
$(\alpha_T, \beta_T, \gamma_T)$ \cite{HVT}, we find a transformation that transforms the space $\rm I\!R^8$ into the direct product
$\rm I\!R^8 = \rm I\!R^5 \otimes S^3$ of the space $\rm I\!R^5({\bf x})$ and the three-dimensional sphere $S^3(\alpha_T, \beta_T,\gamma_T)$S
\cite{M-5}. As a result of such a separation of the space $\rm I\!R^8$, the eight-dimensional isotropic oscillator can be used to construct
a bound system of a charged particle and the five-dimensional $SU(2)$ Yang monopole \cite{YANG-1} with a topological charge $\pm 1$
which we call the  $SU(2)$ Yang--Coulomb monopole \cite{M-9,M-10,M-11}. It should be emphasized that a group of six-dimensional rotations
$SO(6)$ is a group of hidden symmetry $SU(2)$ of the Yang--Coulomb monopole \cite{M-13} and this symmetry is the reason of separation of
variables in the Schr\"{o}dinger equation in the five-dimensional hyperspherical, parabolic, and spheroidal coordinates \cite{M-12}.
The $SU(2)$ Yang-Coulomb monopole is considered in various aspects in \cite{Platyukhov-1,Platyukhov-2,Platyukhov-3}.

And finally, in \cite{Le-1,Le-2,Le-3} it is shown that the generalized Hurwitz transformation, which corresponds to the pair
$(D, d) = (16,9)$, reduces the problem of a sixteen-dimensional isotropic oscillator into the nine-dimensional problem of a hydrogen-like atom
in the field of a non-Abelian  $SO(8)$ monopoles (nine-dimensional MICZ-Kepler problem). A nine-dimensional analogue of the Runge-Lenz vector
is constructed and it is established that the hidden symmetry group of the nine-dimensional MICZ-Kepler system is the $SO(10)$ group.

\section{Hurwitz transformation}
\markboth{CHAPTER 1. COULOMB-OSCILLATOR ANALOGY}{1.2. HURWITZ TRANSFORMATION}

Here we show that the nonbijective bilinear Hurwitz transformation generated by the Kelly matrix \cite{Zhevl}
transforms the eight-dimensional isotropic oscillator problem into a five-dimensional Kepler-Coulomb problem.

Let us introduce a nonbijective quadratic transformation connecting the Cartesian coordinates
$(x_{0}, \dots, x_{4})$ of the space $\rm I\!R^5$ with the coordinates $(u_{0},\dots,u_{7})$
of the space $\rm I\!R^8$ \cite{DMPST}
\begin{eqnarray}
\left(\begin{array}{c}
x_0\\x_1\\x_2\\x_3\\x_4\\0\\0\\0
\end{array}
\right)=\left(
\begin{array}{cccccccc}
u_0&u_1&u_2&u_3&-u_4&-u_5&-u_6&-u_7\\
u_4&u_5&-u_6&-u_7&u_0&u_1&-u_2&-u_3\\
u_5&-u_4&u_7&-u_6&-u_1&u_0&-u_3&u_2\\
u_6&u_7&u_4&u_5&u_2&u_3&u_0&u_1\\
u_7&-u_6&-u_5&u_4&u_3&-u_2&-u_1&u_0\\
u_1&-u_0&u_3&-u_2&u_5&-u_4&u_7&-u_6\\
u_2&-u_3&-u_0&u_1&-u_6&u_7&u_4&-u_5\\
u_3&u_2&-u_1&-u_0&-u_7&-u_6&u_5&u_4\\
\end{array}
\right) \left(
\begin{array}{c}
u_0\\u_1\\u_2\\u_3\\u_4\\u_5\\u_6\\u_7
\end{array}
\right).
\label{1.2.1}
\end{eqnarray}
Hence, it follows that
\begin{eqnarray}
x_0 &=& u_0^2 + u_1^2 + u_2^2 + u_3^2 -u_4^2 - u_5^2- u_6^2 -
u_7^2, \nonumber \\ [2mm] x_1 &=& 2\left((u_0u_4 + u_1u_5 - u_2u_6 -
u_3u_7\right), \nonumber \\ [2mm] x_2 &=& 2\left(u_0u_5 - u_1u_4 + u_2u_7 -
u_3u_6\right), \label{1.2.2} \\ [2mm]
x_3 &=& 2\left(u_0u_6 + u_1u_7 + u_2u_4+ u_3u_5\right), \nonumber \\ [2mm]
x_4 &=& 2(u_0u_7 - u_1u_6 - u_2u_5 +u_3u_4). \nonumber
\end{eqnarray}
It should be noted that to each element in $\rm I\!R^5$ there corresponds not a
single element but a whole set of elements in $\rm I\!R^8$ called a fibre.
Therein lies the property of nonbijectivity of the transformation $\rm
I\!R^8(\bf{u})\to\rm I\!R^5(\bf{x})$.

The matrix $H(u;8)$ in (\ref{1.2.1}) differs from the well-known Kelly
matrix \cite{Zhevl} by a certain rearrangement of lines (see also \cite{Lambert}).
It is easy to verify that for the matrix $H(u;8)$ the following condition holds:
\bea
H_{\mu \lambda}H_{\lambda \nu}^T = u^2 \delta_{\mu \nu},
\label{1.2.3}
\eea
which guarantees the fulfillment of the Euler identity. Hereafter, unless
otherwise specified, Greek letters will take values $0, 1,...,7$; and Latin
letters, $0,1,...,4$.

Using the explicit form of the matrix $H(u;8)$ we also obtain that
\bea
\frac{\partial H_{\mu \nu}}{\partial u_\nu} &=& 0,
\label{1.2.4}
\\ [3mm]
\frac{\partial x_j}{\partial u_\mu} &=& 2 H_{j \mu}.
\label{1.2.5}
\eea

Let us now deal with the transformation of derivatives. Taking (\ref{1.2.5}) into account, we obtain
\bea
\frac{\partial}{\partial u_\mu} = H_{i \mu}
\frac{\partial}{\partial x_i}.
\label{1.2.6}
\eea
Multiplying (\ref{1.2.6}) by $H_{j \mu}$, summing over $\mu$, and keeping in
mind the condition (\ref{1.2.3}), it is easy to show that
\bea \frac{\partial}{\partial x_i} = \frac{1}{2u^2}H_{i \mu}
\frac{\partial}{\partial u_\mu}.
\label{1.2.7}
\eea
Let us introduce the following notation
\bea
\frac{\partial}{\partial q_\mu} = \frac{1}{2u^2}H_{\mu \nu}
\frac{\partial}{\partial u_\nu},
\label{1.2.8}
\eea
where $q_\mu=(x_j,0,0,0)$. Formula (\ref{1.2.8}) generalizes relation
(\ref{1.2.7}) to the case $\mu=5,6,7$.

Now consider the second derivatives. From (\ref{1.2.8}) it follows that
\begin{eqnarray*}
\frac{\partial^2}{\partial q^2_\mu} = \frac{1}{2}
\frac{\partial}{\partial q_\mu}\left(\frac{1}{u^2}H_{\mu \nu}
\frac{\partial}{\partial u_\nu}\right).
\end{eqnarray*}
Repeatedly applying formula (\ref{1.2.8}) and taking into account condition (\ref{1.2.3}) we obtain
\begin{eqnarray*}
\frac{\partial^2}{\partial q^2_\mu} = \frac{1}{4u^2}H_{\mu
\lambda} \left(\frac{\partial}{\partial
u_\lambda}\frac{1}{u^2}H_{\mu \nu}\right) \frac{\partial}{\partial
u_\nu} + \frac{1}{4u^2} \frac{\partial^2}{\partial u^2_\mu}.
\end{eqnarray*}
In the first term, we can differentiate by parts, use condition (\ref{1.2.3}) and then
identity (\ref{1.2.4}), we as a result obtain
\bea
\frac{\partial^2}{\partial q^2_\mu} = \frac{1}{4u^2}
\frac{\partial^2}{\partial u^2_\mu}.
\label{1.2.9}
\eea
Relation (\ref{1.2.9}) and the Euler identity (\ref{1.1.4}) are the only
"mathematical tools" required for transformation of the Schr\"{o}dinger
equation of the eight-dimensional isotropic oscillator into the Schr\"{o}dinger
equation of the five-dimensional Coulomb problem. Before proceeding to this
transformation let us return to formula (\ref{1.2.7}) and introduce the
operators
\bea
\hat{\cal L}_w = iu^2\frac{\partial}{\partial q_w} =
\frac{i}{2}H_{w \mu}\frac{\partial}{\partial u_\mu},
\label{1.2.10}
\eea
where $w=5,6,7$. From this definition and formula (\ref{1.2.5}) it follows that
\begin{eqnarray*}
\hat{\cal L}_w x_j = iu^2{\delta}_{wj} = 0,
\end{eqnarray*}
as $w=5,6,7$ and $j=0,1,...,4$. Thus, the operators (\ref{1.2.10}) are
independent of the coordinates $x_j$ and, therefore, for an arbitrary function
$f(\bf x)$ the following identity holds:
\begin{eqnarray*}
\hat{\cal L}_w f(\bf x) = 0.
\end{eqnarray*}
Based on formula (\ref{1.2.10}) and the matrix $H(u;8)$, it can be proven that operators
(\ref{1.2.10}) have the following explicit forms:
\begin{eqnarray*}
\hat{\cal L}_5 = \frac{i}{2}\left(u_1\frac{\partial}{\partial u_0}-
u_0\frac{\partial}{\partial u_1}+u_3\frac{\partial}{\partial u_2}-
u_2\frac{\partial}{\partial u_3}+u_5\frac{\partial}{\partial u_4}-
u_4\frac{\partial}{\partial u_5}+u_7\frac{\partial}{\partial u_6}-
u_6\frac{\partial}{\partial u_7}\right),
\\ [3mm]
\hat{\cal L}_6 = \frac{i}{2}\left(u_2\frac{\partial}{\partial u_0}-
u_3\frac{\partial}{\partial u_1}-u_0\frac{\partial}{\partial u_2}+
u_1\frac{\partial}{\partial u_3}-u_6\frac{\partial}{\partial u_4}+
u_7\frac{\partial}{\partial u_5}+u_4\frac{\partial}{\partial u_6}-
u_5\frac{\partial}{\partial u_7}\right),
\\ [3mm]
\hat{\cal L}_7 =
\frac{i}{2}\left(u_3\frac{\partial}{\partial u_0}+
u_2\frac{\partial}{\partial u_1}-u_1\frac{\partial}{\partial u_2}-
u_0\frac{\partial}{\partial u_3}-u_7\frac{\partial}{\partial u_4}-
u_6\frac{\partial}{\partial u_5}+u_5\frac{\partial}{\partial u_6}+
u_4\frac{\partial}{\partial u_7}\right).
\end{eqnarray*}
Now denoting these operators in a different way
\begin{eqnarray*}
{\hat {\cal J}}_1 =  \hat{\cal L}_5, \qquad {\hat {\cal J}}_2 =
\hat{\cal L}_6, \qquad {\hat {\cal J}}_3 = \hat{\cal L}_7
\end{eqnarray*}
and using the explicit form of the operators $\hat{\cal L}_5,$ $\hat{\cal
L}_5,$ and $\hat{\cal L}_5,$ one can prove by direct calculation that the
operators ${\hat {\cal J}}_1,$ ${\hat {\cal J}}_2,$ ${\hat {\cal J}}_3,$
satisfy the commutation relations
\begin{eqnarray*} \left[{\hat {\cal J}}_a, {\hat {\cal J}}_b\right] =
i\epsilon_{abc} {\hat {\cal J}}_c,
\end{eqnarray*}
where $a,b,c=1,2,3$. Formula (\ref{1.2.9}) with taking account of
(\ref{1.2.10}) results in that the Laplacian $\frac{\partial^2}{\partial
x^2_j}$ and $\frac{\partial^2}{\partial u^2_\mu}$ are related by
\bea \frac{\partial^2}{\partial u^2_\mu} =
4r\frac{\partial^2}{\partial x^2_j} -\frac{4}{r}{\hat{\cal J}}^2,
\label{1.2.11}
\eea
in which ${\hat{\cal J}}^2$ is determined as

\begin{eqnarray*}
{\hat{\cal J}}^2 = {\hat{\cal J}}^2_1 + {\hat{\cal J}}^2_2 +
{\hat{\cal J}}^2_3.
\end{eqnarray*}

\section{8D oscillator and 5D Coulomb problem}
\markboth{CHAPTER 1. COULOMB-OSCILLATOR ANALOGY}{1.3. 8D OSCILLATOR AND 5D COULOMB PROBLEM}

Now let's relate the eight-dimensional problem of the isotropic oscillator
\bea
\left(-\frac{\hbar^2}{2\mu_0}\frac{{\partial}^2}{\partial u^2_\mu}
+ \frac{\mu_0\omega^2u^2}{2}\right)\psi({\bf u}) = E\psi({\bf u}),
\label{1.3.1} \\
[3mm] E = \hbar \omega \left(N + 4 \right), \qquad N = 0,1,2,...,
\label{1.3.2} \eea
where $N$ is the principal quantum number of the isotropic oscillator,
with the five-dimensional Coulomb problem.

Since the operators ${\hat {\cal J}}_a$ do not depend on the coordinates $x_j$, let us assume that we can
represent the wave function $\psi(\bf u)$ of an eight-dimensional isotropic oscillator in the
following factorized form:
\bea
\psi({\bf u}) = \psi({\bf x})\Phi(\Omega_a),
\label{1.3.3}
\eea
where $\Omega_a \, (a=1,2,3)$ denote the angles on which the operators
${\hat{\cal J}}_a$ depend, and $\Phi(\Omega_a)$ is an eigenfunction of the operator
${\hat{\cal J}}^2$, i.e.,
\bea
{\hat{\cal J}}^2\,\Phi(\Omega_a) = {\cal J}({\cal
J}+1)\Phi\left(\Omega_a\right).
\label{1.3.4}
\eea
Here ${\cal J}({\cal J}+1)$ are the eigenvalues of the operator ${\hat{\cal J}}^2$.
Now, substituting (\ref{1.2.11}) into equation (\ref{1.3.1}), taking into account
relations (\ref{1.3.3}), (\ref{1.3.4}) and formula (\ref{1.1.3}) we arrive at the equation
\bea
\left[-\frac{\hbar^2}{2\mu_0}\frac{{\partial}^2}{\partial
x^2_j} - \frac{e^2}{r} + \frac{\hbar^2}{2\mu_0r^2}{\cal J}({\cal
J}+1) \right]\psi(\bf x) = \epsilon \psi(\bf x).
\label{1.3.5}
\eea
Thus, we obtained that the eight-dimensional isotropic oscillator is dual to an infinite
system of five-dimensional Coulomb systems with an additional term $1/r^2$ and coupling
constant $\kappa=\hbar^2{\cal J}({\cal J}+1)/2\mu_0$. If ${\hat{\cal J}}_a$ are generators
of the $SU(2)$ group, then the system described by equation (\ref{1.3.5}) in \cite{TRUNK}
was called the $SU(2)$ Kepler problem, the $SO(6)$ hidden symmetry group was found,
and the energy spectrum was calculated using a purely algebraic method.

For ${\cal J}=0$, equation (\ref{1.3.5}) becomes the Schr\"{o}dinger equation for the
five-dimensional Coulomb problem
\bea
\left(-\frac{\hbar^2}{2\mu_0}\frac{{\partial}^2}{\partial x^2_j}
- \frac{e^2}{r}\right)\psi(\bf x) = \epsilon \psi(\bf x).
\label{1.3.6}
\eea
The condition ${\cal J}=0$ is equivalent to the requirement
\begin{eqnarray*}
{\hat{\cal J}}_a \psi({\bf x}) = 0.
\end{eqnarray*}
Moreover, it follows from (\ref{1.2.2}) that $\psi(\bf x)$ is an even function
of variables $u$:
\begin{eqnarray*} \psi\left(\bf x(-\bf u)\right) = \psi\left(\bf
x(\bf u)\right).
\end{eqnarray*}
Therefore, any solution of equation (\ref{1.3.6}) $\psi(\bf x)$ can be
expanded over a complete system of even solutions $\psi_{N\alpha}({\bf u})$
($\alpha$ are the remaining quantum numbers) of equation (\ref{1.3.1}), i.e.,
\begin{eqnarray*}
\psi_n({\bf x}) = \sum_{\alpha} C_{n \alpha}\psi_{N\alpha}({\bf u}),
\end{eqnarray*}
where
\bea
 N = 2n.
 \label{1.3.7}
 \eea
It is easy to verify that $n$ coincides with the principal quantum number of the five-dimensional
Coulomb problem. Indeed, substituting the relations $E=4e^2$ and (\ref{1.3.7}) into (\ref{1.3.2}), we obtain

\bea
\omega_n = \frac{2e^2}{\hbar (n+2)}.
\label{1.3.8}
\eea
Thus, in our case, the oscillator energy is fixed and the frequency $\omega$ is quantized.
Now substituting (\ref{1.3.8}) into the condition  $\varepsilon =-\mu_0\omega^2/8$,
we arrive at the expression
\begin{eqnarray}
\varepsilon_n = - \frac{\mu_0e^4}{2\hbar^2 (n+2)^2},
\label{1.3.9}
\end{eqnarray}
which determines the energy spectrum of the five-dimensional Coulomb problem \cite{ALLILUEV}.

\section{Euler parameterization}
\markboth{CHAPTER 1. COULOMB-OSCILLATOR ANALOGY}{1.4. EULER PARAMETERIZATION}

In this section, we are interested in the structure of the Hurwitz transformation and the
reduction mechanism induced by the Hurwitz transformation in the space of state vectors.

Instead of the Cartesian coordinates $u_j$ we introduce the coordinates $u_T,\,
u_K,$ ${\alpha}_T,\, {\beta}_T,\, {\gamma}_T,$ ${\alpha}_K,$ ${\beta}_K,$
${\gamma}_K$ as follows \cite{M-5}:
\bea
u_0 + iu_1 &=& u_T \sin\frac{{\beta}_T}{2}
e^{-i\frac{{\alpha}_T-{\gamma}_T}{2}}, \qquad  u_2 + iu_3 = u_T
\cos\frac{{\beta}_T}{2} e^{i\frac{{\alpha}_T+{\gamma}_T}{2}},
\nonumber \\
\label{1.4.1} \\
u_4 + iu_5 &=& u_K \sin\frac{{\beta}_K}{2}
e^{i\frac{{\alpha}_K-{\gamma}_K}{2}}, \qquad  u_6 + iu_7 = u_K
\cos\frac{{\beta}_K}{2} e^{-i\frac{{\alpha}_K+{\gamma}_K}{2}}.
\nonumber
\eea
New coordinates are defined in areas
\begin{eqnarray*}
0\leq u_T,u_K < \infty, \quad 0 \leq {\beta}_T, {\beta}_K \leq
\pi,\quad 0 \leq {\alpha}_T, {\alpha}_K < 2\pi, \quad 0 \leq
{\gamma}_T, {\gamma}_K < 4\pi.
\end{eqnarray*}
To determine types of coordinates (hyperspherical, cylindrical, parabolic,
spheroidal, etc.), $u_T$ and $u_K$ will be determined in addition. In this
case, $u_T$ and $u_K$ may be arbitrary  but relation to the coordinates $u_j$
is fixed
\begin{eqnarray*}
u_T = (u_0^2+u_1^2+u_2^2+u_3^2)^{1/2},  \qquad
u_K = (u_4^2+u_5^2+u_6^2+u_7^2)^{1/2}.
\end{eqnarray*}
The differential elements of length, volume and the Laplace operator in
coordinates (\ref{1.4.1}) have the form
\begin{eqnarray*} dl_8^2 =
du_T^2+du_K^2+\frac{u_T^2}{4}dl_T^2+\frac{u_K^2}{4}dl_K^2, \qquad
dV_8   &=& u_T^3u_K^3du_Tdu_Kd{\Omega_T}d{\Omega_K},
\end{eqnarray*}
\bea \Delta_{8} = \frac{1}{u_T^3} \frac{\partial}{\partial u_T}
\left(u_T^3\frac{\partial}{\partial u_T} \right)+
\frac{1}{u_K^3}\frac{\partial}{\partial u_K}
\left(u_K^3\frac{\partial}{\partial u_K}\right)-
\frac{4}{u_T^2}{\hat {\bf T}}^2- \frac{4}{u_K^2}{\hat {\bf K}}^2,
\label{1.4.2}\eea
where
\bea
dl_a^2=d{\alpha_a}^2+d{\beta_a}^2+d{\gamma_a}^2+2\cos\beta_ad{\alpha_a}
d{\gamma_a}, \qquad d{\Omega_a} =
\frac{1}{8}\sin\beta_ad{\beta_a}d{\alpha_a}d{\gamma_a},
\label{1.4.3}
\eea
\bea
{\hat {\bf T}}^2 = -\left[\frac{\partial^2}{\partial
\beta_T^2}+ \cot \beta_T\frac{\partial}{\partial \beta_T}+
\frac{1}{\sin^2\beta_T}\left(\frac{\partial^2}{\partial
{\alpha_T}^2}- 2\cos \beta_T \frac{\partial^2}{\partial \alpha_T
\partial \gamma_T}+ \frac{\partial^2}{\partial
{\gamma_T}^2}\right)\right].
\label{1.4.4}
\eea
Index $a=T,\,K$, and the operator ${\hat {\bf K}}^2$ can be derived from the
operator ${\hat {\bf T}}^2$ by substituting $\left(\alpha_T, \beta_T, \gamma_T\right)$ for
$\left(\alpha_K, \beta_K, \gamma_K\right)$.

Now let's find to which coordinates the Hurwitz transformation (\ref{1.2.1}) converts
coordinates (\ref{1.4.1}). After substituting (\ref{1.4.1}) into (\ref{1.2.2}), we obtain
\bea
x_0 = u_T^2 - u_K^2, \qquad x_j = 2u_Tu_K{\hat x}_j.
\label{1.4.5}
\eea
\begin{eqnarray*}
{\hat x}_1 &=& \sin\frac{\beta_T}{2}\sin\frac{\beta_K}{2}
\cos\frac{\alpha_T+\alpha_K-\gamma_T-\gamma_K}{2}-
\cos\frac{\beta_T}{2}\cos\frac{\beta_K}{2}
\cos\frac{\alpha_T+\alpha_K+\gamma_T+\gamma_K}{2},  \\
[3mm] {\hat x}_2 &=& \sin\frac{\beta_T}{2}\sin\frac{\beta_K}{2}
\sin\frac{\alpha_T+\alpha_K-\gamma_T-\gamma_K}{2} -
\cos\frac{\beta_T}{2}\cos\frac{\beta_K}{2}
\sin\frac{\alpha_T+\alpha_K+\gamma_T+\gamma_K}{2}, \\
[3mm] {\hat x}_3 &=& \sin\frac{\beta_T}{2}\cos\frac{\beta_K}{2}
\cos\frac{\alpha_T-\alpha_K-\gamma_T-\gamma_K}{2}+
\cos\frac{\beta_T}{2}\sin\frac{\beta_K}{2}
\cos\frac{\alpha_T-\alpha_K+\gamma_T+\gamma_K}{2}, \\
[3mm] {\hat x}_4 &=& \sin\frac{\beta_T}{2}\cos\frac{\beta_K}{2}
\sin\frac{\alpha_T-\alpha_K-\gamma_T-\gamma_K}{2}+
\cos\frac{\beta_T}{2}\sin\frac{\beta_K}{2}
\sin\frac{\alpha_T-\alpha_K+\gamma_T+\gamma_K}{2}.
\end{eqnarray*}
Now we enter the following coordinates
\bea
x_0 = u_T^2 - u_K^2, \quad x_2 + ix_1 = 2u_T u_K \sin
\frac{\beta}{2} {\rm e}^{i \frac{\alpha-\gamma}{2}}, \quad x_4 +
ix_3 = 2 u_T u_K \cos \frac{\beta}{2}{\rm e}^{i
\frac{\alpha+\gamma}{2}}.
\label{1.4.6}
\eea
In these coordinates we have
\begin{eqnarray*}
dl_5^2 = \frac{\mu+\nu}{4\mu}d{\mu}^2+
\frac{\mu+\nu}{4\nu}d{\nu}^2+\frac{\mu \nu}{4}dl^2, \qquad dV_5 =
\frac{\mu \nu}{4}(\mu+\nu)d{\mu}d{\nu}d{\Omega},
\end{eqnarray*}
\bea
\Delta_{5} = \frac{4}{\mu+\nu}
\left[\frac{1}{\mu}\frac{\partial}{\partial \mu}
\left({\mu}^2\frac{\partial}{\partial \mu}\right)+
\frac{1}{\nu}\frac{\partial}{\partial \nu}
\left({\nu}^2\frac{\partial}{\partial \nu}\right)\right]-
\frac{4}{\mu \nu}{\hat {\bf J}}^2,
\label{1.4.7}
\eea
where $\mu=2u_T^2,\, \nu=2u_K^2$, and $dl^2, d{\Omega}$, and ${\bf J}^2$ can be found
from relations (\ref{1.4.3}) and (\ref{1.4.4}) using the substitution
$\left(\alpha_a,\beta_a,\gamma_a\right) \rightarrow \left(\alpha,\beta,\gamma\right)$. Having identified the coordinates
(\ref{1.4.6}) and (\ref{1.4.5}) we arrive at the following system of trigonometric equations
After solving these equations we have
\bea
\cot \left(\alpha + \alpha_T - \pi\right) = \cos \left(\beta_T - \pi\right) \cot
\left(\gamma_T+\gamma_K\right)+ \cot \beta_K \frac{\sin\left(\beta_T - \pi\right)}{\sin
\left(\gamma_T+\gamma_K\right)},
\nonumber\\[3mm]
\cos \beta = \cos \left(\beta_T - \pi\right)\cos \beta_K- \sin\left(\beta_T -
\pi\right)\sin \beta_K \cos \left(\gamma_T+\gamma_K\right), \label{1.4.8}
\\ [3mm]
\cot \left(\gamma-\alpha_K\right) = \cos \beta_K \cot \left(\gamma_T+\gamma_K\right)+
\cot\left(\beta_T - \pi\right)\frac{\sin \beta_K}{\sin
\left(\gamma_T+\gamma_K\right)}.\nonumber
\eea
Thus, the Hurwitz transformation consists of the conformal Levi--Cevita
transformation $\left(u_T+iu_K\right) \to \left(u_T+iu_K\right)^2$ and the Euler addition of angles
\begin{eqnarray*}
\left(\pi - \alpha_T, \beta_T - \pi, \gamma_T\right) \oplus \left(\gamma_K,
\beta_K, \alpha_K\right) = \left(\alpha, \beta, \gamma\right).
\end{eqnarray*}
Therefore, we call parameterization (\ref{1.4.1}) Eulerian.

From (\ref{1.4.2}) and (\ref{1.4.7}) we see that the Laplace operators $\Delta_{8}$ and $\Delta_{5}$
contain the square of angular momentum operators. If the potential in the Schr\"{o}dinger equation allows for
the separation of variables associated with the operators ${\hat {\bf T}}^2$, ${\hat {\bf K}}^2$ and
${\hat {\bf J}}^2$, then the solutions have a universal dependence on the angles included in these operators.

Fixing the coordinates $u_T$ and $u_K$ we can consider the hyperspheres $S^6$
and $S^3$ in the spaces $\rm I\!R^8(\bf{u})$ and $\rm I\!R^5(\bf{x})$,
respectively. The solutions are of the form
\begin{eqnarray*}
 \psi_6 = \left(\frac{2T+1}{2\pi^2}\right)^{1/2}
\left(\frac{2K+1}{2\pi^2}\right)^{1/2} D_{t,
t'}^{T}\left(\alpha_T,\beta_T,\gamma_T\right) D_{k,
k'}^{K}\left(\alpha_K,\beta_K,\gamma_K\right),
\end{eqnarray*}
\begin{eqnarray*}
\psi_3 = \left(\frac{2L+1}{2\pi^2}\right)^{1/2}
D_{m,m'}^L\left(\alpha,\beta,\gamma\right).
\end{eqnarray*}
For the angles included in (\ref{1.4.8}), the following addition theorem holds \cite{VAR}:
\begin{eqnarray}
D_{m,m'}^L\left(\alpha,\beta,\gamma\right)= {\rm e}^{im\pi} \sum_{k=-L}^L\,
(-1)^{m+m'-L-k}\, D_{m,k}^L\left(\alpha_T,\beta_T,\gamma_T\right)\,
D_{k,m'}^L\left(\alpha_K,\beta_K,\gamma_K\right).
\label{1.4.9}
\end{eqnarray}
Thus, in the Eulerian parametrization, the correspondence between solutions in the spaces
$\rm I\!R^8(\bf{u})$ and $\rm I\!R^5(\bf{x})$ is satisfied according to the addition theorem
of Wigner $D$-function (see also \cite{SISTER1,SISTER2}).

\section{Hyperspherical and Cylindrical Bound States 8D Oscillator}
\markboth{CHAPTER 1. COULOMB-OSCILLATOR ANALOGY}
{1.5. HYPERSPHERICAL AND CYLINDRICAL BOUND STATES 8D OSCILLATOR}

Let us consider the hypespherical and cylindrical bases of the eight-dimensional isotropic
oscillator \cite{KM-1}. Due to the $SU(8)$ hidden symmetry of the eight-dimensional isotropic oscillator
\cite{DEMKOV1}, the energy eigenvalues (\ref{1.3.2}) are degenerate and the degeneracy multiplicity
is equal to \cite{BAKER}:
\begin{eqnarray}
g_{N}= \frac{(N+7)!}{7!\,N!}.
\label{1.5.1}
\end{eqnarray}

According to (\ref{1.4.1}) we define eight-dimensional hyperspherical coordinates as follows
\bea
u_0 + iu_1 &=& u\cos\frac{\theta}{2} \sin\frac{{\beta}_T}{2}
e^{-i\frac{{\alpha}_T-{\gamma}_T}{2}}, \qquad
u_2 + iu_3 = u\cos\frac{\theta}{2}
\cos\frac{{\beta}_T}{2} e^{i\frac{{\alpha}_T+{\gamma}_T}{2}},
\nonumber \\
\label{1.5.2} \\
u_4 + iu_5 &=& u\sin\frac{\theta}{2} \sin\frac{{\beta}_K}{2}
e^{i\frac{{\alpha}_K-{\gamma}_K}{2}}, \qquad  u_6 + iu_7 = u\sin\frac{\theta}{2}
\cos\frac{{\beta}_K}{2} e^{-i\frac{{\alpha}_K+{\gamma}_K}{2}}.
\nonumber
\eea
where $0 \leq u < \infty, 0 \leq \theta \pi$. In coordinates (\ref{1.5.2}), the differential
elements of length, volume and the Laplace operator have the form
\begin{eqnarray*}
dl_{8}^{2} = du^{2} + \frac{u^{2}}{4}\left(d{\theta}^{2} + \cos^{2}\frac{\theta}{2}dl_{T}^{2}
+\sin^{2}\frac{\theta}{2}dl_{K}^{2}\right), \qquad
dV_{8} = u^{7}\sin^{3}\theta du d \theta d\Omega_{T} d\Omega_{K},
\end{eqnarray*}
\begin{eqnarray*}
\Delta_{8} = \frac{1}{u^{7}}\frac{\partial}{\partial u} \left(u^{7}\frac{\partial}{\partial u}\right)
+ \frac{4}{u^{2}\sin^{3}\theta}\frac{\partial}{\partial \theta}
\left(\sin^{3}\theta \frac{\partial}{\partial \theta} \right)-
\frac{4}{u^{2}\cos^{2}\theta/2}\hat{T}^{2}
-\frac{4}{u^{2}\cos^{2}\theta/2}\hat{K}^{2},
\end{eqnarray*}
Where $dl_{T}^{2}, dl_{K}^{2}, d\Omega_{T} d\Omega_{K}, \hat{T}^{2}$
and $\hat{K}^{2}$ are given in formulas (\ref{1.4.3}) and (\ref{1.4.4}).
The components of the operator $\hat{T}$ have the form
\begin{eqnarray}
\hat{T}_{1} &=& i\left(\cos\alpha_{T}\cot\beta_{T}\frac{\partial}{\partial \alpha_{T}}+
\sin\alpha_{T}\frac{\partial}{\partial \beta_{T}}- \frac{\cos\alpha_{T}}{\sin\beta_{T}}
\frac{\partial}{\partial \gamma_{T}}\right),
\nonumber \\ [2mm]
\hat{T}_{2} &=& i\left(\sin\alpha_{T}\cot\beta_{T}\frac{\partial}{\partial \alpha_{T}}-
\cos\alpha_{T}\frac{\partial}{\partial \beta_{T}}-\frac{\sin\alpha_{T}}{\sin\beta_{T}}
\frac{\partial}{\partial \gamma_{T}}\right),
\label{1.5.3} \\ [2mm]
\hat{T}_{3} &=&-i\frac{\partial}{\partial \alpha_{T}}, \qquad
\hat{T}_{3'} =-i\frac{\partial}{\partial \gamma_{T}},
\nonumber
\end{eqnarray}
and the components of the operator $\hat{K}$ are obtained from the components of the operator
$\hat{T}$ by the substitution $\left(\alpha_{T}, \beta_{T}, \gamma_{T}\right)$ by 
$\left(\alpha_{K}, \beta_{K}, \gamma_{K}\right)$.

Since both the operators $\hat{T}^{2}, \hat{T}_{3}$ and $\hat{T}_{3'}$ and $\hat{K}^{2} \hat{K}_{3}$,
and $\hat{K}_{3'}$ commute mutually, then in the coordinates (\ref{1.5.2}) the scheme for separating variables
in the Schr\"{o}dinger equation for an eight-dimensional isotropic oscillator (\ref{1.2.11}) corresponds to the following factorization
\begin{eqnarray}
\Psi^{hsp} = R(u) Z(\theta) D_{tt'}^{T}\left(\alpha_{T}, \beta_{T}, \gamma_{T}\right)
D_{kk'}^{K}\left(\alpha_{K}, \beta_{K}, \gamma_{K}\right),
\label{1.5.4}
\end{eqnarray}
where $D_{mm'}^{j}$ are the Wigner $D$-function.
Now substituting (\ref{1.5.4}) into the Schr\"{o}dinger equation (\ref{1.2.11}), and taking into account that
\begin{eqnarray*}
\hat{T}^{2}D_{tt'}^{T}\left(\alpha_{T}, \beta_{T}, \gamma_{T}\right) &=&
T(T+1)D_{tt'}^{T}\left(\alpha_{T}, \beta_{T}, \gamma_{T}\right), \\ [2mm]
\hat{K}^{2}D_{kk'}^{K}\left(\alpha_{K}, \beta_{K}, \gamma_{K}\right)
&=& K(K+1)D_{kk'}^{K}\left(\alpha_{K}, \beta_{K}, \gamma_{K}\right)
\end{eqnarray*}
we arrive at the following pair of differential equations
\begin{eqnarray}
\left[\frac{1}{\sin^{3}\theta}\frac{d}{d\theta}\left(\sin^{3}\theta \frac{d}{d\theta}\right)
- \frac{T(T+1)}{\cos^{2}\theta/2}-\frac{K(K+1)}{\sin^{2}\theta/2} + J(J+3)\right]Z(\theta)=0,
\label{1.5.5} \\ [2mm]
\left[\frac{1}{u^{7}}\frac{d}{du}\left(u^{7}\frac{d}{du}\right) -
\frac{4J(J+3)}{u^{2}}+\frac{2\mu_{0}E}{\hbar^{2}}-a^{4}u^{2}\right]R(u)=0,
\label{1.5.6}
\end{eqnarray}
where $a=\left(\mu_{0}\omega \hbar \right)^{1/2}$ and $J(J+3)$  is a non-negative separation constant
is the eigenvalue of the operator
\begin{eqnarray}
\hat{J}^{2} = - \frac{1}{\sin^{3}\theta}\frac{\partial}{\partial \theta}
\left(\sin^{3}\theta \frac{\partial}{\partial \theta}\right) +
\frac{1}{\cos^{2}\theta/2}\hat{T}^{2} + \frac{1}{\sin^{2}\theta/2}\hat{K}^{2}.
\label{1.5.7}
\end{eqnarray}
Normalized by the conditions
\begin{eqnarray}
\frac{1}{16} \int_{0}^{\pi} Z_{JKT}(\theta)Z_{J'KT}(\theta)\sin^{3}\theta d\theta = \delta_{JJ'},
\nonumber \\
\label{1.5.8} \\ [2mm]
\int_{0}^{\infty} u^{7} R_{NJ}(u)R_{N'J}(u)du = \delta_{NN'} \nonumber
\end{eqnarray}
the solutions of the equations (\ref{1.5.5}) and (\ref{1.5.6}) have the form
\begin{eqnarray}
Z_{JKT}(\theta) = \sqrt{\frac{2(2J+3)(J-K-T)!(J+K+T+2)!}{(J+K-T+1)!(J-K+T+1)!}} \times \nonumber \\ [2mm]
\label{1.5.9} \\
\times \left(\sin\frac{\theta}{2}\right)^{2K} \left(\cos\frac{\theta}{2}\right)^{2T}
P_{J-K-T}^{(2K+1, 2T+1)}(\cos\theta), \nonumber
\end{eqnarray}
\begin{eqnarray}
R_{NJ}(u) = a^{4}\sqrt{\frac{2\left(\frac{N}{2}+J+3\right)!}{\left(\frac{N}{2}-J\right)!}}
\frac{(au)^{2J}}{(2J+3)!}e^{-a^{2}u^{2}/2} F\left(-\frac{N}{2}+J; 2J+4; a^{2}u^{2}\right),
\label{1.5.10}
\end{eqnarray}
where $P_{n}^{(a,b)}(x)$ is the Jacobi polynomials, and $F(a;b;x)$ is a confluent hypergeometric function. The lower
index of the Jacobi polynomial and the first parameter of the confluent hypergeometric function
indicate that the quantum number J runs through the values $K+T \leq J \leq N/2$, i.e. it can take both
non-negative integer and positive half-integer values.

Normalized per unit
\begin{eqnarray*}
\int \left|\Psi^{hsp}\right|^{2}dV_{8} = 1
\end{eqnarray*}
The complete hyperspherical wave function of an eight-dimensional isotropic oscillator can be written as
\begin{eqnarray}
\Psi^{hsp} =\sqrt{\frac{(2K+1)(2T+1)}{4\pi^{4}}} R_{NJ}(u)Z_{JKT}(\theta)
D_{tt'}^{T}\left(\alpha_{T}, \beta_{T}, \gamma_{T}\right)
D_{kk'}^{K}\left(\alpha_{K}, \beta_{K}, \gamma_{K}\right).
\label{1.5.11}
\end{eqnarray}
Note that when calculating the full normalization factor, the formula was used \cite{VAR}
\begin{eqnarray}
\int D_{m_{2}m_{2}'}^{j_{2}*}\left(\alpha, \beta, \gamma\right)
D_{m_{1}m_{1}'}^{j_{1}}\left(\alpha, \beta, \gamma\right) d\Omega
=\frac{16\pi^{2}}{2j_{1}+1}\delta_{j_{1}j_{2}}\delta_{m_{1}m_{2}}
\delta_{m_{1}'m_{2}'}.
\label{1.5.12}
\end{eqnarray}
Thus, it can be stated that the hyperspherica wave functions (\ref{1.5.11})
are eigenfunctions of the following commuting operators
$\left\{\hat{H}, \hat{J^{2}}, \hat{T^{2}}, \hat{K^{2}}, \hat{T_{3}}, \hat{T_{3'}},
\hat{K_{3}}, \hat{K_{3'}}\right\}$, and
\begin{eqnarray}
\hat{J^{2}} \Psi^{hsp} = J(J+3) \Psi^{hsp}.
\label{1.5.13}
\end{eqnarray}
In the Cartesian coordinates the operator $\hat{J^{2}}$ (\ref{1.5.7}) has the form
\begin{eqnarray}
\hat{J^{2}} = -\frac{u^{2}}{4}\Delta_{8} + \frac{1}{4}u_{\mu}u_{\nu}
\frac{\partial^{2}}{\partial u_{\mu}u_{\nu}} + \frac{7}{4}u_{\mu}\frac{\partial}{\partial u_{\mu}}.
\label{1.5.14}
\end{eqnarray}

In $\rm I\!R^8(\bf{u})$ we define the eight-dimensional cylindrical coordinates as follows:
\begin{eqnarray}
u_{0}+iu_{1} = \rho_{1} \sin\frac{\beta_{T}}{2}e^{-i\frac{\alpha_{T}-\gamma_{T}}{2}}, \qquad
u_{2}+iu_{3} = \rho_{1} \cos\frac{\beta_{T}}{2}e^{i\frac{\alpha_{T}+\gamma_{T}}{2}},  \nonumber \\
\label{1.5.15} \\
u_{4}+iu_{5} = \rho_{2} \sin\frac{\beta_{K}}{2}e^{i\frac{\alpha_{K}-\gamma_{K}}{2}}, \qquad
u_{6}+iu_{7} = \rho_{2} \cos\frac{\beta_{K}}{2}e^{-i\frac{\alpha_{K}+\gamma_{K}}{2}},
\nonumber
\end{eqnarray}
where $0 \leq \rho_{1}, \rho_{2} \leq \infty$. In these coordinates, the potential,
differential elements of length, volume and the Laplace operator have the form
\begin{eqnarray*}
V = \frac{\mu_{0}\omega^{2}}{2}\left(\rho_{1}^{2}+ \rho_{2}^{2}\right),
\end{eqnarray*}
\begin{eqnarray*}
dl_{8}^{2}=d\rho_{1}^{2} + d\rho_{2}^{2} + \frac{\rho_{1}^{2}}{4}dl_{T}^{2} +
\frac{\rho_{2}^{2}}{4}dl_{K}^{2}, \qquad
dV_{8} = \rho_{1}^{3} \rho_{2}^{3}d\rho_{1} d\rho_{2} d\Omega_{T} d\Omega_{K}, \\ [4mm]
\Delta_{8} = \frac{1}{\rho_{1}^{3}}\frac{\partial}{\partial \rho_{1}}
\left(\rho_{1}^{3}\frac{\partial}{\partial \rho_{1}}\right) +
\frac{1}{\rho_{2}^{3}}\frac{\partial}{\partial \rho_{2}}
\left(\rho_{2}^{3}\frac{\partial}{\partial \rho_{2}}\right) -
\frac{4}{\rho_{1}^{2}}\hat{T^{2}} - \frac{4}{\rho_{2}^{2}}\hat{K^{2}}.
\end{eqnarray*}
After the substitution
\begin{eqnarray*}
\Psi^{hsp} = f_{1}(\rho_{1}) f_{2}(\rho_{})
D_{tt'}^{T}\left(\alpha_{T}, \beta_{T}, \gamma_{T}\right)
D_{kk'}^{K}\left(\alpha_{K}, \beta_{K}, \gamma_{K}\right)
\end{eqnarray*}
the variables in the Schr\"{o}dinger equation (\ref{1.2.11}) are separated
and we arrive at the following system of differential equations
\begin{eqnarray}
x_{1}\frac{d^{2}f_{1}}{dx_{1}^{2}} + 2\frac{df_{1}}{dx_{1}} -
\left[\frac{T(T+1)}{x_{1}}+\frac{x_{1}}{4} - \frac{E_{1}}{2\hbar \omega}\right]f_{1}=0,  \nonumber \\
\label{1.5.16} \\
x_{2}\frac{d^{2}f_{2}}{dx_{2}^{2}} + 2\frac{df_{2}}{dx_{2}} -
\left[\frac{K(K+1)}{x_{2}}+\frac{x_{2}}{4} - \frac{E_{2}}{2\hbar \omega}\right]f_{1}=0,
\nonumber
\end{eqnarray}
where $x_{i}=a^{2}\rho_{i}^{2}, a=\left(\mu_{0}\omega \hbar \right)^{1/2}$,
and $E_{1}+E_{2}=E$. Solutions to equations (\ref{1.5.16}) are sought in the form
\begin{eqnarray*}
f_{i}(x_{i}) = e^{-x_{i}/2}x_{i}^{J_{i}}W(x_{i}),
\end{eqnarray*}
where $J_{1}=T$, and $J_{2}=K$. then for W(x) we obtain the equation for the
confluent hypergeometric function
\begin{eqnarray}
z\frac{d^{2}f}{dz^{2}} + (\gamma -z)\frac{df}{dz} - \alpha f =0,
\label{1.5.17}
\end{eqnarray}
with $\alpha = J_{i}+1-E/2\hbar \omega$ and $\gamma = 2J_{i}+2$.
Further, introducing cylindrical quantum numbers
\begin{eqnarray*}
N_{1} = -T-1+\frac{E_{1}}{2\hbar \omega}, \qquad
N_{2} = -K-1+\frac{E_{2}}{2\hbar \omega},
\end{eqnarray*}
and taking into account (\ref{1.3.2}) we find that they are related to the principal
quantum number $N$ as follows:
\begin{eqnarray}
N = 2N_{1} + 2N_{2} +2T + 2K.
\label{1.5.18}
\end{eqnarray}

Normalized by condition
\begin{eqnarray*}
\int \left|\Psi^{cyl}\right|^{2}dV_{8} = 1
\end{eqnarray*}
the cylindrical basis of an eight-dimensional isotropic oscillator can be written as
\begin{eqnarray}
\Psi^{cyl} = \sqrt{\frac{(2K+1)(2T+1)}{4\pi^{4}}} f_{N_{1}T}(\rho_{1})
f_{N_{2}T}(\rho_{2})
D_{tt'}^{T}\left(\alpha_{T}, \beta_{T}, \gamma_{T}\right)
D_{kk'}^{K}\left(\alpha_{K}, \beta_{K}, \gamma_{K}\right),
\label{1.5.19}
\end{eqnarray}
where
\begin{eqnarray*}
f_{N_{i}J_{i}}(\rho_{i})=\frac{a^{2}}{(2J_{i}+1)!}
\sqrt{\frac{2(N_{i}+2J_{i}+1)!}{(N_{i})!}}\left(a \rho_{i}\right)^{2J_{i}}
e^{-a^{2} \rho_{i}^{2}/2}F \left(-N_{i}; 2J_{i}+2; a^{2} \rho_{i}^{2}\right).
\end{eqnarray*}

At the end of this section, we note that the cylindrical wave functions (\ref{1.5.19})
are eigenfunctions of both the operators
$\left\{\hat{H}, \hat{J^{2}}, \hat{T^{2}}, \hat{K^{2}}, \hat{T_{3}}, \hat{T_{3'}},
\hat{K_{3}}, \hat{K_{3'}}\right\}$ and the operator
\begin{eqnarray}
{\hat {\cal P}} = \frac{\hbar}{2\mu_{0}\omega}
\left(-\frac{\partial^{2}}{\partial u_{0}^{2}}-\frac{\partial^{2}}{\partial u_{1}^{2}}
-\frac{\partial^{2}}{\partial u_{2}^{2}}-\frac{\partial^{2}}{\partial u_{3}^{2}}
+\frac{\partial^{2}}{\partial u_{4}^{2}}+\frac{\partial^{2}}{\partial u_{5}^{2}}
+\frac{\partial^{2}}{\partial u_{6}^{2}}+\frac{\partial^{2}}{\partial u_{7}^{2}}\right)
\label{1.5.20}
\end{eqnarray}
in this case
\begin{eqnarray}
{\hat {\cal P}}\Psi^{cyl} = \left(2N_{1}-2N_{2}+2T-2K\right)\Psi^{cyl}.
\label{1.5.21}
\end{eqnarray}

\section{Hidden symmetry and orthogonality in orbital momentum}
\markboth{CHAPTER 1. COULOMB-OSCILLATOR ANALOGY}
{1.6. HIDDEN SYMMETRY AND ORTHOGONALITY IN ORBITAL MOMENTUM}

It is well known that the radial wave functions of a particle moving in a centrally
symmetric field for a given value of the orbital momentum $l$ are orthogonal in the principal
quantum number $n$:
\begin{eqnarray}
\int\limits_{0}^{\infty} R_{nl}(r)  R_{n'l}(r)dr = \delta_{nn'}.
\label{1.6.1}
\end{eqnarray}
Let us prove that for the hydrogen atom, along with (\ref{1.6.1}), the following
"additional" condition of orthogonality with respect to the orbital momentum
$l$ is satisfied \cite{M-6}:
\begin{eqnarray}
J_{ll'} = \int\limits_0^\infty  R_{nl}(r) R_{nl'}(r) dr
= \frac{2}{n^3}\frac{\delta_{ll'}}{2l+1}.
\label{1.6.2}
\end{eqnarray}
where normalized by condition (\ref{1.6.1}) the radial wave function of the hydrogen
atom in units $\mu_{0}=e=\hbar =1$ has the form \cite{LL}
\begin{eqnarray}
R_{nl}(r) = \frac{2}{n^2(2l+1)!} \sqrt{\frac{(n+l)!}{(n-l-1)!}}
\left(\frac{2r}{n}\right)^{l}e^{-r/n}F\left(-n+l+1; 2l+1; \frac{2r}{n}\right).
\label{1.6.3}
\end{eqnarray}
Let us substitute the normalized radial wave function of the hydrogen atom
(\ref{1.6.3}) into (\ref{1.6.2}) and write the confluent hypergeometric function in
$R_{nl'}(r)$ as a polynomial
\begin{eqnarray}
F\left(-n+l'+1; 2l'+2; \frac{2r}{n}\right) = \sum_{s=0}^{n-l'-1}\,
\frac{(-n+l'+1)_s}{s!(2l'+2)_s}\,\left(\frac{2r}{n}\right)^s,
\label{1.6.4}
\end{eqnarray}
integrating by formula \cite{LL}
\bea
\int \limits_{0}^{\infty} e^{-\lambda x} x^\nu F(\alpha,
\gamma; kx) \, dx = \frac{\Gamma(\nu+1)}{\lambda^{\nu+1}} \,
{_2F}_1 \left( \alpha, \nu+1, \gamma ; \frac{k}{\lambda} \right)
\label{1.6.5}
\eea
and taking account of that \cite{BE1}
\bea
{_2F}_1 \left(a, b; c; 1\right) =
\frac{\Gamma(c)\Gamma(c-a-b)} {\Gamma(c-a)\Gamma(c-b)},
\label{1.6.6}
\eea
we have
\bea
J_{ll'} = \frac{2}{n^3}\,\frac{\Gamma\left(l+l'+1\right)}{(2l+1)!}
\sqrt{\frac{(n+l')!}{(n+l)!(n-l-1)!(n-l'-1)!}}\times \nonumber
\\
\label{1.6.7}\\
\times \sum_{s=0}^{n-l'-1}\,
\frac{(-n+l'+1)_s}{s!}\frac{\left(l+l'+1\right)_s}{
(2l'+2)_s}\,\frac{\Gamma\left(n-l'-s\right)}
{\Gamma\left(l-l'-s+1\right)}.
\nonumber
\eea
Next, applying to the gamma functions under the summation sign the formula \cite{BE1}
\bea
\frac{\Gamma(z)}{\Gamma(z-n)} = (-1)^n\frac{\Gamma(-z+n+1)}{\Gamma(-z+1)},
\label{1.6.8}
\eea
we note that the sum over $s$ in (\ref{1.6.7}) turns into a hypergeometric
function of type (\ref{1.6.6}) and, therefore,
\bea
J_{ll'} = \frac{2}{n^3}\,\frac{\left(l+l'\right)!}{\left(l+l'+1\right)!}
\sqrt{\frac{(n+l')!(n-l'-1)!}{(n+l)!(n-l-1)!}}\frac{1}
{\Gamma\left(l-l'+1\right)\Gamma\left(l'-l+1\right)}.
\label{1.6.9}
\eea
The last relation vanishes when $l'\neq l$ due to the product of the gamma functions of
$(l-l'+1)$ and $(l'-l+1)$, which leads to the orthogonality condition (\ref{1.6.2}).

Following the scheme described above, it is easy to verify that orthogonality in the
orbital momentum $l$ also occurs for the radial wave function of an isotropic oscillator:
\begin{eqnarray}
\int\limits_0^\infty  R_{nl'}(r) R_{nl}(r) dr= \frac{2\delta_{ll'}}{2l+1}.
\label{1.6.10}
\end{eqnarray}

Here $R_{nl}(r)$ is the radial wave function of an isotropic oscillator,
which has the form $\left(\mu_{0}=e=\hbar =1\right)$:
\begin{eqnarray}
R_{nl}(r) = \left(\frac{1}{\pi}\right)^{1/4} \frac{2}{(2l+1)!!}
\sqrt{\frac{2^{l}(n+l+1)!!}{(n-l)!!}}r^{l}e^{-r^{2}/2}
F\left(-\frac{n-l}{2}; l+\frac{3}{2}; r^{2}\right).
\label{1.6.11}
\end{eqnarray}

Now we prove that the orthogonality conditions (\ref{1.6.2}) and (\ref{1.6.10}) are a
consequence of the accidental degeneracy of the energy spectrum.

Let us write the radial Schr\"{o}dinger equation in the field $U(r)$ in the form
\begin{eqnarray}
\hat {\cal H} R_{EL}(r) = ER_{EL}(r),
\label{1.6.12}
\end{eqnarray}
where $\hat {\cal H}$ denotes the Hermitian operator
\begin{eqnarray}
\hat {\cal H} = - \frac{1}{2r^2}\frac{d}{dr}
\left(r^2\frac{d}{dr}\right) - \frac{l(l+1)}{r^2} + U(r).
\label{1.6.13}
\end{eqnarray}
From the relations (\ref{1.6.12}) and (\ref{1.6.13}) for a discrete spectrum we obtain
\begin{eqnarray}
(l-l')(l+l'+1)\int \limits_{0}^{\infty}\,R_{E'L'}(r)R_{EL}(r)
\,dr = 2(E-E')\,\int \limits_{0}^{\infty}\,
R_{E'l'}(r)R_{El}(r)\,r^2\,dr.
\label{1.6.14}
\end{eqnarray}
If the spectrum is degenerate in $l$, then for $E'=E$ and $l'\neq l$ we have
\begin{eqnarray}
\int \limits_{0}^{\infty}\,R_{EL'}(r)R_{EL}(r)\,dr = 0.
\label{1.6.15}
\end{eqnarray}
It is known \cite{LL} that for a Hermitian operator ${\hat F}$ depending on some
parameter $\lambda$, the identity holds
\begin{eqnarray}
\left(\frac{\partial \hat F(\lambda)}{\partial \lambda}\right)_{nn}
= \frac{\partial F_n}{\partial \lambda},
\label{1.6.16}
\end{eqnarray}
where the averaging is carried out over the eigenfunctions of the operator ${\hat F}$.
Applying this identity to the operator (\ref{1.6.13}) and choosing the orbital
momentum $l$ as the parameter $\lambda$, we obtain
\begin{eqnarray}
\int \limits_{0}^{\infty}\,R_{nl}(r)R_{nl}(r)\,dr =
\left(\frac{\partial E_n}{\partial l}\right)_{n_r} \frac{2}{2l+1},
\label{1.6.17}
\end{eqnarray}
where a derivative with respect to $l$ is taken at a fixed radial number $n_r$.
Combining (\ref{1.6.15}) with (\ref{1.6.17}) we finally obtain
\begin{eqnarray}
\int \limits_{0}^{\infty}\,R_{nl}(r)R_{nl'}(r)\,dr =
\left(\frac{\partial E_n}{\partial l}\right)_{n_r}
\frac{2\delta_{ll'}}{2l+1}.
\label{1.6.18}
\end{eqnarray}
It is obvious that this formula contains conditions (\ref{1.6.2}) and (\ref{1.6.10}).

For the multidimensional analog of the hydrogen atom \cite{ALLILUEV} and the isotropic
oscillator \cite{BAKER}, the role of operator (\ref{1.6.13}) is played by the
Hermitian operator
\begin{eqnarray}
\hat {\cal H} = - \frac{1}{2r^{D-1}}
\frac{d}{dr}\left(r^{D-1}\frac{d}{dr}\right) +
\frac{L(L+D-2)}{2r^2} + U(r),
\label{1.6.19}
\end{eqnarray}
in which $D \geq 3$ is an integer that determines the dimension of space, and $L$ is
the so-called global moment \cite{VKS} which can take both integer and half-integer
non-negative values.

The reasoning analogous to that given above shows that in the multi-dimensional
case the orthogonality condition holds
\begin{eqnarray}
\int \limits_{0}^{\infty}\,R_{nL}(r)R_{nL'}(r)\,r^{D-3}dr =
\left(\frac{\partial E_n}{\partial L}\right)_{n_r}
\frac{2\delta_{LL'}}{2L+D-2}.
\label{1.6.20}
\end{eqnarray}
The validity of this condition is also confirmed by direct calculations based on the use of the
explicit form of radial wave functions of a multidimensional hydrogen atom \cite{ALLILUEV}
and an isotropic oscillator \cite{POGOSYAN2} (in the next section we will demonstrate this
orthogonality using the example of an eight-dimensional isotropic oscillator).

At the end of this section, we also note that the same (\ref{1.6.20}) additional orthogonality
condition also holds for the Coulomb problem in a curved space of constant curvature,
for example, on a three-dimensional sphere $S^{3}$, \cite{M-6}. This condition explicitly looks like this:
case the orthogonality condition holds
\begin{eqnarray}
J_{ll'} = \int \limits_{0}^{\pi}\,\psi_{nl'\sigma}^{*}(\chi)\psi_{nl\sigma}(\chi)\,d\chi =
\frac{2\left(n^{2}+\sigma^{2}\right)}{nR^{3}} \frac{2\delta_{ll'}}{2l+1}.
\label{1.6.21}
\end{eqnarray}
"Radial" wave function $\psi_{nl\sigma}(\chi)$ normalized by the condition
\begin{eqnarray*}
R^{3}\, \int \limits_{0}^{\pi}\,\sin^{2}\chi\,\psi_{n'l\sigma}^{*}(\chi)\psi_{nl\sigma}(\chi)\,d\chi =
\delta_{nn'}
\end{eqnarray*}
has the form \cite{STEV,BOR,VMPSS}
\begin{eqnarray}
\psi_{nl\sigma}(\chi) &=& C_{nl\sigma} \left(\sin\chi \right)^{l}
\exp\left[-i\chi (n-l-i\sigma -2)\right] \times  \nonumber \\ \label{1.6.22}
\\
&\times& _{2}F_{1}\left(-n+l+1, l+i\sigma +1; 2l+1; 1-e^{2i\chi}\right), \nonumber
\end{eqnarray}
where
\begin{eqnarray}
C_{nl\sigma} = \frac{2^{l+1}e^{\pi \sigma/2}\left|\Gamma\left(l-i\sigma +1\right)\right|}
{R(2l+1)!}\,\sqrt{\frac{\left(n^{2}+\sigma^{2}\right)\,(n+l)!}{2\pi nR\,(n-l-1)!}}, \qquad
\sigma  = \frac{\mu_{0}Ze^{2}R}{n\hbar^{2}}.
\label{1.6.23}
\end{eqnarray}
The "radial" wave function (\ref{1.6.22}) is a solution to the equation
\cite{SCHR-1,SCHR-2,SCHR-3}
\begin{eqnarray}
\left\{\frac{\hbar^{2}}{2\mu_{0}R^{2}}
\left[-\frac{1}{\sin^{2}\chi} \frac{d}{d\chi} \left(\sin^{2}\chi \frac{d}{d\chi}\right)
+\frac{l(l+1)}{\sin^{2}\chi}\right]
-\frac{Ze^{2}}{R}\cot\chi - E\right\}\psi(\chi) = 0,
\label{1.6.24}
\end{eqnarray}
where $V\left(\chi;R\right)=\left(-Ze^{2}/R\right)\,\cot\chi$ is the harmonic potential on a three-dimensional sphere with radius $R$,
i.e. solution of the Laplace equation. The energy spectrum of the Coulomb problem on $S^{3}$ has the form
\begin{eqnarray}
E = \frac{\hbar^{2}}{2\mu_{0}R^{2}}\left(n^{2}-1\right)-\frac{\mu_{0}Ze^{4}}{2\hbar^{2}n^{2}}.
\label{1.6.25}
\end{eqnarray}

Now let's move on to the proof of the orthogonality condition (\ref{1.6.21}). Let us substitute the
normalized "radial" wave function of the Coulomb problem (\ref{1.6.22}) into (\ref{1.6.21}), write the
hypergeometric functions in $\psi_{nl\sigma}(\chi)$ and $\psi_{nl'\sigma}^{*}(\chi)$
in the form of polynomials, and carry out the integration according to the formula \cite{BE1}
\begin{eqnarray*}
\int \limits_{0}^{\pi}\,(\sin t)^{\alpha}\,e^{i\beta t}\,dt =
\frac{\pi \Gamma(1+\alpha)\,e^{i\pi \beta /2}}
{2^{\alpha}\Gamma\left(1+\frac{\alpha + \beta}{2}\right)
\Gamma\left(1+\frac{\alpha - \beta}{2}\right)}, \qquad \Re e \alpha > -1
\end{eqnarray*}
and taking into account formula (\ref{1.6.6}), for the integral $J_{ll'}$ we obtain the expression
\begin{eqnarray*}
J_{ll'} &=& \frac{2(n^{2}+\sigma^{2})\exp\left[i\pi (l-l')/2\right]}
{nR^{3}(2l+1)!}\,\sum \limits_{s=0}^{n-l-1}\,
\frac{(-n+l+1)_{s}\Gamma(l+l'+s+1)\Gamma(n-l-s)}{s!\,(2l+2)_{s}\,\Gamma(l'-l-s+1)}\times \\ [3mm]
&\times&\,\left[\frac{(n+l)!\,\Gamma(l-i\sigma +1)\,\Gamma(l'+i\sigma +1)}
{(n+l')!\,(n-l-1)!\,(n-l'-1)!\,\Gamma(l+i\sigma +1)\,\Gamma(l'-i\sigma +1)}\right]^{1/2}.
\end{eqnarray*}
Next, applying formula (\ref{1.6.8}) to the gamma-functions under the summation sign,
we obtain the expression for the integral $J_{ll'}$
\begin{eqnarray*}
J_{ll'} &=&
\left[\frac{(n+l)!\,(n-l-1)!\,\Gamma(l-i\sigma +1)\,\Gamma(l'+i\sigma +1)}
{(n+l')1\,(n-l'-1)!\,\Gamma(l+i\sigma +1)\,\Gamma(l'-i\sigma +1)}\right]^{1/2} \times \\ [3mm]
&\times& \, \frac{2(n^{2}+\sigma^{2})\exp\left[i\pi (l-l')/2\right]}
{nR^{3}(l+l'+1)}\,\frac{1}{\Gamma(l-l'+1)\,\Gamma(l'-l+1)}.
\end{eqnarray*}
The resulting expression vanishes when $l'?l$ due to the product of gamma functions of
$(l-l'+1)$ and $(l'-l+1)$, which leads to the orthogonality condition (\ref{1.6.21}).

If we now proceed in the same way as in the plane case when obtaining the formulas in (\ref{1.6.20}), we obtain
\begin{eqnarray}
\int \limits_{0}^{\pi}\,\psi_{nl'\sigma}^{*}(\chi)\psi_{nl\sigma}(\chi)\,d\chi =
\frac{2\mu_{0}}{R\hbar^{2}}\,\left(\frac{\partial E_{n}}{\partial l}\right)_{n_{r}}\,
\frac{\delta_{ll'}}{2l+1}.
\label{1.6.26}
\end{eqnarray}
where the derivative with respect to $l$ is taken at a fixed radial quantum number
$n_{r}=n-l-1$. We also note that in the limit to flat space, $R \to \infty$, all the
formulas we obtained in curved space transform into the corresponding formulas of flat space.

\section{Interbasis expansions in the 8D oscillator}
\markboth{CHAPTER 1. COULOMB-OSCILLATOR ANALOGY}
{1.7. INTERBASIS EXPANSIONS IN THE 8D OSCILLATOR}

Before moving on to the problem of interbasis expansions in an eight-dimensional isotropic oscillator,
we prove (as promised in the previous paragraph) that for the radial wave function (\ref{1.5.10}),
along with the orthonormalization condition (\ref{1.5.8}), the following additional condition of
orthogonality with respect to the hypermoment holds:
\begin{eqnarray}
I_{JJ'}=\int \limits_{0}^{\infty}\,u^{5}\,R_{NJ}(u)\,R_{NJ'}(u)\,du =
\frac{a^{2}}{2J+3}\,\delta_{JJ'}.
\label{1.7.1}
\end{eqnarray}
Substituting the normalized radial wave function of the eight-dimensional isotropic
oscillator (\ref{1.5.10}) into (\ref{1.7.1}), we write the confluent hypergeometric
function in $R_{NJ}(u)$ as a polynomial
\begin{eqnarray*}
F\left(-\frac{N}{2}+J;\,2J+4;\,a^{2}u^{2}\right)=
\sum \limits_{0}^{\frac{N}{2}-J}\,\frac{\left(-\frac{N}{2}+J\right)_{s}}
{s!\,(2J+4)_{s}}\,\left(a^{2}u^{2}\right)^{s}\,,
\end{eqnarray*}
let us perform the integration according to formula (\ref{1.6.5}) and taking into
account (\ref{1.6.6}) for the integral (\ref{1.7.1}) we obtain
\begin{eqnarray*}
I_{JJ'}&=&\frac{a^{2}\,\Gamma(J+J'+3)}{(2J+3)!}\,
\sqrt{\frac{\left(\frac{N}{2}+J+3\right)!}{\left(\frac{N}{2}-J\right)!\,
\left(\frac{N}{2}-J'\right)!\,\left(\frac{N}{2}+J'+3\right)!}} \times \\ [3mm]
&\times& \sum \limits_{0}^{\frac{N}{2}-J}\,\frac{\left(-\frac{N}{2}+J\right)_{s}}
{s!}\,\frac{(J+J'+3)_{s}}{(2J+4)_{s}}\,\frac{\Gamma\left(\frac{N}{2}-J-s+1\right)}
{\Gamma\left(J'-J-s+1\right)}.
\end{eqnarray*}
Next, applying formula (\ref{1.6.8}) to the gamma-functions under the summation sign,
we notice that the sum over s is convolved into a hypergeometric function of
type (\ref{1.6.6}) and therefore
\begin{eqnarray*}
I_{JJ'}=\frac{a^{2}}{\Gamma(J-J'+1)\,\Gamma(J'-J+1)}\,
\sqrt{\frac{\left(\frac{N}{2}-J\right)!\,\left(\frac{N}{2}+J+3\right)!}
{\left(\frac{N}{2}-J'\right)!\,\left(\frac{N}{2}+J'+3\right)!}}.
\end{eqnarray*}
From the last expression it is clear that the gamma-functions provide the orthogonality
condition (\ref{1.7.1}).

Now let's return to the problem of interbasis expansion in an 8D oscillator.
For a fixed energy value, we can write the cylindrical bound states (\ref{1.5.19})
as a coherent quantum mixture of hyperspherical bound states (\ref{1.5.11})
\begin{eqnarray}
\Psi^{cyl} = \sum \limits_{J=K+T}^{\frac{N}{2}}\,W_{NN_{1}KT}^{J}\,\Psi^{hsp}.
\label{1.7.2}
\end{eqnarray}
Our purpose is to find an explicit form for the coefficients of
$W_{NN_{1}KT}^{J}$. First, we note that from comparing (\ref{1.5.2}) with (\ref{1.5.15}) we have
\begin{eqnarray}
\rho_{1}=u\,\cos\frac{\theta}{2}, \qquad \rho_{2}=u\,\sin\frac{\theta}{2}.
\label{1.7.3}
\end{eqnarray}
According to (\ref{1.7.3}) in relation (\ref{1.7.2}) we move from cylindrical coordinates
to hyperspherical ones. Further, substituting $\theta =0$, and taking into account that
\begin{eqnarray}
P_{n}^{(a,b)}(1) = \frac{(a+1)_{n}}{n!},
\label{1.7.4}
\end{eqnarray}
and using the condition of orthogonality of the radial wave functions of the eight-dimensional
isotropic oscillator with respect to the hypermoment $J$ (\ref{1.7.1}) for the coefficient
$W_{NN_{1}KT}^{J}$ we obtain the following integral representation
\begin{eqnarray}
W_{NN_{1}KT}^{J} = \frac{\sqrt{(2J+3)\,(J-K-T)!}}{(2J+3)!\,(2T+1)!}\,
A_{NN_{1}N_{2}}^{JKT}\,\,B_{JKT}^{NN_{1}}.
\label{1.7.5}
\end{eqnarray}
Here
\begin{eqnarray}
A_{NN_{1}N_{2}}^{JKT} = \left[\frac{(J-K+T+1)!\,(N_{1}+2T+1)!\,(N_{2}+2K+1)!\,\left(\frac{N}{2}+J+3\right)!}
{(N_{1})!\,(N_{2})!\,(J+K-T+1)!\,(J+K+T+2)!\,\left(\frac{N}{2}-J\right)!}\right]^{1/2},
\label{1.7.6}
\end{eqnarray}
\begin{eqnarray}
B_{JKT}^{NN_{1}} = \int \limits_{0}^{\infty}\,x^{J+K+T+2}\,e^{-x}\,
F\left(-N_{1};\,2T+2;\,x\right)
F\left(-\frac{N}{2}+J;\,2J+4;\,x\right)\,dx,
\label{1.7.7}
\end{eqnarray}
where $x=a^{2}u^{2}$. Then, in (\ref{1.7.7}) writing the confluent hypergeometric function \\
$F\left(-N_{1};\,2T+2;\,x\right)$ in the form of a polynomial, we perform integration according
to formula (\ref{1.6.5}) and using relation (\ref{1.6.6}) we obtain
\begin{eqnarray}
B_{JKT}^{NN_{1}} &=& \frac{(2J+3)!\,(J+K+T+2)!\,\left(\frac{N}{2}-K-T\right)!}
{(J-K-T)!\,\left(\frac{N}{2}+J+3\right)!}\times  \nonumber \\
\label{1.7.8}
\\
&\times &{_3F_2}\left\{\matrix{ -N_{1},\,\,-J+K+T,\,\, J+K+T+2\cr \cr 2T+2,\,\,-\frac{N}{2}+K+T
\cr}\Biggr|1\right\}\,. \nonumber
\end{eqnarray}

Now returning to the integral representation (\ref{1.7.7}), taking into account
(\ref{1.7.6}) and (\ref{1.7.8}) for $W_{NN_{1}KT}^{J}$ we obtain the expression
\begin{eqnarray}
W_{NN_{1}KT}^{J} &=&
\left[\frac{(2J+3)\,(J-K+T+1)!\,(J+K+T+2)!\,(N_{1}+2T+1)!\,(N_{2}+2K+1)!}
{(N_{1})!\,(N_{2})!\,(J-K-T)!\,(J+K-T+1)!\,\left(\frac{N}{2}-J\right)!\,
\left(\frac{N}{2}+J+3\right)!}\right]^{1/2}\times \nonumber \\
\label{1.7.9}
\\
&\times & \frac{\left(\frac{N}{2}-K-T\right)!}{(2T+1)!}
{_3F_2}\left\{\matrix{ -N_{1},\,\,-J+K+T,\,\, J+K+T+2\cr \cr 2T+2,\,\,-\frac{N}{2}+K+T
\cr}\Biggr|1\right\}\,. \nonumber
\end{eqnarray}

It is known that the Clebsch-Gordan coefficients of the $SU(2)$ group have the following
representation \cite{VAR}
\begin{eqnarray}
&& C_{a \alpha; b \beta}^{c\gamma} = (-1)^{a-\alpha}\,
\left[\frac{((2c+1)\,(a+\alpha)!\,(c+\gamma)!}
{(a-\alpha)!\,(c-\gamma)!\,(a+b+c+1)!\,(a+b-c)!\,(a-b+c)!\,(b-a+c)!}\right]^{1/2}\times
\nonumber  \\
\label{1.7.10}
\\
&\times& \delta_{\gamma, \alpha+\beta}
\frac{(a+b-\gamma)!\,(b+c-\alpha)!}{\sqrt{(b-\beta)!\,(b+\beta)!}}
{_3F_2}\left\{\matrix{ -a-b-c-1,\,\,-a+\alpha,\,\, -c+\gamma\cr \cr
-a-b+\gamma,\,\, -b-c+\alpha \cr}\Biggr|1\right\}\,.
\nonumber
\end{eqnarray}
Using the formula \cite{BE1}
\begin{eqnarray}
{_3F_2}\left\{\matrix{ s,\,\,s',\,\, -N\cr \cr
t',\,\, 1-N-t \cr}\Biggr|1\right\}=\frac{(t+s)_{N}}{(t)_{N}}\,
{_3F_2}\left\{\matrix{ s,\,\,t'-s',\,\, -N\cr \cr
t',\,\, t+s \cr}\Biggr|1\right\}
\label{1.7.11}
\end{eqnarray}
relation (\ref{1.7.10}) can be written in the form
\begin{eqnarray}
&& C_{a \alpha; b \beta}^{c\gamma} = \delta_{\gamma, \alpha+\beta}\,
\left[\frac{(2c+1)\,\,(b-a+c)!\,(a+\alpha)!\,(b+\beta)!\,(c+\gamma)!}
{(b-\beta)!\,(c-\gamma)!\,(a+b-c)!\,(a-b+c)!(a+b+c+1)!}\right]^{1/2}\times
\nonumber  \\
\label{1.7.12}
\\
&\times &
\frac{(-1)^{a-\alpha}}{\sqrt{(a-\alpha)!}}\frac{(a+b-\gamma)!}{(b-a+\gamma)!}\,\,
{_3F_2}\left\{\matrix{ -a+\alpha,\,\, c+\gamma +1\,\, -c+\gamma\cr \cr
\gamma -a-b,\,\, b-a+\gamma +1 \cr}\Biggr|1\right\}\,.
\nonumber
\end{eqnarray}

Finally, comparing (\ref{1.7.10}) with (\ref{1.7.12}) we arrive at the following
representation for the coefficient $W_{NN_{1}KT}^{J}$:
\begin{eqnarray}
W_{NN_{1}KT}^{J} =
(-1)^{N_{1}}\,C_{\frac{N_{1}+N_{2}+2K+1}{2},\,\frac{N_{2}-N_{1}+2K+1}{2};\,
\frac{N_{1}+N_{2}+2T+1}{2},\,\frac{N_{1}-N_{2}+2T+1}{2}}^{J+1,\,K+T+1}.
\label{1.7.13}
\end{eqnarray}

The inverse transformation has the form
\begin{eqnarray}
\Psi^{hsp} = \sum \limits_{N_{1}=0}^{\frac{N}{2}K-T}\,\tilde{W}_{NJKT}^{N_{1}}\,\Psi^{cyl}.
\label{1.7.14}
\end{eqnarray}
The expansion coefficients in (\ref{1.7.14}) are given by the expression
\begin{eqnarray}
\tilde{W}_{NJKT}^{N_{1}} =
(-1)^{N_{1}}\,C_{\frac{N-2T+2K+2}{4},\,\frac{N-2T+2K+2}{4}-N_{1};\,
\frac{N+2T-2K+2}{4},\,N_{1}+2T-\frac{N+2T-2K-2}{4}}^{J+1,\,K+T+1}.
\label{1.7.15}
\end{eqnarray}

\section{Spheroidal basis of the 8D oscillator}
\markboth{CHAPTER 1. COULOMB-OSCILLATOR ANALOGY}
{1.8. SPHEROIDAL BASIS OF THE 8D OSCILLATOR}

Due to $SU(8)$ hidden symmetry, the variables in the Schr\"{o}dinger equation for an eight-dimensional
isotropic oscillator are separated not only in hyperspherical and cylindrical coordinates,
but also in eight-dimensional spheroidal coordinates \cite{KM-3}.

We define eight-dimensional spheroidal coordinates as follows:
\begin{eqnarray}
u_{0}+iu_{1}&=& \frac{d}{2}\,\sqrt{(\xi +1)(1+\eta)}\sin\frac{\beta_{T}}{2}\,e^{-i(\alpha_{T}-\beta_{T})/2}, \nonumber \\
u_{2}+iu_{3}&=& \frac{d}{2}\,\sqrt{(\xi +1)(1+\eta)}\cos\frac{\beta_{T}}{2}\,e^{i(\alpha_{T}+\beta_{T})/2},\nonumber \\
\label{1.8.1}
\\
u_{4}+iu_{5}&=& \frac{d}{2}\,\sqrt{(\xi - 1)(1-\eta)}\sin\frac{\beta_{K}}{2}\,e^{i(\alpha_{K}-\beta_{K})/2}, \nonumber \\
u_{6}+iu_{7}&=& \frac{d}{2}\,\sqrt{(\xi - 1)(1-\eta)}\cos\frac{\beta_{K}}{2}\,e^{-i(\alpha_{K}+\beta_{K})/2},
\nonumber
\end{eqnarray}
where $\xi \in [1,\infty ),\,\eta \in [-1,1]$. Parameter $d$ is the interfocal distance and within
$d \to 0$ and $d \to \infty $ the spheroidal coordinates (\ref{1.8.1}) turn into hyperspherical
and cylindrical coordinates, respectively.

In a spheroidal coordinate system, the oscillatory potential has the form
\begin{eqnarray*}
V = \frac{\mu_{0}\omega^{2}d^{2}}{2}\, (\xi + \eta).
\end{eqnarray*}
In coordinates (\ref{1.8.1}), the differential elements of length, volume and the Laplace operator are
written in the following form:
\begin{eqnarray*}
dl^{2}&=& \frac{d^{2}}{8}\,(\xi - \eta)\,\left(\frac{d\xi^{2}}{\xi^{2}-1}+
\frac{d\eta^{2}}{1-\eta^{2}}\right)+\frac{d^{2}}{16}\,(\xi +1)(1+\eta)\,dl_{T}^{2}\,+
\frac{d^{2}}{16}\,(\xi- 1)(1-\eta)\,dl_{K}^{2}, \\ [3mm]
dV_{8}&=&\frac{d^{8}}{512}\,(\xi- \eta)\,d\xi\,d\eta\,d\Omega_{T}\,d\Omega_{K},\\ [3mm]
\Delta_{8}&=& \frac{8}{d^{2}\,(\xi - \eta)}\,\left[\frac{1}{\xi^{2}- 1}\frac{\partial}{\partial \xi}
\left(\xi^{2}- 1\right)^{2}\,\frac{\partial}{\partial \xi} + \frac{1}{1-\eta^{2}}\,
\frac{\partial}{\partial \eta}\,\left(1-\eta^{2}\right)^{2}\,\frac{\partial}{\partial \eta}\right] - \\ [3mm]
&-&\frac{16{\hat T}^{2}}{d^{2}(\xi +1)(1+\eta)} - \frac{16{\hat K}^{2}}{d^{2}(\xi -1)(1-\eta)}.
\end{eqnarray*}
After substitution
\begin{eqnarray*}
\Psi^{spheroidal} = f_{1}(\xi)\,f_{2}(\eta)\,D_{tt'}^{T}\left(\alpha_{T},\,\beta_{T},\,\gamma_{T}\right)\,
D_{kk'}^{K}\left(\alpha_{K},\,\beta_{K},\,\gamma_{K}\right)
\end{eqnarray*}
the variables in the Schr\"{o}dinger equation are separated and we arrive at the following equations
\begin{eqnarray}
\Biggl[\frac{1}{\xi^{2}-1}\,\frac{d}{d\xi}\,\left(\xi^{2}-1\right)^{2}\,\frac{d}{d\xi} +
\frac{2T(T+1)}{\xi +1} - \frac{2K(K+1)}{\xi -1} + \frac{\mu_{0}d^{2}E}{4\hbar^{2}}-
\frac{a^{4}d^{4}}{16}\,\left(\xi^{2}-1\right)\Biggr]\,f_{1}(\xi) &=& \nonumber \\
= X(d)\,f_{1}(\xi), \nonumber \\
\label{1.8.2}
\\
\Biggl[\frac{1}{1 -\eta^{2}}\,\frac{d}{d\eta}\,\left(1-\eta^{2}\right)^{2}\,\frac{d}{d\eta} -
\frac{2T(T+1)}{1+\eta} - \frac{2K(K+1)}{1-\eta} - \frac{\mu_{0}d^{2}E}{4\hbar^{2}}-
\frac{a^{4}d^{4}}{16}\,\left(1-\eta^{2}\right)\Biggr]\,f_{2}(\eta) &=& \nonumber \\
= - X(d)\,f_{2}(\eta), \nonumber
\nonumber
\end{eqnarray}
where $X(d)$ is the separation constant in spheroidal coordinates.
Now, excluding the energy $E$ from the system of equations (\ref{1.8.2}),
we obtain the spheroidal integral of motion
\begin{eqnarray*}
\hat{X} &=& - \frac{1}{\xi - \eta}\,\left[\frac{\eta}{\xi^{2}-1}\,
\frac{\partial}{\partial \xi}\,\left(\xi^{2}-1\right)^{2}\,
\frac{\partial}{\partial \xi} + \frac{\xi}{1 -\eta^{2}}\,
\frac{\partial}{\partial \eta}\,\left(1-\eta^{2}\right)^{2}\,\frac{\partial}{\partial \eta}\right] + \\ [3mm]
&+& \frac{2(\xi + \eta +1)}{(\xi +1)\,(1+\eta)}\,\hat{T}^{2} -
\frac{2(\xi + \eta -1)}{(\xi -1)\,(1-\eta)}\,\hat{K}^{2}
+ \frac{a^{4}d^{4}}{16}\,(\xi \eta +1),
\end{eqnarray*}
whose eigenvalues are the separation constant $X(d)$, and the eigenfunctions are
$\Psi^{spheroidal}$. Writing down the operator $\hat{X}$ in Cartesian coordinates, we get
\begin{eqnarray}
\hat{X} = \hat{J}^{2} + \frac{a^{4}d^{4}}{4}\,{\hat {\cal P}}.
\label{1.8.3}
\end{eqnarray}
Explicit expressions for the operators $\hat{J}^{2}$ and ${\hat {\cal P}}$ in Cartesian
coordinates are given in the formulas (\ref{1.5.14}) and (\ref{1.5.20}).

So, we have the following spectral problem
\begin{eqnarray}
\hat{X}\,\Psi^{spheroidal} = X_{p}(d)\, \Psi^{spheroidal},
\label{1.8.4}
\end{eqnarray}
where $0 \leq p \leq \frac{N}{2}-K-T-1$ numbers the eigenvalues of the operator
$\hat{X}$.

Now construct the spheroidal basis of the $8D$ oscillator using the following expansions:
\begin{eqnarray}
\Psi^{spheroidal} = \sum \limits_{J=K+T}^{N/2}\,V_{NpKT}^{J}(d)\,\Psi^{hsp},
\label{1.8.5}
\end{eqnarray}
\begin{eqnarray}
\Psi^{spheroidal} = \sum \limits_{N_{1}=0}^{N/2 - K-T}\,U_{NpKT}^{N_{1}}(d)\,\Psi^{cyl}.
\label{1.8.6}
\end{eqnarray}

First, consider expansion (\ref{1.8.5}). Acting with the operator $\hat{X}$ on both sides of equation
(\ref{1.8.5}), using relations (\ref{1.8.3}), (\ref{1.8.4}), (\ref{1.5.13}) and the orthonormalization condition
for the hyperspherical basis with respect to the quantum number $J$, we obtain the following algebraic equation:
\begin{eqnarray}
\frac{4\hbar}{a^{2}\,d^{2}}\,\left[X_{p}(d) - J(J+3)\right] = \sum \limits_{J'}\,
V_{NpKT}^{J}(d)\, \left({\hat {\cal P}}\right)_{JJ'},
\label{1.8.7}
\end{eqnarray}
where
\begin{eqnarray}
 \left({\hat {\cal P}}\right)_{JJ'}=\int \,\Psi_{J}^{hsp*}\,
 {\hat {\cal P}}\,\Psi_{J'}^{hsp}\,dV_{8}.
\label{1.8.8}
\end{eqnarray}
The calculation of the matrix elements $ \left({\hat {\cal P}}\right)_{JJ'}$ can be
carried out using the expansion of the hyperspherical basis into a cylindrical one
(\ref{1.7.14}) and taking into account the eigenvalue equation of the operator
$\hat {\cal P}$ (\ref{1.5.21}). Then instead of relation (\ref{1.8.8}) we get
\begin{eqnarray}
 \left({\hat {\cal P}}\right)_{JJ'}= \sum \limits_{N_{1}=0}^{N/2-K-T}\,
 (4N_{1}-N+4T)\,\tilde{W}_{NJKT}^{N_{1}}\,\tilde{W}_{NJ'KT}^{N_{1}}.
\label{1.8.9}
\end{eqnarray}
Next, using equation (\ref{1.7.15}) together with the recurrence relation \cite{VAR}
\begin{eqnarray}
&&C_{a\,\alpha \,;b\,\beta}^{c\,\gamma} = -\left[\frac{4c^{2}\,(2c+1)\,(2c-1)}
{(c+\gamma)\,(c-\gamma)\,(b-a+c)\,(a-b+c)\,(a+b-c+1)\,(a+b+c+1)}\right]^{1/2} \,\times
\nonumber \\
&\times&\,
\Biggl\{
\left[\frac{(c-\gamma -1)\,(c+\gamma -1)\,(b-a+c-1)\,(a-b+c-1)\,(a+b-c+2)}
{4\,(c-1)^{2}\,(2c-3)\,(2c-1)}\right]^{1/2} \times \label{1.8.10} \\
&\times&\, \sqrt{a+b+c}\, C_{a\,\alpha \,;b\,\beta}^{c-2,\,\gamma}
- \frac{(\alpha-\beta)\,c\,(c-1) - \gamma a(a+1) + \gamma b(b+1)}{2\,c\,(c-1)}\,
C_{a\,\alpha \,;b\,\beta}^{c-1,\,\gamma}\Biggr\} \nonumber
\end{eqnarray}
and the orthogonality condition for the Clebsch-Gordan coefficients
\begin{eqnarray}
\sum \limits_{\alpha + \beta = \gamma}\,\,C_{a\,\alpha \,;b\,\beta}^{c\,\gamma}
\,\, C_{a\,\alpha \,;b\,\beta}^{c',\,\gamma'} = \delta_{c,c'}\,\,\delta_{\gamma, \gamma'}
\label{1.8.11}
\end{eqnarray}
for matrix elements $\left({\hat {\cal P}}\right)_{JJ'}$ we obtain the expression
\begin{eqnarray}
\left({\hat {\cal P}}\right)_{JJ'} = A_{J+1}\,\,\delta_{J',\,J+1} +
 B_{J}\,\,\delta_{J',\,J} +  A_{J}\,\,\delta_{J',\,J-1}
\label{1.8.12}
\end{eqnarray}
where
\begin{eqnarray*}
A_{J} &=& - \sqrt{(J-T-K)\,\,(J+T+K+2)} \times \\ [3mm]
&\times& \left[\frac{(J+T-K+1)\,(J-T+K+1)\,(2N-2J+2)\,(N+2J+6)}
{(J+1)^{2}\,\,(2J+1)\,\,(2J+3)}\right]^{1/2}, \\ [3mm]
B_{J} &=& \frac{(N+4)\,\,(T-K)\,\,(T-K+1)}{(J+1)\,\,(J+2)}.
\end{eqnarray*}
Now substituting (\ref{1.8.12}) into (\ref{1.8.7}) we arrive at the following three-term
recurrence relation for the coefficients $V_{NpKT}^{J}(d)$:
\begin{eqnarray}
A_{J+1}\,\,V_{NpKT}^{J+1} + \left\{B_{J} - \frac{4}{a^{2}\,d^{2}}
\left[X_{p}(d)-J(J+3)\right]\right\}\,\,V_{NpKT}^{J} + A_{J}\,\,V_{NpKT}^{J-1}=0.
\label{1.8.13}
\end{eqnarray}
Recurrence relations (\ref{1.8.13}) are $N/2-K-T+1$ linear homogeneous equations that can be
solved together with the normalization condition
\begin{eqnarray*}
\sum \limits_{J=K+T}^{N/2}\,\,\left|V_{NpKT}^{J}(d)\right|^{2}=1.
\end{eqnarray*}
The eigenvalues $X_{p}(d)$ of the operator $\hat{X}$ can be found from the condition that the
determinant of the system (\ref{1.8.13}) is equal to zero.

Now consider the expansion of the spheroidal basis of an isotropic oscillator into a cylindrical
one (\ref{1.8.6}). Using a technique similar to finding (\ref{1.8.7}), we obtain
\begin{eqnarray}
\left[X_{p}(d) - \frac{a^{2}\,d^{2}}{2}\,\,(N_{1}-N_{2} +T-K)\right]\,\,
U_{NpKT}^{N_{1}}(d) = \sum \limits_{N_{1}'}\,\,U_{NpKT}^{N_{1}'}(d)\,\,
\left(\hat{J}^{2}\right)_{N_{1},N_{1}'},
\label{1.8.14}
\end{eqnarray}
where
\begin{eqnarray*}
\left(\hat{J}^{2}\right)_{N_{1},N_{1}'} =
\int\,\,\Psi^{cyl}_{N_{1}}\,\,\hat{J}^{2}\,\,\Psi^{cyl}_{N_{1}'}\,\,d\,\,V_{8}.
\end{eqnarray*}
The matrix elements of $\left(\hat{J}^{2}\right)_{N_{1},N_{1}'}$ are calculated in the
same way as $\left({\hat {\cal P}}\right)_{JJ'}$, only now we will use the relation \cite{KIBLER}
\begin{eqnarray}
&&\left[c(c+1) -a(a+1) -b(b+1) - 2\alpha \beta\right]\,\,C_{a\,\alpha\,;\,b\,\beta}^{c\,\gamma} = \nonumber \\
&=& \sqrt{(a+\alpha)\,(a-\alpha +1)\,(b-\beta)\,(b+\beta +1)}\,\,
C_{a\,\alpha-1\,;\,b\,\beta+1}^{c\,\gamma} + \label{1.8.15} \\
&+& \sqrt{(a-\alpha)\,(a+\alpha +1)\,(b+\beta)\,(b-\beta +1)}\,\,
C_{a\,\alpha+1\,;\,b\,\beta-1}^{c\,\gamma},
\nonumber
\end{eqnarray}
and orthonormalization condition
\begin{eqnarray}
\sum \limits_{c=|\gamma|}^{a+b}\,\,
C_{a\,\alpha\,;\,b\,\beta}^{c\,\gamma}\,\,
C_{a\,\alpha'\,;\,b\,\beta'}^{c\,\gamma}  =
\delta_{\alpha\,\alpha'}\,\,\delta_{\beta\,\beta'}
\label{1.8.16}
\end{eqnarray}
then for the matrix elements $\left(\hat{J}^{2}\right)_{N_{1},N_{1}'}$ we obtain the expression
\begin{eqnarray}
\left(\hat{J}^{2}\right)_{N_{1},N_{1}'} = C_{N_{1}+1}\,\,\delta_{N_{1}',\,N_{1}+1} +
D_{N_{1}}\,\,\delta_{N_{1}',\,N_{1}} +
C_{N_{1}}\,\,\delta_{N_{1}',\,N_{1}-1},
\label{1.8.17}
\end{eqnarray}
where
\begin{eqnarray*}
C_{N_{1}} &=& -\sqrt{N_{1}\,\,(N_{2}+1)\,\,(N_{1}+2T+1)\,\,(N_{2}+2K+2)}\,\,, \\ [3mm]
D_{N_{1}} &=&N_{2}\,\,(N_{1}+1) + (N_{1}+2T+1)\,\,(N_{2}+2K+2) +
(T-K)(T-K-1) -2\,.
\end{eqnarray*}
Finally, substituting (\ref{1.8.17}) into (\ref{1.8.14}) for the coefficients
$U_{NpKT}^{N_{1}}(d)$ we obtain the following three-term recurrence relation
\begin{eqnarray}
&&\left[D_{N_{1}}-X_{p}(d)+\frac{a^{2}d^{2}}{2}\left(N_{1}-N_{2}+T-K\right)\right]\,\,
U_{NpKT}^{N_{1}}(d) + \nonumber \\
\label{1.8.18}
\\
&+&C_{N_{1}+1}\,\,U_{NpKT}^{N_{1}+1}(d) +
C_{N_{1}}\,\,U_{NpKT}^{N_{1}-1}(d) = 0. \nonumber
\end{eqnarray}
This system of homogeneous equations, as in the previous case, must be solved
together with the normalization condition
\begin{eqnarray*}
\sum\limits_{N_{1}=0}^{N/2-K-T}\,\,\left|U_{NpKT}^{N_{1}}(d)\right|^{2}=1.
\end{eqnarray*}
Here again the eigenvalues of $X_{p}(d)$ can be found by solving a system
of $N/2-K-T+1$ linear homogeneous equations.

Finally, we present the following four limit relations:
\begin{eqnarray*}
\lim\limits_{d\to 0} V_{NpKT}^{J}(d)=\delta_{p,J}, \qquad
\lim\limits_{d\to \infty0} V_{NpKT}^{J}(d)=W_{NN_{1}KT}^{J}, \\ [3mm]
\lim\limits_{d\to 0} U_{NpKT}^{N_{1}}(d) = \tilde{W}_{NJKT}^{N_{1}}, \qquad
\lim\limits_{d\to \infty} U_{NpKT}^{N_{1}}(d) = \delta_{p,N_{1}},
\end{eqnarray*}
which are a good test to check the calculations carried out in sections 6 and 7.

\section{Hyperspherical basis of the 5D Coulomb problem}
\markboth{CHAPTER 1. COULOMB-OSCILLATOR ANALOGY}
{1.9. HYPERSPHERICAL BASIS OF THE 5D COULOMB PROBLEM}

As was shown in Section 3 of the Hurwitz transformation (\ref{1.2.1}), the eight-dimensional
oscillator with the coupling ${\hat{\cal J}}_a \psi({\bf x}) = 0$ translates into the
five-dimensional Coulomb problem (\ref{1.3.6}). The energy eigenvalues of the five-dimensional Coulomb
problem (\ref{1.3.9}) are degenerate, and the multiplicity of degeneracy is equal to
\begin{eqnarray}
g_{n}=\frac{1}{12}\,\,(n+1)\,\,(n+2)^{2}\,\,(n+3).
\label{1.9.1}
\end{eqnarray}
Due to $SO(6)$ hidden symmetry of the $5D$ Coulomb problem \cite{ALLILUEV},
the variables in equation (\ref{1.3.6}) are separated in hyperspherical, parabolic
and spheroidal coordinate systems \cite{KM-2,KM-3}.

We introduce five-dimensional hyperspherical coordinates
$r \in [0,\infty), \theta \in [0, \pi], \alpha \in [0, 2\pi), \beta \in [0, \pi], \gamma \in [0, 4\pi)$
in $\rm I\!R^5(\bf{x})$ as follows:
\begin{eqnarray}
x_{0}=r\,\cos\theta, \quad x_{2}+ix_{1}=r\sin\theta\,\sin\frac{\beta}{2}\,e^{i\frac{\alpha - \gamma}{2}},
\quad x_{4}+ix_{3}=r\sin\theta\,\cos\frac{\beta}{2}\,e^{i\frac{\alpha + \gamma}{2}}.
\label{1.9.2}
\end{eqnarray}
The differential elements of length, volume and the Laplace operator in coordinates (\ref{1.9.2}) have the form
\begin{eqnarray*}
dl_{5}^{2} &=& dr^{2} + r^{2}\,d\theta^{2} + \frac{r^{2}}{4}\sin^{2}\theta
\left(d\alpha^{2} + d\beta^{2} + d\gamma^{2} + 2\cos\beta\,d\alpha\,d\gamma\right), \\ [3mm]
dV_{5} &=& \frac{r^{2}}{8}\,\sin^{3}\theta\,\sin\beta\,dr\,d\theta\,d\alpha\,d\beta\,d\gamma\,, \\ [3mm]
\Delta_{5} &=& \frac{1}{r^{4}}\frac{\partial}{\partial r}\left(r^{4}\,\frac{\partial}{\partial r}\right)
+ \frac{1}{r^{2}\,\sin^{3}\theta}\frac{\partial}{\partial \theta}
\left(\sin^{3}\theta \,\frac{\partial}{\partial \theta}\right)
- \frac{4}{r^{2}\,\sin^{2}\theta}\,\hat{L^{2}},
\end{eqnarray*}
where
\begin{eqnarray}
\hat{L^{2}} = - \left[\frac{\partial^{2}}{\partial \beta^{2}} + \cot\beta\,
\frac{\partial}{\partial \beta} + \frac{1}{\sin^{2}\beta}\,
\left(\frac{\partial^{2}}{\partial \alpha^{2}} - 2 \cos\beta \,
\frac{\partial^{2}}{\partial \alpha\,\partial \gamma} +
\frac{\partial^{2}}{\partial \gamma^{2}}\right)\right].
\label{1.9.3}
\end{eqnarray}
The components of the operator $\hat{L}$ can be obtained from the components of the operator
$\hat{T}$ (\ref{1.5.3}) by replacing $\left(\alpha_{T}, \beta_{T}, \gamma_{T}\right)\to
\left(\alpha, \beta, \gamma\right)$.

After substitution into equations (\ref{1.3.6})
\begin{eqnarray*}
\Psi^{hsp} = R(r)\,Z(\theta)\,D_{mm'}^{L}\left(\alpha, \beta, \gamma\right)
\end{eqnarray*}
we arrive at the following pair of differential equations
\begin{eqnarray}
\frac{1}{\sin^{3}\theta}\frac{d}{d \theta}\left(\sin^{3}\theta \,\frac{dZ}{d \theta}\right)
- \frac{4L(L+1)}{\sin^{2}\theta}\,Z + \lambda (\lambda +3)Z = 0,
\label{1.9.4}
\end{eqnarray}
\begin{eqnarray}
\frac{1}{r^{4}}\frac{d}{d r}\left(r^{4} \,\frac{dR}{d r}\right)
- \frac{\lambda (\lambda +3)}{r^{2}}\,R +
\left[\frac{2}{r_{0}\,r} - \frac{1}{r_{0}^{2}\,(n+2)^{2}}\right]R = 0.
\label{1.9.5}
\end{eqnarray}
Here $\lambda (\lambda +3)$ is the separation constant ($\lambda$ is a non-negative integer),
$r_{0}$ - Bohr radius and
\begin{eqnarray*}
n+2 = \sqrt{-\frac{\mu_{0}e^{4}}{2\hbar^{2}\,\varepsilon}}, \qquad
r_{0} = \frac{\hbar^{2}}{\mu_{0}e^{2}}.
\end{eqnarray*}
The solution to equation (\ref{1.9.4}) is the Gegenbauer polynomials \cite{BE2}
\begin{eqnarray}
C_{n}^{\nu}(x) = \frac{\Gamma(n+2\nu)}{N!\,\Gamma(2\nu)}\,\,
_{2}F_{1}\left(-n,\,\,\,\,n+2\nu;\,\,\nu + \frac{1}{2};\,\,\frac{1-x}{2}\right)
\label{1.9.6}
\end{eqnarray}
multiplied by the factor $(\sin\theta)^{2L}$. Normalizing by the condition
\begin{eqnarray*}
\int\limits_{0}^{\pi}\,\,\left|Z_{\lambda, L}(\theta)\right|^{2}
\,\,\sin^{3}\theta \,\,d\,\theta = 1
\end{eqnarray*}
we find
\begin{eqnarray}
Z_{\lambda, L}(\theta) = 2^{2L+1}\,\Gamma\left(2L + \frac{3}{2}\right)\,\,
\sqrt{\frac{(2\lambda + 3)\,(\lambda - 2L)!}{2\,\pi\,(\lambda +2L+2)!}}\,\,
(\sin\theta)^{2L}\,\,C_{\lambda - 2L}^{2L+3/2}(\cos\theta).
\label{1.9.7}
\end{eqnarray}

Normalized by the condition
\begin{eqnarray*}
\int\limits_{0}^{\infty}\,\,r^{4}\,\,R_{n,\lambda}^{2}(r)
\,\,d\,r = 1
\end{eqnarray*}
the radial wave function can be expressed in terms of the confluent
hypergeometric function as follows
\begin{eqnarray}
R_{n,\lambda}(r) = \frac{4}{r_{0}^{5/2}\,(n+2)^{3}}\,
\sqrt{\frac{(n+\lambda + 3)!}{(n-\lambda)!}}\,
\frac{e^{-\kappa r}\,(2\kappa r)^{\lambda}}{(2\lambda + 3)!}\,
F\left(-n+\lambda;\,2\lambda + 4;\,2\kappa r\right),
\label{1.9.8}
\end{eqnarray}
where $\kappa=\sqrt{-2\mu_{0}\varepsilon}/\hbar = 1/r_{0}(n+2)$.
Quantum numbers $\lambda$ and $n$ take the values
\begin{eqnarray*}
\lambda = 2L, 2L+2,...; \qquad n=\lambda, \lambda+1,....
\end{eqnarray*}
So, the normalized hyperspherical wave functions of the $5D$ Coulomb problem can be written as
\begin{eqnarray}
\Psi^{hsp} = \sqrt{\frac{2L+1}{2\pi^{2}}}\,\,R_{n,\lambda}(r)\,\,
Z_{\lambda, L}(\theta)\,\,D_{mm'}^{L}\left(\alpha, \beta, \gamma\right).
\label{1.9.9}
\end{eqnarray}

The hyperspherical basis (\ref{1.9.9}) is an eigenfunction of the following commuting operators
$\left\{\hat{H},\,\,\hat{L}^{2},\,\,\hat{L}_{z},\,\,\hat{L}_{z'}\right\}$
and the square of the global angular momentum
\begin{eqnarray*}
\hat{\Lambda}^{2} = -\,\frac{1}{\sin^{3}\theta}\,\,\frac{\partial}{\partial \theta}\,\,
\left(\sin^{3}\theta\,\,\frac{\partial}{\partial \theta}\right)
+ \frac{4}{\sin^{2}\theta}\,\,\hat{L^{2}},
\end{eqnarray*}
moreover
\begin{eqnarray}
\hat{\Lambda}^{2}\,\,\Psi^{hsp} = \lambda\,(\lambda + 3)\,\,\Psi^{hsp}.
\label{1.9.10}
\end{eqnarray}

In Cartesian coordinates, the operator $\hat{\Lambda}^{2}$ has the form
\begin{eqnarray}
\hat{\Lambda}^{2} = -r^{2}\,\Delta_{5} + x_{i}\,x_{j}\,\,
\frac{\partial^{2}}{\partial x_{i}\,\partial x_{j}} +
4\,x_{i}\,\,\frac{\partial^{2}}{\partial x_{i}},
\label{1.9.11}
\end{eqnarray}
where $i,j=0,1,2,3,4$.

\section{5D Parabolic bound states}
\markboth{CHAPTER 1. COULOMB-OSCILLATOR ANALOGY}
{1.10. 5D PARABOLIC BOUND STATES}

Now consider the five-dimensional Coulomb problem in parabolic coordinates.

In $\rm I\!R^5(\bf{x})$ we define parabolic coordinates as follows:
\begin{eqnarray}
x_{0}=\frac{1}{2}(\mu - \nu), \quad x_{2}+ix_{1}=\sqrt{\mu \nu}\,\sin\frac{\beta}{2}e^{i\frac{\alpha - \gamma}{2}},
\quad x_{4}+ix_{3}=\sqrt{\mu \nu}\,\cos\frac{\beta}{2}e^{i\frac{\alpha + \gamma}{2}}.
\label{1.10.1}
\end{eqnarray}
where $\mu\,,\nu \in [0,\,\infty)$.
In coordinates (1.10.1), the Coulomb potential, the differential elements of length,
volume and the Laplace operator have the form
\begin{eqnarray*}
V=- \frac{2e^{2}}{\mu + \nu},
\end{eqnarray*}
\begin{eqnarray*}
dl^{5} &=& \frac{\mu + \nu}{4}\,\left(\frac{1}{\mu}\,d\mu^{2} +
\frac{1}{\nu}\,d\nu^{2}\right) + \frac{\mu \nu}{4}\,
\left(d\alpha^{2} + d\beta^{2} + d\gamma^{2} + 2\cos\beta\,d\alpha\,d\gamma\right), \\ [3mm]
dV_{5} &=& \frac{\mu \nu}{4}\,(\mu + \nu)\,\sin\beta\,d\mu\,d\nu\,d\alpha\,d\beta\,d\gamma\,, \\ [3mm]
\Delta_{5} &=& \frac{4}{\mu + \nu}\,\left[\frac{1}{\mu}\,\frac{\partial}{\partial \mu}
\left(\mu^{2}\,\frac{\partial}{\partial \mu}\right) +
\frac{1}{\nu}\,\frac{\partial}{\partial \nu}
\left(\nu^{2}\,\frac{\partial}{\partial \nu}\right)\right]
-\frac{4}{\mu \nu}\,\hat{L^{2}}.
\end{eqnarray*}

Starting from the product
\begin{eqnarray*}
\Psi^{par} =\Phi_{1}(\mu)\,\Phi_{2}(\nu)\,D_{mm'}^{L}\left(\alpha, \beta, \gamma\right)
\end{eqnarray*}
we obtain equations
\begin{eqnarray}
\frac{1}{\mu}\,\frac{d}{d \mu}\,\left(\mu^{2}\,\frac{d\Phi_{1}}{d \mu}\right) +
\left[\frac{\mu_{0}\,\varepsilon}{2\hbar^{2}}\,\mu - \frac{L(L+1)}{\mu} +
\frac{\sqrt{\mu_{0}}}{2\hbar}\,\Omega + \frac{\mu_{0}\,e^{2}}{2\hbar^{2}}\right]\,\Phi_{1} =0,
\nonumber \\
\label{1.10.2}
\\
\frac{1}{\nu}\,\frac{d}{d \nu}\,\left(\nu^{2}\,\frac{d\Phi_{2}}{d \nu}\right) +
\left[\frac{\mu_{0}\,\varepsilon}{2\hbar^{2}}\,\nu - \frac{L(L+1)}{\nu} -
\frac{\sqrt{\mu_{0}}}{2\hbar}\,\Omega + \frac{\mu_{0}\,e^{2}}{2\hbar^{2}}\right]\,\Phi_{2} =0,
\nonumber
\end{eqnarray}
where $\Omega$ is the parabolic separation constant. We seek solutions to equations (\ref{1.10.2}) in the form
\begin{eqnarray*}
\Phi_{1}(y_{1}) = e^{-y_{1}/2}\,\,y_{1}^{L}\,\,f_{1}(y_{1}), \qquad
\Phi_{2}(y_{2}) = e^{-y_{2}/2}\,\,y_{2}^{L}\,\,f_{2}(y_{2}),
\end{eqnarray*}
where $y_{1} = \kappa \mu, \,y_{2} = \kappa \nu$. Then for the functions
$f_{1}(y_{1})$ and $f_{2}(y_{2})$ we obtain the equation for the confluent
hypergeometric function
\begin{eqnarray*}
y_{1}\,\frac{d^{2}\,f_{1}}{d\,y_{1}^{2}} +
(2L+2-y_{1})\,\,\frac{d\,f_{1}}{d\,y_{1}} -
\left(L+1-\frac{\sqrt{\mu_{0}}}{2\kappa\,\hbar}\,\Omega -
\frac{\mu_{0}\,e^{2}}{2\kappa \hbar^{2}}\right)f_{1} = 0, \\ [3mm]
y_{2}\,\frac{d^{2}\,f_{2}}{d\,y_{2}^{2}} +
(2L+2-y_{2})\,\,\frac{d\,f_{2}}{d\,y_{2}} -
\left(L+1 + \frac{\sqrt{\mu_{0}}}{2\kappa\,\hbar}\,\Omega -
\frac{\mu_{0}\,e^{2}}{2\kappa \hbar^{2}}\right)f_{2} = 0.
\end{eqnarray*}
Now demanding that
\begin{eqnarray*}
n_{1} = -L-1 + \frac{\sqrt{\mu_{0}}}{2\kappa\,\hbar}\,\Omega -
\frac{\mu_{0}\,e^{2}}{2\kappa \hbar^{2}}, \qquad
n_{2} = -L-1 - \frac{\sqrt{\mu_{0}}}{2\kappa\,\hbar}\,\Omega -
\frac{\mu_{0}\,e^{2}}{2\kappa \hbar^{2}},,
\end{eqnarray*}
where $n_{1}$ and $n_{2}$ are non-negative integers, taking into account (\ref{1.3.9})
from the last relations we obtain that the parabolic quantum numbers
$n_{1}$, $n_{2}$ and L are related to the principal quantum number n as follows:
\begin{eqnarray*}
n= n_{1} + n_{2} +2L.
\end{eqnarray*}
Normalized by condition
\begin{eqnarray*}
\int\,\left|\Psi^{par}\right|^{2}\,d\,V_{5} = 1
\end{eqnarray*}
the parabolic basis has the form
\begin{eqnarray}
\Psi^{par} = \frac{1}{(n+2)^{3}}\,\,\sqrt{\frac{2L+1}{\pi^{2}\,r_{0}^{5}}}\,\,
\Phi_{n_{1}, L}(\mu)\,\,\Phi_{n_{2}, L}(\nu)\,\,
D_{mm'}^{L}\left(\alpha, \beta, \gamma\right),
\label{1.10.3}
\end{eqnarray}
where
\begin{eqnarray*}
\Phi_{p, q}(x)= \frac{1}{(2q+1)!}\,\,\sqrt{\frac{(p+2q+1)!}{P!}}\,\,
e^{-\kappa x/2}\,(\kappa x)^{q}\,\,F(-p;\,\,2q+2;\,\,\kappa x).
\end{eqnarray*}

Now excluding the energy $\varepsilon$ from equations (\ref{1.10.2})
we obtain the following integral of motion
\begin{eqnarray*}
\it{\hat{A}_{0}} = \frac{\hbar}{\sqrt{\mu_{0}}}\,\left\{\frac{2}{\mu + \nu}
\left[\frac{\mu}{\nu}\,\frac{\partial}{\partial \nu}\left(\nu^{2}\,\frac{\partial}{\partial \nu}\right)
-\frac{\nu}{\mu}\,\frac{\partial}{\partial \mu}\left(\mu^{2}\,\frac{\partial}{\partial \mu}\right)\right]
- \frac{2(\mu - \nu)}{\mu \nu}\,\hat{L}^{2} + \frac{\mu_{0}\,e^{2}}{\hbar^{2}}\,\frac{\mu - \nu}{\mu + \nu}\right\}.
\end{eqnarray*}
It is easy to notice that $\it{\hat{A}_{0}}$ is a component of the five-dimensional analogue
of the Runge-Lenz vector \cite{revai}
\begin{eqnarray}
\it{\hat{A}_{k}} = \frac{1}{2\sqrt{\mu_{0}}}\,\left(\hat{p}_{i}\,\hat{L}_{i k} +
\hat{L}_{i k}\,\hat{p}_{i} + \frac{2\mu_{0}\,e^{2}}{\hbar^{2}}\,\frac{x_{k}}{r}\right),
\label{1.10.4}
\end{eqnarray}
where
\begin{eqnarray*}
\hat{L}_{i k} = \frac{1}{\hbar}\left(x_{i}\,\hat{p}_{k} - x_{k}\,\hat{p}_{i}\right).
\end{eqnarray*}

Thus the parabolic basis of the five-dimensional Coulomb problem
(\ref{1.10.3}) is an eigenfunction of the operator $\it{\hat{A}_{0}}$,
and the following equation holds:
\begin{eqnarray}
{\it{\hat{A}_{0}}}\,\Psi^{par} = \Omega\,\Psi^{par} = \frac{e^{2}\sqrt{\mu_{0}}}{\hbar}
\,\frac{n_{1} - n_{2}}{n_{1} + n_{2}}\Psi^{par}.
\label{1.10.5}
\end{eqnarray}

In Cartesian coordinates, the operator $\it{\hat{A}_{0}}$ has the form
\begin{eqnarray}
{\it{\hat{A}_{0}}} = \frac{\hbar}{\sqrt{\mu_{0}}}\,\left[x_{0}\,
\frac{\partial^{2}}{\partial x_{\sigma} \partial x_{\sigma}}- x_{0}\,
\frac{\partial^{2}}{\partial x_{0} \partial x_{\sigma}} - 2\frac{\partial}{\partial x_{0}} +
\frac{\mu_{0} e^{2}}{\hbar^{2}}\,\frac{x_{0}}{r}\right],
\label{1.10.6}
\end{eqnarray}
where $\sigma = 1,2,3,4$.

\section{Generalized Park-Tarter expansion}
\markboth{CHAPTER 1. COULOMB-OSCILLATOR ANALOGY}
{1.11. GENERALIZED PARK-TARTER EXPANSION}

Now, for a fixed energy value, we can write parabolic bound states (\ref{1.10.3})
as a coherent quantum mixture of hyperspherical bound states (\ref{1.9.9}):
\begin{eqnarray}
\Psi^{par} = \sum \limits_{\lambda = 2L}^{n}\,\,W^{\lambda}_{n_{1} n_{2} L}\,\,\Psi^{hsp}.
\label{1.11.1}
\end{eqnarray}
Expansion (\ref{1.11.1}) generalizes the Park-Tarter expansion \cite{PARK,TARTER}
from the theory of the hydrogen atom to the $5$-dimensional case.

The purpose of this section is to find an exact expression for the amplitudes
$W^{\lambda}_{n_{1} n_{2} L}$. First, note that the hyperspherical and parabolic
coordinates are related by the relations:
\begin{eqnarray}
\mu = r\,(1 + \cos\theta), \qquad \nu = r\,(1  \sin\theta).
\label{1.11.2}
\end{eqnarray}
Now, in relation (\ref{1.11.1}) according to (\ref{1.11.2}), we move from parabolic
coordinates to spherical ones. Then substituting $\theta = 0$, taking into account that
\begin{eqnarray}
C_{n}^{\nu}(1) = \frac{(2\nu)!}{n!}
\label{1.11.3}
\end{eqnarray}
and using the condition of orthogonality of the radial wave functions of the
five-dimensional Coulomb problem (\ref{1.9.9}) with respect to the hypermoment (\ref{1.6.20})
\begin{eqnarray}
\int\limits_{0}^{\infty}\,\,r^{2}\,R_{n \lambda}(r)\,R_{n \lambda'}(r)\,dr
= \frac{2}{r_{0}^{2}(n+2)^{3}}\,\,\frac{\delta_{\lambda \lambda'}}{2\lambda + 3},
\label{1.11.4}
\end{eqnarray}
we arrive at the following integral representation for the generalized
Park-Tarter amplitude
\begin{eqnarray}
W^{\lambda}_{n_{1} n_{2} L} = \frac{E_{n_{1} n_{2}}^{\lambda L}\,\,
K_{\lambda L}^{n n_{1}}}{(2L+1)!\,(2\lambda + 3)!}.
\label{1.11.5}
\end{eqnarray}
Here
\begin{eqnarray}
E_{n_{1} n_{2}}^{\lambda L} = \left[\frac{(2\lambda + 3)\,(\lambda - 2L)!\,
(n + \lambda + 3)!\,(n_{1} + 2L +1)!\,(n_{2} + 2L +1)!}
{(n_{1})!\,(n_{2})!\,(n - \lambda)!\,(\lambda +2L + 2)!}\right]^{1/2},
\label{1.11.6}
\end{eqnarray}
\begin{eqnarray*}
K_{\lambda L}^{n n_{1}} =\int\limits_{0}^{\infty}\,\,e^{-x}\,x^{\lambda +2L + 2}\,
F\left(-n_{1};\,2L + 2;\,x\right)\,F\left(-n + \lambda;\,2\lambda + 4;\,x\right)\,dx,
\end{eqnarray*}
where $x=2\kappa r$. Now writing $F\left(-n_{1};\,2L + 2;\,x\right)$
as a polynomial and using formulas (\ref{1.6.5}) and (\ref{1.6.6}) we get
\begin{eqnarray*}
K_{\lambda L}^{n n_{1}} = \frac{(2\lambda + 3)!\,(\lambda +2L + 2)!\,(n - \lambda)!}
{(\lambda - 2L)!\,(n + \lambda + 3)!}\,\,
{_3F_2}\left\{\matrix{ -n_{1},\,\,-\lambda +2L,\,\, \lambda +2L + 3\cr \cr
2L + 2,\,\,-n + 2L
\cr}\Biggr|1\right\}.
\end{eqnarray*}
Next, substituting the resulting expression for the integral
$K_{\lambda L}^{n n_{1}}$ and relation (\ref{1.11.6}) into (\ref{1.11.5})
we obtain a generalized version of the Tarter representation \cite{TARTER}:
\begin{eqnarray}
W^{\lambda}_{n_{1} n_{2} L} &=& \frac{(n-2L)!}{(2L + 1)!}
\left[\frac{(2\lambda + 3)\,(\lambda +2L + 2)!\,(n_{1} + 2L +1)!\,(n_{2} + 2L +1)!}
{(n_{1})!\,(n_{2})!\,(\lambda - 2L)!\,(n - \lambda)!(n + \lambda + 3)!}\right]^{1/2}\,\times
\nonumber \\
\label{1.11.7}
\\
&\times& {_3F_2}\left\{\matrix{ -n_{1},\,\,-\lambda +2L,\,\, \lambda +2L + 3\cr \cr
2L + 2,\,\,-n + 2L
\cr}\Biggr|1\right\}.
\nonumber
\end{eqnarray}
And finally, using formula (\ref{1.7.12}), the amplitude $W^{\lambda}_{n_{1} n_{2} L}$
can be expressed in terms of the Clebsch-Gordan coefficients of the $SU(2)$ group, namely
\begin{eqnarray}
W^{\lambda}_{n_{1} n_{2} L} = (-1)^{n_{1}}\,
C_{\frac{n+1}{2},\,\frac{2L+n_{2}-n_{1}+1}{2};\,\frac{n+1}{2},\,\frac{2L+n_{1}-n_{2}+1}{2}}
^{\lambda + 1,\,2L+1},
\label{1.11.8}
\end{eqnarray}
which is a generalization of Park's result \cite{PARK}.

Reverse expansion
\begin{eqnarray}
\Psi^{hsp} = \sum \limits_{n_{1} = 0}^{n-2L}\,\,\tilde{W}^{n_{1}}_{n \lambda L}\,\,\Psi^{par}
\label{1.11.9}
\end{eqnarray}
directly obtained from (\ref{1.11.1}) using the orthonormality condition for the
Clebsch-Gordan coefficients (\ref{1.8.16}). The inverse expansion coefficients (\ref{1.11.9})
have the form:
\begin{eqnarray}
\tilde{W}^{n_{1}}_{n \lambda L} = (-1)^{n_{1}}\,
C_{\frac{n+1}{2},\,\frac{n+1}{2}-n_{1};\,\frac{n+1}{2},\,n_{1}+2L-\frac{n-1}{2}}
^{\lambda + 1,\,2L+1}.
\label{1.11.10}
\end{eqnarray}

\section{Generalized amplitudes of Coulson-Joseph}
\markboth{CHAPTER 1. COULOMB-OSCILLATOR ANALOGY}
{1.12. GENERALIZED AMPLITUDES OF COULSON-JOSEPH}

The variables in the Schr\"{o}dinger equation (\ref{1.3.6}) for the five-dimensional
Coulomb problem are also separated in spheroidal coordinates. In $\rm I \!R^{5}(\bf x)$
we define spheroidal coordinates as follows \cite{KM-3}:
\bea
x_{0} &=& \frac{R}{2}\left(\xi \eta +1\right),  \nonumber \\
x_{2}+ix_{1} &=& \frac{R}{2}\sqrt{\left(\xi^{2}-1\right)\left(1- \eta^{2}\right)}
\sin\frac{\beta}{2}e^{i(\alpha-\gamma)/2},
\label{1.12.1}
\\
x_{4}+ix_{3} &=& \frac{R}{2}\sqrt{\left(\xi^{2}-1\right)\left(1- \eta^{2}\right)}
\cos\frac{\beta}{2}e^{i(\alpha+\gamma)/2},
\nonumber
\eea
where $\xi \in [1, \infty), \eta \in [-1, 1]$. The Coulomb potential, differential
elements of length, volume and the Laplace operator in these coordinates have the form
\begin{eqnarray*}
V = - \frac{2e^{2}}{R(\xi + \eta},
\end{eqnarray*}
\begin{eqnarray*}
dl_{5}^{2} &=& \frac{R^{2}}{4}\left(\xi^{2} - \eta^{2}\right)
\left(\frac{d\xi^{2}}{\xi^{2} -1} +
\frac{d\eta^{2}}{1-\eta^{2}}\right) + \frac{R^{2}}{16}
\left(\xi^{2}-1\right)\left(1- \eta^{2}\right)dl^{2}, \\ [3mm]
dV_{5} &=& \frac{R^{5}}{256}\left(\xi^{2} - \eta^{2}\right)
\left(\xi^{2}-1\right)\left(1- \eta^{2}\right)d\xi\, d\eta\, d\alpha\, d\beta\, d\gamma,
\end{eqnarray*}
\begin{eqnarray*}
\Delta_{5} &=& \frac{4}{R^{2}\left(\xi^{2} - \eta^{2}\right)}
\left\{\frac{1}{\xi^{2} -1}\frac{\partial}{\partial \xi}
\left[\left(\xi^{2} - 1\right)^{2}\frac{\partial}{\partial \xi}\right] +
\frac{1}{1-\eta^{2}}\frac{\partial}{\partial \eta}
\left[\left(1- \eta^{2}\right)^{2}\frac{\partial}{\partial \eta}\right]\right\} - \\ [3mm]
&-& \frac{16 \hat{L}^{2}}{R^{2}\left(\xi^{2}-1\right)\left(1- \eta^{2}\right)},
\end{eqnarray*}
where $dl^{2}=d\alpha^{2} + d\beta^{2} + d\gamma^{2} + 2\cos\beta\, d\alpha\, d\beta\, d\gamma$
and R – interfocal distance. Within $R \to 0$ and $R \to \infty$ we return to hyperspherical
and parabolic coordinates, respectively.

Separation of variables in spheroidal coordinates
\begin{eqnarray*}
\Psi^{spheroidal} = \sqrt{\frac{2L+1}{2\pi^{2}}}\,f_{1}(\xi)\,f_{2}(\eta)\,
D_{mm'}^{L}(\alpha, \beta, \gamma)
\end{eqnarray*}
generates the separation constant $\Pi (R)$ and leads to two ordinary differential equations:
\bea
\Biggl[\frac{1}{\xi^{2}-1}\frac{d}{d \xi}\left(\xi^{2}-1\right)^{2}
\frac{d}{d \xi} - \frac{4L(L+1)}{\xi^{2}-1} + \frac{R}{r_{0}}\,\xi
+ \frac{\mu_{0}R^{2}\varepsilon}{2\hbar^{2}}\,\left(\xi^{2}-1\right)\Biggr]
f_{1}(\xi) = \Pi(R)\,f_{1}(\xi), \nonumber \\
\label{1.12.2}
\\
\Biggl[\frac{1}{1- \eta^{2}}\frac{d}{d \eta}\left(1- \eta^{2}\right)^{2}
\,\frac{d}{d \eta} - \frac{4L(L+1)}{1- \eta^{2}} - \frac{R}{r_{0}}\,\eta
+ \frac{\mu_{0}R^{2}\varepsilon}{2\hbar^{2}}\,\left(1- \eta^{2}\right)\Biggr]\,
f_{2}(\eta) = -\Pi(R)\,f_{2}(\eta). \nonumber
\eea
Eliminating the energy $\varepsilon$ from equations (\ref{1.12.2}) we find the Hermitian operator
\begin{eqnarray*}
\hat{\Pi} &=& \frac{1}{R^{2}\left(\xi^{2} - \eta^{2}\right)}\,
\left[\frac{1-\eta^{2}}{\xi^{2}-1}\,\frac{\partial}{\partial \xi}
\left(\xi^{2}-1\right)^{2}\,\frac{\partial}{\partial \xi} -
\frac{\xi^{2}-1}{1-\eta^{2}}\,\frac{\partial}{\partial \eta}\,
\left(1- \eta^{2}\right)^{2}\,\frac{\partial}{\partial \eta}\right] + \\ [3mm]
&+& \frac{4\left(\xi^{2} + \eta^{2} - 2\right)}{\left(\xi^{2}-1\right)\left(1- \eta^{2}\right)}\,
\hat{L}^{2} + \frac{R}{r_{0}}\,\frac{\xi \eta + 1}{\xi + \eta},
\end{eqnarray*}
whose eigenvalues are $\Pi (R)$, and whose eigenfunctions are $\Psi^{spheroidal}$, i.e.
\begin{eqnarray}
\hat{\Pi}\,\Psi^{spheroidal} = \Pi_{q}(R)\,\Psi^{spheroidal},
\label{1.12.3}
\end{eqnarray}
where the integer index $q$ numbers the eigenvalues of the operator $\hat{\Pi}$ and takes
values within the range $0 \leq q \leq n-2L$. Returning to Cartesian coordinates and after
long calculations for operator $\hat{\Pi}$, we obtain that
\begin{eqnarray}
\hat{\Pi} = \hat{\Lambda}^{2} + \frac{R\,\sqrt{\mu_{0}}}{\hbar}\,{\it{\hat{A}_{0}}}.
\label{1.12.4}
\end{eqnarray}
Thus, operator $\hat{\Pi}$ is a simple linear combination of the hyperspherical and parabolic integrals of motion.

Our next steps are as follows:

\noindent (a) consider the expansions
\begin{eqnarray}
\Psi^{spheroidal} = \sum\limits_{\lambda =2L}^{n}\,V_{nqL}^{\lambda}(R)\,\Psi^{hsp},
\label{1.12.5}
\end{eqnarray}
\begin{eqnarray}
\Psi^{spheroidal} = \sum\limits_{n_{1} = 0}^{n - 2L}\,U_{nqL}^{n_{1}}(R)\,\Psi^{par};
\label{1.12.6}
\end{eqnarray}
(b) we act with operator $\hat{\Pi}$ on both sides of (\ref{1.12.5}) and (\ref{1.12.6}),
taking into account relation (\ref{1.12.4});

\noindent (c) Using equations (\ref{1.9.10}) and (\ref{1.10.6}) we obtain two systems of
linear homogeneous equations:
\begin{eqnarray}
\left[\Pi_{q}(R) - \lambda(\lambda + 3)\right]\,V_{nqL}^{\lambda}(R) =
\frac{R\,\sqrt{\mu_{0}}}{\hbar}\,\sum\limits_{\lambda =2L}^{n}\,V_{nqL}^{\lambda'}(R)\,
\left({\it{\hat{A}_{0}}}\right)_{\lambda \lambda'},
\label{1.12.7}
\end{eqnarray}
\begin{eqnarray}
\left[\Pi_{q}(R) - \frac{R}{r_{0}}\,\frac{n_{1} - n_{2}}{n + 2}\right] =
\sum\limits_{n_{1} = 0}^{n - 2L}\,U_{nqL}^{n_{1}'}(R)\,
\left(\hat{\Lambda}^{2}\right)_{n_{1} n_{1}'}.
\label{1.12.8}
\end{eqnarray}
Here
\begin{eqnarray*}
\left({\it{\hat{A}_{0}}}\right)_{\lambda \lambda'} =
\int\,\Psi_{\lambda}^{hsp*}\,{\it{\hat{A}_{0}}}\,\Psi_{\lambda'}^{hsp}\,
d\,V_{5}, \\ [3mm]
\left(\hat{\Lambda}^{2}\right)_{n_{1} n_{1}'} = \int\,\Psi_{n_{1}}^{par*}\,
\hat{\Lambda}^{2}\,\Psi_{n_{1}'}^{par}\,d\,V_{5}.
\end{eqnarray*}

Now using expansions (\ref{1.11.1}), (\ref{1.11.9}) and relations (\ref{1.8.10}), (\ref{1.8.11}),
(\ref{1.8.15}) and (\ref{1.8.16}) for matrix elements $\left({\it{\hat{A}_{0}}}\right)_{\lambda \lambda'}$
and $\left(\hat{\Lambda}^{2}\right)_{n_{1} n_{1}'}$ we obtain the following expressions:
\begin{eqnarray}
\left({\it{\hat{A}_{0}}}\right)_{\lambda \lambda'} =
\frac{e^{2}\,\sqrt{\mu_{0}}}{\hbar\,(n+2)}
\left(B_{n L}^{\lambda +1}\,\delta_{\lambda', \lambda +1} +
B_{n L}^{\lambda}\,\delta_{\lambda', \lambda -1}\right),
\label{1.12.9}
\end{eqnarray}
\begin{eqnarray}
&&\left(\hat{\Lambda}^{2}\right)_{n_{1} n_{1}'} = \left[\left(n_{1}+1\right)\,
\left(n-n_{1}-2L\right) + \left(n-n_{1}+2\right)\,\left(n_{1}+2L+1\right) - 2\right]\,
\delta_{n_{1}', n_{1}} - \nonumber \\ [2mm]
&-& \sqrt{\left(n_{1}+1\right)\,\left(n-n_{1}+1\right)\,\left(n_{1}+2L+2\right)\,
\left(n-n_{1}-2L\right)}\,\delta_{n_{1}', n_{1}+1} -
\label{1.12.10}
\\ [2mm]
&-& \sqrt{n_{1}\,\left(n-n_{1}+2\right)\,\left(n_{1}+2L+1\right)\,
\left(n-n_{1}-2L+1\right)}\,\delta_{n_{1}', n_{1}-1},
\nonumber
\end{eqnarray}
where
\begin{eqnarray*}
B_{n L}^{\lambda} = - \sqrt{\frac{\left(\lambda - 2L\right)\,\left(\lambda + 2L + 2\right)\,
\left(n - \lambda + 1\right)\,\left(n + \lambda + 3\right)}
{\left(2 \lambda + 1\right)\,\left(2 \lambda + 3\right)}}.
\end{eqnarray*}
Substituting (\ref{1.12.9}) and (\ref{1.12.10}) into (\ref{1.12.7}) and (\ref{1.12.8}),
respectively, we obtain the following three-term recurrence relations for the coefficients
$V_{nqL}^{\lambda}(R)$ and $U_{nqL}^{n_{1}}(R)$:
\begin{eqnarray}
\left[\Pi_{q}(R) - \lambda(\lambda + 3)\right]\,V_{nqL}^{\lambda}(R) =
\frac{R}{r_{0}\,(n+2)}
\left[B_{n L}^{\lambda +1}\,V_{nqL}^{\lambda + 1}(R) +
B_{n L}^{\lambda}\,V_{nqL}^{\lambda - 1}(R)\right],
\label{1.12.11}
\end{eqnarray}
\begin{eqnarray}
&&\left[n_{2}\left(n_{1}+1\right) + \left(n-n_{1}+2\right)\,\left(n_{1}+2L+1\right) - 2 +
\frac{R}{r_{0}}\,\frac{n_{1} - n_{2}}{n + 2} - \Pi_{q}(R)\right]\,
U_{nqL}^{n_{1}}(R) = \nonumber \\ [3mm]
&=&  \sqrt{\left(n_{1}+1\right)\,\left(n-n_{1}+1\right)\,\left(n_{1}+2L+2\right)\,
\left(n-n_{1}-2L\right)}\,U_{nqL}^{n_{1}+1}(R) + \label{1.12.12} \\ [3mm]
&+& \sqrt{n_{1}\,\left(n-n_{1}+2\right)\,\left(n_{1}+2L+1\right)\,
\left(n-n_{1}-2L+1\right)}\,U_{nqL}^{n_{1}-1}(R). \nonumber
\end{eqnarray}
These equations must be solved together with the normalization conditions
\begin{eqnarray*}
\sum\limits_{\lambda = 2L}^{n}\,\left|V_{nqL}^{\lambda}(R)\right|^{2} = 1, \qquad
\sum\limits_{n_{1} = 0}^{n - 2L}\,\left|U_{nqL}^{n_{1}}(R)\right|^{2} = 1.
\end{eqnarray*}

Note that the matrices $V_{nqL}^{\lambda}(R)$ and $U_{nqL}^{n_{1}}$ generalize
the well-known Coulson-Joseph amplitudes \cite{COJO,COULSON1} from the
three-dimensional problem of two Coulomb centers.

Now note that from the definitions of coordinates (\ref{1.5.2}), (\ref{1.5.15}),
(\ref{1.8.1}), (\ref{1.9.2}), (\ref{1.10.1}), (\ref{1.12.1}) and the Hurwitz
transformation (\ref{1.2.2}) follows that
\begin{eqnarray*}
r = u^{2}, \qquad \mu = 2\rho_{1}^{2}, \qquad \nu = 2\rho_{2}^{2}, \qquad
R =d^{2}.
\end{eqnarray*}
Next, using the addition theorem for Wigner $D$-functions (\ref{1.3.9}) and the
orthonormality of the wave functions of the $8D$ oscillator and the $5D$ Coulomb problem, we obtain:
\begin{eqnarray}
\Psi_{c}^{hsp} = \frac{\pi}{4(n+2)}\frac{1}{\sqrt{r_{0}\,(2L+1)}}
\delta_{N,2n}\,\delta_{J,\lambda}\,\delta_{T,L}\,\delta_{K,L}\,\delta_{t,m}\,\delta_{t',m'}\,
\delta_{k',m'}\,\sum\limits_{k=-L}^{L}\,\Psi_{os}^{hsp},
\label{1.12.13}
\end{eqnarray}
\begin{eqnarray}
\Psi_{c}^{par} = \frac{\pi}{4(n+2)}\frac{1}{\sqrt{r_{0}\,(2L+1)}}
\delta_{N_{1},n_{1}}\,\delta_{N_{2},n_{2}}\,\delta_{T,L}\,\delta_{K,L}\,\delta_{t,m}\,\delta_{t',m'}\,
\delta_{k',m'}\,\sum\limits_{k=-L}^{L}\,\Psi_{os}^{cyl}.
\label{1.12.14}
\end{eqnarray}
Now, using the relations between quantum numbers following from (\ref{1.12.13}) and (\ref{1.12.14}),
it is easy to show that the three-term recurrence relations (\ref{1.8.13}) and (\ref{1.8.18}) turn
into relations (\ref{1.12.11}) and (\ref{1.12.12}) respectively, if the spheroidal separation constant
of the $8D$ isotropic oscillator $X_{p}(d)$ transforms into the spheroidal separation constant of
the $5D$ Coulomb problem $\Pi_{q}(R)$. Thus we have
\begin{eqnarray*}
\Psi_{c}^{spheroidal} = \frac{\pi}{4(n+2)}\frac{1}{\sqrt{r_{0}\,(2L+1)}}
\delta_{N,2n}\,\delta_{p,q}\,\delta_{T,L}\,\delta_{K,L}\,\delta_{t,m}\,\delta_{t',m'}\,
\delta_{k',m'}\,\sum\limits_{k=-L}^{L}\,\Psi_{os}^{spheroidal}.
\end{eqnarray*}
Here $\Psi_{os}$ and $\Psi_{c}$ are the bases of the $8D$ isotropic oscillator and the
$5D$ Coulomb problem, respectively.

\section{Scattering problem in a 5D Coulomb field}
\markboth{CHAPTER 1. COULOMB-OSCILLATOR ANALOGY}
{1.13. SCATTERING PROBLEM IN A 5D COULOMB FIELD}

Before moving on to the problem of scattering charged particles in a $5D$ Coulomb field,
we define the hyperspherical and parabolic bases of the five-dimensional Coulomb problem
in the continuous spectrum.

The solution of equation (\ref{1.3.6}) in hyperspherical coordinates (\ref{1.9.2}) can be written as \cite{M-8}
\begin{eqnarray}
\Psi^{hsp} = \sqrt\frac{(2L+1)}{2\,\pi^{2}}\,
R_{k \lambda}\,(r)\,Z_{\lambda L}\,(\theta)\,D_{m m'}^{L}(\alpha, \beta, \gamma),
\label{1.13.1}
\end{eqnarray}
where $D_{m m'}^{L}(\alpha, \beta, \gamma)$ is the Wigner function and function
$Z_{\lambda L}\,(\theta)$ has the form (\ref{1.9.7}). The radial wave function
$R_{k \lambda}\,(r)$ of the continuous spectrum has the form
\begin{eqnarray}
R_{k \lambda}\,(r) = C_{k \lambda}\,\frac{(2ikr)^{\lambda}}{2\lambda + 3)!}\,
e^{-ikr}\,F\left(\lambda + 2 + \frac{i}{k r_{0}};\,2\lambda + 4;\,2ikr\right).
\label{1.13.2}
\end{eqnarray}
Here $k=\sqrt{2\,\mu_{0}\,\varepsilon}/\hbar$, $C_{k \lambda}$ is the normalization
constant to be determined. For this purpose, using the following representation of the
confluent hypergeometric function \cite{LL}
\begin{eqnarray}
F\left(a;\,c;\,z\right) &=& \frac{\Gamma(c)}{\Gamma(c-a)}\,(-z)^{-a}\,
G\left(a;\,a-c+1;\,-z\right) + \nonumber \\
\label{1.13.3}
\\
&+& \frac{\Gamma(c)}{\Gamma(a)}\,e^{z}\,z^{a-c}\,
G\left(c-a;\,1-a;\,z\right),
\nonumber
\end{eqnarray}
where
\begin{eqnarray*}
G\left(a;\,c;\,z\right) = 1 + \frac{a c}{1!\,z} +
\frac{a(a+1)\,c(c+1)}{2!\,z^{2}} + ...,
\end{eqnarray*}
for the radial wave function $R_{k \lambda}\,(r)$ we obtain the expression
\begin{eqnarray}
&&R_{k \lambda}\,(r) = C_{k \lambda}\,\frac{(-i)^{\lambda}}{2k^{2}r^{2}}\,e^{-\pi/2k r_{0}}\times
\nonumber \\
\label{1.13.4}
\\
&\times& \Re e \left\{\frac{e^{-i\left[kr-\frac{\pi}{2}\,(\lambda +2)+\frac{1}{kr_{0}}\ln 2kr\right]}}
{\Gamma\left(\lambda + 2 -\frac{i}{k r_{0}}\right)}\,
G\left(\lambda + 2 - \frac{i}{k r_{0}};\, \frac{i}{k r_{0}} - \lambda - 1;\, -2ikr\right)\right\}.
\nonumber
\end{eqnarray}
If the radial wave function is normalized by the condition
\begin{eqnarray*}
\int\limits_{0}^{\infty}\,r^{4}\,R^{*}_{k' \lambda}\,(r)\,R_{k \lambda}\,(r)\,dr =
2\pi\,\delta(k-k'),
\end{eqnarray*}
then the normalization coefficient $C_{k \lambda}$ is equal to
\begin{eqnarray}
C_{k \lambda} = (-i)^{\lambda}\,4k^{2}\,e^{\pi/2k r_{0}}\,
\left|\Gamma\left(\lambda + 2 -\frac{i}{k r_{0}}\right)\right|.
\label{1.13.5}
\end{eqnarray}
Asymptotic expression for $R_{k \lambda}\,(r)$ for large $r$
(first term of expansion (\ref{1.13.4}))
\begin{eqnarray*}
R_{k \lambda}\,(r) \approx \frac{2}{r^{2}}\,
\sin\left[kr+\frac{1}{kr_{0}}\ln 2kr-\frac{\pi}{2}\,(\lambda +1) + \delta_{\lambda}\right],
\end{eqnarray*}
where
\begin{eqnarray*}
\delta_{\lambda} = arg\,\Gamma\left(\lambda + 2 -\frac{i}{k r_{0}}\right).
\end{eqnarray*}

Normalized by condition
\begin{eqnarray*}
\frac{1}{4}\,\int\,\Psi^{*}_{k'\Omega' L m m'}\,\,\Psi_{k \Omega L m m'}
\,\mu\,\nu\,(\mu + \nu)\,d\mu\,d\nu = 2\pi\,\delta(k-k')\,\delta\left(\Omega - \Omega'\right),
\end{eqnarray*}
recall that $\Omega$ is the parabolic separation constant, the parabolic
basis of the continuous spectrum of the $5D$ Coulomb problem has the form:
\begin{eqnarray}
\Psi_{k \Omega L m m'}(\mu\,\nu\,\alpha,\,\beta,],\gamma) =
\sqrt{\frac{2L+1}{2\pi^{2}}}\,C_{k \Omega L}\,\Phi_{k \Omega L}(\mu)\,
\Phi_{k, -\Omega, L}(\nu)\,D_{mm'}^{L}(\alpha, \beta, \gamma),
\label{1.13.6}
\end{eqnarray}
where
\begin{eqnarray*}
\Phi_{k \Omega L}(x) = \frac{(ikx)^{L}}{(2L+1)!}\,e^{-ikx/2}\,
F\left(L+1+\frac{i}{2k r_{0}}+i\frac{\sqrt{\mu_{0}}}{2\hbar k}\Omega;\,
2L+2;\,ikx\right), \\ [3mm]
C_{k \Omega L} = (-1)^{L}\,\sqrt{\frac{\hbar^{2} k^{3}}{2\pi \mu_{0}}}
e^{\pi/2kr_{0}}\left|\Gamma\left(L+1-\frac{i}{2k r_{0}}-i\frac{\sqrt{\mu_{0}}}{2\hbar k}\Omega\right)
\,\Gamma\left(L+1-\frac{i}{2k r_{0}}+i\frac{\sqrt{\mu_{0}}}{2\hbar k}\Omega\right)\right|.
\end{eqnarray*}
Let us note that when calculating the normalization factor $C_{k \Omega L}$
we used the representation of the confluent hypergeometric function (\ref{1.13.3}).

Now let's consider the problem of scattering a charged particle in a $5$-dimensional Coulomb field.
Since motion in a Coulomb field of $d > 3$ dimensions is a two-dimensional problem, the wave function
does not depend on the angles $\alpha, \beta$ and $\gamma$, i.e. on quantum numbers
$L, m$ and $m^{'}$. Substituting $L=0$ into equation (\ref{1.10.2}), we get
\begin{eqnarray}
\frac{1}{\mu}\,\frac{d}{d \mu}\,\left(\mu^{2}\,\frac{d\Phi_{1}}{d \mu}\right) +
\left[\frac{k^{2}}{4}\,\mu  + \frac{\sqrt{\mu_{0}}}{2\hbar}\,\Omega +
\frac{1}{r_{0}}\right]\,\Phi_{1} =0, \nonumber \\
\label{1.13.7} \\
\frac{1}{\nu}\,\frac{d}{d \nu}\,\left(\nu^{2}\,\frac{d\Phi_{2}}{d \nu}\right) +
\left[\frac{k^{2}}{4}\,\nu - \frac{\sqrt{\mu_{0}}}{2\hbar}\,\Omega + \frac{1}{r_{0}}\right]\,\Phi_{2} =0.
\nonumber
\end{eqnarray}
We must find solutions to equations (\ref{1.13.7}) such that the solution to the
Schr\"{o}dinger equation for negative $x_{0} \in (-\infty, 0)$ and large
$r \to \infty$ has the form of a plane wave
\begin{eqnarray}
\Psi_{k \Omega}\,(\mu, \nu) \approx e^{ikx_{0}} = e^{ik(\mu - \nu)/2}.
\label{1.13.8}
\end{eqnarray}
This condition can be satisfied if we assume that the parabolic separation constant is
\begin{eqnarray*}
\Omega  = - \frac{\hbar}{r_{0}\,\sqrt{\mu_{0}}}-
i\,\frac{2\hbar k}{\sqrt{\mu_{0}}}.
\end{eqnarray*}
Substituting the last relation into equation (\ref{1.13.7}), we find the following
solution to the Schr\"{o}dinger equation, which describes the scattering of a charged
particle in a $5$-dimensional Coulomb field
\begin{eqnarray}
\Psi_{k \Omega}\,(\mu, \nu) = C_{k}\,e^{ik(\mu - \nu)/2}\,
F\left(\frac{i}{k r_{0}};\,2;\,ik\nu\right),
\label{1.13.9}
\end{eqnarray}
where $C_{k}$ is the normalization constant. In order to distinguish the incident
and scattered waves in function (\ref{1.13.9}), it is necessary to consider its behavior
at large distances from the Coulomb scattering center. Using the first two terms of
representation (\ref{1.13.3}) for the confluent hypergeometric function, for large $\nu$ we obtain
\begin{eqnarray*}
F\left(\frac{i}{k r_{0}};\,2;\,ik\nu\right) &\approx& e^{-\pi/2kr_{0}}\,
\Biggl[\frac{e^{-\frac{i}{k r_{0}}\ln k\nu}}{\Gamma\left(2-i/k r_{0}\right)}\left(1 + \frac{1+ik r_{0}}
{ik^{3}r_{0}^{2}\nu}-\frac{1+k^{2} r_{0}^{2}}
{2k^{6}r_{0}^{4}\nu^{2}}\right) - \\ [3mm]
&-& \frac{i(k r_{0}+i)}{\Gamma\left(2+i/k r_{0}\right)}
\frac{e^{ik\nu}}{k^{4} r_{0}^{2}\nu^{2}}e^{\frac{i}{k r_{0}}\ln k\nu}
\Biggr].
\end{eqnarray*}
Now substituting the last relation into the wave function (\ref{1.13.9}),
choosing a constant normalization of $C_{k}$ in the form
\begin{eqnarray*}
C_{k} = e^{-\pi/2kr_{0}}\,\Gamma\left(2-i/k r_{0}\right),
\end{eqnarray*}
so that the incident plane wave has a unit amplitude, and passing to hyperspherical
coordinates according to formula (\ref{1.11.2}), we obtain
\begin{eqnarray*}
\Psi_{k}\,(\mu, \nu) &=& \left[1 + \frac{k r_{0} - i}
{2k^{3}r_{0}^{2}\,r\,\sin^{2}\theta/2}\right]\exp\left[ikx_{0} - \frac{i}{kr_{0}}
\ln\left(2kr\sin^{2}\frac{\theta}{2}\right)\right] + \\ [3mm]
&+& \frac{f(\theta)}{r^{2}}\,\exp\left[ikr + \frac{i}{kr_{0}}
\ln\left(2kr\right)\right],
\end{eqnarray*}
where $f(\theta)$ – the scattering amplitude has the form
\begin{eqnarray}
f(\theta) = \frac{1-ik r_{0}}{4k^{4}\,r_{0}^{2}\,\sin^{4}\theta/2}\,
\frac{\Gamma\left(2-i/k r_{0}\right)}{\Gamma\left(2+i/k r_{0}\right)}\,
\exp\left(\frac{2i}{k r_{0}}\,\ln\,\frac{\theta}{2}\right).
\label{1.13.10}
\end{eqnarray}
Thus, for the scattering cross section $d\sigma =|f(\theta)|^{2}\, d\Omega$ ($d\Omega$
is the differential element of the solid angle) we obtain the formula
\begin{eqnarray}
d\sigma = \frac{1+k^{2} r_{0}^{2}}{16k^{4}\,r_{0}^{2}\,\sin^{8}\theta/2}\,
d\Omega,
\label{1.13.10}
\end{eqnarray}
which generalizes Rutherford's formula in the five-dimensional case.

\newpage

\chapter{Dyon-oscillator duality}
\markboth{CHAPTER 2. DYON-OSCILLATOR DUALITY}{}

\section{$8D$ oscillator in space $\rm I \!R^{5} \bigotimes S^{3}$}
\markboth{CHAPTER 2. DYON-OSCILLATOR DUALITY}{2.1. $8D$ OSCILLATOR IN SPACE $\rm I \!R^{5} \bigotimes S^{3}$}

In the previous chapter, we showed that the eight-dimensional isotropic oscillator, when imposing the
condition ${\hat {\cal J}}_a\,\psi (\bf x) = 0$, is dual to the five-dimensional Coulomb system and the duality
transformation is the nonbijective quadratic Hurwitz transformation (\ref{1.2.2}).
Now let's find out into what system the problem of an eight-dimensional isotropic oscillator is translated into
by the bijective, i.e. a one-to-one, bilinear transformation, which is obtained by adding three angles to the Hurwitz
transformation, independent of the coordinates $x_{j}$ of the space $\rm I \!R^{5}(\bf x)$.

By adding angles \cite{HVT} to the transformations (\ref{1.2.2})
\begin{eqnarray}
\alpha_{T} &=& \frac{i}{2}\,\ln\,\frac{\left(u_{0} + iu_{1}\,\right)\,\left(u_{2} - iu_{3}\,\right)}
{\left(u_{0} - iu_{1}\,\right)\,\left(u_{2} + iu_{3}\,\right)}\,\in [0, 2\pi), \nonumber \\
\beta_{T} &=& 2\,\arctan \left(\frac{u_{0}^{2} + u_{1}^{2}},
{u_{2}^{2} + u_{3}^{2}}\right)^{1/2}\,\in [0, \pi],
\label{2.1.1}
\\
\gamma_{T} &=& \frac{i}{2}\,\ln\,\frac{\left(u_{0} - iu_{1}\,\right)\,\left(u_{2} - iu_{3}\,\right)}
{\left(u_{0} + iu_{1}\,\right)\,\left(u_{2} + iu_{3}\,\right)}\,\in [0, 4\pi)
\nonumber
\end{eqnarray}
we obtain a transformation that takes $\rm I \!R^{8}$ into the direct product
$\rm I \!R^{5} \bigotimes S^{3}$ of the space $\rm I \!R^{5}(\bf x)$ and the three-dimensional
sphere $S^{3}\,(\alpha_{T}, \beta_{T}, \gamma_{T})$.

Now using the definition of the Laplace operator
\begin{eqnarray*}
\Delta = \frac{1}{\sqrt{g}}\,\frac{\partial}{\partial x_{\mu}}
\left(\sqrt{g}\,g^{\mu \nu}\,\frac{\partial}{\partial x_{\nu}}\right),
\end{eqnarray*}
where $\mu, \nu = 0,1...,7$, $g^{\mu \nu}$ are contravariant components of the metric tensor,
$g$ is the determinant of the metric tensor $g_{\mu \nu}$g, the eight-dimensional operator
in coordinates $x_{j} (j = 0,1...,4)$ and $\alpha_{T}, \beta_{T}, \gamma_{T}$ will be written as
\begin{eqnarray}
\Delta_{8} = 4\,r\,\left[\Delta_{5} - 2i\,A_{j}^{a(+)}\,\frac{\partial}{\partial x_{j}}
- \frac{2}{r(r+x_{0})}{\hat T}^{2}\right].
\label{2.1.2}
\end{eqnarray}
Here $r = \left(x_{j} x_{j}\right)^{1/2} = u^{2}$, and the operators ${\hat T}_{a} (a=1, 2, 3)$
are generators of the group $SU(2)$ and their explicit form is given in formula (\ref{1.4.3}).

The triplet of five-dimensional vectors ${\bf A}^{a(+)}$ have the form
\cite{M-9,M-10}:
\begin{eqnarray}
{\bf A}^{1(+)} &=& \frac{2}{r(r+x_{0})}\,\left(0, x_{4}, x_{3}, -x_{2}, -x_{1}\right),
\nonumber \\
{\bf A}^{2(+)} &=& \frac{2}{r(r+x_{0})}\,\left(0, -x_{3}, x_{4}, x_{1}, -x_{2}\right),
\label{2.1.3}
\\
{\bf A}^{3(+)} &=& \frac{2}{r(r+x_{0})}\,\left(0, x_{2}, -x_{1}, x_{4}, -x_{3}\right).
\nonumber
\end{eqnarray}
It is easy to notice that vectors ${\bf A}^{a(+)}$ are orthogonal to each other
\begin{eqnarray}
A_{j}^{a(+)}\,A_{j}^{b(+)}\, = \frac{1}{r^{2}}\,\frac{r - x_{0}}{r+x_{0}}\,\delta_{a b},
\label{2.1.4}
\end{eqnarray}
the same applies to the vector $r=\left(x_{0}, x_{1}, x_{2}, x_{3}, x_{4}\right)$.

Now, using formula (\ref{2.1.2}) and the orthogonality condition (\ref{2.1.4}), the
Schr\"{o}dinger equation (\ref{1.3.1}) for an eight-dimensional isotropic oscillator
can be written in the form:
\begin{eqnarray}
\frac{1}{2\mu_{0}}\,\left(-i\hbar\,\frac{\partial}{\partial x_{j}} -
\hbar\,A_{j}^{a(+)}\,{\hat T}_{a}\right)^{2}\psi +
\frac{\hbar^{2}}{2\mu_{0}r^{2}}\,{\hat T}^{2}\psi - \frac{e^{2}}{r}\,\psi
= \varepsilon \psi,
\label{2.1.5}
\end{eqnarray}
where $\varepsilon = -\mu_{0} \omega^{2}/8$, and $e^{2} = E/4$.

Thus, we have obtained an equation identical to the Pauli equation and therefore we can give
the meaning of a tripet of five-dimensional vectors ${\bf A}^{a(+)}$ to the meaning of vector
potentials with the singularity axis directed along the negative part of the $x_{0}$ axis.

For further calculations, it is convenient to write the vector potentials ${\bf A}^{a(+)}$
in the form \cite{M-13}:
\begin{eqnarray}
A_{i}^{a(+)} = \frac{2i}{r (r+x_{0})}\,\tau_{i j}^{a}\,x_{j}.
\label{2.1.6}
\end{eqnarray}
Here $\tau_{i j}^{a}$ is a $5 \times 5$ matrix and has the form
\begin{eqnarray*}
\tau^{1} = \frac{1}{2}\,
\left(
\begin{array}{ccc}
0&0&0\\
0&0&-i\sigma^{1}\\
0&i\sigma^{1}&0\\
\end{array}
\right), \qquad
\tau^{2} = \frac{1}{2}\,
\left(
\begin{array}{ccc}
0&0&0\\
0&0&i\sigma^{3}\\
0&i-\sigma^{3}&0\\
\end{array}
\right), \qquad
\tau^{3} = \frac{1}{2}\,
\left(
\begin{array}{ccc}
0&0&0\\
0&\sigma^{2}&0\\
0&0&\sigma^{1}\\
\end{array}
\right),
\end{eqnarray*}
where $sigma^{a}$ of the Pauli matrix:
\begin{eqnarray*}
\sigma^{1} =
\left(
\begin{array}{cc}
0&1\\
1&0\\
\end{array}
\right), \qquad
\sigma^{2} =
\left(
\begin{array}{cc}
0&-i\\
i&0\\
\end{array}
\right), \qquad
\sigma^{2} =
\left(
\begin{array}{cc}
1&0\\
0&-1\\
\end{array}
\right).
\end{eqnarray*}
For $\tau^{a}$ matrices the following relations hold:
\begin{eqnarray}
\left[\tau^{a}, \tau^{b}\right] = i \epsilon_{a b c} \tau^{c}, \qquad
\left\{\tau^{a}, \tau^{b}\right\} = \frac{1}{2}\,\delta_{a b}, \nonumber \\
\label{2.1.7}
\\
4 \tau^{a}_{i j} \tau^{b}_{jk} = \delta_{a b} \left(\delta_{i k} -
\delta_{i 0} \delta_{k 0}\right) + 2i \epsilon_{a b c} \tau^{c}_{i k},
\nonumber
\end{eqnarray}
\begin{eqnarray}
\epsilon_{a b c}\tau^{b}_{i j} \tau^{c}_{k m} &=& \frac{i}{2} \Bigl[
\left(\delta_{i 0} \delta_{k 0} - \delta_{i k}\right)\tau^{a}_{j m}  -
\left(\delta_{i 0} \delta_{m 0} - \delta_{i m}\right)\tau^{a}_{j k} +
\nonumber \\
\label{2.1.8}
\\
&+& \left(\delta_{j 0} \delta_{m 0} - \delta_{j m}\right)\tau^{a}_{i k}-
\left(\delta_{j 0} \delta_{k 0} - \delta_{j k}\right)\tau^{a}_{i m}\Bigr].
\nonumber
\end{eqnarray}
Here the brackets $[,]$ and $\{,\}$ denote the commutator and anticommutator, respectively,
and $\epsilon_{a b c}$ is the antisymmetric unit tensor of the third rank.

Now we present the proof of formula (\ref{2.1.8}).

Since $\epsilon_{a b c}\tau^{b}_{i j} \tau^{c}_{k m}$ is a pseudovector on $S^{3}$ and a
tensor on $\rm I\!R^{5}$, we can construct the following general expression:
\begin{eqnarray}
\epsilon_{a b c}\tau^{b}_{i j} \tau^{c}_{k m} &=& A \left(\delta_{i k} - \delta_{i 0} \delta_{k 0}\right)\tau^{a}_{j m}  +
B \left(\delta_{i m} - \delta_{i 0} \delta_{m 0}\right)\tau^{a}_{j k} +
C \left(\delta_{j k} - \delta_{j 0} \delta_{k 0}\right)\tau^{a}_{i m} + \nonumber \\
\label{2.1.9}
\\
&+& D \left(\delta_{j m} - \delta_{j 0} \delta_{m 0}\right)\tau^{a}_{i k} +
E \left(\delta_{i j}-\delta_{i 0} \delta_{j 0}\right)\tau^{a}_{k m} +
F \left(\delta_{k m}-\delta_{k 0} \delta_{m 0}\right)\tau^{a}_{i j}.  \nonumber
\end{eqnarray}
Here the left side of relation (\ref{2.1.9}) is symmetric with respect to the indices
$i, j$ and $k, m$, respectively. Therefore $E=F=0, C=B, D=A$ and
\begin{eqnarray*}
\epsilon_{a b c}\tau^{b}_{i j} \tau^{c}_{k m} &=&
A \left[\left(\delta_{i k} - \delta_{i 0} \delta_{k 0}\right)\tau^{a}_{j m}  +
\left(\delta_{j m} - \delta_{j 0} \delta_{m 0}\right)\tau^{a}_{i k}\right] + \\ [4mm]
&+& B \left[\left(\delta_{i m} - \delta_{i 0} \delta_{m 0}\right)\tau^{a}_{j k} +
\left(\delta_{j k} - \delta_{j 0} \delta_{k 0}\right)\tau^{a}_{i m}\right].
\end{eqnarray*}
Next, $B=-A$, because after $i \leftrightarrow j$ the left side changes sign,
and the right side is  changed places by brackets nd therefore:
\begin{eqnarray*}
\epsilon_{a b c}\tau^{b}_{i j} \tau^{c}_{k m} &=&
A \Bigl[\left(\delta_{i k} - \delta_{i 0} \delta_{k 0}\right)\tau^{a}_{j m}  +
\left(\delta_{j m} - \delta_{j 0} \delta_{m 0}\right)\tau^{a}_{i k} - \\ [4mm]
&-& \left(\delta_{i m} - \delta_{i 0} \delta_{m 0}\right)\tau^{a}_{j k} -
\left(\delta_{j k} - \delta_{j 0} \delta_{k 0}\right)\tau^{a}_{i m}\Bigr].
\end{eqnarray*}
Summing over the indices $j$ and $k$, and using formula (\ref{2.1.7}), we get that $A=-i/2$.

Now consider the field tensor
\begin{eqnarray*}
F_{i k}^{a(+)} =
\frac{\partial A_{k}^{a(+)}}{\partial x_{i}} -
\frac{\partial A_{i}^{a(+)}}{\partial x_{k}} +
\epsilon_{a b c}A_{i}^{b(+)} A_{k}^{c(+)}.
\end{eqnarray*}
Using formulas (\ref{2.1.6}) and (\ref{2.1.8}) for the field tensor we obtain the expression:
\begin{eqnarray}
F_{i k}^{a(+)} = \frac{1}{r^{2}}\left[\left(x_{k} + r\delta_{k 0}\right)\,A_{i}^{a(+)} -
\left(x_{i} + r\delta_{i 0}\right)\,A_{k}^{a(+)} - 2i\,\tau^{a}_{i k}\right].
\label{2.1.10}
\end{eqnarray}
Now, taking into account the orthogonality condition $x_{i} A_{i}^{a(+)}=0$ and the explicit form
of the field tensor (\ref{2.1.10}), it is easy to show that $x_{i} F_{i k}^{a(+)}=0$,
i.e. both the vector potential and the field tensor are orthogonal to the radius vector $\bf r$.

Since $A_{0}^{a(+)}=0$, we have
\begin{eqnarray*}
F_{0 \mu}^{a(+)} = - \frac{r + x_{0}}{r^{2}}\,A_{\mu}^{a(+)} =
- \frac{2i}{r^{3}}\,\tau^{a}_{\mu \nu}\,x_{\nu}.
\end{eqnarray*}
Here $\mu, \nu = 1,2,3,4$. In other cases, i.e. when not one of the field tensor indices is
equal to zero, then from (\ref{2.1.6}) and (\ref{2.1.10}) we have
\begin{eqnarray*}
F_{\mu \nu}^{a(+)} = \frac{1}{r^{2}}\,\left(x_{\nu}\,A_{\mu}^{a(+)} -
x_{\mu}\,A_{\nu}^{a(+)} - 2i \tau^{a}_{\mu \nu}\right).
\end{eqnarray*}
Finally, we note that from formulas (\ref{2.1.4}), (\ref{2.1.6}), (\ref{2.1.7})
and (\ref{2.1.10}) it follows that
\begin{eqnarray}
F_{i j}^{a(+)} F^{b(+)jk} = F_{i j}^{a(+)} F^{b(+)}_{jk} =
\frac{1}{r^{6}}\,\left(x_{i} x_{k} - r^{2}\,\delta_{i 0}\right)\,\delta_{a b}
+  \frac{1}{r^{2}}\,\epsilon_{a b c}\,F_{i k}^{c(+)},
\label{2.1.11}
\end{eqnarray}
or
\begin{eqnarray}
F_{i j}^{a(+)} F^{b(+)jk} = F_{i j}^{a(+)} F^{b(+)}_{jk} =
\frac{4}{r^{4}}\,\delta_{a b}.
\label{2.1.12}
\end{eqnarray}
Let us give the following useful commutation relation
\begin{eqnarray}
\left[\pi_{i}, F_{i k}^{a(+)}\right] = 0,
\label{2.1.13}
\end{eqnarray}
where
\begin{eqnarray*}
\pi_{i} = -i\hbar\,\frac{\partial}{\partial x_{i}} -
\hbar\,A_{i}^{a(+)}{\hat T}_{a}.
\end{eqnarray*}
We need formulas (\ref{2.1.10}) – (\ref{2.1.13}) to calculate commutation relations
when finding the hidden symmetry group of this system.

From potentials $A_{i}^{a(+)}$ using gauge transformation
\begin{eqnarray*}
B_{j}^{(+)} = S_{+}\,A_{j}^{(+)}\, S_{+}^{-1} +
i\, S_{+}\,\frac{\partial}{\partial x_{j}}S_{+}^{-1},
\end{eqnarray*}
where $A_{j}^{(+)} = A_{j}^{a(+)}{\hat T}_{a}$,
$B_{j}^{(+)} = B_{j}^{a(+)}{\hat T}_{a}$, and
$S_{+} = e^{-i\gamma T_{3}}\,e^{-i\beta T_{2}}\,e^{-i\alpha T_{3}}$,
it is possible to construct potentials $B_{j}^{(+)}$, the singularity of which is directed
along the positive part of the $x_{0}$ axis and they have the form
\begin{eqnarray}
{\bf B}^{1(+)} &=& \frac{2}{r(r-x_{0})}\,\left(0, -x_{4}, x_{3}, -x_{2}, x_{1}\right), \nonumber \\
{\bf B}^{2(+)} &=& \frac{2}{r(r-x_{0})}\,\left(0, -x_{3}, -x_{4}, x_{1}, x_{2}\right),
\label{2.1.14} \\
{\bf B}^{3(+)} &=& \frac{2}{r(r-x_{0})}\,\left(0, x_{2}, -x_{1}, -x_{4}, x_{3}\right).
\nonumber
\end{eqnarray}
Euler angles $\alpha, \beta$ and $\gamma$ according to (\ref{1.9.2}) are defined as follows:
\begin{eqnarray*}
\alpha &=& \frac{i}{2}\,\ln\,\frac{\left(x_{2} - ix_{1}\,\right)\,\left(x_{4} - ix_{3}\,\right)}
{\left(x_{2} + ix_{1}\,\right)\,\left(x_{4} + ix_{3}\,\right)}\,\in [0, 2\pi),  \\ [5mm]
\beta &=& 2\,\arctan \left(\frac{x_{1}^{2} + x_{2}^{2}}{x_{3}^{2} + x_{4}^{2}}\right)^{1/2}\,\in [0, \pi], \\ [5mm]
\gamma &=& \frac{i}{2}\,\ln\,\frac{\left(x_{2} + ix_{1}\,\right)\,\left(x_{4} - ix_{3}\,\right)}
{\left(x_{2} - ix_{1}\,\right)\,\left(x_{4} + ix_{3}\,\right)}\,\in [0, 4\pi).
\end{eqnarray*}
The field tensor for potential $B_{j}^{a(+)}$ will be denoted by $\tilde{F}_{i j}$.

\section{Topological charge}
\markboth{CHAPTER 2. DYON-OSCILLATOR DUALITY}{2.2. TOPOLOGICAL CHARGE}

Now, knowing the explicit form of the field tensor, we need to find out which physical system is
described by equation (\ref{2.1.5}). To do this, first of all, you need to calculate the topological
charge, which is one of the sources of the field described by the tensor $F_{i j}$.
It is convenient to calculate the amount of topological charge in five-dimensional hyperspherical
coordinates (\ref{1.9.2}). Before this, for completeness, we present explicit expressions for the
Cartesian components of the tensors $F_{i j}^{a(+)}$ and $\tilde{F}_{i j}^{a(+)}$:
\begin{eqnarray*}
F_{0 1}^{1(+)} &=& -\frac{x_{4}}{r^{3}}, \qquad  F_{0 2}^{1(+)} = -\frac{x_{3}}{r^{3}}, \qquad
F_{0 3}^{1(+)} = \frac{x_{2}}{r^{3}}, \qquad F_{0 4}^{1(+)} = \frac{x_{1}}{r^{3}},  \\ [5mm]
F_{1 2}^{1(+)} &=& \frac{x_{2}x_{4} - x_{1}x_{2}}{r^{3} (r + x_{0})}, \qquad
F_{1 3}^{1(+)} = \frac{x_{1}x_{2} + x_{3}x_{4}}{r^{3} (r + x_{0})}, \qquad
F_{1 4}^{1(+)} =  \frac{1}{r^{2}}\,\left[\frac{x_{1}^{2} + x_{4}^{2}}{r (r + x_{0})} - 1\right], \\ [5mm]
F_{2 3}^{1(+)} &=&  \frac{1}{r^{2}}\,\left[\frac{x_{2}^{2} + x_{3}^{2}}{r (r + x_{0})} - 1\right], \qquad
F_{2 4}^{1(+)} = \frac{x_{1}x_{2} + x_{3}x_{4}}{r^{3} (r + x_{0})}, \qquad
F_{3 4}^{1(+)} = \frac{x_{1}x_{3} - x_{2}x_{4}}{r^{3} (r + x_{0})};
\end{eqnarray*}
\vspace{7mm}
\begin{eqnarray*}
F_{0 1}^{2(+)} &=& \frac{x_{3}}{r^{3}}, \qquad  F_{0 2}^{2(+)} = -\frac{x_{4}}{r^{3}}, \qquad
F_{0 3}^{2(+)} = -\frac{x_{1}}{r^{3}}, \qquad F_{0 4}^{2(+)} = \frac{x_{2}}{r^{3}},  \\ [5mm]
F_{1 2}^{2(+)} &=& -\frac{x_{1}x_{4} + x_{2}x_{3}}{r^{3} (r + x_{0})}, \qquad
F_{1 3}^{2(+)} = -\frac{1}{r^{2}}\,\left[\frac{x_{1}^{2} + x_{3}^{2}}{r (r + x_{0})} - 1\right],  \qquad
F_{1 4}^{2(+)} =  \frac{x_{1}x_{2} - x_{3}x_{4}}{r^{3} (r + x_{0})}, \\ [5mm]
F_{2 3}^{2(+)} &=& \frac{x_{3}x_{4} - x_{1}x_{2}}{r^{3} (r + x_{0})},  \qquad
F_{2 4}^{2(+)} = \frac{1}{r^{2}}\,\left[\frac{x_{2}^{2} + x_{4}^{2}}{r (r + x_{0})} - 1\right],  \qquad
F_{3 4}^{2(+)} = \frac{x_{1}x_{4} + x_{2}x_{3}}{r^{3} (r + x_{0})};
\end{eqnarray*}
\vspace{7mm}
\begin{eqnarray*}
F_{0 1}^{3(+)} &=& -\frac{x_{2}}{r^{3}}, \qquad  F_{0 2}^{3(+)} = \frac{x_{1}}{r^{3}}, \qquad
F_{0 3}^{3(+)} = -\frac{x_{4}}{r^{3}}, \qquad F_{0 4}^{3(+)} = \frac{x_{3}}{r^{3}},  \\ [5mm]
F_{1 2}^{3(+)} &=& \frac{1}{r^{2}}\,\left[\frac{x_{1}^{2} + x_{2}^{2}}{r (r + x_{0})} - 1\right],  \qquad
F_{1 3}^{3(+)} =  \frac{x_{2}x_{3} - x_{1}x_{4}}{r^{3} (r + x_{0})}, \qquad
F_{1 4}^{3(+)} =  \frac{x_{1}x_{3} + x_{2}x_{4}}{r^{3} (r + x_{0})}, \\ [5mm]
F_{2 3}^{3(+)} &=& -\frac{x_{1}x_{3} + x_{2}x_{4}}{r^{3} (r + x_{0})},  \qquad
F_{2 4}^{3(+)} = \frac{x_{2}x_{3} - x_{1}x_{4}}{r^{3} (r + x_{0})},  \qquad
F_{3 4}^{3(+)} = \frac{1}{r^{2}}\,\left[\frac{x_{3}^{2} + x_{4}^{2}}{r (r + x_{0})} - 1\right];
\end{eqnarray*}
\vspace{12mm}
\begin{eqnarray*}
\tilde{F}_{0 1}^{1(+)} &=& \frac{x_{4}}{r^{3}}, \qquad  \tilde{F}_{0 2}^{1(+)} = -\frac{x_{3}}{r^{3}}, \qquad
\tilde{F}_{0 3}^{1(+)} = \frac{x_{2}}{r^{3}}, \qquad \tilde{F}_{0 4}^{1(+)} = -\frac{x_{1}}{r^{3}},  \\ [5mm]
\tilde{F}_{1 2}^{1(+)} &=& -\frac{x_{1}x_{3} + x_{2}x_{4}}{r^{3} (r - x_{0})},    \qquad
\tilde{F}_{1 3}^{1(+)} =  \frac{x_{1}x_{2} - x_{3}x_{4}}{r^{3} (r - x_{0})}, \qquad
\tilde{F}_{1 4}^{1(+)} = \frac{1}{r^{2}}\,\left[1 - \frac{x_{1}^{2} + x_{4}^{2}}{r (r - x_{0})}\right],  \\ [5mm]
\tilde{F}_{2 3}^{1(+)} &=& \frac{1}{r^{2}}\,\left[\frac{x_{2}^{2} + x_{3}^{2}}{r (r - x_{0})} + 1\right],  \qquad
\tilde{F}_{2 4}^{1(+)} = -\frac{x_{1}x_{2} - x_{3}x_{4}}{r^{3} (r - x_{0})},  \qquad
\tilde{F}_{3 4}^{1(+)} = -\frac{x_{1}x_{3} + x_{2}x_{4}}{r^{3} (r - x_{0})};
\end{eqnarray*}
\vspace{7mm}
\begin{eqnarray*}
\tilde{F}_{0 1}^{2(+)} &=& \frac{x_{3}}{r^{3}}, \qquad  \tilde{F}_{0 2}^{2(+)} = \frac{x_{4}}{r^{3}}, \qquad
\tilde{F}_{0 3}^{2(+)} = -\frac{x_{1}}{r^{3}}, \qquad \tilde{F}_{0 4}^{2(+)} = -\frac{x_{2}}{r^{3}},  \\ [5mm]
\tilde{F}_{1 2}^{2(+)} &=& \frac{x_{1}x_{4} - x_{2}x_{3}}{r^{3} (r - x_{0})},    \qquad
\tilde{F}_{1 3}^{2(+)} =  \frac{1}{r^{2}}\,\left[1 - \frac{x_{1}^{2} + x_{3}^{2}}{r (r - x_{0})}\right],  \qquad
\tilde{F}_{1 4}^{2(+)} = -\frac{x_{1}x_{2} + x_{3}x_{4}}{r^{3} (r - x_{0})},  \\ [5mm]
\tilde{F}_{2 3}^{2(+)} &=& -\frac{x_{1}x_{2} + x_{3}x_{4}}{r^{3} (r - x_{0})},   \qquad
\tilde{F}_{2 4}^{2(+)} =  \frac{1}{r^{2}}\,\left[1 - \frac{x_{2}^{2} + x_{4}^{2}}{r (r - x_{0})}\right], \qquad
\tilde{F}_{3 4}^{2(+)} = \frac{x_{1}x_{4} - x_{2}x_{3}}{r^{3} (r - x_{0})};
\end{eqnarray*}
\vspace{7mm}
\begin{eqnarray*}
\tilde{F}_{0 1}^{3(+)} &=& -\frac{x_{2}}{r^{3}}, \qquad  \tilde{F}_{0 2}^{3(+)} = \frac{x_{1}}{r^{3}}, \qquad
\tilde{F}_{0 3}^{3(+)} = \frac{x_{4}}{r^{3}}, \qquad \tilde{F}_{0 4}^{3(+)} = -\frac{x_{3}}{r^{3}},  \\ [5mm]
\tilde{F}_{1 2}^{3(+)} &=& \frac{1}{r^{2}}\,\left[\frac{x_{1}^{2} + x_{2}^{2}}{r (r - x_{0})} + 1\right],      \qquad
\tilde{F}_{1 3}^{3(+)} = \frac{x_{1}x_{4} + x_{2}x_{3}}{r^{3} (r - x_{0})},   \qquad
\tilde{F}_{1 4}^{3(+)} = \frac{x_{2}x_{4} - x_{1}x_{3}}{r^{3} (r - x_{0})},  \\ [5mm]
\tilde{F}_{2 3}^{3(+)} &=& \frac{x_{2}x_{4} - x_{1}x_{3}}{r^{3} (r - x_{0})},   \qquad
\tilde{F}_{2 4}^{3(+)} =  -\frac{x_{1}x_{4} + x_{2}x_{3}}{r^{3} (r - x_{0})}, \qquad
\tilde{F}_{3 4}^{3(+)} = \frac{1}{r^{2}}\,\left[1 - \frac{x_{3}^{2} + x_{4}^{2}}{r (r - x_{0})}\right].
\end{eqnarray*}

It is known that the components of a second-rank tensor in different coordinate systems are
related to each other by the formula
\begin{eqnarray*}
\bar{f}_{i k} = \frac{\partial x_{m}}{\partial \bar{x}_{i}}\,
\frac{\partial x_{n}}{\partial \bar{x}_{k}}\,f_{m n}.
\end{eqnarray*}
In our case $\bar{x}_{0}=r, \bar{x}_{1}=\theta, \bar{x}_{2}=\beta, \bar{x}_{3}=\alpha, \bar{x}_{4}=\gamma$.
Direct calculations show that all $F_{i k}^{a(+)} \equiv 0$, where
$k=r, \theta, \beta, \alpha, \gamma$, and non-zero components have the form
\begin{eqnarray*}
F_{\theta \beta}^{1(+)} = \frac{1}{2}\,\sin\theta \sin\alpha, \qquad
F_{\theta \alpha}^{1(+)} = 0, \qquad
F_{\theta \gamma}^{1(+)} = -\frac{1}{2}\,\sin\theta\,\sin\beta\,\cos\alpha,   \\ [3mm]
F_{\beta \alpha}^{1(+)} = -\frac{1}{4}\,\sin^{2}\theta\,\cos\alpha, \qquad
F_{\beta \gamma}^{1(+)} = -\frac{1}{4}\sin^{2}\theta\,\cos\beta\,\cos\alpha,  \qquad
F_{\alpha \gamma}^{1(+)} = \frac{1}{4}\sin^{2}\theta\,\sin\beta\,\sin\alpha;
\end{eqnarray*}
\vspace{1mm}
\begin{eqnarray*}
F_{\theta \beta}^{2(+)} = \frac{1}{2}\,\sin\theta \cos\alpha, \qquad
F_{\theta \alpha}^{2(+)} = 0, \qquad
F_{\theta \gamma}^{2(+)} = \frac{1}{2}\,\sin\theta\,\sin\beta\,\sin\alpha,   \\ [3mm]
F_{\beta \alpha}^{2(+)} = \frac{1}{4}\,\sin^{2}\theta\,\sin\alpha, \qquad
F_{\beta \gamma}^{2(+)} = \frac{1}{4}\sin^{2}\theta\,\cos\beta\,\sin\alpha,  \qquad
F_{\alpha \gamma}^{2(+)} = \frac{1}{4}\sin^{2}\theta\,\sin\beta\,\cos\alpha;
\end{eqnarray*}
\vspace{1mm}
\begin{eqnarray*}
F_{\theta \beta}^{2(+)} = 0, \qquad
F_{\theta \alpha}^{2(+)} = \frac{1}{2}\,\sin\theta, \qquad
F_{\theta \gamma}^{2(+)} = \frac{1}{2}\,\sin\theta\,\cos\beta,   \\ [3mm]
F_{\beta \alpha}^{2(+)} = 0, \qquad
F_{\beta \gamma}^{2(+)} = -\frac{1}{4}\sin^{2}\theta\,\sin\beta,  \qquad
F_{\alpha \gamma}^{2(+)} = 0.
\end{eqnarray*}
\vspace{4mm}
\begin{eqnarray*}
\tilde{F}_{\theta \beta}^{1(+)} = -\frac{1}{2}\,\sin\theta\,\sin\gamma, \qquad
\tilde{F}_{\theta \alpha}^{1(+)} = \frac{1}{2}\,\sin\theta\,\sin\beta\,\cos\gamma, \qquad
\tilde{F}_{\theta \gamma}^{1(+)} = 0,   \\ [3mm]
\tilde{F}_{\beta \alpha}^{1(+)} = -\frac{1}{4}\,\sin^{2}\theta\,\cos\beta\,\cos\gamma, \qquad
\tilde{F}_{\beta \gamma}^{1(+)} = -\frac{1}{4}\sin^{2}\theta\,\cos\gamma,  \qquad
\tilde{F}_{\alpha \gamma}^{1(+)} = -\frac{1}{4}\sin^{2}\theta\,\sin\beta\,\sin\gamma;
\end{eqnarray*}
\vspace{1mm}
\begin{eqnarray*}
\tilde{F}_{\theta \beta}^{2(+)} = \frac{1}{2}\,\sin\theta\,\cos\gamma, \qquad
\tilde{F}_{\theta \alpha}^{2(+)} = \frac{1}{2}\,\sin\theta\,\sin\beta\,\sin\gamma, \qquad
\tilde{F}_{\theta \gamma}^{2(+)} = 0,   \\ [3mm]
\tilde{F}_{\beta \alpha}^{2(+)} = -\frac{1}{4}\,\sin^{2}\theta\,\cos\beta\,\sin\gamma, \qquad
\tilde{F}_{\beta \gamma}^{2(+)} = -\frac{1}{4}\sin^{2}\theta\,\sin\gamma,  \qquad
\tilde{F}_{\alpha \gamma}^{2(+)} = \frac{1}{4}\sin^{2}\theta\,\sin\beta\,\cos\gamma;
\end{eqnarray*}
\vspace{1mm}
\begin{eqnarray*}
\tilde{F}_{\theta \beta}^{3(+)} = 0, \qquad
\tilde{F}_{\theta \alpha}^{3(+)} = \frac{1}{2}\,\sin\theta\,\cos\beta, \qquad
\tilde{F}_{\theta \gamma}^{3(+)} = \frac{1}{2}\,\sin\theta,   \\ [3mm]
\tilde{F}_{\beta \alpha}^{3(+)} = \frac{1}{4}\,\sin^{2}\theta\,\sin\beta, \qquad
\tilde{F}_{\beta \gamma}^{3(+)} = 0  \qquad
\tilde{F}_{\alpha \gamma}^{3(+)} = 0.
\end{eqnarray*}

Now using the definition of the dual tensor
\begin{eqnarray*}
^{*}f^{\mu \nu} = \frac{\sqrt{g}}{2}\,\epsilon^{\mu \nu \lambda \sigma}\,
f_{\lambda \sigma},
\end{eqnarray*}
where $\epsilon^{1 2 3 4} = 1$, and with exact expressions for the hyperspherical components
of the field tensor $F^{a(+)\mu \nu}$ we obtain that the tensor $F^{a(+)\mu \nu}$ is self-dual, i.e.
\begin{eqnarray}
^{*}F^{a(+)\mu \nu} = F^{a(+)\mu \nu}.
\label{2.2.1}
\end{eqnarray}
Further, using the definition of topological charge
\begin{eqnarray*}
q = \frac{1}{32\,\pi^{2}}\,\oint\,^{*}F^{a(+)\mu \nu}\,
F^{a(+)}_{\mu \nu}\,dS,
\end{eqnarray*}
where
\begin{eqnarray*}
dS = \frac{r^{4}}{8}\,\sin^{3}\theta\,d\theta\,d\beta\,d\alpha\,d\gamma,
\end{eqnarray*}
taking into account the self-duality equation (\ref{2.2.1}) and the orthogonality
condition (\ref{2.1.12}) we obtain that in our case the topological charge $q=+1$ \cite{M-11}.

Thus, equation (\ref{2.1.5}) describes a system consisting of a charged particle and an
$SU(2)$ Yang monopole \cite{YANG-1} with topological charge $q=+1$, which we will further
call the SU(2) Yang-Coulomb monopole.

If now in the first equation of relation (\ref{1.2.2}) we replace
$x_{0} \to - x_{0}$, i.e.
\begin{eqnarray*}
x_{0} = u_{4}^{2} + u_{5}^{2} + u_{6}^{2} + u_{7}^{2} -
u_{0}^{2} -u_{1}^{2} - u_{2}^{2} - u_{3}^{2}
\end{eqnarray*}
then from the potentials $B_{j}^{a(+)}$ and $A_{j}^{a(+)}$ we obtain the potentials
$A_{j}^{a(-)}$ and $B_{j}^{a(-)}$, respectively. Naturally, the latter are also interconnected
by a gauge transformation and the transition function $S_{-}$ has the form:
\begin{eqnarray*}
S_{-} = e^{i\alpha T_{3}}\,e^{i\beta T_{2}}\,e^{i\gamma T_{3}},
\end{eqnarray*}
i.e. $S_{-} = S_{+}^{-1}$.

The field tensors $F_{\mu \nu}^{a(-)}$ and $\tilde{F}_{\mu \nu}^{a(-)}$ and the corresponding
potentials $A_{j}^{a(-)}$ and $B_{j}^{a(-)}$ are anti-self-dual, i.e.
\begin{eqnarray*}
^{*}F^{a(-)\mu \nu} = - F^{a(-)\mu \nu}, \qquad
^{*}\tilde{F}^{a(-)\mu \nu} = - \tilde{F}^{a(-)\mu \nu}.
\end{eqnarray*}
and therefore they describe a five-dimensional $SU(2)$ Yang-Coulomb monopole (YCM)with topological
charge $q = -1$ and singularities directed along the negative and positive values of the
$x_{0}$ coordinate, respectively.

The components of the tensors $F_{\mu \nu}^{a(-)}$ and $\tilde{F}_{\mu \nu}^{a(-)}$ can be
obtained from the components $\tilde{F}_{\mu \nu}^{a(+)}$ and  $F_{\mu \nu}^{a(+)}$
respectively using the substitution $\theta \to \theta + \pi$.

\section{$LT$-Interaction}
\markboth{CHAPTER 2. DYON-OSCILLATOR DUALITY}{2.3. $LT$-INTERACTION}

Now let's proceed to solving the Schr\"{o}dinger equation (\ref{2.1.5}), which describes
$SU(2)$ YCM \cite{M-12}. Using the orthogonality condition (\ref{2.1.4}) for vectors
$A_{j}^{a(+)}$, we can transform equation (\ref{2.1.5}) into the equation
\begin{eqnarray}
\left[\Delta_{5} -2i\,A_{j}^{a(+)}\,\hat{T}_{a}\,\frac{\partial}{\partial x_{j}} -
\frac{2}{r(r + x_{0})}\hat{\bf{T}}^{2}\right]\,\psi + \frac{2 \mu_{0}}{\hbar^{2}}\,
\left(\varepsilon + \frac{e^{2}}{r}\right)\,\psi = 0.
\label{2.3.1}
\end{eqnarray}
Further, taking into account that
\begin{eqnarray*}
i\,A_{j}^{a(+)}\,\frac{\partial}{\partial x_{j}} =
\frac{2}{r(r+x_{0})}\,\hat{L}_{a},
\end{eqnarray*}
where
\begin{eqnarray*}
\hat{L}_{1} &=& \frac{i}{2}\,\left(x_{4}\,\frac{\partial}{\partial x_{1}} +
x_{3}\,\frac{\partial}{\partial x_{2}} - x_{2}\,\frac{\partial}{\partial x_{3}}
-x_{1}\,\frac{\partial}{\partial x_{4}}\right), \\ [4mm]
\hat{L}_{2} &=& \frac{i}{2}\,\left(-x_{3}\,\frac{\partial}{\partial x_{1}} +
x_{4}\,\frac{\partial}{\partial x_{2}} + x_{1}\,\frac{\partial}{\partial x_{3}}
-x_{2}\,\frac{\partial}{\partial x_{4}}\right), \\ [4mm]
\hat{L}_{3} &=& \frac{i}{2}\,\left(x_{2}\,\frac{\partial}{\partial x_{1}} -
x_{1}\,\frac{\partial}{\partial x_{2}} + x_{4}\,\frac{\partial}{\partial x_{3}}
-x_{3}\,\frac{\partial}{\partial x_{4}}\right)
\end{eqnarray*}
we write equation (\ref{2.3.1}) in the form
\begin{eqnarray}
\left[\Delta_{5} -\frac{4}{r(r + x_{0})}\hat{\bf{L}}\hat{\bf{T}} -
\frac{2}{r(r + x_{0})}\hat{\bf{T}}^{2}\right]\,\psi + \frac{2 \mu_{0}}{\hbar^{2}}\,
\left(\varepsilon + \frac{e^{2}}{r}\right)\,\psi = 0.
\label{2.3.2}
\end{eqnarray}
We see that equation (\ref{2.3.2}) contains an LT interaction term, which indicates that we
cannot divide the wave function into the product of two functions depending on the coordinates
$\rm I\!R^{5}$ and $S^{3}$, respectively.

Now we introduce the following operators $\hat{J}_{a}=\hat{L}_{a}+\hat{T}_{a}$.
Since $\hat{\bf{J}}^{2} = \hat{\bf{L}}^{2} + \hat{\bf{T}}^{2} + 2\hat{L}_{a}\hat{T}_{a}$,
then equation (\ref{2.3.2}) in hyperspherical coordinates (\ref{1.9.2}) will have the form:
\begin{eqnarray}
\left(\Delta_{r \theta} -\frac{\hat{\bf{L}}^{2}}{r^{2}\,\sin^{2}\theta/2} -
\frac{\hat{\bf{J}}^{2}}{r^{2}\,\cos^{2}\theta/2}\right)\,\psi + \frac{2 \mu_{0}}{\hbar^{2}}\,
\left(\varepsilon + \frac{e^{2}}{r}\right)\,\psi = 0.
\label{2.3.3}
\end{eqnarray}
Here
\begin{eqnarray*}
\Delta_{r \theta} = \frac{1}{r^{4}}\,\frac{\partial}{\partial r}
\left(r^{4}\,\frac{\partial}{\partial r}\right) +
\frac{1}{r^{2}\,\sin^{3}\theta}\,\frac{\partial}{\partial \theta}\,
\left(\sin^{3}\theta\,\frac{\partial}{\partial \theta}\right).
\end{eqnarray*}
Note also that the following commutation relations hold:
\begin{eqnarray*}
\left[\hat{L}_{a}, \hat{L}_{b}\right] = i\epsilon_{a b c}\,\hat{L}_{c}, \qquad
\left[\hat{J}_{a}, \hat{J}_{b}\right] = i\epsilon_{a b c}\,\hat{J}_{c}.
\end{eqnarray*}

Let us represent the wave function in the form
\begin{eqnarray}
\psi = \Phi(r, \theta)\, \cal{D}\left(\alpha, \beta, \gamma, \alpha_{T}, \beta_{T}, \gamma_{T}\right),
\label{2.3.4}
\end{eqnarray}
where $\cal{D}$ are the eigenfunctions of the operators $\hat{\bf{L}}^{2},
\hat{\bf{T}}^{2}$ and $\hat{\bf{J}}^{2}$ with eigenvalues $L(L+1), T(T+1)$
and $J(J+1)$, respectively. If we substitute the last ansatz into equation
(\ref{2.3.3}), then for the function $\Phi(r, \theta)$ we obtain the following differential equation:
\begin{eqnarray}
\left[\Delta_{r \theta} -\frac{L(L+1)}{r^{2}\,\sin^{2}\theta/2} -
\frac{J(J+1)}{r^{2}\,\cos^{2}\theta/2}\right]\,\Phi + \frac{2 \mu_{0}}{\hbar^{2}}\,
\left(\varepsilon + \frac{e^{2}}{r}\right)\,\Phi = 0.
\label{2.3.5}
\end{eqnarray}
Because of the $LT$ - interaction, orthonormalized by the condition
\begin{eqnarray*}
\int\,{\cal{D}}_{LTm't'}^{JM*}\,{\cal{D}}_{L_{1}T_{1}m_{1}'t_{1}'}^{J_{1}M_{1}}\,
d\,\Omega\,d\,d\,\Omega_{T} = \delta_{JJ_{1}}\,\delta_{LL_{1}}\,\delta_{TT_{1}}\,
\delta_{MM_{1}}\,\delta_{m'm_{1}'}\,\delta_{t't_{1}'}
\end{eqnarray*}
function $\cal{D}$ has the form
\begin{eqnarray}
{\cal{D}} = \sqrt{\frac{(2L+1) (2T+1)}{4 \pi^{4}}}\sum\limits_{M=m+t}
C_{Lm;Tt}^{JM}\,D_{mm'}^{L}\left(\alpha, \beta, \gamma\right)\,
D_{tt'}^{T}\left(\alpha_{T}, \beta_{T}, \gamma_{T}\right),
\label{2.3.6}
\end{eqnarray}
where $C_{Lm;Tt}^{JM}$ are the Clebsch-Gordan coefficients, and $D_{mm'}^{L}$
and $D_{tt'}^{T}$ are Wigner functions. When calculating the normalization factor,
formula (\ref{1.5.12}) was used and it was taken into account that
\begin{eqnarray*}
d\,\Omega = \frac{1}{8}\,\sin\beta\,d\beta\,d\alpha\,d\gamma, \qquad
d\,\Omega_{T} = \frac{1}{8}\,\sin\beta_{T}\,d\beta_{T}\,d\alpha_{T}\,d\gamma_{T}.
\end{eqnarray*}
Now let us prove that in formula (\ref{2.3.6}) the coefficients of the expansion
$C_{Lm;Tt}^{JM}$ are indeed the Clebsch-Gordan coefficients of the $SU(2)$ group.

Let us write the expansion in the form
\begin{eqnarray*}
{\cal{D}}_{LTm't'}^{JM}\left(\alpha, \beta, \gamma, \alpha_{T}, \beta_{T}, \gamma_{T}\right)
= \sum\limits_{M=m+t}\,\left(JM|Lm;Tt\right)
\,D_{mm'}^{L}\left(\alpha, \beta, \gamma\right)\,
D_{tt'}^{T}\left(\alpha_{T}, \beta_{T}, \gamma_{T}\right).
\end{eqnarray*}
Let us act on this expansion with the operator
$\hat{\bf{J}}^{2} = \hat{\bf{L}}^{2} + \hat{\bf{T}}^{2} + 2\hat{L}_{a}\hat{T}_{a}$,
remember that ${\cal{D}}_{LTm't'}^{JM}$ is an eigenfunction of the operators
$\hat{\bf{L}}^{2}$, $\hat{\bf{T}}^{2}$ and $\hat{\bf{J}}^{2}$ simultaneously and using
the orthonormalization integral for the $D$ - Wigner function (\ref{1.5.12}), we obtain
\begin{eqnarray}
&&\frac{2\pi^{4}}{(2L+1)(2J+1)}\left[J(J+1) - L(L+1) - T(T+1)\right]\,
\left(JM|Lm;Tt\right)\,\delta_{m'm_{1}'}\,\delta_{t't_{1}'} = \nonumber \\
\label{2.3.7} \\
&=& \sum\limits_{M=m+t} \left(JM|L_{1}m_{1}';T_{1}t_{1}'\right)\,
\left\langle Lmm';Ttt' \left|\hat{L}_{a} \hat{T}_{a}\right| L_{1}m_{1}m_{1}';T_{1}t_{1}t_{1}'\right\rangle,
\nonumber
\end{eqnarray}
where
\begin{eqnarray}
&&\left\langle Lmm';Ttt' \left|\hat{L}_{a} \hat{T}_{a}\right| L_{1}m_{1}m_{1}';T_{1}t_{1}t_{1}'\right\rangle =
\nonumber \\
\label{2.3.8}
\\
&=& \int\,D_{mm'}^{L*}\left(\alpha, \beta, \gamma\right)\,D_{tt'}^{T*}\left(\alpha_{T}, \beta_{T}, \gamma_{T}\right)
\hat{L}_{a} \hat{T}_{a}\,D_{mm'}^{L}\left(\alpha, \beta, \gamma\right)\,D_{tt'}^{T}\left(\alpha_{T}, \beta_{T}, \gamma_{T}\right)
d\,\Omega\,d\,d\,\Omega_{T}.
\nonumber
\end{eqnarray}
Now using the definition of the cyclic components of the vector \cite{VAR}
\begin{eqnarray*}
A_{x} = \frac{1}{\sqrt{2}}\left(A_{-1} - A_{+1}\right), \qquad
A_{y} = \frac{i}{\sqrt{2}}\left(A_{+1} - A_{-1}\right), \qquad
A_{z} = A_{0}
\end{eqnarray*}
we have that
\begin{eqnarray*}
\hat{L}_{a} \hat{T}_{a} = \hat{L}_{0} \hat{T}_{0} - \hat{L}_{-1} \hat{T}_{+1}
\hat{L}_{+1} \hat{T}_{-1}.
\end{eqnarray*}
Next, taking into account the formula \cite{VAR}
\begin{eqnarray*}
\hat{J}_{\nu} D_{MM'}^{J}\left(\alpha, \beta, \gamma\right)=
\Bigg\{
\begin{array}{cc}
-M D_{MM'}^{J}(\alpha, \beta, \gamma)&\nu = 0,\\ [7mm]
\pm \sqrt{\frac{J(J+1) - M(M+1)}{2}}D_{M\mp 1 M'}^{J}(\alpha, \beta, \gamma)&\nu = \pm 1\\
\end{array}
\end{eqnarray*}
and the orthonormalization integral (\ref{1.5.12}), for the matrix element (\ref{2.3.8}) we obtain
\begin{eqnarray*}
\left\langle Lmm';Ttt' \left|\hat{L}_{a} \hat{T}_{a}\right| L_{1}m_{1}m_{1}';T_{1}t_{1}t_{1}'\right\rangle =
\frac{2\pi^{4}}{(2L+1)(2J+1)}\,\delta_{L L_{1}}\,\delta_{T T_{1}}\,\delta_{m'm_{1}'}\,\delta_{t't_{1}'} \times  \\ [4mm]
\times \Bigl[2m t\,\delta_{m_{1} m}\,\delta_{t_{1} t} +
\sqrt{(L+m)(L-m+1)(T-t)(T+t+1)}\delta_{m_{1}, m-1}\,\delta_{t_{1}, t+1} +   \\ [4mm]
+ \sqrt{(L-m)(L+m+1)(T+t)(T-t+1)}\delta_{m_{1}, m+1}\,\delta_{t_{1}, t-1}\Bigr].
\end{eqnarray*}
Substituting the obtained expression for the matrix element into formula (\ref{2.3.7}) we have
\begin{eqnarray*}
\left[J(J+1) - L(L+1) - T(T+1)- 2m t\right]\,
\left(JM|Lm;Tt\right) =  \\ [4mm]
= \sqrt{(L+m)(L-m+1)(T-t)(T+t+1)}\,\left(JM|L,m-1;T,t+1\right) + \\ [4mm]
+ \sqrt{(L-m)(L+m+1)(T+t)(T-t+1)}\,\left(JM|L,m+1;T,t-1\right).
\end{eqnarray*}
Comparison of the last formula with the three-term recurrence relation (\ref{1.8.15})
shows that indeed the expansion coefficients (\ref{2.3.6}) are the Clebsch-Gordan
coefficients of the SU(2) group.

\section{Hyperspherical basis $SU(2)$ YCM}
\markboth{CHAPTER 2. DYON-OSCILLATOR DUALITY}{2.4. HYPERSPHERICAL BASIS $SU(2)$ YCM}

Now, after establishing the explicit dependence of the wave function of the $SU(2)$ YCM
on the coordinates associated with the "isospin-orbital" interaction, we will find the complete wave
function in hyperspherical coordinates. To do this, we represent the function $\Phi(r,\theta)$
in relation (\ref{2.3.4}) in the form:
\begin{eqnarray*}
\Phi(r,\theta) = R(r)\,Z(\theta).
\end{eqnarray*}
Then the variables in equation (\ref{2.3.5}) are separated and we arrive at a system of two ordinary
differential equations:
\begin{eqnarray}
\frac{1}{\sin^{3}\theta}\,\frac{d}{d \theta}\,\left(\sin^{3}\theta\,
\frac{d Z}{d \theta}\right) - \frac{2L(L+1)}{1-\cos\theta} -
\frac{2J(J+1)}{1+\cos\theta}Z + \lambda(\lambda + 3)\,Z = 0,
\label{2.4.1}
\end{eqnarray}
\begin{eqnarray}
\frac{1}{r^{4}}\,\frac{d}{d r}\,\left(r^{4}\,\frac{d R}{d r}\right) -
\frac{\lambda(\lambda + 3)}{r^{2}}R + \frac{2\mu_{0}}{\hbar^{2}}\,
\left(\varepsilon + \frac{e^{2}}{r}\right)R = 0.
\label{2.4.2}
\end{eqnarray}
Here the non-negative separation constant $\lambda(\lambda + 3)$ is the eigenvalues
of the square of the hypermoment operator:
\begin{eqnarray*}
{\hat{\Lambda}}^{2} = - \frac{1}{\sin^{3}\theta}\,\frac{\partial}{\partial \theta}\,
\left(\sin^{3}\theta\,\frac{\partial}{\partial \theta}\right)
+ \frac{4}{\sin^{2}\theta}\,{\hat{\bf L}}^{2} +
\frac{2}{1+\cos\theta}\,{\hat{\bf T}}^{2} + \frac{4}{1+\cos\theta}\,
\hat{L}_{a}\,\hat{T}_{a}.
\end{eqnarray*}

In equation (\ref{2.4.1}) it is convenient to go to the new variable
$y=(1-\cos\theta)/2$ and write
\begin{eqnarray*}
Z(y) = y^{L}\,(1-y)^{J}\,W(y).
\end{eqnarray*}
Substituting the last relation into (\ref{2.4.1}) we obtain the hypergeometric equation
\begin{eqnarray*}
y\,(1 - y)\,\frac{d^{2}W}{dy^{2}} + \left[c - (a + b + 1)y\right]\,\frac{d W}{d y}
- ab\,W = 0
\end{eqnarray*}
with $a = -\lambda + L + J, b = \lambda + L + J + 3, c = 2L + 2$.

Thus we find that
\begin{eqnarray*}
Z(\theta) = \left(1 - \cos\theta\right)^{L}\,\left(1 + \cos\theta\right)^{J}\,
_{2}F_{1}\,\left(-\lambda + L + J, \lambda + L + J + 3; 2L + 2;
\frac{1 - \cos\theta}{2}\right).
\end{eqnarray*}
This solution behaves well for $\theta = \pi$, if the series
$_{2}F_{1}$ is finite, i.e.
\begin{eqnarray*}
-\lambda + L + J = -n_{\theta}, \qquad \rm{where} \qquad n_{\theta} = 0, 1, 2,....
\end{eqnarray*}
Now using the formula \cite{BE2}
\begin{eqnarray*}
_{2}F_{1}\,\left(-n, n + a + b + 1; a + 1; \frac{1 - y}{2}\right)
= \frac{n!\,\Gamma(a + 1)}{\Gamma(n + a + 1)}\,P_{n}^{a,b}(y),
\end{eqnarray*}
where $P_{n}^{a,b}(y)$ is a Jacobi polynomial, and taking into account the integral
\cite{Ryzhik}
\begin{eqnarray*}
\int\limits_{-1}^{1}\,(1-y)^{a}\,(1+y)^{b}\,\left[P_{n}^{a,b}(y)\right]^{2}\,d(y)
= \frac{2^{a+b+1}}{2n+a+b+1}\,\frac{\Gamma(n + a + 1)\,\Gamma(n + b + 1)}
{n!\,\Gamma(n + a + b + 1)},
\end{eqnarray*}
normalized by the condition
\begin{eqnarray}
\int\limits_{0}^{\pi}\,Z_{\lambda' L J}(\theta)\,Z_{\lambda L J}(\theta)\,
\sin^{3}\theta\,d\theta = \delta_{\lambda \lambda'}
\label{2.4.3}
\end{eqnarray}
function $Z_{\lambda L J}(\theta)$ can be written in the form
\begin{eqnarray*}
Z_{\lambda L J}(\theta) = C_{LJT}^{\lambda}\,\left(1 - \cos\theta\right)^{L}\,
\left(1 + \cos\theta\right)^{J}\,P_{\lambda -L-J}^{2L+1,2J+1}(\cos\theta),
\end{eqnarray*}
where
\begin{eqnarray*}
C_{LJT}^{\lambda}\sqrt{\frac{(2\lambda +3)\,(\lambda -J-L)!\,
\Gamma(\lambda +J+L + 3)}{2^{2J+2L+3}\,\Gamma(\lambda +J-L + 2)\,
\Gamma(\lambda -J+L + 2)}}.
\end{eqnarray*}
Function
\begin{eqnarray*}
\Phi^{\lambda L J}_{LTm't'}\,\left(\theta, \alpha, \beta, \gamma,
\alpha_{T}, \beta_{T}, \gamma_{T}\right) = Z_{\lambda L J}(\theta) \,
{\cal {D}}^{JM}_{LTm't'\,}\left(\alpha, \beta, \gamma,
\alpha_{T}, \beta_{T}, \gamma_{T}\right)
\end{eqnarray*}
are $SU(2)$ monopole harmonics \cite{YANG-2}.

Now let's return to the radial equation (\ref{2.4.2}). After substitution in (\ref{2.4.2})
\begin{eqnarray*}
R(r) = e^{-\kappa r}\,r^{\lambda}\,f(r)
\end{eqnarray*}
for the function $f(r)$ we obtain an equation for the confluent hypergeometric function
(\ref{1.5.17}) from the argument $z=2\kappa r$, and with parameters
$\alpha =\lambda +2-1/\kappa r_{0}, \gamma = 2\lambda +4$, where
$\kappa =\sqrt{-2\mu_{0}\varepsilon}/\hbar$, and $r_{0}=\hbar^{2}/\mu_{0}e^{2}$ is the Bohr radius.
For bound states $(\varepsilon < 0)$, we have
\begin{eqnarray*}
\lambda + 2 - \frac{1}{\kappa r_{0}} = - n_{r} = 0, -1, -2,...,
\end{eqnarray*}
whence it follows that the energy spectrum of the $SU(2)$ YCM has the form
\begin{eqnarray}
\varepsilon_{N}^{T} = -\frac{\mu_{0}e^{4}}{2\hbar^{2} \left(\frac{N}{2} + 2\right)^{2}},
\label{2.4.4}
\end{eqnarray}
where $N=2\left(n_{r} + \lambda\right)=2\left(n_{r} + n_{\theta} + J + L\right)$ is the principal quantum number.

Solution of the radial equation (\ref{2.4.2}) normalized by the condition
\begin{eqnarray}
\int\limits_{0}^{\infty}\,r^{4}\,R_{N \lambda}(r)\,R_{N \lambda}(r)\,dr=
\delta_{NN'}
\label{2.4.5}
\end{eqnarray}
has the appearance.
Solution of the radial equation (\ref{2.4.2}) normalized by the condition
\begin{eqnarray}
R_{N \lambda}(r) = \frac{32}{r_{0}^{5/2}\,(N+4)^{3}}\,
\sqrt{\frac{\left(\frac{N}{2}+\lambda +3\right)!}{\left(\frac{N}{2}-\lambda \right)!}}\,
\frac{(2\kappa r)^{\lambda}\,e^{-\kappa r}}{(2\lambda +3)!}\,
F\left(-\frac{N}{2}+\lambda; 2\lambda +4; 2\kappa r\right).
\label{2.4.6}
\end{eqnarray}

The complete wave function can be written as
\begin{eqnarray}
\psi^{hsp} = R_{N \lambda}(r)\,
Z_{\lambda L J}(\theta) \,
{\cal {D}}^{JM}_{LTm't'\,}\left(\alpha, \beta, \gamma,
\alpha_{T}, \beta_{T}, \gamma_{T}\right).
\label{2.4.7}
\end{eqnarray}
It is normalized by the condition
\begin{eqnarray*}
\int\,\left|\psi^{hsp}\right|^{2}\,dv =1, \qquad {\rm where} \quad
dv = r^{4}\,\sin^{3}\theta\,dr\,d\theta\,d\Omega\,d\Omega_{T}.
\end{eqnarray*}

Thus we solved the following spectral problem
\begin{eqnarray*}
\hat{H}\,\psi^{hsp} &=& \varepsilon_{N}^{T}\,\psi^{hsp}, \qquad
\hat{\Lambda}^{2}\,\psi^{hsp} = \lambda(\lambda+3)\,\psi^{hsp}, \qquad
\hat{\bf J}^{2}\,\psi^{hsp} = J(J+1)\,\psi^{hsp}, \\ [5mm]
\hat{\bf L}^{2}\,\psi^{hsp} &=& L(L+1)\,\psi^{hsp}, \qquad
\hat{\bf T}^{2}\,\psi^{hsp} = T(T+1)\,\psi^{hsp}, \\ [5mm]
\hat{J}_{3}\,\psi^{hsp} &=& M\,\psi^{hsp}, \qquad
\hat{L}_{3'}\,\psi^{hsp} = m'\,\psi^{hsp}, \qquad
\hat{T}_{3'}\,\psi^{hsp} = t'\,\psi^{hsp},
\end{eqnarray*}
where $\hat{L}_{3'} = -i\,\partial/\partial \gamma$, and
$\hat{T}_{3'} = -i\,\partial/\partial \gamma_{T}$.

It should be noted that in Cartesian coordinates the square of the hypermoment operator has the form
\begin{eqnarray}
\hat{\Lambda}^{2} = -r^{2}\,\Delta_{5} + x_{i}x_{j}\,\frac{\partial^{2}}{\partial x_{i} \partial x_{j}}
+ 4x_{i}\,\frac{\partial}{\partial x_{i}} + \frac{2r}{r+x_{2}}\,\hat{\bf T}^{2} +
2ir^{2}\,A_{i}^{a(+)}\,\hat{T}_{a}\,\frac{\partial}{\partial x_{i}}.
\label{2.4.8}
\end{eqnarray}

From the formula (\ref{2.4.4}) it follows that the energy spectrum of the $SU(2)$ YCM is
degenerate in the quantum number $\lambda$. Now let's calculate the multiplicity of this degeneracy.

For a fixed value of the quantum number $T$, the energy levels $\varepsilon_{N}^{T}$ do not depend
on the quantum numbers $L, J$ and $\lambda$, i.e. they are degenerate. The total multiplicity of
degeneracy for a fixed value of $T$ is equal to
\begin{eqnarray*}
g_{N}^{T} = (2T+1)\,\sum\limits_{\lambda}\,\sum\limits_{L}\,(2L+1)\,\sum\limits_{J}\,(2J+1).
\end{eqnarray*}
Since $\lambda = n_{\theta}+J+L$ and $N=2(n_{r}+ \lambda)$, it follows that (for fixed $N$ and $T$)
$\lambda =  T+ T+1,…, N/2$. Then $L_{max}= \lambda- J_{min}$ ($L_{max}$ is fixed) and therefore
$L_{max}= \lambda - \left(L_{max} - T\right)$ or $L_{max}=(\lambda + T)/2$. Thus
\begin{eqnarray*}
g_{N}^{T} = (2T+1)\,\sum\limits_{\lambda = T}^{N/2}\,
\sum\limits_{L=0,1/2}^{(\lambda - T)/2}\,(2L+1)\,\sum\limits_{J}\,(2J+1).
\end{eqnarray*}
Now from comparing the relations $|L-T| \leq J \leq L+T$ and
$J \leq \lambda - L$ we find out that
\begin{eqnarray*}
&&(a)\,\,\, J = |L-T|, |L-T|+1,..., L+T, \qquad {\rm for}
\quad L=1,\frac{1}{2},..., \frac{\lambda -T}{2}, \\ [5mm]
&&(b)\,\,\, J = |L-T|, |L-T|+1,..., \lambda - L, \qquad {\rm for} \quad
L=\frac{\lambda -T+1}{2},..., \frac{\lambda +T}{2},
\end{eqnarray*}
and we can rewrite the formula for g in a more defined form
\begin{eqnarray*}
g_{N}^{T} = (2T+1)\,\sum\limits_{\lambda = T}^{N/2}\,\left\{
\sum\limits_{L=0,\frac{1}{2}}^{\frac{\lambda -T}{2}}\,(2L+1)\,
\sum\limits_{J=|L-T|}^{L+T}\,(2J+1) +
\sum\limits_{L=\frac{\lambda -T+1}{2}}^{\frac{\lambda +T}{2}}\,(2L+1)\,
\sum\limits_{J=|L-T|}^{L+T}\,(\lambda -T)\right\}.
\end{eqnarray*}
Finally, after some tedious calculations we obtain the following result:
\begin{eqnarray}
g_{N}^{T} &=& \frac{1}{12}(2T+1)^{2}\,\left(\frac{N}{2} - T + 1\right)\,
\left(\frac{N}{2} - T + 2\right) \times \nonumber \\
\label{2.4.9}
\\
&\times& \left[\left(\frac{N}{2} - T + 2\right)^{2}\,
\left(\frac{N}{2} - T + 3\right) + 2T(N+5)\right].
\nonumber
\end{eqnarray}
For $T=0$ and $N=2n$ (even) we have the multiplicity of degeneracy of the Coulomb
levels (\ref{1.9.1}). Further, $T=0,1,…,N/2$ and $T=1/2,3/2…,N/2$, for even and odd
$N$, respectively, we have
\begin{eqnarray*}
g_{N} = \sum\limits_{T=0,1/2}^{N/2}\,g_{N}^{T} =
\frac{(N+7)!}{7!\,N!},
\end{eqnarray*}
i.e. multiplicity of degeneracy of energy levels for an eight-dimensional isotropic
oscillator (\ref{1.5.1}).

\section{Parabolic basis of the $SU(2)$ YCM}
\markboth{CHAPTER 2. DYON-OSCILLATOR DUALITY}{2.5. PARABOLIC BASIS OF THE $SU(2)$
YCM}

In parabolic coordinates (\ref{1.10.1}) equation (\ref{2.1.5}) has the form
\begin{eqnarray}
\left[\Delta_{\mu \nu} - \frac{4{\hat{\bf J}}^{2}}{\mu (\mu + \nu)}
- \frac{4{\hat{\bf L}}^{2}}{\nu (\mu + \nu)}\right]\,\psi^{par} +
\frac{2\mu_{0}}{\hbar^{2}}\left(\varepsilon + \frac{2e^{2}}{\mu + \nu}
\right)\,\psi^{par} = 0,
\label{2.5.1}
\end{eqnarray}
where
\begin{eqnarray*}
\Delta_{\mu \nu} = \frac{4}{\mu + \nu}\,\left[\frac{1}{\mu}\,\frac{\partial}{\partial \mu}\,
\left(\mu^{2}\frac{\partial}{\partial \mu}\right) + \frac{1}{\nu}\,\frac{\partial}{\partial \nu}\,
\left(\nu^{2}\frac{\partial}{\partial \nu}\right)\right].
\end{eqnarray*}
After substitution
\begin{eqnarray*}
\psi^{par} = f_{1}(\mu)\,f_{2}(\nu) \,
{\cal {D}}^{JM}_{LTm't'\,}(\alpha, \beta, \gamma,
\alpha_{T}, \beta_{T}, \gamma_{T}).
\end{eqnarray*}
The variables in equation (\ref{2.5.1}) are separated and we arrive at the following system of equations
\begin{eqnarray}
\frac{1}{\mu}\,\frac{d}{d \mu}\,
\left(\mu^{2}\frac{d f_{1}}{d \mu}\right) +
\left[\frac{\mu_{0}\varepsilon}{2\hbar^{2}}\mu - \frac{J(J+1)}{\mu} +
\frac{\sqrt{\mu_{0}}}{2\hbar}\Omega + \frac{\mu_{0}e^{2}}{2\hbar^{2}}\right]\,
f_{1} = 0, \nonumber \\
\label{2.5.2}
\\
\frac{1}{\nu}\,\frac{d}{d \nu}\,
\left(\nu^{2}\frac{d f_{2}}{d \nu}\right) +
\left[\frac{\mu_{0}\varepsilon}{2\hbar^{2}}\nu - \frac{L(L+1)}{\nu} -
\frac{\sqrt{\mu_{0}}}{2\hbar}\Omega + \frac{\mu_{0}e^{2}}{2\hbar^{2}}\right]\,
f_{2} = 0, \nonumber
\end{eqnarray}
where $\Omega$ is the separation constant. At $T=0$ (i.e. $J=L$) these equations coincide
with equations (\ref{1.10.2}) for the five-dimensional Coulomb problem, and consequently
\begin{eqnarray}
\psi^{par} =\kappa^{3}\,\sqrt{2r_{0}}\, f_{n_{1} J}(\mu)\,f_{n_{2} L}(\nu) \,
{\cal {D}}^{JM}_{LTm't'\,}(\alpha, \beta, \gamma,
\alpha_{T}, \beta_{T}, \gamma_{T}),
\label{2.5.3}
\end{eqnarray}
\begin{eqnarray*}
f_{p q}(x) = \frac{1}{(2q+1)!}\,\sqrt{\frac{(p+2q+1)!}{p!}}\,
\exp\left(-\frac{\kappa x}{2}\right)\,(\kappa x)^{q}\,
F\left(-p; 2q+2; \kappa x\right).
\end{eqnarray*}
Here $n_{1}$ and $n_{2}$ are non-negative integers, and
\begin{eqnarray*}
n_{1} = -J-1+\frac{\sqrt{\mu_{0}}}{2\hbar \kappa}\Omega + \frac{1}{2\kappa r_{0}},
\qquad
n_{2} = -L-1-\frac{\sqrt{\mu_{0}}}{2\hbar \kappa}\Omega + \frac{1}{2\kappa r_{0}}.
\end{eqnarray*}
From the last relations and from formula (\ref{2.4.4}) it follows that the parabolic
quantum numbers $n_{1}$ and $n_{2}$ are related to the principal quantum number $N$ as follows:
\begin{eqnarray*}
N = 2\left(n_{1} + n_{2} + J + L\right).
\end{eqnarray*}
Eliminating energy from equations (\ref{2.5.2}) we obtain the integral of motion
\begin{eqnarray*}
\hat{M}_{0} &=& \frac{\hbar}{\sqrt{\mu_{0}}}\,\Biggl\{\frac{2}{\mu + \nu}\,
\left[\frac{\mu}{\nu}\,\frac{\partial}{\partial \nu}\left(\nu^{2}\,
\frac{\partial}{\partial \nu}\right) - \frac{\nu}{\mu}\,\frac{\partial}{\partial \mu}\left(\mu^{2}\,
\frac{\partial}{\partial \mu}\right)\right] - \\ [5mm]
&-& \frac{2(\mu - \nu)}{\mu \nu}{\hat{\bf L}}^{2} +
\frac{4\nu}{\mu (\mu + \nu)}\hat{L_{a}}\hat{T}_{a} + \frac{2\nu}{\mu (\mu + \nu)}
{\hat{\bf T}}^{2} + \frac{\mu_{0}e^{2}}{\hbar^{2}}\,\frac{\mu - \nu}{\mu + \nu}\Biggr\}
\end{eqnarray*}
eigenvalues of which is:
\begin{eqnarray*}
\frac{\hbar^{2}}{\mu_{0}}\,\Omega = \frac{2e^{2}\sqrt{\mu_{0}}}{\hbar}\,\,
\frac{n_{1} - n_{2} + J - L}{N+4}.
\end{eqnarray*}
In Cartesian coordinates, the operator $\hat{M}_{0}$ will be written in the form
\begin{eqnarray}
\hat{M}_{0} &=& \Biggl[x_{0}\,\frac{\partial^{2}}{\partial x_{\sigma} \partial x_{\sigma}} -
x_{\sigma}\,\frac{\partial^{2}}{\partial x_{0} \partial x_{\sigma}} + i(r - x_{0})\,
A_{j}^{a(+)}\,\hat{T}_{a}\,\frac{\partial}{\partial x_{j}} - \nonumber \\
\label{2.5.4}
\\
&-& 2\,\frac{\partial}{\partial x_{0}}
+ \frac{r - x_{0}}{r(r - x_{0})}{\hat{\bf T}}^{2} + \frac{\mu_{0}e^{2}}{\hbar^{2}}\,
\frac{x_{0}}{r}\Biggr],
\nonumber
\end{eqnarray}
where $\sigma = 1,2,3,4$.

So, the parabolic basis (\ref{2.5.3}) is simultaneously an eigenfunction of the following commuting operators:
\begin{eqnarray*}
\hat{H}\,\psi^{par} &=& \varepsilon_{N}^{T}\,\psi^{par}, \qquad
\hat{M}_{0}\,\psi^{par} = \frac{2e^{2}\sqrt{\mu_{0}}}{\hbar}\,\,
\frac{n_{1} - n_{2} + J - L}{N+4}\,\psi^{par}, \\ [5mm]
\hat{\bf J}^{2}\,\psi^{par} &=& J(J+1)\,\psi^{par}, \qquad
\hat{\bf L}^{2}\,\psi^{par} = L(L+1)\,\psi^{par}, \qquad
\hat{\bf T}^{2}\,\psi^{par} = T(T+1)\,\psi^{par}, \\ [5mm]
\hat{J}_{3}\,\psi^{par} &=& M\,\psi^{par}, \qquad
\hat{L}_{3'}\,\psi^{par} = m'\,\psi^{par}, \qquad
\hat{T}_{3'}\,\psi^{par} = t'\,\psi^{par}.
\end{eqnarray*}

Now we calculate the multiplicity of degeneracy of energy levels using parabolic quantum numbers.

For a fixed value of T, complete degeneracy can be written as
\begin{eqnarray*}
g_{N}^{T} = (2T+1)\,\sum\limits_{n_{1} + n_{2}}\,\sum\limits_{L}\,(2L+1)\,\sum\limits_{J}\,(2J+1).
\end{eqnarray*}
Since $n_{1} + n_{2} = \frac{N}{2} - J - L$, it follows that (for fixed $N$ and $T$)
$n_{1} + n_{2} = 0,1,...,\frac{N}{2} - T$. Then $L_{max} = \frac{N}{2} - n_{1} - n_{2} - J_{min}$
($L_{max}$ is fixed) and therefore $L_{max} = \frac{N}{2} - n_{1} - n_{2} -\left(L_{max} - T\right)$ or
$L_{max} = \frac{1}{2}\,\left(\frac{N}{2} - n_{1} - n_{2} + T\right)$. Thus
\begin{eqnarray*}
g_{N}^{T} = (2T+1)\,\sum\limits_{n_{1} + n_{2}=0}^{\frac{N}{2} - T}\,
\sum\limits_{L=0,1/2}^{\frac{1}{2}\,\left(\frac{N}{2} - n_{1} - n_{2} + T\right)}\,(2L+1)\,
\sum\limits_{J}\,(2J+1).
\end{eqnarray*}
Now comparing the relations $|L-T| \leq J \leq L+T$ and
$J \leq \frac{N}{2} - n_{1} - n_{2} - L$ we calculate that
\begin{eqnarray*}
&& (a)\,\,\, J = |L-T|,|L-T|+1,...,L+T, \quad \rm{for}\,\,\,
L=0,\frac{1}{2},...,\frac{1}{2}\,\left(\frac{N}{2} - n_{1} - n_{2} - T\right), \\ [5mm]
&& (b)\,\,\, J = |L-T|,|L-T|+1,...,\frac{N}{2} - n_{1} - n_{2} - L, \quad \rm{for} \\ [5mm]
&& L = \frac{1}{2}\,\left(\frac{N}{2} - n_{1} - n_{2} - T + 1\right),...,
\frac{1}{2}\,\left(\frac{N}{2} - n_{1} - n_{2} + T\right)
\end{eqnarray*}
and rewrite the formula for $g_{N}^{T}$ in the form
\begin{eqnarray*}
g_{N}^{T} &=& (2T+1)\,\sum\limits_{n_{1} + n_{2}=0}^{\frac{N}{2} - T}\Biggl\{
\sum\limits_{L=0,1/2}^{\frac{1}{2}\,\left(\frac{N}{2} - n_{1} - n_{2} - T\right)}\,(2L+1)\,
\sum\limits_{J=|L-T|}^{L+T}\,(2J+1) + \\ [7mm]
&+& \sum\limits_{L =\left(\frac{N}{2} - n_{1} - n_{2} - T + 1\right)}^{\frac{1}{2}\,\left(\frac{N}{2} - n_{1} - n_{2} + T\right)}\,(2L+1)\,
\sum\limits_{J=|L-T|}^{\frac{N}{2} - n_{1} - n_{2} - L}\,(2J+1)\Biggr\}.
\end{eqnarray*}
Finally, after long and tedious calculations we obtain formula (\ref{2.4.9}).

\section{Analogue of the Park-Tarter expansion}
\markboth{CHAPTER 2. DYON-OSCILLATOR DUALITY}{2.6. ANALOGUE OF THE
PARK-TARTER EXPANSION}

First, we note that for the radial wave function $R_{N \lambda}(r)$ (\ref{2.4.6}),
along with (\ref{2.4.5}), the following "additional" condition of orthogonality with
respect to the hypermoment $\lambda$ is satisfied:
\begin{eqnarray}
I_{\lambda \lambda'} = \int\limits_{0}^{\infty}\,r^{2}\,R_{N \lambda}(r)\,
R_{N \lambda'}(r)\,dr = \frac{16}{r_{0}^{2} (N+4)^{3}}\,\frac{1}{2\lambda + 3}\,
\delta_{\lambda \lambda'},
\label{2.6.1}
\end{eqnarray}
which is easy to prove if you apply the calculation technique given in paragraph $1.6$.

Now let's move on to finding the coefficients of the expansion of parabolic wave functions of the
$SU(2)$ YCM (\ref{2.5.3}) in hyperspherical wave functions (\ref{2.4.7}).
We write the desired expansion for fixed energy values in the form \cite{M-12}:
\begin{eqnarray}
\psi^{par} = \sum\limits_{\lambda = T}^{N/2}\,
W^{\lambda}_{NJLn_{1}}\,\psi^{hsp}.
\label{2.6.2}
\end{eqnarray}
The calculation of the coefficients $W^{\lambda}_{NJLn_{1}}$ is carried out in the
same way as in the case of the five-dimensional Coulomb problem, only now we will
use formulas (\ref{1.7.4}) and (\ref{2.6.1}).
As a result, we obtain the following integral representation for the coefficients
$W^{\lambda}_{NJLn_{1}}$:
\begin{eqnarray}
W^{\lambda}_{NJLn_{1}} = \frac{\sqrt{(2\lambda + 3)\left(\frac{N}{2}+\lambda +3\right)!}}
{(2J+1)!\,(2\lambda + 3)!}\,E_{JL\lambda}^{Nn_{1}}\,K_{JL\lambda}^{Nn_{1}}.
\label{2.6.3}
\end{eqnarray}
Here
\begin{eqnarray}
E_{JL\lambda}^{Nn_{1}} = \left[\frac{(\lambda - J - L)!\,(\lambda +J - L + 1)!\,
(n_{1} + 2J + 1)!\,(n_{2} + 2L + 1)!}{(n_{1})!\,(n_{2})!\,\left(\frac{N}{2} - \lambda\right)!\,
(\lambda - J + L + 1)!\,(\lambda + J + L + 2)!}\right]^{1/2},
\label{2.6.4}
\end{eqnarray}
\begin{eqnarray*}
K_{JL\lambda}^{Nn_{1}} = \int\limits_{0}^{\infty}\,e^{-x}\,x^{\lambda + J + L + 2}\,
F\left(-n_{1}; 2J+2; x\right)\,F\left(-\frac{N}{2} + \lambda; 2\lambda+4; x\right)\,dx,
\end{eqnarray*}
where $x=2\kappa r$. In the integral $K_{JL\lambda}^{Nn_{1}}$, writing the confluent hypergeometric function \\
$F\left(-n_{1}; 2J+2; x\right)$ as a polynomial, integrating according to formula (\ref{1.6.5}) and
taking into account (\ref{1.6.6}) we obtain
\begin{eqnarray}
K_{JL\lambda}^{Nn_{1}} &=& \frac{(2\lambda + 3)!\,)\left(\frac{N}{2}- J - L\right)!\,
(\lambda + J + L + 2)!}{(\lambda - J - L)!\,\left(\frac{N}{2}+\lambda +3\right)!}\times
\nonumber \\
\label{2.6.5} \\
&\times& {_3F_2}\left\{\matrix{ -n_{1},\,\,-\lambda+J+L,\,\, \lambda+J+L+3\cr \cr
2J+2,\,\,-\frac{N}{2}+J+L
\cr}\Biggr|1\right\}.
\nonumber
\end{eqnarray}
After substituting (\ref{2.6.4}) and (\ref{2.6.5}) into (\ref{2.6.3}) we have
\begin{eqnarray}
W^{\lambda}_{NJLn_{1}} &=& \left[\frac{(2\lambda + 3)\,(\lambda + J + L + 2)!\,
(\lambda + J - L + 1)!\,(n_{1} + 2J + 1)!\,(n_{2} + 2L + 1)!}
{(n_{1})!\,(n_{2})!\,(\lambda - J - L)!\,(\lambda - J + L + 1)!\,\left(\frac{N}{2} - \lambda\right)!\,
\left(\frac{N}{2}+\lambda +3\right)!}\right]^{1/2}\times \nonumber \\
\label{2.6.6}
\\
&\times& \frac{\left(\frac{N}{2}- J - L\right)!}{(2J+1)!}\,
{_3F_2}\left\{\matrix{ -n_{1},\,\,-\lambda+J+L,\,\, \lambda+J+L+3\cr \cr
2J+2,\,\,-\frac{N}{2}+J+L
\cr}\Biggr|1\right\}.
\nonumber
\end{eqnarray}
Comparing expression (\ref{2.6.6}) with the representation of the Clebsch-Gordan
coefficients of the $SU(2)$ group (\ref{1.7.12}) gives
\begin{eqnarray}
W^{\lambda}_{NJLn_{1}} = (-1)^{n_{1}}\,
C_{\frac{N-2J+2L+2}{4},\,L+\frac{n_{2}-n_{1}+1}{2};\,
\frac{N+2J-2L+2}{4},\,J+\frac{n_{1}-n_{2}+1}{2}}^{\lambda + 1,\,J+L+1}.
\label{2.6.7}
\end{eqnarray}

Reverse expansion, i.e. the expansion of a hyperspherical basis into a parabolic one has the form:
\begin{eqnarray}
\psi^{hsp} = \sum\limits_{n_{1} = 0}^{\frac{N}{2}-J-L}\,
\tilde{W}^{n_{1}}_{NJL\lambda}\,\psi^{par}.
\label{2.6.8}
\end{eqnarray}
Using the orthonormalization condition for the Clebsch-Gordan coefficients of the
$SU(2)$ group  (\ref{1.8.16}), for the expansion coefficients (\ref{2.6.8}) we obtain
\begin{eqnarray}
\tilde{W}^{n_{1}}_{NJL\lambda} = (-1)^{n_{1}}\,
C_{\frac{N-2J+2L+2}{4},\,\frac{N-2J+2L+2}{4}-n_{1};\,
\frac{N+2J-2L+2}{4},\,n_{1} + J - \frac{N - 2J - 2L - 2}{4}}^{\lambda + 1,\,J+L+1}.
\label{2.6.9}
\end{eqnarray}

Finally, we note that at $T=0$ (i.e. $J=L$) formulas (\ref{2.6.6}) – (\ref{2.6.9}) transform
into the corresponding formulas (\ref{1.11.7}) – (\ref{1.11.10}) for the five-dimensional Coulomb problem.

\section{Spheroidal basis of the $SU(2)$ YCM}
\markboth{CHAPTER 2. DYON-OSCILLATOR DUALITY}{2.7. SPHEROIDAL BASIS OF THE
$SU(2)$ YCM}

In five-dimensional spheroidal coordinates (\ref{1.12.1}), equation (\ref{2.1.5}) will be written as
\begin{eqnarray}
\left\{\Delta_{\xi \eta} - \frac{8}{R^{2}(\xi + \eta)}\left[\frac{{\hat{\bf J}}^{2}}{(\xi + 1)(1 + \eta)}
- \frac{{\hat{\bf L}}^{2}}{(\xi - 1)(1 - \eta)}+\frac{\mu_{0}e^{2}R}{2\hbar^{2}}\right] +
\frac{2\mu_{0}\varepsilon}{\hbar^{2}}\right\}\psi^{sphd} = 0,
\label{2.7.1}
\end{eqnarray}
where
\begin{eqnarray*}
\Delta_{\xi \eta} = \frac{4}{R^{2}(\xi^{2} - \eta^{2})\,}
\left\{\frac{1}{\xi^{2} - 1}\,\frac{\partial}{\partial \xi}\,
\left[(\xi^{2} - 1)^{2}\,\frac{\partial}{\partial \xi}\right] +
\frac{1}{1 - \eta^{2}}\,\frac{\partial}{\partial \eta}\,
\left[(1 - \eta^{2})^{2}\,\frac{\partial}{\partial \eta}\right]\right\}.
\end{eqnarray*}
The following factorization corresponds to the scheme for separating variables in the
Schr\"{o}dinger equation (\ref{2.7.1}):
\begin{eqnarray}
\psi^{sphd} = \phi_{1}(\xi)\,\phi_{2}(\eta) \,
{\cal {D}}^{JM}_{LTm't'\,}(\alpha, \beta, \gamma,
\alpha_{T}, \beta_{T}, \gamma_{T}).
\label{2.7.2}
\end{eqnarray}
Now, substituting (\ref{2.7.2}) into (\ref{2.7.1}) we arrive at the following equations
\begin{eqnarray}
\Biggl[\frac{1}{\xi^{2} - 1}\,\frac{d}{d \xi}\,(\xi^{2} - 1)^{2}\,\frac{d}{d \xi} -
\frac{2L(L+1)}{\xi - 1} + \frac{2J(J+1)}{\xi + 1} + \frac{R}{r_{0}}\,\xi + \cr
 + \frac{\mu_{0}R^{2}\varepsilon}{2\hbar^{2}}\,(\xi^{2} - 1) - A\Biggr]\,\phi_{1} = 0,
\nonumber \\ [2mm]
\label{2.7.3} \\ [2mm]
\Biggl[\frac{1}{1 - \eta^{2}}\,\frac{d}{d \eta}\,(1 - \eta^{2})^{2}\,\frac{d}{d \eta} -
\frac{2L(L+1)}{1 - \eta} - \frac{2J(J+1)}{1 + \eta} - \frac{R}{r_{0}}\,\eta + \cr
+ \frac{\mu_{0}R^{2}\varepsilon}{2\hbar^{2}}\,(1 - \eta^{2}) + A\Biggr]\,\phi_{2} = 0.
\nonumber
\end{eqnarray}
The spheroidal basis (\ref{2.7.2}) and the separation constant $A(R)$ are eigenfunctions and
eigenvalues of the operator $\hat{A}$, i.e. we have the following equation
\begin{eqnarray}
\hat{A}\,\psi^{spheroidal} = A_{q}(R)\,\psi^{spheroidal}.
\label{2.7.4}
\end{eqnarray}
The index $q$ enumerates the eigenvalues of the operator $\hat{A}$ and takes the following
values $0 \leq q \leq \frac{N}{2} - L - J$. The explicit form of the operator $\hat{A}$ in
spheroidal coordinates is obtained from equations (\ref{2.7.3}) when energy is eliminated:
\begin{eqnarray*}
\hat{A} &=& \frac{1}{\xi^{2} - \eta^{2}}\,\left[\frac{1 - \eta^{2}}{\xi^{2} - 1}\,
\frac{\partial}{\partial \xi}\,(\xi^{2} - 1)^{2}\,\frac{\partial}{\partial \xi} -
\frac{\xi^{2} - 1}{1 - \eta^{2}}\,\frac{\partial}{\partial \eta}\,
(1 - \eta^{2})^{2}\,\frac{\partial}{\partial \eta}\right] + \\ [5mm]
&+& \frac{4(\xi^{2} - \eta^{2} - 2)}{(\xi^{2} - 1)\,(1 - \eta^{2})}\,{\hat{\bf L}}^{2} +
+ i\,\frac{R^{2}}{2}\,(\xi^{2} + \eta^{2} + \xi \eta + \xi + \eta- 1)\,
A_{j}^{a(+)}\,\hat{T}_{a}\,\frac{\partial}{\partial x_{j}} + \\ [5mm]
&+& \frac{2(\xi^{2} + \eta^{2} + \xi \eta + \xi + \eta - 1)}{(\xi + \eta)(\xi + 1)\,(1 + \eta)}\,
{\hat{\bf T}}^{2} + \frac{R(\xi \eta + 1)}{r_{0}(\xi + \eta)}.
\end{eqnarray*}
In the last expression, passing to Cartesian coordinates and taking into account
(\ref{2.4.8}) and (\ref{2.5.4}) we obtain
\begin{eqnarray}
\hat{A} = {\hat{\Lambda}}^{2} + \frac{R\sqrt{\mu_{0}}}{\hbar}\,{\hat{M}}_{0}.
\label{2.7.5}
\end{eqnarray}

Now we construct a spheroidal basis of the $SU(2)$ YCM using the
hyperspherical (\ref{2.4.7}) and parabolic (\ref{2.5.3}) bases. We write the required
expansions in the form:
\begin{eqnarray}
\psi^{spheroidal} = \sum\limits_{\lambda = J + L}^{N/2}\,U_{NqJL}^{\lambda}(R)\,\psi^{hsp},
\label{2.7.6}
\end{eqnarray}
\begin{eqnarray}
\psi^{spheroidal} = \sum\limits_{n_{1} = 0}^{\frac{N}{2}-J-L}\,V_{NqJL}^{n_{1}}(R)\,\psi^{par}.
\label{2.7.7}
\end{eqnarray}

Acting with operator $\hat{A}$ on both sides of expansion (\ref{2.7.6}), (\ref{2.7.7}), and taking
into account relations (\ref{2.7.4}), (\ref{2.7.5}) and
\begin{eqnarray*}
{\hat{\Lambda}}^{2}\,\psi^{hsp} = \lambda(\lambda + 3)\,\psi^{hsp}, \qquad
{\hat{M}}_{0}\,\psi^{par} = \frac{2e^{2}\sqrt{\mu_{0}}}{\hbar}\,\,
\frac{n_{1} - n_{2} + J - L}{N+4}\,\psi^{par},
\end{eqnarray*}
we obtain the following systems of linear homogeneous equations:
\begin{eqnarray}
\left[A_{q}(R) - \lambda(\lambda + 3)\right]\,
U_{NqJL}^{\lambda} (R)= \frac{R\sqrt{\mu_{0}}}{\hbar}\,
\sum\limits_{\lambda' = J + L}^{N/2}\,U_{NqJL}^{\lambda'}(R)\,\left(\hat{M}_{0}\right)_{\lambda \lambda'},
\label{2.7.8}
\end{eqnarray}
\begin{eqnarray}
\left[A_{q}(R) - \frac{2R}{r_{0}(N+4)}\,\left(n_{1} - n_{2} + J - L\right)\right]\,
V_{NqJL}^{n_{1}}(R) = \sum\limits_{n_{1}' = 0}^{\frac{N}{2}-J-L}\,V_{NqJL}^{n_{1}'}(R)\,
\left({\hat{\Lambda}}^{2}\right)_{n_{1} n_{1}'}.
\label{2.7.9}
\end{eqnarray}
Here
\begin{eqnarray*}
\left(\hat{M}_{0}\right)_{\lambda \lambda'} = \int\,\psi_{\lambda}^{hsp*}\,
{\hat{M}}_{0}\,\psi_{\lambda'}^{hsp}\,dv, \qquad
\left({\hat{\Lambda}}^{2}\right)_{n_{1} n_{1}'} =
\int\,\psi_{n_{1}}^{par*}\,{\hat{\Lambda}}^{2}\,\psi_{n_{1}'}^{par}\,dv.
\end{eqnarray*}
Then, using expansions (\ref{2.6.1}), (\ref{2.6.7}) and formulas (\ref{1.8.10}),
(\ref{1.8.11}), (\ref{1.8.15}) and (\ref{1.8.16}) for the Clebsch-Gordan coefficients, we obtain
\begin{eqnarray*}
\left(\hat{M}_{0}\right)_{\lambda \lambda'} = -\frac{e^{2}\sqrt{\mu_{0}}}{\hbar (N+4)}\,
\left[\frac{(L-J)(L+J+1)(N+4)}{(\lambda + 1)(\lambda + 2)}\,\delta_{\lambda \lambda'} +
B_{NJL}^{\lambda + 1}\,\delta_{\lambda', \lambda + 1} +
B_{NJL}^{\lambda}\,\delta_{\lambda', \lambda - 1}\right],
\end{eqnarray*}
\begin{eqnarray*}
\left({\hat{\Lambda}}^{2}\right)_{n_{1} n_{1}'} &=&
\left[n_{2}(n_{1}+1) + (n_{1}+2J+1)(n_{2}+2L+1) - (L-J)(L-J+1) - 2\right]\,
\delta_{n_{1} n_{1}'} - \\ [4mm]
&-& \left[n_{1}(n_{1}+2J+1)(n_{2}+1)(n_{2}+2L+2)\right]^{1/2}\,
\delta_{n_{1}', n_{1}-1} - \\ [4mm]
&-& \left[n_{2}(n_{2}+2L+1)(n_{1}+1)(n_{1}+2J+2)\right]^{1/2}\,
\delta_{n_{1}', n_{1}+1},
\end{eqnarray*}
where
\begin{eqnarray*}
B_{NJL}^{\lambda} = \frac{\sqrt{(N-2\lambda + 2)(N+2\lambda + 6)}}{\lambda + 1}\,\times \\ [3mm]
\times \left[\frac{(\lambda + J + L + 2)(\lambda - J - L)(\lambda - J + L + 1)
(\lambda + J - L + 1)}{(2\lambda + 1)(2\lambda + 3)}\right]^{1/2}.
\end{eqnarray*}
After substituting the expressions of the matrix elements into (\ref{2.7.8}) and
(\ref{2.7.9}), we finally obtain the following three-term recurrence relations:
\begin{eqnarray*}
&&\left[A_{q}(R) - \lambda(\lambda + 3) + \frac{R}{r_{0}}\,
\frac{(L-J)(L+J+1)}{(\lambda + 1)(\lambda + 2)}\right]\,
U_{NqJL}^{\lambda}(R) + \\ [5mm]
&+& \frac{R}{r_{0}(N+4)}\,
\left[B_{NJL}^{\lambda + 1}\,U_{NqJL}^{\lambda + 1}(R)
+ B_{NJL}^{\lambda}\,U_{NqJL}^{\lambda - 1}(R)\right] = 0,
\end{eqnarray*}
\begin{eqnarray*}
&&\Bigl[n_{2}(n_{1}+1) + (n_{1}+2J+1)(n_{2}+2L+1) - (L-J)(L-J+1) - 2 + \\ [3mm]
&+&\frac{2R}{r_{0}(N+4)}(4n_{1} - N + 4J)- A_{q}(R)\Bigr]\,
V_{NqJL}^{n_{1}}(R) =  \\ [5mm]
&=& \left[n_{1}(n_{1}+2J+1)(n_{2}+1)(n_{2}+2L+2)\right]^{1/2}\,V_{NqJL}^{n_{1}+1}(R) + \\ [3mm]
&+& \left[n_{2}(n_{2}+2L+1)(n_{1}+1)(n_{1}+2J+2)\right]^{1/2}\,V_{NqJL}^{n_{1}-1}(R),
\end{eqnarray*}
which must be solved together with the normalization conditions:
\begin{eqnarray*}
\sum\limits_{\lambda = J + L}^{N/2}\,\left|U_{NqJL}^{\lambda}(R)\right|^{2} =1, \qquad
\sum\limits_{n_{1} = 0}^{\frac{N}{2}-J-L}\,\left|V_{NqJL}^{n_{1}}(R)\right|^{2} = 1.
\end{eqnarray*}
The following limit formulas hold:
\begin{eqnarray*}
\lim_{R \to 0}\,U_{NqJL}^{\lambda}(R) &=& \delta_{\lambda q}, \qquad
\lim_{R \to \infty}\,U_{NqJL}^{\lambda}(R) =\tilde{W}_{NJL\lambda}^{n_{1}}, \\ [5mm]
\lim_{R \to 0}\,V_{NqJL}^{n_{1}} &=& W_{NJLn_{1}}^{\lambda}, \qquad
\lim_{R \to \infty}\,V_{NqJL}^{n_{1}} = \delta_{q n_{1}}.
\end{eqnarray*}
Note that at $T=0$, i.e. at $J=L$, all the formulas we obtained in section 1.11 are reproduced.

\section{Hidden symmetry of the $SU(2)$ YCM}
\markboth{CHAPTER 2. DYON-OSCILLATOR DUALITY}{2.8. HIDDEN SYMMETRY OF THE $SU(2)$ YCM}

In this section we analyze the properties of symmetry $SU(2)$ YCM \cite{M-13}.

Before moving on to consider the issue of hidden symmetry, we prove that the group $SO(5)$
is the kinematic symmetry group of the $SU(2)$ YCM.

Let us write the Hamiltonian of this system in the form
\begin{eqnarray}
\hat{H} = \frac{1}{2\mu_{0}}\,\hat{\pi}^{2} + \frac{\hbar^{2}}{2\mu_{0}r^{2}}\,
\hat{\bf T}^{2} - \frac{e^{2}}{r}.
\label{2.8.1}
\end{eqnarray}
Here $\hat{\pi}^{2} =  \hat{\pi}_{j} \hat{\pi}_{j}$, and
\begin{eqnarray*}
\hat{\pi}_{j} = -i\hbar\,\frac{\partial}{\partial x_{j}} - \hbar\,A_{j}^{a(+)}\hat{T}_{a}
\end{eqnarray*}
and the following fundamental commutation relations are valid:
\begin{eqnarray}
\left[\hat{\pi}_{i},\,x_{j}\right] = -i\hbar\,\delta_{i j}, \qquad
\left[\hat{\pi}_{i},\,\hat{\pi}_{j}\right] = i\hbar^{2}\,F_{i j}^{a(+)}\,\hat{T}_{a},
\qquad
\left[\hat{\pi}_{i},\,\hat{\pi}^{2}\right] = 2i\hbar^{2}\,F_{i j}^{a(+)}\,\hat{T}_{a}\,\hat{\pi}_{j}.
\label{2.8.2}
\end{eqnarray}
Now consider the operator
\begin{eqnarray*}
\hat{L}_{i k} = \frac{1}{\hbar}\,\left(x_{i}\,\hat{\pi}_{k} - x_{k}\,\hat{\pi}_{i}\right) -
r^{2}\,F_{i k}^{a(+)}\,\hat{T}_{a}.
\end{eqnarray*}
It's easy to verify that
\begin{eqnarray}
\left[\hat{L}_{i k},\,x_{j}\right] = i\,\delta_{i j}\,x_{k} - i\,\delta_{k j}\,x_{i}, \qquad
\left[\hat{L}_{i k},\,\hat{\pi}_{j}\right] = i\delta_{i j}\,\hat{\pi}_{k} - i\delta_{k j}\,\hat{\pi}_{i}.
\label{2.8.3}
\end{eqnarray}
Next, using relations (\ref{2.8.3}) we obtain the commutation rule for generators of the group $SO(5)$
\begin{eqnarray}
\left[\hat{L}_{i j},\,\hat{L}_{m n}\right] = i\,\delta_{i m}\,\hat{L}_{j n} - i\,\delta_{j m}\,\hat{L}_{i n}
- i\,\delta_{i n}\,\hat{L}_{j m} + i\,\delta_{j n}\,\hat{L}_{i m}.
\label{2.8.4}
\end{eqnarray}
In addition, from (\ref{2.8.3}) it follows that the tensor L commutes with the Hamiltonian $\hat{H}$.
The group $SO(5)$ was previously proposed by Yang \cite{YANG-1} as a dynamic symmetry group for
the Hamiltonian, which includes only the monopole - isospin interaction.

Now, by analogy with the multidimensional Runge-Lenz vector \cite{revai}, consider the operator
\begin{eqnarray}
\hat{M}_{k} = \frac{1}{2\sqrt{\mu_{0}}}\,\left(\hat{\pi}_{i}\,\hat{L}_{i k} +
\hat{L}_{i k}\,\hat{\pi}_{i} + \frac{2\mu_{0}e^{2}}{\hbar}\,\frac{x_{k}}{r}\right).
\label{2.8.5}
\end{eqnarray}
Using formulas (\ref{2.1.10}) – (\ref{2.1.13}), (\ref{2.8.2}) and (\ref{2.8.3}), the following
commutation relations can be calculated:
\begin{eqnarray*}
\left[\hat{M}_{k},\,\hat{\pi}_{j}\right] = \frac{1}{\sqrt{\mu_{0}}}\,\left[i\,
\hat{\pi}_{k}\,\hat{\pi}_{j} - i\delta_{k j}\,\hat{\pi}^{2} -
i\hbar\,x_{k}\,F_{i j}^{a(+)}\,\hat{T}_{a}\,\hat{\pi}_{i} +
i\,\left(\frac{\hbar^{2}}{r}\,\hat{\bf T}^{2} - \mu_{0}e^{2}\right)\,
\left(\frac{x_{k}\,x_{j}}{r^{3}} - \frac{1}{r}\,\delta_{k j}\right)\right],
\end{eqnarray*}
\begin{eqnarray*}
\left[\hat{M}_{k},\,\frac{1}{r}\right] = \frac{\hbar}{\sqrt{\mu_{0}}}\,
\left(i\,\frac{x_{i}}{r^{3}}\,\hat{L}_{i k} -\frac{2x_{k}}{r^{3}}\right),
\end{eqnarray*}
which make it possible to establish that $\left[\hat{H},\,\hat{M}_{k}\right] = 0$, i.e.
$\hat{M}_{k}$ – integral of motion. Now using relations (\ref{2.8.3}) and (\ref{2.8.4}) we can show that
\begin{eqnarray*}
\left[\hat{L}_{i j},\,\hat{M}_{k}\right] =
i\,\delta_{i k}\,\hat{M}_{j} - i\,\delta_{j k}\,\hat{M}_{i}.
\end{eqnarray*}
More complex calculations lead to the formula
\begin{eqnarray*}
\left[\hat{M}_{i},\,\hat{M}_{k}\right] = -2i\,\hat{H}\,\hat{L}_{i k} - \frac{i}{\mu_{0}}\,
x_{i}\,x_{k}\,F_{mn}^{a(+)}\,\hat{T}_{a}\,\hat{\pi}_{m}\,\hat{\pi}_{n} -
\frac{2\hbar^{2}}{\mu_{0}}\,\frac{x_{i}\,x_{k}}{r^{4}}\,\hat{\bf T}^{2}.
\end{eqnarray*}
Taking into account (\ref{2.1.12}) and (\ref{2.8.3}) it is easy to check that
the last two terms cancel each other and, therefore,
\begin{eqnarray}
\left[\hat{M}_{i},\,\hat{M}_{k}\right] = -2i\,\hat{H}\,\hat{L}_{i k}.
\label{2.8.6}
\end{eqnarray}
This commutation rule generalize relations known from the theory of the Coulomb
problem \cite{LL}. For the operator $\hat{M}_{j}^{'} = \left(-2\,\hat{H}\right)^{-1/2}\,\hat{M}_{j}$
we have
\begin{eqnarray}
\left[\hat{M}_{i}^{'},\,\hat{M}_{k}^{'}\right] = i\,\hat{L}_{i k}.
\label{2.8.7}
\end{eqnarray}
Finally, let us introduce the $6\times 6$ matrix
\begin{eqnarray*}
\hat{D} =
\left(
\begin{array}{cc}
\hat{L}_{i j}&-\hat{M}_{j}^{'}\\
\hat{M}_{j}^{'}&0\\
\end{array}
\right),
\end{eqnarray*}
The components $\hat{D}_{\mu \nu}$ (here $\mu, \nu = 0,1,...,5$) satisfy
the commutations relations
\begin{eqnarray*}
\left[\hat{D}_{\mu \nu},\,\hat{D}_{\lambda \rho}\right] = i\,\delta_{\mu \lambda}\,\hat{D}_{\nu \rho} -
i\,\delta_{\nu \lambda}\,\hat{D}_{\mu \rho}
- i\,\delta_{\mu \rho}\,\hat{D}_{\nu \lambda} + i\,\delta_{\nu \rho}\,\hat{D}_{\mu \lambda},
\end{eqnarray*}
i.e. $\hat{D}_{\mu \nu}$ are the generators of the group $SO(6)$. Since
$\left[\hat{H},\,\hat{D}_{\mu \nu}\right] = 0$, one concludes that $SU(2)$
YCM is provided by the $SO(6)$ group of hidden symmetry.

Now knowing the group of hidden symmetry, we can algebraically calculate the energy eigenvalues.

It is known \cite{BARUT-RON} that the Casimir operators for the group $SO(6)$ have the form
\begin{eqnarray*}
\hat{C}_{2} &=& \frac{1}{2}\,\hat{D}_{\mu \nu}\,\hat{D}_{\mu \nu}, \\ [2mm]
\hat{C}_{3} &=& \epsilon_{\mu \nu \rho \sigma \tau \lambda}\,\hat{D}_{\mu \nu}\,\hat{D}_{\rho \sigma}\,
\hat{D}_{\tau \lambda}, \\ [2mm]
\hat{C}_{4} &=& \frac{1}{2}\,\hat{D}_{\mu \nu}\,\hat{D}_{\nu \rho}
\hat{D}_{\rho \tau}\,\hat{D}_{\tau \mu}.
\end{eqnarray*}
According to \cite{PER-POP2}, the eigenvalues of these operators can be represented as
\begin{eqnarray*}
C_{2} &=& \mu_{1}\,(\mu_{1} + 4) + \mu_{2}\,(\mu_{2} + 2) + \mu_{3}^{2}, \\ [2mm]
C_{3} &=& 48\,(\mu_{1} + 2)\,(\mu_{2} + 1)\,\mu_{3},  \\ [2mm]
C_{4} &=& \mu_{1}^{2}\,(\mu_{1} + 4)^{2} +6\mu_{1}\,(\mu_{1} + 4) +
\mu_{2}^{2}\,(\mu_{2} + 2)^{2} + \mu_{3}^{4} -2\mu_{3}^{2},
\end{eqnarray*}
where $\mu_{1}, \mu_{2}$ and $\mu_{3}$ are positive integers or half-integers
and $\mu_{1}\geq \mu_{2} \geq \mu_{3}$.

Direct calculations lead to the representation
\begin{eqnarray}
\hat{C}_{2} &=& -\frac{\mu_{0}e^{4}}{2\hbar^{2}\hat{H}} + 2\hat{\bf T}^{2} - 4, \nonumber \\[2mm]
\hat{C}_{3} &=& 48\,\left(-\frac{\mu_{2}e^{4}}{2\hbar^{2}\hat{H}}\right)^{1/2}\,\hat{\bf T}^{2}, \\ [2mm]
\hat{C}_{4} &=& \hat{C}_{2}^{2} + 6\hat{C}_{2} - 4\,\hat{C}_{2}\,\hat{\bf T}^{2} -
12\hat{\bf T}^{2} + 6\hat{\bf T}^{4}. \nonumber
\label{2.8.8}
\end{eqnarray}
From the last equations we can find another expression for the eigenvalues of $\hat{C}_{4}$
\begin{eqnarray*}
C_{4} = \left[C_{2} - 2T(T+1)\right]^{2} + 6\left[C_{2} - 2T(T+1)\right] +
2T^{2}(T+1)^{2}
\end{eqnarray*}
and calculate that
\begin{eqnarray}
C_{2} - 2T(T+1) = \mu_{1}\,(\mu_{1} + 4),
\label{2.8.9}
\end{eqnarray}
\begin{eqnarray}
\mu_{2}^{2}\,(\mu_{2} + 2)^{2} + \mu_{3}^{4} -2\mu_{3}^{2} = 2T^{2}(T+1)^{2}.
\label{2.8.10}
\end{eqnarray}
The energy levels of the $SU(2)$ YCM can be found from (\ref{2.8.8}) and (\ref{2.8.9})
\begin{eqnarray}
\varepsilon_{N}^{T} = - \frac{\mu_{0}e^{4}}{2\hbar^{2}\,\left(\frac{N}{2} + 2\right)^{2}},
\label{2.8.11}
\end{eqnarray}
where $\mu_{1} = N/2$, $N$ is a non-negative integer.
The substitution of the eigenvalues of $\hat{H}$ and $T^{2}$ into the equation for
$\hat{C}_{3}$ gives more formula for $C_{3}$
\begin{eqnarray*}
C_{3} = 48\,(\mu_{1} + 4)\,T(T+1).
\end{eqnarray*}
Now, we have two expressions for $C3$ and comparing them leads to the relation
\begin{eqnarray}
T(T+1) = (\mu_{2} + 1)\,\mu_{3}.
\label{2.8.12}
\end{eqnarray}
Comparing (\ref{2.8.12}) with (\ref{2.8.10}) we obtain the equation
\begin{eqnarray*}
\left(\mu_{2}^{2} - \mu_{3}^{2}\right)\,\left[(\mu_{2} + 1)^{2} - \mu_{3}^{2}\right] = 0.
\end{eqnarray*}
Since $\mu_{3} \leq \mu_{2}$, we have that $\mu_{3} = \mu_{2}$. Then from
(2.8.12) it follows that $\mu_{2} = T$. Thus, $N$ in formula (\ref{2.8.11})
can only take the values $N/2=T,T+1,T+2,…$.

The quantum number $T$ fixes the type of the YCS: the bosonic type for $T =
0, 1, 2, \dots$, and the fermionic type for $T = 1/2, 3/2, \dots$.
The bosonic type of YCS with $T = 0$ is the five–dimensional Coulomb system.

Let us complete the present consideration of YCM by three remarks:

$\bullet$ It is know \cite{ALLILUEV} that the five--dimensional Coulomb problem is provided
with the hidden $SO(6)$ symmetry. From this position it is argued that the inclusion of the
$SU(2)$ gauge field does not violate the hidden symmetry of the original system.
The same phenomenon takes place for the Abelian particle - charge system \cite{ZWANZIG}.

$\bullet$ The elegant approach presented above to solve the YCM eigenvalue problem has  been used for
the same purpose in connection with the so--called $SU(2)$ Kepler problem \cite{TRUNK}.

$\bullet$ At the end of this section, we note that for a continuous spectrum
$(\varepsilon > 0)$ from formula (\ref{2.8.6}) instead of the commutation relation
(\ref{2.8.7}) we obtain
\begin{eqnarray*}
\left[\hat{M}_{i}^{''},\,\hat{M}_{k}^{''}\right] = -i\,\hat{L}_{i k}, \qquad
{\rm where} \quad
\hat{M}_{i}^{''} = \left(2\,\hat{H}\right)^{-1/2}\,\hat{M}_{i},
\end{eqnarray*}
which indicates that in the continuous spectrum the $SU(2)$ YCM has a hidden
symmetry group $SO(5,1)$.

\section{Scattering problem in the field of $SU(2)$ YCM}
\markboth{CHAPTER 2. DYON-OSCILLATOR DUALITY}{2.9. SCATTERING PROBLEM IN
THE FIELD $SU(2)$-INTERACTION}

It is convenient to consider the problem of scattering of charged particles on the
$SU(2)$ YCM for vector potentials $B_{i}^{a(+)}$ (\ref{2.1.14}). First, we present
explicit expressions for the hyperspherical and parabolic bases of the continuous
spectrum \cite{M-14}.

The hyperspherical basis of the $SU(2)$ YCM in a continuous spectrum can be represented as:
\begin{eqnarray}
\psi^{hsp} = R_{k \lambda}(r)\,Z_{\lambda LJ}(\theta) \,
{\cal {D}}^{JM}_{LTm't'\,}(\alpha, \beta, \gamma,
\alpha_{T}, \beta_{T}, \gamma_{T}).
\label{2.9.1}
\end{eqnarray}
Here, normalized by the condition
\begin{eqnarray*}
\int\limits_{0}^{\infty}\,r^{4}\,R_{k^{'} \lambda}(r)\,R_{k \lambda}(r)\,
dr = 2\pi\,\delta(k - k'),
\end{eqnarray*}
the radial wave function has the form
\begin{eqnarray*}
R_{k \lambda}(r) = C_{k \lambda}\,\frac{(2ikr)^{\lambda}}{(2\lambda + 3)!}\,
e^{-ikr}\,F\left(\lambda + 2 + \frac{i}{kr_{0}};\, 2\lambda + 4;\, 2ikr\right),
\end{eqnarray*}
where
\begin{eqnarray*}
 C_{k \lambda} = (-i)^{\lambda} \,4k^{2}\,e^{\pi/2kr_{0}}\,
\left|\Gamma\left(\lambda + 2 - \frac{i}{kr_{0}}\right)\right|.
\end{eqnarray*}

The parabolic wave function has the form
\begin{eqnarray}
\psi^{par} = (-i)^{L+J}\,C_{k \Omega}^{JL}\,f_{k \Omega}^{L}(\mu)\,f_{k, -\Omega}^{J}(\nu)\,
{\cal {D}}^{JM}_{LTm't'\,}(\alpha, \beta, \gamma,
\alpha_{T}, \beta_{T}, \gamma_{T}),
\label{2.9.2}
\end{eqnarray}
where
\begin{eqnarray*}
f_{k \Omega}^{q}(x) = \frac{(ikx)^{q}}{(2q+1)!}\,e^{-ikx/2}\,
F\left(q + 1 + \frac{i}{2kr_{0}} + i\frac{\sqrt{\mu_{0}}}{2\hbar k}\,\Omega;\, 2q + 2;\, ikx\right),
\end{eqnarray*}
\begin{eqnarray*}
C_{k \Omega}^{JL} = \sqrt{\frac{\hbar k^{3}}{2\pi \sqrt{\mu_{0}}}}\,
e^{\pi/2kr_{0}}\,
\left|\Gamma\left(J + 1 - \frac{i}{2kr_{0}} - i\frac{\sqrt{\mu_{0}}}{2\hbar k}\,\Omega\right)\,
\Gamma\left(L + 1 - \frac{i}{2kr_{0}} + i\frac{\sqrt{\mu_{0}}}{2\hbar k}\,\Omega\right)\right|.
\end{eqnarray*}
When finding the normalization factor $C_{k \Omega}^{JL}$, we used the following normalization condition:
\begin{eqnarray*}
\frac{1}{4}\,\int\,\Phi_{k' \Omega'}^{JL*}(\mu,\nu)\,\Phi_{k \Omega}^{JL}(\mu,\nu)\,
\mu \nu\,(\mu + \nu)\,d\mu\,d\nu = 2\pi\,\delta(k - k')\,\,\delta(\Omega - \Omega'),
\end{eqnarray*}
where $\Phi_{k \Omega}^{JL}(\mu,\nu) = f_{k \Omega}^{L}(\mu)\,f_{k, -\Omega}^{J}(\nu)$.

Now consider the scattering of charged particles in the field of the $SU(2)$ YCM \cite{M-15}.
Since the $SU(2)$ YCM is a Coulomb-like system, and motion in a Coulomb field of arbitrary
dimension $d \geq 3$ is a two-dimensional problem, the wave function of a charged particle
should not depend on the angles $\alpha, \beta$ and $\gamma$, i.e. $L=0$, and $J=T$.
Then the system of equations (\ref{2.5.2}) takes the form
\begin{eqnarray}
\frac{1}{\mu}\,\frac{d}{d \mu}\,
\left(\mu^{2}\frac{d f_{1}}{d \mu}\right) +
\left[\frac{k^{2}}{4}\mu  +
\frac{\sqrt{\mu_{0}}}{2\hbar}\Omega + \frac{1}{2r_{0}}\right]\,
f_{1} = 0, \cr
\frac{1}{\nu}\,\frac{d}{d \nu}\,
\left(\nu^{2}\frac{d f_{2}}{d \nu}\right) +
\left[\frac{k^{2}}{4}\nu - \frac{T(T+1)}{\nu} -
\frac{\sqrt{\mu_{0}}}{2\hbar}\Omega + \frac{1}{2r_{0}}\right]\,
f_{2} = 0,
\label{2.9.3}
\end{eqnarray}

Further, as in the case of a five-dimensional Coulomb field, we must find solutions to equation
(\ref{2.9.3}) such that the solution to the Schr\"{o}dinger equation for negative
$x_{0}\in (-\infty; 0)$ and  $r \to \infty$ looks like a plane wave
\begin{eqnarray*}
\Phi(\mu,\nu) \approx e^{-ikx_{0}} = e^{ik(\mu - \nu)/2}.
\end{eqnarray*}
This condition can be satisfied if we assume that the parabolic separation constant is equal to
\begin{eqnarray}
\Omega = - \frac{\hbar}{r_{0}\sqrt{\mu_{0}}}
-i\frac{2\hbar k}{\sqrt{\mu_{0}}}.
\label{2.9.4}
\end{eqnarray}

Substituting the last relation in (\ref{2.9.3}) we find the following solution to the Schr\"{o}dinger
equation, which describes the scattering of charged particles in the field of the $SU(2)$ YCM:
\begin{eqnarray*}
\psi = \Phi_{k T}(\mu,\nu)\,{\cal {D}}^{T}_{t t'\,}(\alpha_{T}, \beta_{T}, \gamma_{T}).
\end{eqnarray*}
Here
\begin{eqnarray}
\Phi_{k T}(\mu,\nu) = C_{kT}\,e^{ik(\mu - \nu)/2}\,(ik\nu)^{T\,}
F\left(T + \frac{i}{kr_{0}};\, 2T + 2;\, ik\nu\right),
\label{2.9.5}
\end{eqnarray}
where $C_{kT}$ is the normalization constant. In order to distinguish the incident and
scattered waves in function (\ref{2.9.5}), it is necessary to consider its behavior at
large distances from the scattering center.
Using the representation of the confluent hypergeometric function (\ref{1.13.3}) and
limiting ourselves to the first two terms for large $\nu$, we obtain
\begin{eqnarray*}
F\left(T + \frac{i}{kr_{0}};\, 2T + 2;\, ik\nu \right) \approx
e^{-\pi/2kr_{0}}\,(ik\nu)^{T}\,\Biggl\{\frac{e^{i\pi T/2}}{\Gamma\left(T + 2 - \frac{i}{kr_{0}}\right)}\times \\ [2mm]
\times \left[1 + \frac{1 + ikr_{0} + k^{2}r_{0}^{2}\,T(T+1)}{ik^{3}r_{0}^{2}\,\nu} -
\frac{1 + k^{2}r_{0}^{2}\,(T+1)^{2}}{2k^{6}r_{0}^{4}\,\nu^{2}}\right]\exp\left(-i\,\frac{\ln\,k\nu}{kr_{0}}\right)- \\ [2mm]
- \frac{\left(kr_{0}\,T + i\right)\left[kr_{0}\left(T + 1\right) + i\right]}
{\Gamma\left(T + 2 + \frac{i}{kr_{0}}\right)}\,\frac{e^{ik\nu}}{k^{4}r_{0}^{2}\,\nu^{2}}
\,\exp\left(i\,\frac{\ln\,k\nu}{kr_{0}}\right)\Biggr\}.
\end{eqnarray*}
Now substituting the last relation into the wave function (\ref{2.9.5}),
choosing the normalization constant $C_{kT}$ in the form
\begin{eqnarray*}
C_{kT} = \frac{\Gamma\left(T + 2 - \frac{i}{kr_{0}}\right)}{\Gamma\left(2T + 2\right)}\,
\exp\left[\pi\,\left(\frac{1}{2kr_{0}} - i\,\frac{T}{2}\right)\right],
\end{eqnarray*}
so that the incident plane wave has a unit amplitude, and passing to hyperspherical coordinates
(\ref{1.9.2}) according to formulas (\ref{1.11.2}), we obtain
\begin{eqnarray*}
\Phi_{k T}(\mu,\nu) =
\left\{1 - \frac{1 + ikr_{0} + k^{2}r_{0}^{2}\,T(T+1)}{2k^{3}r_{0}^{2}\,r\,\sin^{2}\theta/2} -
\frac{\left(1 + k^{2}r_{0}^{2}\,T^{2}\right)\left[1 + k^{2}r_{0}^{2}\,(T+1)^{2}\right]}
{8k^{6}r_{0}^{4}\,r^{2}\,\sin^{2}\theta/2}\right\}\,\times \\ [2mm]
\times \exp\left[i\,k\,x_{0} - \frac{i}{kr_{0}}\,\ln\,\left(2kr\sin^{2}\theta\right)\right]
+ \frac{f(\theta)}{r^{2}}\,
\exp\left[ikr + \frac{i}{k\,r_{0}}\,\ln\,\left(2kr\right)\right],
\end{eqnarray*}
where the scattering amplitude $f(\theta)$ has the form
\begin{eqnarray*}
f(\theta) = - \frac{\left(kr_{0}\,T + i\right)\left[kr_{0}\left(T + 1\right) + i\right]}
{4k^{4}r_{0}^{2}\,r\,\sin^{4}\theta/2}\,\,\frac{\Gamma\left(T + 2 - \frac{i}{kr_{0}}\right)}
{\Gamma\left(T + 2 + \frac{i}{kr_{0}}\right)}\,\,
\exp\left(\frac{2i}{kr_{0}}\,\ln\,\sin\frac{\theta}{2} - i\frac{\pi\,T}{2}\right).
\end{eqnarray*}
So, for the scattering cross section we obtain the following formula:
\begin{eqnarray*}
d\,\sigma = \left|f(\theta)\right|^{2}\,d\,\Omega =
\frac{\left(1 + k^{2}r_{0}^{2}\,T^{2}\right)\left[1 + k^{2}r_{0}^{2}\,(T+1)^{2}\right]}
{16k^{8}r_{0}^{2}\,\sin^{8}\theta/2}\,d\,\Omega.
\end{eqnarray*}

At the end, we note that all the formulas obtained in this section at $T = 0$
go into the corresponding formulas in paragraph $1.12$.

\section{Interbasis expansions in the continuous spectrum}
\markboth{CHAPTER 2. DYON-OSCILLATOR DUALITY}{2.10. INTERBASIS EXPANSIONS IN THE CONTINUOUS SPECTRUM}

First, let us consider the expansion of the parabolic basis of the continuous spectrum of the $SU(2)$ YCM
in terms of the hyperspherical basis. Since the hypermoment $\lambda$ takes both integer and half-integer
values and is bounded from below by the inequality $\lambda \geq T$, the required expansion can be written in the form
\begin{eqnarray}
\psi^{par} = \sum\limits_{\lambda = T}^{\infty}\,W_{k \Omega J L}^{\lambda}\,\psi^{hsp}.
\label{2.10.1}
\end{eqnarray}
Let us multiply both sides of the expansion (\ref{2.10.1}) by $\sin^{3}\theta\,Z_{\lambda' J L}(\theta)$,
integrate over the angle $\theta$ and use the orthonormalization condition (\ref{2.4.3}).
Further, we present the confluent hypergeometric functions included in the parabolic wave functions in the form of series
\begin{eqnarray}
F(a; c; z) = \sum\limits_{s = 0}^{\infty}\,\frac{(a)_{s}\,z^{s}}{s!\,(c)_{s}},
\label{2.10.2}
\end{eqnarray}
where
\begin{eqnarray*}
(a)_{s} = \frac{\Gamma(a+s)}{\Gamma(a)}
\end{eqnarray*}
is the Pochhammer's symbol, and let us move from parabolic coordinates to hyperspherical
ones according to relation (\ref{1.11.2}). Then instead of transformation (\ref{2.10.1}) we will have
\begin{eqnarray*}
W_{k \Omega J L}^{\lambda}\,F\left(\lambda + 2 +\frac{i}{kr_{0}}; 2\lambda + 4; 2ikr\right) =
\frac{(2\lambda + 3)!}{2^{\lambda}\,(2J + 1)!\, (2L + 1)!}\,\frac{C_{k \Omega}^{J L}}{C_{k \lambda}}\times \\ [2mm]
\times \sum\limits_{s = 0}^{\infty}\,\sum\limits_{t = 0}^{\infty}\,\frac{(u)_{s}\,(v)_{t}}{(2J + 2)_{s}\,(2L + 2)_{t}}\,
\frac{i^{s+t+J+L-\lambda}}{s!\,t!}\,(kr)^{s+t+J+L-\lambda}\,Q_{\lambda J L}^{s t},
\end{eqnarray*}
where
\begin{eqnarray*}
u = J + 1 + \frac{i}{2kr_{0}} + i\,\frac{\sqrt{\mu_{0}}}{2\hbar k}\,\Omega, \qquad
v = L + 1 + \frac{i}{2kr_{0}} - i\,\frac{\sqrt{\mu_{0}}}{2\hbar k}\,\Omega,
\end{eqnarray*}
and
\begin{eqnarray}
Q_{\lambda J L}^{s t} = \int\limits_{0}^{\pi}\,\sin^{3}\theta\,
(1 + \cos\theta)^{J+s}\,(1 - \cos\theta)^{L+t}\,Z_{\lambda J L}(\theta)\,d\theta.
\label{2.10.3}
\end{eqnarray}
Now using Rodrigues' formula for Jacobi polynomials \cite{SEGO}
\begin{eqnarray}
P_{n}^{(a,b)} = \frac{(-1)^{n}}{2^{n}\,n!}\,(1-x)^{-a}\,(1+x)^{-b}\,
\frac{d^{n}}{d x^{n}}\,(1-x)^{a+n}\,(1+x)^{b+n}
\label{2.10.4}
\end{eqnarray}
and sequentially integrating by parts we are convinced that $Q_{\lambda J L}^{s t}$
is nonzero only under the condition $s+t+J+L- \lambda \leq 0$ and therefore all terms
of the series contain r to a non-negative power, so in the limit $r \to 0$ we obtain
\begin{eqnarray}
&&W_{k \Omega J L}^{\lambda} =
\frac{(2\lambda + 3)!}{2^{\lambda}\,(2J + 1)!\, (2L + 1)!}\,\frac{C_{k \Omega}^{J L}}{C_{k \lambda}}\times
\nonumber \\
\label{2.10.5}
\\
&\times&\,\sum\limits_{s = 0}^{\lambda - J - L}\,\frac{(u)_{s}\,(v)_{\lambda - J - L - s}}{(2J + 2)_{s}\,
(2L + 2)_{\lambda - J - L - s}}\,\frac{Q_{\lambda J L}^{s, \lambda - J - L - s}}{s!\,(\lambda - J - L - s)!}.
\nonumber
\end{eqnarray}
The integral $Q_{\lambda J L}^{s t}$ at $t = \lambda - J - L - s$ becomes a closed expression
\begin{eqnarray*}
&&Q_{\lambda J L}^{s, \lambda - J - L - s} = (-1)^{\lambda - J - L + s}\,
\frac{2^{\lambda + 2}}{(2\lambda + 3)!}\,\sqrt{\frac{2\lambda + 3}{2}} \times \\ [2mm]
&\times& \sqrt{(\lambda - J - L)!\,(\lambda + J + L + 2)!\,(\lambda + J - L + 1)!\,
(\lambda - J + L + 1)!}.
\end{eqnarray*}
Substituting it into formula (\ref{2.10.5}) and taking into account the auxiliary equalities
\begin{eqnarray*}
(v)_{\lambda - J - L - s} &=& (-1)^{s}\,\frac{(v)_{\lambda - J - L}}
{(1 - \lambda + J + L - v)_{s}}, \\ [2mm]
(2L + 2)_{\lambda - J - L - s} &=& (-1)^{s}\,\frac{(\lambda - J + L +1)!}
{(2L + 1)!\,(-\lambda + J - L - 1)_{s}}, \\ [2mm]
(\lambda - J - L - s)! &=& (-1)^{s}\,
\frac{\lambda - J - L)!}{(-\lambda + J + L)_{s}}
\end{eqnarray*}
we will get
\begin{eqnarray*}
W_{k \Omega J L}^{\lambda} &=& \frac{(-1)^{\lambda - J - L}}{(2J + 1)!}\,
\frac{C_{k \Omega}^{J L}}{C_{k \lambda}}\,\sqrt{8\,(2\lambda + 3)
\frac{(\lambda + J + L + 2)!\,(\lambda + J - L + 1)!}{(\lambda - J - L)!\,
(\lambda - J + L + 1)!}}\times \\ [2mm]
&\times& (v)_{\lambda - J - L}\,
{_3F_2}\left\{\matrix{ -\lambda + J + L,\,\,-\lambda + J - L - 1,\,\, u\cr \cr
2J+2,\,\,1-\lambda + J - L - v
\cr}\Biggr|1\right\}.
\end{eqnarray*}
Further, using formula (\ref{1.7.11}), the last relation can be written in the form
\begin{eqnarray}
&&W_{k \Omega J L}^{\lambda} = \frac{(-1)^{\lambda - J - L}}{(2J + 1)!}\,
\frac{\Gamma\left(\lambda + 2 + \frac{i}{kr_{0}}\right)}
{\Gamma\left(J + L+ 2 + \frac{i}{kr_{0}}\right)}\,
\left[\frac{(\lambda + J + L + 2)!\,(\lambda + J - L + 1)!}
{(\lambda - J - L)!\,(\lambda - J + L + 1)!}\right]^{1/2}\times \nonumber \\
\label{2.10.6}
\\
&\times& \frac{4C_{k \Omega}^{J L}}{C_{k \lambda}}\,\sqrt{\lambda + \frac{3}{2}}\,
{_3F_2}\left\{\matrix{ -\lambda + J + L,\,\,\lambda + J + L + 3,\,\,
J + 1 + \frac{i}{2kr_{0}} + i\,\frac{\sqrt{\mu_{0}}}{2\hbar k}\,\Omega\cr \cr
2J+2,\,\, J + L + 2 + \frac{i}{2kr_{0}}
\cr}\Biggr|1\right\}.
\nonumber
\end{eqnarray}
Let us note that up to now we have not used an explicit forms of normalization constants
$C_{k \lambda}$ and $C_{k \Omega}^{J L}$. In this sense, formula (\ref{2.10.6}) is correct
for an arbitrary method of the wave function normalization. In our case, we have
\begin{eqnarray}
&&W_{k \Omega J L}^{\lambda} =(-i)^{\lambda - J - L}\,\frac{(2\lambda + 3)}{(2J + 1)!}\,
\sqrt{\frac{(\lambda + J + L + 2)!\,(\lambda + J - L + 1)!}
{(\lambda - J - L)!\,(\lambda - J + L + 1)!}}\,\times \nonumber \\ [2mm]
&\times& \sqrt{\frac{8\hbar}{\pi k\sqrt{\mu_{0}}}}\,
\frac{\left|\Gamma\left(J + 1 - \frac{i}{2kr_{0}} - i\,\frac{\sqrt{\mu_{0}}}{2\hbar k}\,\Omega\right)\,
\Gamma\left(L + 1 - \frac{i}{2kr_{0}} + i\,\frac{\sqrt{\mu_{0}}}{2\hbar k}\,\Omega\right)\right|}
{\Gamma\left(J + L+ 2 + \frac{i}{kr_{0}}\right)}\,\times \\ [2mm]
\label{2.10.7}
&\times& e^{-\delta_{\lambda}}\,\frac{4C_{k \Omega}^{J L}}{C_{k \lambda}}\,
{_3F_2}\left\{\matrix{ -\lambda + J + L,\,\,\lambda + J + L + 3,\,\,
J + 1 + \frac{i}{2kr_{0}} + i\,\frac{\sqrt{\mu_{0}}}{2\hbar k}\,\Omega\cr \cr
2J+2,\,\, J + L + 2 + \frac{i}{2kr_{0}}
\cr}\Biggr|1\right\},
\nonumber
\end{eqnarray}
where
\begin{eqnarray*}
\delta_{\lambda} = \arg\,\Gamma\left(\lambda + 2 + \frac{i}{kr_{0}}\right).
\end{eqnarray*}
Using formula (\ref{1.7.11}) we can verify that the coefficients
$W_{k \Omega J L}^{\lambda}$ are real.

Now consider the inverse expansion, i.e. expansion of the hyperspherical basis of the
$SU(2)$ YCM in a parabolic basis. Since the parabolic separation constant $\Omega$
can also take complex values, for example, in the scattering problem of charged particles on the $SU(2)$ YCM
(see formula (\ref{2.9.4})), it is certainly unclear over which $\Omega$ should be integrated
in the inverse expansion.

Consider the integral
\begin{eqnarray*}
Q_{\lambda \lambda'} = \int\limits_{-\infty}^{\infty}\,
W_{k \Omega J L}^{\lambda}\,W_{k \Omega J L}^{\lambda' *}\,d\Omega.
\end{eqnarray*}
Let us substitute expression (\ref{2.10.7}) into the integral instead of
$W_{k \Omega J L}^{\lambda}$ and move from the generalized hypergeometric function
$_3F_2$ to the finite sum through which it is expressed. Then, also replacing
$\Omega$ with $z=i\,\frac{\sqrt{\mu_{0}}}{2\hbar k}\,\Omega$, we get
\begin{eqnarray*}
Q_{\lambda \lambda'} = \frac{(-i)^{\lambda}\,i^{\lambda'}}{\left[(2J + 1)!\right]^{2}}\,
\frac{e^{i(\delta_{\lambda}' - \delta_{\lambda})}}
{\left|\Gamma\left(J + L + 2 - \frac{i}{kr_{0}}\right)\right|^{2}}\,
\sqrt{(2\lambda + 3)\,(2\lambda' + 3)}\,\times \\[3mm]
\times \left[\frac{(\lambda + J + L + 2)!\,(\lambda + J - L + 1)!\,
(\lambda' + J + L + 2)!\,(\lambda' + J - L + 1)!}
{(\lambda - J - L)!\,(\lambda' - J - L)!\,(\lambda - J + L + 1)!\,
(\lambda' - J + L + 1)!}\right]^{1/2}\,\times \\ [3mm]
\sum\limits_{s=0}^{\lambda - J - L}\,\frac{(-\lambda + J + L)_{s}
(\lambda + J + L + 3)_{s}}{s!\,(2J + 2)_{s}\,\left(J + L + 2 + \frac{i}{kr_{0}}\right)_{s}}\,
\sum\limits_{t=0}^{\lambda' - J - L}\,\frac{(-\lambda' + J + L)_{t}
(\lambda' + J + L + 3)_{t}}{t!\,(2J + 2)_{t}\,\left(J + L + 2 - \frac{i}{kr_{0}}\right)_{t}}\,
B_{s t},
\end{eqnarray*}
where
\begin{eqnarray*}
B_{s t}, &=& \frac{1}{2\pi i}\,\int\limits_{-i\infty}^{i\infty}\,
\Gamma\left(J + 1 + s + \frac{i}{2kr_{0}} + z\right)\,
\Gamma\left(J + 1 + t - \frac{i}{2kr_{0}} - z\right)\,\times \\ [3mm]
&\times& \Gamma\left(L + 1 - \frac{i}{2kr_{0}} + z\right)\,
\Gamma\left(L + 1 + \frac{i}{2kr_{0}} - z\right)\,dz.
\end{eqnarray*}
According to Barnes' lemma \cite{WITWAT}
\begin{eqnarray}
&&\frac{1}{2\pi i}\,\int\limits_{-i\infty}^{i\infty}\,
\Gamma\left(\alpha + s\right)\,\Gamma\left(\beta + s\right)\,\Gamma\left(\gamma - s\right)\,
\Gamma\left(\delta - s\right)\,ds = \nonumber \\
\label{2.10.8} \\
&=& \frac{\Gamma\left(\alpha + \gamma\right)\,\Gamma\left(\alpha + \delta\right)\,
\Gamma\left(\beta + \gamma\right)\,\Gamma\left(\beta + \delta\right)}
{\Gamma\left(\alpha + \beta + \gamma + \delta\right)},
\nonumber
\end{eqnarray}
if the poles of the expression $\Gamma\left(\gamma - s\right)\,\Gamma\left(\delta - s\right)$
lie to the right of the integration path, and the poles of the expression
$\Gamma\left(\alpha + s\right)\,\Gamma\left(\beta + s\right)$ lie to the left,
and none of the poles of the first set coincides with either one pole of the second set.
In our case, the requirements of this lemma are met, and therefore
\begin{eqnarray*}
Q_{\lambda \lambda'} = (-i)^{\lambda}\,i^{\lambda'}\frac{(2L + 1)!}{(2J + 1)!}\,
\sqrt{(2\lambda + 3)\,(2\lambda' + 3)\frac{(\lambda + J + L + 2)!\,(\lambda + J - L + 1)!}
{(\lambda - J - L)!\,(\lambda' - J - L)!}}\,\times \\ [3mm]
\sqrt{\frac{(\lambda' + J + L + 2)!\,(\lambda' + J - L + 1)!}
{(\lambda - J + L + 1)!\,(\lambda' - J + L + 1)!}}\,
\sum\limits_{s=0}^{\lambda - J - L}\,
\frac{(-\lambda + J + L)_{s}(\lambda + J + L + 3)_{s}}{s!\,\Gamma\left(2J + 2L + s + 4\right)}\,\times \\ [3mm]
e^{i(\delta_{\lambda}' - \delta_{\lambda})}\,
\times {_3F_2}\left\{\matrix{ -\lambda' + J + L,\,\,\lambda' + J + L + 3,\,\,
2J + 2 \cr \cr
2J+2,\,\, 2J + 2L + s + 4
\cr}\Biggr|1\right\}.
\end{eqnarray*}
Using now the Saalsch\"{u}tz theorem \cite{BE1}
\begin{eqnarray}
{_3F_2}\left\{\matrix{ a,\,\,b,\,\, -n\cr \cr c,\,\,1+a+b-c-n
\cr}\Biggr|1\right\} = \frac{(c-a)_{n}\,(c-b)_{n}}{(c)_{n}\,(c-a-b)_{n}},
\label{2.10.9}
\end{eqnarray}
\begin{eqnarray*}
Q_{\lambda \lambda'} = \frac{(-1)^{\lambda' - J - L}\,(-i)^{\lambda}\,i^{\lambda'}\,
e^{i(\delta_{\lambda'} - \delta_{\lambda})}\,\sqrt{(2\lambda + 3)\,(2\lambda' + 3)}}
{(\lambda' + J + L + 3)!\,\Gamma(-\lambda' + J + L + 1)}\,\times \\ [3mm]
\times \sqrt{\frac{(\lambda + J + L + 2)!\,(\lambda + J - L + 1)!\,(\lambda' + J + L + 2)!\,
(\lambda' - J + L + 1)!}{(\lambda - J - L)!\,(\lambda' - J - L)!\,
(\lambda' + J - L + 1)!\,(\lambda + J - L + 1)!}}\,\times \\ [3mm]
\times {_3F_2}\left\{\matrix{ -\lambda + J + L,\,\,\lambda + J + L + 3,\,\,
1 \cr \cr
-\lambda' + J + L + 1,\,\, \lambda' + J + L + 4
\cr}\Biggr|1\right\}.
\end{eqnarray*}
Using the Saalsch\"{u}tz theorem (\ref{2.10.9}) for the second time, we finally obtain
\begin{eqnarray*}
Q_{\lambda \lambda'} = \frac{(-i)^{\lambda}\,i^{\lambda'}\,
e^{i(\delta_{\lambda'} - \delta_{\lambda})}}{\Gamma(\lambda - \lambda' + 1)
\Gamma(\lambda' - \lambda + 1)}\,\frac{\sqrt{(2\lambda + 3)\,(2\lambda' + 3)}}
{\lambda - \lambda' + 3}\,\times \\ [3mm]
\times \sqrt{\frac{(\lambda' - J + L)!\,(\lambda + J + L + 2)!\,(\lambda + J - L + 1)!\,
(\lambda' - J + L + 1)!}{(\lambda - J - L)!\,(\lambda' + J + L + 2)!\,
(\lambda - J + L + 1)!\,(\lambda' + J - L + 1)!}}.
\end{eqnarray*}
Since the numbers $\lambda$ and $\lambda'$ are integers or half-integers at the same time,
the last expression vanishes for $\lambda \neq \lambda'$ due to the product of the gamma
functions of $(\lambda - \lambda' + 1)$ and $(\lambda' - \lambda + 1)$, and in unity – at
$\lambda = \lambda' $, i.e.
\begin{eqnarray}
\int\limits_{-\infty}^{\infty}\,
W_{k \Omega J L}^{\lambda}\,W_{k \Omega J L}^{\lambda' *}\,d\Omega
= \delta_{\lambda \lambda'}.
\label{2.10.10}
\end{eqnarray}
Then, taking into account (\ref{2.10.10}) for the inverse expansion we obtain
\begin{eqnarray}
\psi^{hsp} = \int\limits_{-\infty}^{\infty}\,
W_{k \Omega J L}^{\lambda}\,\psi^{par}\,d\Omega,
\label{2.10.11}
\end{eqnarray}
where integration is carried out along the real axis.

\newpage

\chapter{The MIC-Kepler problem and the Coulomb-oscillator correspondence}
\markboth{CHAPTER 3. THE MIC-KEPLER PROBLEM}{}

\section{MIC-Kepler problem}
\markboth{CHAPTER 3. THE MIC-KEPLER PROBLEM}{3.1. MIC-KEPLER PROBLEM}

The superintegrable MIC-Kepler system, or charge-dyon system, describing the motion of a
charged particle in a Dirac dyon, was constructed by Zwanziger \cite{ZWANZIG} and then
rediscovered by McIntosh and Cisneros \cite{MIC}. This system is described by the Hamiltonian
\begin{eqnarray}
\hat{\mathcal{H}}_{0} = \frac{1}{2\mu_0}\left(-i\hbar{\bf{\nabla}} -
\frac{e}{c}{\bf A}^{(\pm)}\right)^2 +\frac{\hbar^2{s}^2}{2\mu_0
r^2}-\frac{e^2}{r},
\label{3.1.1}
\end{eqnarray}
where
\begin{eqnarray}
{\bf A}^{(\pm)} = \frac{g}{r(r \mp z)}\left(\pm y, \mp x.
0\right).
\label{3.1.2}
\end{eqnarray}
The vector potentials ${\bf A}^{(\pm)}$ correspond to the Dirac monopole
\cite{DIRAC} with a magnetic charge $g=\frac{\hbar cs}{e}$ ($s=0,\pm
\frac{1}{2},\pm 1,\dots$) and the singularity axis at $z
> 0$ and $z < 0$, respectively.

It is easy to notice that the vector potentials $A^{(+)}_i$ and $A^{(-)}_i$
are related to each other by the gauge transformation
\begin{eqnarray}
A^{(-)}_i = A^{(+)}_i + \frac{\partial f}{\partial x_i},
\label{3.1.2-1}
\end{eqnarray}
where $f=2g\arctan y/x$, and the intensity of a magnetic field created by the
dyon has the form
\begin{eqnarray*}
{\bf B} = {\bf \nabla} \times {\bf A}^{(\pm)} = g\frac{{\bf r}}
{r^3}.
\end{eqnarray*}
Further, we will carry out all calculations only for the vector potential
$A^{(+)}_i$  and for brevity we will omit the sign $(+)$ in what follows.

A specific feature of the MIC--Kepler problem is hidden symmetry inherent in
the Coulomb problem. A direct calculation can really show that the following six
operators commute with Hamiltonian (\ref{3.1.1}):
\begin{eqnarray}
{\hat {\bf J}}&=& {\bf r}\times \left(-i{\bf\nabla}
-\frac{e}{\hbar c}{\bf A}\right)
- s\frac{{\bf r}}{r},
\nonumber 
\\
\label{3.1.3}
\\
{\hat {\bf I}}&=&\frac{1}{2\sqrt{\mu_0}}\left[{\hat {\bf J}}\times
\left(-i\hbar{\bf \nabla} - \frac{e}{c}{\bf
A}\right) - \left(-i\hbar{\bf \nabla} + \frac{e}{c}{\bf A}\right)
\times{\hat{\bf J}}\right] -\frac{e^2\sqrt{\mu_0}}{\hbar}\frac{\bf r}{r}.
\nonumber
\end{eqnarray}
The operators ${\hat {\bf J}}$ determine the rotational moment of the system
and the operator ${\hat{\bf I}}$ are an analog of the Runge-Lenz vector.

At fixed negative values of energy the integrals of motion (\ref{3.1.3}) form
an algebra isomorphic to the algebra $so(4)$; whereas at positive values of
energy, $so(3.1)$ \cite{ZWANZIG}. By virtue of hidden symmetry the MIC--Kepler
problem splits up not only in the spherical and parabolic but also prolate
spheroidal coordinates. Thus, the MIC--Kepler system can be considered a
natural generalization of the Coulomb problem in the presence of the Dirac
monopole.

At first, consider the MIC-Kepler problem in spherical coordinates \cite{MONO}.

In the spherical coordinates
\begin{eqnarray}
x = r\,\sin\theta\,\cos\varphi, \qquad y = r\,\sin\theta\,\sin\varphi, \qquad
z = r\,\cos\theta
\label{3.1.4}
\end{eqnarray}
a solution of the spectral problem
\begin{eqnarray}
\hat{\mathcal{H}}_{0}\,\psi=E^{(0)}\,\psi,\qquad {\hat{\bf J}}^2\,\psi=j(j+1)\,\psi, \qquad
{\hat J}_z\,\psi=\left(s - i\frac{\partial}{\partial \varphi}\right)\,\psi= m\,\psi,
\label{3.1.5}
\end{eqnarray}
which describes a bound MIC-Kepler system with energy $E^{(0)}$,
rotational moment $j$ and $z$-component of the rotational moment $m$, can be
written as follows \cite{TN,MONO}:
\begin{eqnarray}
\psi_{njm}^{(s)}({\bf r}) =
\left(\frac{2j+1}{4\pi}\right)^{1/2}R_{nj}^{(s)}(r)
d^j_{ms}(\theta)e^{i(m-s)\varphi},
\label{3.1.6}
\end{eqnarray}
where $d^j_{ms}$ is the Wigner $d$-function \cite{VAR}, and $R_{nj}^{(s)}(r)$
is determined by the expression
\begin{eqnarray}
R_{nj}^{(s)}(r) = \frac{2e^{-\frac{r}{r_0n}}}{n^2r_0^{3/2}(2j+1)!}
\sqrt{\frac{(n+j)!}{(n-j-1)!}}\,\left(\frac{2r}{r_0n}\right)^j
\, F\left(j-n+1, 2j+2,\frac{2r}{r_0n}\right).
\label{3.1.7}
\end{eqnarray}
Now taking into account formula \cite{VAR}
\begin{eqnarray}
d^{J}_{MM'}(\beta) = (-1)^{\frac{M-M'+|M-M'|}{2}}\,
\left[\frac{k!(k+a+b)!}{(k+a)!(k+b)!}\right]^{\frac{1}{2}}\,
\left(\sin\frac{\beta}{2}\right)^{a}\left(\cos\frac{\beta}{2}\right)^{b}
P^{(a,b)}_{k}(\cos\beta),
\label{3.1.8}
\end{eqnarray}
where the indices of the Jacobi polynomial $k, a, b$ are related to $J, M, M'$ by
\begin{eqnarray*}
a= |M-M'|, \qquad b= |M+M'|,  \qquad k = J - \frac{1}{2}(a+b),
\end{eqnarray*}
the angular wave function can be written in the form
\begin{eqnarray}
&&Z_{jm}^{(s)}(\theta,
\varphi) = \left(\frac{2j+1}{4\pi}\right)^{1/2}d^j_{ms}(\theta)
e^{i(m+s)\varphi}= \left[\frac{(2j+1)(j-m_+)!(j+m_+)!}
{4\pi(j-m_-)!(j+m_-)!}\right]^{1/2}\times
\nonumber \\[3mm]
\label{3.1.9} \\[3mm]
&\times& (-1)^{\frac{m-s+|m-s|}{2}}\,
\left(\sin\frac{\theta}{2}\right)^{|m-s|}
\left(\cos\frac{\theta}{2}\right)^{|m+s|}\,
P^{(|m-s|,|m+s|)}_{j-m_+}(\cos\theta)\,e^{i(m-s)\varphi}.
\nonumber
\end{eqnarray}
Here we introduced the notation
\begin{eqnarray}
m_{\pm} = \frac{|m+s| \pm |m-s|}{2}.
\label{3.1.10}
\end{eqnarray}
The angular wave function $Z_{jm}^{(s)}(\theta, \varphi)$ is called the Tamm monopole
harmonics \cite{TAMM}.

Now the total wave function (\ref{3.1.6}) of the coupled MIC-Kepler system can be rewritten
in the following form
\begin{eqnarray}
\psi_{njm}^{(s)}({\bf r}) = R_{nj}^{(s)}(r)\,Z_{jm}^{(s)}(\theta,
\varphi).
\label{3.1.11}
\end{eqnarray}
The system spectrum is determined by the conditions:
\begin{eqnarray}
&\varepsilon_n& = -\frac{\mu_0 e^4}{2\hbar^2n^2},\qquad
n=|s|+1,|s|+2, \ldots ,
\label{3.1.12}
\end{eqnarray}
\begin{eqnarray*}
&j&= |{s}|, |{s}|+1,\ldots, n-1;\quad  m =-j, -j+1,\ldots, j-1, j.
\end{eqnarray*}
Quantum numbers $j$ and $m$ define the total moment of the system and its
projection onto the axis $z$. At integer $s$ the system has an
integer spin; and at half-integer, $s$ half-integer. At ${s}=0$ it turns into a
hydrogen-like system.

Under the identity transformation $\varphi\to\varphi+2\pi$ the wave function of
the system is unambiguous at integer ${s}$ and changes the sign at half-integer
${s}$.In the second case, the ambiguity of the wave function can be
interpreted as the presence of a magnetic field of an infinitely thin solenoid
(directed along the axis $z$) generating spin $1/2$ in the system \cite{ARTUR,NTA}.

Now consider the MIC-Kepler system in the parabolic basis.
In parabolic coordinates $\mu, \nu  \in [0, \infty), \varphi \in [0, 2\pi)$,
which are defined by the formulas \cite{LL}
\begin{eqnarray}
x = \sqrt{\mu \nu}\,\cos\varphi, \qquad y = \sqrt{\mu \nu}\,\sin\varphi, \qquad
z = \frac{1}{2}\,(\mu - \nu),
\label{3.1.13}
\end{eqnarray}
differential elements of length and volume have the form
\begin{eqnarray*}
d\,l^{2} = \frac{\mu + \nu}{4}\,\left(\frac{d\,\mu^{2}}{\mu} + \frac{d\,\nu^{2}}{\nu}\right) +
\mu \nu\,d\,\varphi^{2},  \qquad d\,V = \frac{1}{4}\,(\mu + \nu)\,d\,\mu\,d\,\nu\,d\,\varphi,
\end{eqnarray*}
and the Laplace operator is as follows
\begin{eqnarray}
\Delta = \frac{4}{\mu + \nu}\,\left[\frac{\partial}{\partial \mu}\,
\left(\mu\,\frac{\partial}{\partial \mu}\right) + \frac{\partial}{\partial \nu}\,
\left(\nu\,\frac{\partial}{\partial \nu}\right)\right] + \frac{1}{\mu \nu}\,
\frac{\partial^{2}}{\partial \varphi^{2}}.
\label{3.1.14}
\end{eqnarray}
After substitution
\begin{eqnarray*}
\psi(\mu,\nu,\varphi) = \Phi_1(\mu)
\Phi_2(\nu)\,\frac{e^{i(m-s)\varphi}}{\sqrt{2\pi}}
\end{eqnarray*}
the variables in the Schr\"{o}dinger equation are separated and we arrive at the
following system of equations \cite{M-16}:
\begin{eqnarray}
\frac{d}{d \mu}\left(\mu \frac{d\Phi_1}{d \mu}\right) +
\left[\frac{\mu_0E^{(0)}}{2\hbar^2}\mu - \frac{(m-s)^2}{4\mu} +
\frac{\sqrt{\mu_0}}{2\hbar}\Omega + \frac{1}{2r_0}\right]\Phi_1
&=& 0,
\label{3.1.15}
\\ [5mm]
\frac{d}{d \nu}\left(\nu
\frac{d\Phi_2}{d \nu}\right) + \left[\frac{\mu_0
E^{(0)}}{2\hbar^2}\nu - \frac{(m+s)^2}{4\nu}-
\frac{\sqrt{\mu_0}}{2\hbar}\Omega + \frac{1}{2r_0}\right]\Phi_2
&=& 0,
\label{3.1.16}
\end{eqnarray}
where $\Omega$ is the separation constant that is an eigenvalue of the
$z$-component of the Runge--Lenz vector ${\hat{\bf I}}$.

At $s=0$ these equations coincide with those for the hydrogen atom in the
parabolic coordinates \cite{LL}. Hence, we have
\begin{eqnarray}
\psi_{n_1n_2 m}^{(s)}(\mu, \nu, \varphi) = \frac{1}{n^2r_0^{3/2}}
\Phi_{n_1, m-s}(\mu)\,\Phi_{n_2, m+s}(\nu)\,
\frac{e^{i(m-s)\varphi}}{\sqrt{\pi}},
\label{3.1.17}
\end{eqnarray}
where
\begin{eqnarray*}
\Phi_{p q}(x) = \frac{1}{|q|!} \sqrt{\frac{(p+|q|)!}{p!}}
e^{-x/2r_0n}\left(\frac{x}{r_0n}\right)^{|q|/2} F\left(-p;|q|+1;
\frac{x}{r_0n}\right).
\end{eqnarray*}
Here $n_1$ and $n_2$ are non-negative integers
\begin{eqnarray*}
n_1 = - \frac{|m-s|+1}{2} + \frac{\sqrt{\mu_0}}{2\kappa
\hbar}\,\Omega + \frac{1}{2r_0\kappa}, \qquad n_2 = -
\frac{|m+s|+1}{2} - \frac{\sqrt{\mu_0}}{2\kappa \hbar}\,\Omega +
\frac{1}{2r_0\kappa},
\end{eqnarray*}
and $\kappa=\sqrt{-2\mu_0 E^{(0)}}/\hbar$.
From the last relations, taking into account (\ref{3.1.12}), we obtain that the parabolic
quantum numbers $n_{1}$ and $n_{2}$ are related to the principal quantum number $n$ as follows:
\begin{eqnarray}
n = n_1 + n_2 + m_{+} + 1.
\label{3.1.18}
\end{eqnarray}
Thus, we solved the spectral problem
\begin{eqnarray}
\hat{\mathcal{H}}_{0}\,\psi=E^{(0)}\,\psi, \qquad {\hat I}_z\,\psi=
\frac{e^2\sqrt{\mu_0}}{\hbar n}\left(n_{1} -n_{2} - m_{-}\right)\,\psi,
\qquad {\hat J}_z\,\psi= m\psi,
\label{3.1.19}
\end{eqnarray}
where $\hat{\mathcal{H}}_{0}, {\hat I}_z, {\hat J}_z$ are determined by expressions
(\ref{3.1.1}), (\ref{3.1.3}).

\section{Generalization of the Park-Tarter expansion}
\markboth{CHAPTER 3. THE MIC-KEPLER PROBLEM}{3.2. GENERALIZATION OF THE PARK-TARTER EXPANSION}

First, we note that both for the radial wave functions of the $8D$ oscillator,
the $5D$ Coulomb problem and the $SU(2)$ Yang-Coulomb monopole, and for the radial wave
function of the coupled MIC-Kepler system, along with the orthonormalization condition
\begin{eqnarray*}
\int\limits_{0}^{\infty}\,r^{2}\,R_{n'j}^{(s)}(r)R_{nj}^{(s)}(r)dr
= \delta_{nn'}
\end{eqnarray*}
there is a similar condition of orthogonality with respect to the orbital momentum $j$:
\begin{eqnarray}
\int\limits_{0}^{\infty}\,R_{nj'}^{(s)}(r)R_{nj}^{(s)}(r)dr
=\frac{2}{r_0^2n^3} \frac{\delta_{jj'}}{2j+1}.
\label{3.2.1}
\end{eqnarray}

Now let's return to the problem of interbasis expansion of a parabola - sphere.
The sought expansion at a fixed energy value can be written as follows \cite{M-16}:
\begin{eqnarray}
\psi_{n_1n_2m}^{{(s)}}(\mu, \nu, \varphi) =
\sum_{j=m_+}^{n-1}\,W^{j}_{nn_1ms}\psi_{njm}^{(s)}(r, \theta,
\varphi).
\label{3.2.2}
\end{eqnarray}
Passing in the left-hand side of relation (\ref{3.2.2}) from the parabolic
coordinates to the spherical ones, according to formulae
\begin{eqnarray}
\mu = r\left(1+\cos\theta\right), \qquad \nu = r\left(1-\cos\theta\right),
\label{3.2.3}
\end{eqnarray}
putting $\theta =0$ and taking into account (\ref{1.7.4}), we establish an equality
that includes only the variable $r$. Then using the orthogonality condition for the
radial wave function in the angular momentum (\ref{3.2.1}), we get
\begin{eqnarray}
W_{nn_1ms}^j = (-1)^{\frac{m-s+|m-s|}{2}}\frac{\sqrt{(2j +1)\,
(n+j)!}} {|m-s|!\,(2j +1)!}\, E_{nn_1n_2}^{jms}\,K_{jms}^{nn_1}\,,
\label{3.2.4}
\end{eqnarray}
where
\begin{eqnarray}
E_{nn_1n_2}^{jms} = \left[\frac{(n_1+|m-s|)!(n_2+|m+s|)!
\left(j - m_{-}\right)! \left(j - m_{+}\right)!} {(n_1)!(n_2)!(n-j -1)!
\left(j + m_{+}\right)! \left(j + m_{-}\right)!}\right]^{1/2},
\label{3.2.5}
\end{eqnarray}
\begin{eqnarray*}
K_{jms}^{nn_1}=\int \limits_{0}^{\infty} e^{-x} x^{j+m_{+}}F(-n_1,
|m-s|+1; x) F(-n+j+1, 2j +2; x)dx. \nonumber
\end{eqnarray*}
Here $x=2r/r_0n$. For calculation of the integral $K_{jms}^{nn_1}$ we write down the confluent
hypergeometric $F(-n_1, |m+s|+1; x)$ as a polynomial, perform integration by
formula (\ref{1.6.5}) and with relation (\ref{1.6.6}) taken into account we get
\begin{eqnarray}
K_{jms}^{nn_1}= \frac{(2j +1)!\,(j+m_{+})!\,\left(n + m_{-} -1\right)!}
{(j-m_+)!\,(n+j)!} {_3F}_2 \left\{
\begin{array}{l}
-n_1,  -j +m_{+}, j +m_+  +1
\\ [3mm] |m-s|+1,  -n+m_+ +1
\end{array}
\biggr| 1 \right\}.
\label{3.2.6}
\end{eqnarray}
Substitution of (\ref{3.2.5}) and (\ref{3.2.6}) into (\ref{3.2.4}) gives
\begin{eqnarray}
W_{nn_1ms}^j
&=& \sqrt{\frac{(2j+1)(n_1-|m+s|)!(n_2+|m+s|)! \left(j
+ m_{+}\right)! \left(j - m_{-}\right)!} {(n_1)!(n_2)!(n-j -1)!
(n+j)!\left(j - m_{+}\right)! \left(j + m_{-} \right)!}}\,
\frac{\left(n - m_{+} -1\right)!}{|m-s|!}\times
\nonumber \\ [3mm]
\label{3.2.7}
\\ [3mm]
&\times&
(-1)^{n_1+\frac{m-s+|m-s|}{2}}
{_3F}_2 \left\{
\begin{array}{l}
-n_1,  -j +m_{+}, j+ m_{+} +1
\\ [3mm] |m-s|+1,   -n+ m_{+} +1
\end{array}
\biggr| 1 \right\}.
\nonumber
\end{eqnarray}
Finally, comparing (\ref{3.2.7}) with (\ref{1.7.12}), we arrive at the following representation:
\begin{eqnarray}
W_{nn_1ms}^j = (-1)^{n_1+\frac{m-s+|m-s|}{2}}
C_{\frac{n_1+n_2+|m+s|}{2},\frac{n_2-n_1+|m+s|}{2};
\frac{n_1+n_2+|m-s|}{2},\frac{n_1-n_2+|m-s|}{2}}^ {j,\,m_+}.
\label{3.2.8}
\end{eqnarray}

Inverse transformation, i.e., expansion of the spherical basis over the
parabolic one, has the form
\begin{eqnarray}
\psi_{n j m}^{(s)}(r,\theta,\varphi) = \sum_{n_1=0}^{n-m_+ -1}
{\tilde W}_{n j ms}^{n_1} \psi_{n_1n_2m}^{(s)}(\mu,\nu,\varphi).
\label{3.2.9}
\end{eqnarray}
Then using the orthonormalization condition of the Clebsch--Gordan coefficients
of the $SU(2)$ group (\ref{1.8.16}), we get
\begin{eqnarray}
{\tilde W}_{n j ms}^{n_1} = (-1)^{n_1+\frac{m-s+|m-s|}{2}}
C_{\frac{n + m_{-} -1}{2},\frac{n + m_{-} -1}{2}-n_1; \frac{n - m_{-} - 1}{2},
\frac{m_{+} +|m-s|-n+1}{2}+n_1}^ {j, m_+}.
\label{3.2.10}
\end{eqnarray}
Taking into account formula (\ref{1.7.12}) ${\tilde W}_{n j ms}^{n_1}$ can be
written in terms of the functions ${_3}F_2$. At $s=0$ formulae (\ref{3.2.7}) and
(\ref{3.2.8}) turn into the corresponding relations,obtained by Tarter
\cite{TARTER} and Park \cite{PARK}.

\section{Spheroidal basis of the MIC-Kepler problem}
\markboth{CHAPTER 3. THE MIC-KEPLER PROBLEM}{3.3. SPHEROIADAL BASIS OF THE MIC-KEPLER PROBLEM}

Spheroidal coordinates are a natural tool for investigation of many problems of
mathematical physics \cite{KOMPOSLA}. In quantum mechanics these coordinates
are used to describe the behavior of a charged particle in the field of two
Coulomb centers. The distance $R$ between the centers is taken as a dimensional
parameter specifying spheroidal coordinates and has a dynamic meaning, i.e.,
enters into the energy--spectrum expression. If the charge of one of the
centers is put as zero, one arrives at a one-center problem and the parameter
$R$ becomes purely kinematic. This simplifies the problem considerably. At the
same time the mathematical structure of the spheroidal equations remains the
same as the energy enters into both the radial and angular equations.
Consequently, the spheroidal analysis of a MIC-Kepler problem becomes the first step
in the investigation of the two-center MIC-Kepler problem.

In prolate spheroidal coordinates
\begin{eqnarray}
x = \frac{R}{2}\sqrt{(\xi^2-1)(1-\eta^2)}\cos \varphi, \quad y =
\frac{R}{2}\sqrt{(\xi^2-1)(1-\eta^2)}\sin \varphi,  \quad z =
\frac{R}{2}(\xi \eta +1),
\label{3.3.1}
\end{eqnarray}
where $\xi \in [1, \infty)$, $\eta \in [-1, 1]$, $\varphi \in [0, 2\pi)$, $R
\in [0, \infty)$. The differential elements of length, volume, and the Laplace
operator have the form
\begin{eqnarray*}
dl^2 &=& \frac{R^2}{4}\left[(\xi^2 - \eta^2)\left(
\frac{d\xi^2}{\xi^2-1}+\frac{d\eta^2}{1-\eta^2}\right)+
(\xi^2-1)(1-\eta^2)d\varphi^2\right],  \\[4mm]
dV &=& \frac{R^3}{8}(\xi^2 - \eta^2)d\xi d\eta d \varphi, \\[4mm]
\Delta &=& \frac{4}{R^2(\xi^2 {-} \eta^2)}\left[
\frac{\partial}{\partial \xi}(\xi^2-1)\frac{\partial}{\partial
\xi} + \frac{\partial}{\partial
\eta}(1-\eta^2)\frac{\partial}{\partial \eta} \right] +
\frac{4}{R^2(\xi^2{-}1)(1{-}\eta^2)} \frac{\partial^2}{\partial
\varphi^2}.
\end{eqnarray*}
The parameter $R$ is an interfocus distance, and as $R \to 0$ and $R \to
\infty$ the prolate spheroidal coordinates turn into spherical and parabolic
coordinates, respectively \cite{KOMPOSLA}.

After substitution
\begin{eqnarray*}
\psi^{(s)}(\xi, \eta, \varphi; R) = X^{(s)}(\xi;R)Y^{(s)}(\eta;R)
\frac{e^{i(m - s)\varphi}}{\sqrt{2\pi}}.
\end{eqnarray*}
the variables in the Schr\"{o}dinger equation of the MIC-Kepler problem (\ref{3.1.1})
are separated and we arrive at the following system of ordinary differential equations:
\begin{eqnarray}
\left[\frac{d}{d\xi}(\xi^2-1)\frac{d}{d\xi} +
\frac{(m-s)^2}{2(\xi+1)}- \frac{(m+s)^2}{2(\xi -1)} +
\frac{\mu_0R^2E^{(0)}} {2\hbar^2}(\xi^2-1) + \frac{R}{r_0}\xi
-\Lambda(R)\right]X^{(s)} = 0, \nonumber \\ [3mm]
\label{3.3.2}
\\  [3mm]
\left[\frac{d}{d\eta}(1-\eta^2)\frac{d}{d\eta} -
\frac{(m-s)^2}{2(1+\eta)}- \frac{(m+s)^2}{2(1-\eta)} +
\frac{\mu_0R^2E^{(0)}}{2\hbar^2}(1-\eta^2) - \frac{R}{r_0}\eta +
\Lambda(R)\right]Y^{(s)} = 0, \nonumber
\end{eqnarray}
where $\Lambda(R)$ is the separation constant.

Excluding the energy from the system of equations (\ref{3.3.2}) we obtain the
spheroidal integral of motion
\begin{eqnarray*}
\hat \Lambda &=&
\frac{1}{\xi^2-\eta^2}\left[(1-\eta^2)\frac{\partial}{\partial
\xi} (\xi^2-1)\frac{\partial}{\partial \xi}- (\xi^2-1)
\frac{\partial}{\partial \eta}(1-\eta^2)\frac{\partial}{\partial
\eta} \right]-\frac{\xi^2+\eta^2-2}{(\xi^2-1)(1-\eta^2)}
\frac{\partial^2}{\partial \varphi^2} -  \\[5mm]
&-& \frac{2s}{(\xi - 1)(1-\eta)}
\left(\xi+\eta-1-\frac{\xi\eta+1}{\xi+\eta}
\right)\left(s - i\frac{\partial}{\partial \varphi}\right) +
\frac{R}{r_0}\frac{\xi\eta+1}{\xi+\eta}.
\end{eqnarray*}
In the last expression, moving to Cartesian coordinates and taking into account that
the square of the angular momentum operator ${\hat {\bf J}}^2$ and the $z$ - component of
the analogue of the Runge-Lenz vector ${\hat I}_z$ in Cartesian coordinates have the form:
\begin{eqnarray}
{\hat {\bf J}}^2 = - \left(r^{2}\,\Delta - x_{i}\,x_{j}\,\frac{\partial^{2}}{\partial x_{i}
\,\partial x_{j}} - 2_{i}\,\frac{\partial}{\partial x_{i}}\right) +
\frac{2isr}{r - z}\,\left(y\,\frac{\partial}{\partial x} -
x\,\frac{\partial}{\partial y}\right) + \frac{2s^{2}r}{r - z},
\label{3.3.3}
\end{eqnarray}
\vspace{3mm}
\begin{eqnarray}
{\hat I}_z &=& \frac{\hbar}{\sqrt{\mu_{0}}}
\Bigl[z\,\left(\frac{\partial^{2}}{\partial x^{2}} + \frac{\partial^{2}}{\partial y^{2}}\right) -
x\,\frac{\partial^{2}}{\partial x \partial z} - y\,\frac{\partial^{2}}{\partial y \partial z} -
\nonumber \\
\\
&-& is\,\frac{r + z}{r\,(r - z)}\left(y\,\frac{\partial}{\partial x} -
x\,\frac{\partial}{\partial y} - is\right) - \frac{\partial}{\partial z} +
\frac{z}{r_{0}\,r}\Bigr], \nonumber
\label{3.3.4}
\end{eqnarray}
we get
\begin{eqnarray}
\hat \Lambda = {\hat {\bf J}}^2 + \frac{R\sqrt{\mu_0}}{\hbar}
{\hat I}_z.
\label{3.3.5}
\end{eqnarray}
So we have
\begin{eqnarray}
\hat \Lambda \psi^{(s)}_{nqm}(\xi, \eta, \varphi; R)=
\Lambda_q(R)\psi^{(s)}_{nqm}(\xi, \eta, \varphi; R).
\label{3.3.6}
\end{eqnarray}
Index $q$ numbers eigenvalues for the operator ${\hat \Lambda}$ and takes the
following values: $1 \leq q \leq n-m_+$.
Let us now express the spheroidal basis as expansion over the spherical
(\ref{3.1.11}) and parabolic (\ref{3.1.17}) bases:
\begin{eqnarray}
\psi_{nqm}^{(s)}(\xi,\eta,\varphi;R)
&=& \sum_{j=m_+}^{n-1}
V_{nqms}^j \psi_{njm}^{(s)}(r,\theta,\varphi),
\label{3.3.7}
\\[5mm]
\psi_{nqm}^{(s)}(\xi,\eta,\varphi;R)
&=& \sum_{n_1=0}^{n-m_+ -1}
U_{nqms}^{n_1} \psi_{n_1n_2m}^{(s)}(\mu,\nu,\varphi).
\label{3.3.8}
\end{eqnarray}
Acting by operator (\ref{3.3.5}) on (\ref{3.3.7}) and (\ref{3.3.8}) and using
the spectral equations for the operators ${\hat J}^2$ (\ref{3.1.4}), ${\hat
I}_z$ (\ref{3.1.18}), and ${\hat \Lambda}$ (\ref{3.3.6}), we get
\begin{eqnarray}
[\Lambda_q(R)-j(j +1)]V_{nqms}^j(R)
= \frac{R\sqrt{\mu_0}}{\hbar}
\sum_{j'}\,V_{nqms}^{j'}(R) \left({\hat I}_z\right)_{j j'}, \nonumber
\\ [5mm]
\label{3.3.9}
\\ [5mm]
\left[\Lambda_q(R) - \frac{R}{nr_0}\left(n_1-n_2 -
m_-\right)\right]U_{nqms}^{n_1}(R)
=
\sum_{n_1'}\,U_{nqms}^{n_1'}(R)
\left({\hat {\bf J}}^2\right)_{n_1 n_1'},\nonumber
\end{eqnarray}
where
\begin{eqnarray*}
\left({\hat I}_z\right)_{j j'}
&=& \int \psi_{nj
m}^{(s)*}(r,\theta,\varphi){\hat I}_z \psi_{nj'm}^{(s)}
(r,\theta,\varphi) dV,
\\ [5mm]
\left({\hat {\bf J}}^2\right)_{n_1 n_1'}
&=&
\int \psi_{n_1n_2m}^{(s)*}(\mu,\nu,\varphi)
{\hat J}^2 \psi_{n_1' n_2' m}^{(s)}(\mu,\nu,\varphi) dV.
\end{eqnarray*}
Further, doing the same as when finding the spheroidal bases of the $8D$ oscillator,
the $5D$ Coulomb problem and the $SU(2)$ Yang-Coulomb monopole, for the matrix elements
$\left({\hat I}_z\right)_{jj'}$, and $\left({\hat {\bf J}}^2\right)_{n_1{n'}_1}$ we obtain
\begin{eqnarray}
\left({\hat I}_z\right)_{j j'}
&=& \frac{e^2\sqrt{\mu_0}}{\hbar n}
\left(A_{j +1}\delta_{j',j+1} - B_j\delta_{j',j} +
A_j\delta_{j',j-1}\right), \nonumber \\ [5mm]
\label{3.3.10}
\\  [5mm]
\left({\hat {\bf J}}^2\right)_{n_1 n_1'}
&=&
C_{n_1+1}\delta_{n_1',n_1+1}+ D_{n_1}\delta_{n_1',n_1} +
C_{n_1}\delta_{n_1',n_1-1}, \nonumber
\end{eqnarray}
where
\begin{eqnarray*}
A_j &=& -\sqrt{\frac{\left(j^2-m_+^2\right)
\left(j^2-m_-^2\right)(n^2-j^2)} {j^2(2j-1)(2j+1)}}, \qquad B_j =
\frac{nm_+ m_-}{j(j+1)},\\
[7mm] C_{n_1} &=& - \sqrt{n_1(n_1+|m-s|) \left(n-n_1 - m_{-}\right)
\left(n-n_1-m_{+}\right)}, \\ [7mm]
D_{n_1} &=& n(n - 1)
- n_1\left(n_1+|m-s|\right) - n_2\left(n_2+|m+s|\right).
\end{eqnarray*}
Now substituting the matrix elements (\ref{3.3.10}) into
(\ref{3.3.9}) we derive the three-term recurrence relations
\begin{eqnarray}
A_{j+1}V_{nqms}^{j+1}(R)
&+& \left\{B_j -
\frac{nr_0}{R}\left[\Lambda_q(R)-j(j+1)
\right]\right\}V_{nqms}^j(R) + A_jV_{nqms}^{j-1}(R) = 0, \nonumber \\ [5mm]
\label{3.3.11}
\\ [5mm]
C_{n_1+1}U_{nqms}^{n_1+1}(R)
&+& \left[D_{n_1}-\frac{R}{nr_0}
\left(n_1-n_2 - m_{-}\right)-\Lambda_q(R)\right]U_{nqms}^{n_1}(R) +
C_{n_1}U_{nqms}^{n_1-1}(R) = 0, \nonumber
\end{eqnarray}
to be solved simultaneously with the normalization conditions
\begin{eqnarray*}
\sum_{j}\,\left|V_{nqms}^j(R)\right|^2 = 1, \qquad
\sum_{n_1}\,\left|U_{nqms}^{n_1}(R)\right|^2 = 1.
\end{eqnarray*}

To illustrate the method for constructing spheroidal wave functions, let us
consider in detail one particular case. Let $n=2, m=s=0$, then according to
expansion (\ref{3.3.7}) we have
\begin{eqnarray}
\psi_{2q0}^{(0)} = V_{2q00}^{0}\,\psi_{200}^{(0)} + V_{2q00}^{1}\,\psi_{210}^{(0)}.
\label{3.3.12}
\end{eqnarray}
For the coefficients $V_{2q00}^{0}$ and $V_{2q00}^{1}$ from recurrence relations
(\ref{3.3.12}) we obtain the following system of algebraic equations
\begin{eqnarray*}
V_{2q00}^{1} + \frac{2r_{0}}{R}\,\Lambda\,V_{2q00}^{0} = 0, \\ [5mm]
\frac{2r_{0}}{R}\,\Lambda\,(\Lambda - 2)\,V_{2q00}^{1} + V_{2q00}^{0} = 0.
\end{eqnarray*}
Equating the determinant of the system of equations to zero, we obtain the quadratic
equation for the separation constant $\Lambda$
\begin{eqnarray*}
\Lambda\,(\Lambda - 2) = \frac{R^{2}}{4r_{0}^{2}},
\end{eqnarray*}
whence it follows that
\begin{eqnarray*}
\Lambda_{1,2} = 1 \pm \sqrt{1 + \frac{R^{2}}{4r_{0}^{2}}}.
\end{eqnarray*}
After this, taking into account the condition for orthonormalization of the
coefficients $V_{nqms}^{j}(R)$, we obtain
\begin{eqnarray*}
V_{2q00}^{0} = \frac{1}{\sqrt{2}}\,\left(\frac{\Lambda_{1,2} - 2}{\Lambda_{1,2} - 1}\right)^{1/2},
\qquad
V_{2q00}^{1}  = -\frac{1}{\sqrt{2}}\,\left(\frac{\Lambda_{1,2}}{\Lambda_{1,2} - 1}\right)^{1/2}.
\end{eqnarray*}
Next, substituting the obtained expressions for the coefficients
$V_{2q00}^{0}$ and $V_{2q00}^{1}$ into (\ref{3.3.12}) and using the explicit
expression for the spherical basis of the "charge-dyon" system  (\ref{3.1.6}), we have
\begin{eqnarray*}
\psi_{2q0}^{(0)} = \frac{e^{-R(\xi + \eta)/4r_{0}}}{4\sqrt{\pi r_{0}^{3}}}\,\left(\frac{\Lambda_{1,2} - 2}{\Lambda_{1,2} - 1}\right)^{1/2}
\left[1 + \frac{2r_{0}\,\Lambda_{1,2}}{R + 2r_{0}\,\Lambda_{1,2}}\,(\xi - 1)\right]\,
\left[1 + \frac{2r_{0}\,\Lambda_{1,2}}{R - 2r_{0}\,\Lambda_{1,2}}\,(1 + \eta)\right].
\end{eqnarray*}
This result can also be obtained using expansion (\ref{3.3.8}).

\section{Stark effect in the charge-dyon system}
\markboth{CHAPTER 3. THE MIC-KEPLER PROBLEM}{3.4. STARK EFFECT IN CHARGE-DYON SYSTEM }

The Hamiltonian of the MIC-Kepler system in the external constant uniform electric field
is of the form \cite{M-17}
\begin{eqnarray}
\hat{\mathcal{H}} = \frac{1}{2\mu_0}\left(-i\hbar{\bf{\nabla}} -
\frac{e}{c}{\bf A}^{(\pm)}\right)^2 +\frac{\hbar^2{s}^2}{2\mu_0
r^2}-\frac{e^2}{r} + |e|\,\varepsilon\,z.
\label{3.4.1}
\end{eqnarray}
We have assumed that the electric field $\varepsilon$ is directid along positive
$z$-semiaxes, and the force acting the electron is directed along negative $z$-semiaxes.

Since the Hamiltonian (\ref{3.4.1}) possesses axial symmetry, it is convenient to consider the Schr\"{o}dinger
equation of the charge-dyon system in the external constant uniform electric field in the parabolic coordinates.
Thus, for convenience, the parabolic wave functions (\ref{3.1.17}) of the MIC-Kepler system are chosen
as nonperturbed ones for the calculation of the matrix elements of transitions between the mutually
degenerated states.

We are mostly interested the matrix elements of the transitions
$n_{1} n_{2} m \to  n′_{1}n′_{2} m′$ for the fixed value of the principal quantum
number $n$. The perturbation operator in parabolic coordinates reads
\begin{eqnarray*}
\hat{V} = |e|\,\varepsilon\,(\mu - \nu)/2.
\end{eqnarray*}
According to the perturbation theory the first-order corrections to the energy eigenvalue
$E_{n}^{(0)}$ (\ref{3.1.12}) are of the form
\begin{eqnarray*}
E_{n}^{(1)} = \int\,\psi_{n_{1} n_{2} m}^{*}(\mu, \nu, \varphi)\,
\hat{V}\,\psi_{n_{1} n_{2} m}(\mu, \nu, \varphi)\,dv.
\end{eqnarray*}
Hence, taking into account (\ref{3.1.17}), we get
\begin{eqnarray*}
E_{n}^{(1)} = \frac{|e|\,\varepsilon}{4r_{0}^{3}\,n^{4}}\,
\left({\mathcal{I}}_{n_{1},m-s}\,I_{n_{2}, m+s} - I_{n_{1}, m-s}\,{\mathcal{I}}_{n_{2},m+s}\right),
\end{eqnarray*}
where
\begin{eqnarray*}
I_{p q} = \int\limits_{0}^{\infty}\,\left[\Phi_{p q}(x)\right]^{2}\,dx,
\qquad
{\mathcal{I}}_{p q} = \int\limits_{0}^{\infty}\,x^{2\,}\left[\Phi_{p q}(x)\right]^{2}\,dx.
\end{eqnarray*}
Further, using formulas (\ref{1.6.5}) and (\ref{1.6.6}) for the integrals $I_{p q}$
and ${\mathcal{I}}_{p q}$, we obtain the following expressions
\begin{eqnarray*}
I_{p q} = r_{0}\,n, \qquad
{\mathcal{I}}_{p q} = 2\left(r_{0}\,n\right)^{3}\,\left[3p\,(p + |q| + 1) +
\frac{1}{2}\,|q|\,(|q| + 3) + 1\right].
\end{eqnarray*}
Now, using these formulae, we could get the first-order correction to
the energy eigenvalue (\ref{3.1.12})
\begin{eqnarray}
E_{n}^{(1)} = \frac{3\hbar^{2}\,|e|\,\varepsilon}{2\mu_{0}\,e^{2}}\,
\left[n\,\left(n_{1} - n_{2} - m_{-}\right) - \frac{m\,s}{3}\right].
\label{3.4.2}
\end{eqnarray}
Similar to the hydrogen atom, the linear term on $n$ is proportional to the $z$-component
of the Runge-Lenz vector. However, there is an extra correction linear on $m$, which removes
the degeneracy on the $z$-component of the angular momentum.

Thus, in the “charge-Dirac dyon” system there is the linear Stark effect, completely
removing the degeneracy on azimuth quantum number $m$.

For the fixed s, according to the formula (\ref{3.1.18}), the two extreme components of the splited
energy level correspond to the following values of the parabolic quantum numbers:
$n_{1} = n - |s| - 1, n_{2} = 0$ and $n_{1} = 0, n_{2} = n - |s| - 1$.
According (\ref{3.4.2}) the distance between these levels is
\begin{eqnarray*}
\Delta\,E_{n} = \frac{3\hbar^{2}\,|e|\,\varepsilon}{\mu_{0}\,e^{2}}\,
n\,(n - |s| - 1),
\end{eqnarray*}
i.e. the complete splitting of the level is proportional to $n^{2}$,
as on the case of the hydrogen atom.

The presence of the linear Stark effect means that in the unperturbed
state the charge-dyon bound system has a dipole momentum with a mean value
\begin{eqnarray*}
\bar{d}_{z} = -\frac{3\hbar^{2}\,|e|}{2\mu_{0}\,e^{2}}\,
\left[n\,\left(n_{1} - n_{2} - m_{-}\right) - \frac{m\,s}{3}\right].
\end{eqnarray*}
From the expression for the mean dipole momentum it is natural to give the
following definition for the dipole momentum’s operator
\begin{eqnarray*}
\hat{\bf{d}} = -\frac{3\hbar^{2}\,|e|}{2\mu_{0}\,e^{2}}\,
\left(n^{2}\hat{\bf I} - \frac{s}{3}\,\hat{\bf J}\right) =
|e|\,\left(\frac{3e^{2}}{4E_{n}^{(0)}}\,\hat{\bf I} +
\frac{\hbar^{2}\,s}{2\mu_{0}\,e^{2}}\,\hat{\bf J}\right).
\end{eqnarray*}
It is mentioned that all the formulae obtained for $s = 0$ yields the
corresponding formulae for the hydrogen atom.

\section{Scattering of electrons on the dyone}
\markboth{CHAPTER 3. THE MIC-KEPLER PROBLEM}{3.5. SCATTERING OF ELECTRONS ON THE DYON}

First, we present formulae that determine the spherical and parabolic wave functions of
the MIC-Kepler problem in the continuous spectrum \cite{M-18}.

The spherical basis in the continuous spectrum of the MIC-Kepler system can be written in the following form
\begin{eqnarray}
\psi_{k j m}^{(s)}(r, \theta, \varphi) = R_{k j}^{(s)}(r)\,
Z_{j m}^{(s)}(\theta, \varphi).
\label{3.5.1}
\end{eqnarray}
Normalized by the condition
\begin{eqnarray*}
\int\limits\,r^{2}\,R_{k' j}^{(s)*}(r)\,R_{k j}^{(s)}(r)\,dr = 2\pi\,\delta(k - k')
\end{eqnarray*}
the radial wave function has the form
\begin{eqnarray}
R_{k j}^{(s)}(r) = C_{k j}\,\frac{(2ikr)^{j}}{(2j + 1)!}\,e^{-ikr}\,
F\left(\frac{i}{kr_{0}} + j + 1; 2j + 2; 2ikr\right),
\label{3.5.2}
\end{eqnarray}
where the normalization constant is
\begin{eqnarray}
C_{k j} = 2k\,e^{\pi/2kr_{0}}\,
\left|\Gamma\left(j + 1 - \frac{i}{kr_{0}}\right)\right|,
\label{3.5.3}
\end{eqnarray}
and the explicit form of the angular wave function $Z_{j m}^{(s)}(\theta, \varphi)$
is given in formula (\ref{3.1.7}).

Normalized by the condition
\begin{eqnarray*}
\int\,r^{2}\,\psi_{k \Omega m}^{(s)}(\mu, \nu, \varphi)\,
\psi_{k' \Omega' m'}^{(s)*}(\mu, \nu, \varphi)\,dV
= 2\pi\,\delta(k - k')\,\delta(\Omega - \Omega')\,\delta_{m m'}
\end{eqnarray*}
the parabolic basis is given by
\begin{eqnarray}
\psi_{k \Omega m}^{(s)}(\mu, \nu, \varphi)  = C_{k \Omega m}^{(s)}\,
\Phi_{k, \Omega, m-s}^{(s)}(\mu)\,\Phi_{k, -\Omega, m+s}^{(s)}(\nu)\,
\frac{e^{i(m-s)\varphi}}{2\pi},
\label{3.5.4}
\end{eqnarray}
where
\begin{eqnarray*}
\Phi_{k, \Omega, q}^{(s)}(x) = e^{-ikx/2}\,\frac{(ikx)^{|q|/2}}{|q|!}\,
F\left(\frac{|q| + 1}{2} + \frac{i}{2kr_{0}} + i\frac{\sqrt{\mu_{0}}}{2\hbar k}\,\Omega;
|q| + 1; ikx\right)
\end{eqnarray*}
and the normalization constant is
\begin{eqnarray}
C_{k \Omega m}^{(s)} = \sqrt{\frac{k\sqrt{\mu_{0}}}{2\pi \hbar}}\,e^{\pi/2kr_{0}}\,\times \nonumber \\
\label{3.5.5} 
\\  [3mm]
\times \left|\Gamma\left(\frac{|m+s| + 1}{2} - \frac{i}{2kr_{0}} - i\frac{\sqrt{\mu_{0}}}{2\hbar k}\,\Omega\right)\,
\Gamma\left(\frac{|m-s| + 1}{2} - \frac{i}{2kr_{0}} + i\frac{\sqrt{\mu_{0}}}{2\hbar k}\,\Omega\right)\right|.
\nonumber
\end{eqnarray}
We note that to calculate normalization factors (\ref{3.5.3}) and (\ref{3.5.5}),
we use representation (\ref{1.13.3}) for the confluent hypergeometric
function.

We now consider the scattering of an electron on a dyon. Because the charge–dyon system is Coulomb-like,
the wave function is independent of the angle $\varphi$. Substituting $m+s=0$ into equations
(\ref{3.1.15}) and (\ref{3.1.16}), we obtain
\begin{eqnarray}
\frac{d}{d \mu}\left(\mu \frac{d\Phi_1}{d \mu}\right) +
\left[\frac{k^{2}}{4}\,\mu +
\frac{\sqrt{\mu_0}}{2\hbar}\,\Omega + \frac{1}{2r_0}\right]\Phi_1 &=& 0, \nonumber \\ [3mm]
\label{3.5.6}
\\  [3mm]
\frac{d}{d \nu}\left(\nu \frac{d\Phi_2}{d \nu}\right) +
\left[\frac{k^{2}}{4}\,\nu - \frac{s^2}{\nu}-
\frac{\sqrt{\mu_0}}{2\hbar}\Omega + \frac{1}{2r_0}\right]\Phi_2
&=& 0. \nonumber
\end{eqnarray}
We seek solutions of Eqs. (\ref{3.5.6}) such that the solution of the Schr\"{o}odinger
equation for negative $z \in (-\infty, 0)$ and large $r \to \infty$ has the form of a flat wave:
\begin{eqnarray*}
\psi^{(s)} \sim e^{ikz} = e^{ik(\mu - \nu)/2}.
\end{eqnarray*}
This condition can be satisfied if we set the separation constant equal to
\begin{eqnarray*}
\Omega = -i\,\frac{\hbar k}{\sqrt{\mu_0}} - \frac{\hbar}{r_0\,\sqrt{\mu_0}}.
\end{eqnarray*}
Substituting this relation in Eqs. (\ref{3.5.6}), we obtain a solution of the Schr\"{o}dinger
equation that describes the scattering of an electron in the field of a dyon:
\begin{eqnarray}
\psi^{(s)} = C_{k}^{(s)}\,(ik\nu)^{|s|}\,e^{ik(\mu - \nu)/2}\,
F\left(|s| + \frac{i}{kr_0}; 2|s| + 1; ik\nu\right),
\label{3.5.7}
\end{eqnarray}
where $C_k^{(s)}$ is the normalization constant. To separate the incident and the scattered
waves in function (\ref{3.5.7}), we must investigate the behavior of this function at large
distances from the scattering center. Using the first two terms in representation (\ref{1.13.3})
for the confluent hypergeometric function, we obtain
\begin{eqnarray*}
F\left(|s| + \frac{i}{kr_0}; 2|s| + 1; ik\nu\right) \approx e^{-\pi/2kr_0}\,
\Gamma(2|s| + 1)\,(ik\nu)^{-|s|}\times
\\ [7mm]
\times \left[\frac{e^{i\pi |s|}}{\Gamma\left(|s| + 1 - \frac{i}{kr_{0}}\right)}\,
\left(1 + \frac{1 + k^{2}r_{0}^{2}s^{2}}{ik^{3}r_{0}^{2}\nu}\right)
\,e^{-\frac{i}{{kr_0}}\ln k\nu}
+ \frac{|s| + i/(kr_0)}{\Gamma\left(|s| + 1 + \frac{i}{kr_{0}}\right)}
\,\frac{e^{ik\nu}}{ik\nu}\,e^{\frac{i}{{kr_0}}\ln k\nu}\right].
\end{eqnarray*}
for large $\nu$. We now substitute this relation in wave function (\ref{3.5.7})
and select the normalization constant $C_{k}^{(s)}$ in the form
\begin{eqnarray*}
C_{k}^{(s)} = e^{i\pi |s|}\,e^{\pi/2kr_0}\,
\frac{\Gamma\left(|s| + 1 - \frac{i}{kr_{0}}\right)}{\Gamma\left(2|s| + 1\right)},
\end{eqnarray*}
for the incident wave to have a unit amplitude. Using the formulas
\begin{eqnarray*}
r = \frac{1}{2}\,(\mu + \nu), \qquad \nu = r - z = r\,(1 - \cos\theta)
\end{eqnarray*}
to change to spherical coordinates, we obtain
\begin{eqnarray*}
\psi^{(s)} = \left(1 - \frac{i}{2k^{3}r_{0}^{2}\,\sin^{2}\theta/2}\right)\,
\exp\left[ikz - \frac{i}{kr_{0}}\,\ln\left(2kr\,\sin^{2}\frac{\theta}{2}\right)\right]
+ \frac{f(\theta)}{r}\,\exp\left[ikr + \frac{i}{kr_{0}}\,\ln\left(2kr\right)\right],
\end{eqnarray*}
where $f(\theta)$ is the scattering amplitude,
\begin{eqnarray}
f(\theta) = - \frac{e^{-i\pi |s|}(1 - ikr_{0}|s|}{2k^{2}r_{0}\sin^{2}\theta/2}\,
\frac{\Gamma\left(|s| + 1 - \frac{i}{kr_{0}}\right)}
{\Gamma\left(|s| + 1 + \frac{i}{kr_{0}}\right)}\,
\exp\left(\frac{2i}{kr_{0}}\,\ln\,\sin\frac{\theta}{2}\right).
\label{3.5.8}
\end{eqnarray}
Therefore, for the scattering cross section $d\sigma = |f(\theta)|^{2}\,d\Omega$
($d\Omega$ is the element of the solid angle), we obtain the formula
\begin{eqnarray*}
d\sigma = \frac{1 + k^{2}r_{0}^{2}s^{2}}{4k^{4}r_{0}^{2}\,\sin^{4}\theta/2}\,d\Omega,
\end{eqnarray*}
which, as can be easily seen, simplifies to the Rutherford formula for $s = 0$.

For the vector potential $A^{(-)}$ in (\ref{3.1.2}), the solution of Schr\"{o}dinger equation
for the MIC-Kepler system for the problem of scattering of an electron on a dyon is given by
\begin{eqnarray*}
\psi^{(-)} = \psi^{(+)}\,e^{2is\varphi}, \quad {\rm where} \quad \psi^{(+)}\equiv\psi^{(s)}
\end{eqnarray*}
as follows from gauge transformation (\ref{3.1.2-1}).

We return to formula (18). Scattering amplitude (\ref{3.5.8}) obtained here differs from the scattering amplitude
obtained in \cite{ZWANZIG} for the phase factor $e^{2is\varphi}$ of the gauge transformation.

\section{Interbasis expansions in the continuous spectrum}
\markboth{CHAPTER 3. THE MIC-KEPLER PROBLEM}{3.6. INTERBASIS EXPANSIONS IN THE CONTINUOUS SPECTRUM}

In the continuous spectrum the orbital moment $j$  runs all integer values
restricted from below by the inequality $j \geq m_{+}$. In this connection the
expansion of the parabolic basis of the MIC-Kepler problem (\ref{3.5.3}) over the spherical
one (\ref{3.5.1}) has the form
\begin{eqnarray}
\psi_{k \Omega m}^{(s)}(\mu, \nu, \varphi)
= \sum_{j=m_{+}}^{\infty}\,
W_{k\Omega m s}^{j}
\psi_{kjm}^{(s)}(r,\theta,\varphi)\,.
\label{3.6.1}
\end{eqnarray}
Next, to use the orthonormality of the monopole harmonics of Tamm (\ref{3.1.9})
\begin{eqnarray*}
\int\limits_{0}^{\pi}\,\int\limits_{0}^{2\pi}\,\sin\theta\,Z_{j'm'}^{(s)*}(\theta,\varphi)\,
Z_{jm}^{(s)}(\theta,\varphi)\,d\theta\,d\varphi = \delta_{j j'}\,\delta_{m m'}
\end{eqnarray*}
we multiply both sides of expansion (\ref{3.6.1}) by $\sin\theta\,Z_{jm}^{(s)*}(\theta,\varphi)$,
integrate over the angles $\theta$ and $\varphi$. After that, represent the confluent hypergeometric
functions in the parabolic basis as the series (\ref{2.10.2}), and pass to spherical coordinates
according to relation (\ref{3.2.3}). Then, instead of (\ref{3.6.1}) we have
\begin{eqnarray*}
W_{k\Omega m s}^{j}\,F\left(j + 1 + \frac{i}{kr_{0}}; 2j + 2; 2ikr\right) =
(-1)^{\frac{m-s + |m-s|}{2}}\,\frac{(2j + 1)!}{2^{j+m_{+}}\,|m+s|!\,|m-s|!}\,
\frac{C_{k \Omega m}^{(s)}}{C_{k j}}\times \\ [7mm]
\times \left[\frac{(2j + 1)\,(j-m_{+})!\,(j+m_{+})!}{2(j-m_{-})!\,(j+m_{-})!}\right]^{1/2}\,
\sum_{p=0}^{\infty}\,\sum_{t=0}^{\infty}\,\frac{(u)_{p}\,(v)_{t}}{p!\,t!}\,
\frac{(ikr)^{p+t-j+m_{+}}}{(|m-s|+1)_{p}\,(|m+s|+1)_{t}}\,Q_{j m s}^{p t},
\end{eqnarray*}
where
\begin{eqnarray*}
u = \frac{|m-s|+1}{2} + \frac{i}{2kr_{0}} + i\frac{\sqrt{\mu_{0}}}{2\hbar k}\,\Omega,
\qquad
v = \frac{|m+s|+1}{2} + \frac{i}{2kr_{0}} - i\frac{\sqrt{\mu_{0}}}{2\hbar k}\,\Omega,
\end{eqnarray*}
and
\begin{eqnarray*}
Q_{j m s}^{p t} = \int\limits_{0}^{\pi}\,\sin\theta\,(1 + \cos\theta)^{|m-s|+p}\,
(1 - \cos\theta)^{|m+s|+t}\,P_{j-m_{+}}^{(|m+s|,|m-s|)}(\cos\theta)\,d\theta.
\end{eqnarray*}
Now, as in section $2.10$, using the Rodrigue formula for the Jacobi polynomial (\ref{2.10.4}),
and sequentially integrating in parts, we make sure that the integral $Q_{j m s}^{p t}$
is nonzero under the condition $p+t-j+m_{+} \geq> 0$ and therefore all terms of the series
contain $r$ to a non-negative degree, so that in the limit $r \to 0$ we get
\begin{eqnarray}
W_{k\Omega m s}^{j} &=& (-1)^{j+\frac{m-s + |m-s|}{2}}\,\frac{(2j + 1)!}{2^{j+m_{+}}\,|m+s|!\,|m-s|!}\,
\left[\frac{(2j + 1)\,(j-m_{+})!\,(j+m_{+})!}{2(j-m_{-})!\,(j+m_{-})!}\right]^{1/2}\times \nonumber \\ [5mm]
\label{3.6.2}
\\  [5mm]
&\times& \frac{C_{k \Omega m}^{(s)}}{C_{k j}}\,
\sum_{p=0}^{j-m_{+}}\,\frac{(u)_{p}\,(v)_{j-m_{+}-p}}{p!\,(j-m_{+}-p)!}\,
\frac{Q_{j m s}^{p, j-m_{+}-p}}{(|m-s|+1)_{p}\,(|m+s|+1)_{j-m_{+}-p}}\,.\nonumber
\end{eqnarray}
The integral $Q_{j m s}^{p t}$ at $t=j-m_{+}-p$ passes into a closed expression
\begin{eqnarray*}
Q_{j m s}^{p j-m_{+}-p} = (-1)^{\frac{m-s + |m-s|}{2}}\,
\frac{2^{j+m_{+}+1}}{(2j + 1)!}\,(j - m_{-})!\,(j + m_{-})!.
\end{eqnarray*}
Substituting the last relation into the formula (\ref{3.6.2}) and considering the
auxiliary equalities,
\begin{eqnarray*}
(v)_{j-m_{+}-p} = \frac{(-1)^{p}\,(v)_{j-m_{+}}}{(1-j+m_{+}-v)_{p}}, \qquad
(j-m_{+}-p)! = \frac{(-1)^{p}\,(j-m_{+})!}{(-j+m_{+})_{p}},
\end{eqnarray*}
\begin{eqnarray*}
(|m+s|+1)_{j-m_{+}-p} = \frac{(-1)^{p}\,(j + m_{-})!}{|m-s|!\,(-j - m_{-})_{p}}
\end{eqnarray*}
for the expansion coefficients $W_{k\Omega m s}^{j}$, we obtain
\begin{eqnarray*}
W_{k\Omega m s}^{j} = (-1)^{j+\frac{m-s + |m-s|}{2}}\,\frac{C_{k \Omega m}^{(s)}}{C_{k j}}\,
\frac{2}{|m-s|!}\sqrt{\frac{2j+1}{2}\,\frac{(j+m_{+})!\,(j-m_{-})!}
{(j-m_{+})!\,(j+m_{-})!}}\times \\ [5mm]
\times (v)_{j-m_{+}}\,\,
{_3F_2}\left\{\matrix{ -j+m_{+},\,\,-j-m_{-},\,\, u\cr \cr
|m-s| + 1,\,\,1-j + m_{+} - v
\cr}\Biggr|1\right\}.
\end{eqnarray*}
Further, using the formula (\ref{1.7.11}) and explicit expressions of the normalization
constants $C_{k j}$ and $C_{k \Omega m}^{(s)}$, finally obtain
\begin{eqnarray}
W_{k\Omega m s}^{j} &=& (-1)^{j+\frac{m-s + |m+s|}{2}}\,\sqrt{\frac{\mu_{0}^{1/2}}
{\pi \hbar k}}\,\frac{e^{-i\delta_{j}}}{|m+s|!}
\sqrt{\frac{2j+1}{4}\,\frac{(j+m_{+})!\,(j-m_{-})!}
{(j-m_{+})!\,(j+m_{-})!}}\times \nonumber \\ [5mm]
\label{3.6.3}
&\times& \frac{\left|\Gamma\left(\frac{|m+s| + 1}{2} - \frac{i}{2kr_{0}} - i\frac{\sqrt{\mu_{0}}}{2\hbar k}\,\Omega\right)\,
\Gamma\left(\frac{|m-s| + 1}{2} - \frac{i}{2kr_{0}} + i\frac{\sqrt{\mu_{0}}}{2\hbar k}\,\Omega\right)\right|}
{\Gamma\left(m_{+} + 1 + \frac{i}{kr_{0}}\right)}\times \\  [5mm]
&\times& {_3F_2}\left\{\matrix{ -j+m_{+},\,\,j+m_{+}+1,\,\,
\frac{|m-s| + 1}{2} + \frac{i}{2kr_{0}} + i\frac{\sqrt{\mu_{0}}}{2\hbar k}\,\Omega\cr \cr
|m-s| + 1,\,\,m_{+} + 1 + \frac{i}{kr_{0}}
\cr}\Biggr|1\right\}. \nonumber
\end{eqnarray}
where
\begin{eqnarray*}
\delta_{j} = \arg\,\Gamma\left(j + 1 - \frac{i}{kr_{0}}\right)
\end{eqnarray*}
the scattering phase of the charge-dyon system. Using the formula (\ref{1.7.11}),
we can verify that the expansion coefficients (\ref{3.6.3}) are real.

Now let's consider the inverse expansion, i.e. the expansion of the spherical basis
of the charge-dyon system according to the parabolic basis.

Generally speaking, the parabolic constant $\Omega$ can be real or complex (for example, in the problem of
a charged particle scattering in the field of a dyon). Therefore, it is not known beforehand over which
$\Omega$ should we integrate in the inverse transformation.

We consider the integral
\begin{eqnarray*}
Q_{j j'} = \int\limits_{-\infty}^{\infty}\,W_{k\Omega m s}^{j}\,
W_{k\Omega m s}^{j'*}\,d\Omega.
\end{eqnarray*}
We substitute expression (\ref{3.6.3}) for $W_{k\Omega m s}^{j}$ in this integral
and replace the generalized hypergeometric function ${_3F_2}$ with its expression
as a finite sum. If we also replace $\Omega$ with $z = i\frac{\sqrt{\mu_{0}}}{2\hbar k}\,\Omega$,
we obtain
\begin{eqnarray*}
Q_{j j'} = \frac{(-1)^{j+j'+m-s - |m+s|}}{\left|\Gamma\left(m_{+} + 1 + \frac{i}{kr_{0}}\right)\right|^{2}}\,
\frac{e^{-i(\delta_{j'}-\delta_{j})}}{\left(|m-s|!\right)^{2}}\,\sqrt{(2j+1)\,(2j'+1)}\times \\ [7mm]
\times \left[\frac{(j+m_{+})!\,(j-m_{-})!\,(j'+m_{+})!\,(j'-m_{-})!}
{(j-m_{+})!\,(j+m_{-})!\,(j'-m_{+})!\,(j'+m_{-})!}\right]^{1/2}\times \\ [7mm]
\times \sum\limits_{p=0}^{j-m_{+}}\,\frac{(j-m_{+})_{p}\,(j+m_{+}+1)_{p}}
{p!\,\left(m_{+} + 1 + \frac{i}{kr_{0}}\right)_{p}}\,\,
\sum\limits_{t=0}^{j'-m_{+}}\,\frac{(j'-m_{+})_{t}\,(j'+m_{+}+1)_{t}}
{t!\,\left(m_{+} + 1 - \frac{i}{kr_{0}}\right)_{t}}\,\,B_{p t},
\end{eqnarray*}
where
\begin{eqnarray*}
B_{p t} = \frac{1}{2\pi i}\,\int\limits_{-i\infty}^{i\infty}\,
\Gamma\left(\frac{|m-s| + 1}{2} + p + \frac{i}{2kr_{0}} + z\right)\,
\Gamma\left(\frac{|m-s| + 1}{2} + t - \frac{i}{2kr_{0}} - z\right)\times \\ [7mm]
\times \Gamma\left(\frac{|m+s| + 1}{2} - \frac{i}{2kr_{0}} + z\right)\,
\Gamma\left(\frac{|m+s| + 1}{2} + \frac{i}{2kr_{0}} - z\right)\,dz.
\end{eqnarray*}
The integral $B_{p t}$ can be calculated using Barnes' lemma (\ref{2.10.8})
and as a result, for the integral $Q_{j j'}$ we obtain the expression
we obtain
\begin{eqnarray*}
Q_{j j'} = (-1)^{j+j'+m-s - |m+s|}\,\,
\frac{e^{-i(\delta_{j'}-\delta_{j})}\,|m+s|!}{|m-s|!\,(2m_{+} + 1)!}\,\sqrt{(2j+1)\,(2j'+1)}\times \\ [7mm]
\times \left[\frac{(j+m_{+})!\,(j-m_{-})!\,(j'+m_{+})!\,(j'-m_{-})!}
{(j-m_{+})!\,(j+m_{-})!\,(j'-m_{+})!\,(j'+m_{-})!}\right]^{1/2}\times \\ [7mm]
\times \sum\limits_{p=0}^{j-m_{+}}\,\frac{(j-m_{+})_{p}\,(j+m_{+}+1)_{p}}
{p!\,\left(2m_{+} + 2\right)_{p}}\,\,
{_3F_2}\left\{\matrix{ -j'+m_{+},\,\,j'+m_{+}+1,\,\,
|m-s| + p + 1 \cr \cr
|m-s| + 1,\,\,2m_{+} + p + 2
\cr}\Biggr|1\right\}.
\end{eqnarray*}
Now applying Saalschutz's theorem twice (\ref{2.10.9}) we will get
\begin{eqnarray*}
Q_{j j'} = \frac{(-1)^{2(j-j'+ \frac{m-s + |m-s|}{2})}\,e^{-i(\delta_{j'}-\delta_{j})}}
{\Gamma\left(j - j' + 1\right)\,\Gamma\left(j' - j + 1\right)}\,
\frac{\sqrt{(2j+1)\,(2j'+1)}}{j + j' +1}\times \\ [7mm]
\times \left[\frac{(j+m_{+})!\,(j-m_{-})!\,(j'-m_{+})!\,(j'+m_{-})!}
{(j-m_{+})!\,(j+m_{-})!\,(j'+m_{+})!\,(j'-m_{-})!}\right]^{1/2}.
\end{eqnarray*}
Since the numbers $j$ and $j’$ are integers or half-integers at the same time,
the last expression turns to zero at $j \neq j’$ due to gamma functions from
$(j - j’ +1)$ and $(j‘- j +1)$, and to one – at $j = j’$ i.e.
\begin{eqnarray}
Q_{j j'} = \int\limits_{-\infty}^{\infty}\,W_{k\Omega m s}^{j}\,
W_{k\Omega m s}^{j'*}\,d\Omega = \delta_{j j'}.
\label{3.6.4}
\end{eqnarray}
Then, taking into account (\ref{3.6.4}) for the inverse expansion, we obtain
\begin{eqnarray*}
\psi_{kjm}^{(s)}(r,\theta, \varphi) = \int\limits_{-\infty}^{\infty}\,
W_{k\Omega m s}^{j}\,\psi_{k\Omega m}^{(s)}(\mu, \nu, \varphi),
\end{eqnarray*}
where integration is carried out along the real axis.

The results obtained in this section at $s=0$ with the accuracy of the phase
multiplier coincide with the formulas obtained in \cite{POGOSYAN1}.

\section{The MIC-Kepler problem and the 4D oscillator}
\markboth{CHAPTER 3. THE MIC-KEPLER PROBLEM}{3.7. THE MIC-KEPLER PROBLEM AND THE $4D$ OSCILLATOR}

Let us show that the MIC-Kepler problem is dual to the four-dimensional
isotropic oscillator \cite{M-16}. In the spherical coordinates (\ref{3.1.4})
the Schr\"{o}dinger equation (\ref{3.1.1}) has the form
\begin{eqnarray}
&&\frac{1}{r^2}\frac{\partial}{\partial r}\left( r^2\frac{\partial
\Psi^{(s)}}{\partial r}\right)
+ \frac{1}{r^2}\left[\frac{1}{\sin
\theta}\frac{\partial}{\partial \theta}\left( \sin \theta
\frac{\partial \Psi^{(s)}}{\partial \theta}\right) +
\frac{1}{\sin^2\theta}\frac{\partial^2\Psi^{(s)}} {\partial
\varphi^2}\right]- \nonumber \\ [5mm]
\label{3.7.1}
\\ [5mm]
&-&
\frac{2is}{r^2(1-\cos\theta)}\frac{\partial \Psi^{(s)}}{\partial
\varphi}-
 \frac{2s^2}{r^2(1-\cos\theta)}\Psi +
\frac{2\mu_0}{\hbar^2}\left(\varepsilon +
\frac{e^2}{r}\right)\Psi^{(s)} = 0. \nonumber
\end{eqnarray}
Now, if the following substitutions are made in equation (\ref{3.7.1})
\begin{eqnarray*}
\Psi^{(s)}({\bf r}) \to \Psi({\bf r},\gamma) = \Psi^{(s)}({\bf r})
\frac{e^{-i\gamma}}{\sqrt{4\pi}}, \qquad s \to
-i\frac{\partial}{\partial \gamma},\qquad {\rm where} \quad \gamma
\in [0, 4\pi),
\end{eqnarray*}
then equation (\ref{3.7.1}) will go into the equation
\begin{eqnarray}
\left[\frac{1}{r^2}\frac{\partial}{\partial r}\left(r^2
\frac{\partial}{\partial r}\right) - \frac{{\hat {\bf L}
}^2}{r^2}\right] \Psi + \frac{2\mu_0}{\hbar^2}\left(\varepsilon +
\frac{e^2}{r}\right) \Psi = 0,
\label{3.7.2}
\end{eqnarray}
where
\begin{eqnarray*}
{\hat {\bf L}}^2 = -\left[\frac{1}{\sin
\beta}\frac{\partial}{\partial \beta} \left(\sin
\beta\frac{\partial}{\partial \beta}\right) +
\frac{1}{\sin^2\beta}\left(\frac{\partial^2}{\partial \alpha^2} -
2\cos\beta\frac{\partial^2}{\partial \alpha \partial \gamma}+
\frac{\partial^2}{\partial \gamma^2}\right)\right].
\end{eqnarray*}
Here, we have introduced the following notation: $\beta = \theta$ and
$\alpha = \varphi$.

Further, passing from the spherical coordinates ($r, \alpha, \beta, \gamma$)
to the new coordinates ($u_0, u_1, u_2, u_3$) according to
\begin{eqnarray}
u_0 + iu_1 = u\cos{\frac{\beta}{2}} e^{i\frac{\alpha +
\gamma}{2}}, \qquad u_2 + iu_3 = u\sin{\frac{\beta}{2}}
e^{i\frac{\alpha - \gamma}{2}}
\label{3.7.3}
\end{eqnarray}
so that $u^2 = u_0^2+u_1^2+u_2^2+u_3^2 = r$, taking into account the fact that
the Laplace operator in the coordinates $u_{\mu}$ has the form
\begin{eqnarray}
\frac{\partial^{2}}{\partial u_{\mu^{2}}} = \Delta_4 =
\frac{1}{u^3}\frac{\partial}{\partial u}
\left(u^3\frac{\partial}{\partial u}\right) - \frac{4}{u^2}{\hat
{\bf L}}^2, \qquad  {\rm where} \qquad \mu =0,1,2,3
\label{3.7.4}
\end{eqnarray}
\begin{eqnarray}
E = 4e^2, \qquad \varepsilon = -\frac{\mu_0 \omega^2}{8}
\label{3.7.5}
\end{eqnarray}
one can easily see that equation (\ref{3.7.2}) turns exactly into the
Schr\"{o}dinger equation for the four-dimensional isotropic oscillator
\begin{eqnarray}
\left[\Delta_4 +
\frac{2\mu_0}{\hbar}\left(E - \frac{\mu_0\omega^2
u^2}{2}\right)\right] \psi({\bf u}) = 0,
\label{3.7.6}
\end{eqnarray}
the energy spectrum of which is given by the well-known formula
\begin{eqnarray}
E_N = \hbar \omega (N+2), \qquad {\rm where} \qquad
N=0,1,2,....
\label{3.7.7}
\end{eqnarray}
From a comparison of the energy spectra of the MIC--Kepler problem
(\ref{3.1.12}) and the $4D$ oscillator (\ref{3.7.7}), with taking into account
(\ref{3.7.5}) we get the following relation between the principal quantum
numbers $n$ and $N$:
\begin{eqnarray}
N = 2n-2.
\label{3.7.8}
\end{eqnarray}
Establish a relation between the Cartesian coordinates $x, y, z$ and the
coordinates $u_\mu$. From the definition of the coordinates $u_\mu$
(\ref{3.7.3}) with the allowance made for the notation \\
$r=u^2,\qquad \theta = \beta, \qquad \varphi = \alpha$, we obtain the following relations:
\begin{eqnarray*}
\cos\theta &=&
\frac{u_0^2+u_1^2-u_2^2-u_3^2}{u_0^2+u_1^2+u_2^2+u_3^2}, \qquad
\sin\theta =
\frac{2\sqrt{(u_0^2+u_1^2)(u_2^2+u_3^2)}}{u_0^2+u_1^2+u_2^2+u_3^2},
\\ [7mm]
\cos\varphi &=&
\frac{u_0u_2-u_1u_3}{\sqrt{(u_0^2+u_1^2)(u_2^2+u_3^2)}}, \qquad
\sin\varphi
=\frac{u_0u_3+u_1u_2}{\sqrt{(u_0^2+u_1^2)(u_2^2+u_3^2)}}\,.
\end{eqnarray*}
Substituting the last relations in (\ref{3.1.4}) and and recalling that
$r=u^2=u_0^2+u_1^2+u_2^2+u_3^2$, we have
\begin{eqnarray}
x = 2(u_0u_2 - u_1u_3), \qquad y &=& 2(u_0u_3 + u_1u_2), \qquad
z = u_0^2 + u_1^2 - u_2^2 - u_3^2,  \nonumber \\ [3mm]
\label{3.7.9}
\\  [3mm]
\gamma &=& \frac{i}{2}\ln{\frac{(u_0-iu_1)(u_2+iu_3)}
{(u_0+iu_1)(u_2-iu_3)}}. \nonumber
\end{eqnarray}
The first three equations in (\ref{3.7.9}) are the transformation
${\rm I\!R}^4 \to {\rm I \!R}^3$ used by Kustaanheimo and Stiefel for regularization
of equations of the celestial mechanics \cite{KS}, and are called the
Kustaanheimo--Stiefel transformation (KS-transformation).

Thus, we have proved that the bound MIC--Kepler system is dual to the
four-dimensional isotropic oscillator and that the generalized version of the
KS-transformation (\ref{3.7.9}) represent the duality transformation.
rep
Let us discuss the problem of degeneracy multiplicity of energy levels
(\ref{3.1.11}) and (\ref{3.7.7}). It follows from formula (\ref{3.1.17}) that
at fixed quantum numbers $n, m$ and $s$ the degeneracy multiplicity of energy
levels (\ref{3.1.11}) equals
\begin{eqnarray*}
g_{nm}^s = n - m_{+}.
\end{eqnarray*}
For $s \geq 0$, the multiplicity of degeneracy of energy levels (\ref{3.1.12}) at
fixed $n$ and $s$ will be
\begin{eqnarray*}
g_{n}^s = \sum_{|m|\geq s}g_{nm}^s + \sum_{|m|\leq s-1}g_{nm}^s\,,
\end{eqnarray*}
where the upper limit of summation is determined from the condition $g_{nm}^s
\geq 1$, i.е.,
\begin{eqnarray*}
m_{+} \leq 2n - 2.
\end{eqnarray*}
Consequently,
\begin{eqnarray}
g_{n}^s = \sum_{m=-s+1}^{s-1}(n - s) + 2\sum_{m=s}^{n-1}(n - m) =
(n-s)(n+s)\,.
\label{3.7.10}
\end{eqnarray}
The same result follows from a similar calculation for $s < 0$.
Since the quantum numbers $n$ and $s$ simultaneously take either integer or half-integer values,
summing up the formula (\ref{3.7.10}), we obtain
\begin{eqnarray*}
g_n = \sum_{s=-n+1}^{n-1} g_{n}^s = \frac{1}{3}n(2n-1)(2n+1),
\end{eqnarray*}
where $g_n$ is the multiplicity of the degeneracy of the energy levels of the
$4D$ oscillator (\ref{3.7.7}). Since, according to (\ref{3.7.8}),
$N = 2n - 2$, we arrive at the well-known result \cite{BAKER}:
\begin{eqnarray}
g_N = \frac{1}{6}(N+1)(N+2)(N+3).
\label{3.7.11}
\end{eqnarray}

\section{Fundamental bases of the $4D$ oscillator}
\markboth{CHAPTER 3. THE MIC-KEPLER PROBLEM}{3.8. FUNDAMENTAL BASES OF THE $4D$ OSCILLATOR}

We consider the problem of four-dimensional isotropic oscillator only in the
following five systems of coordinates: 1) Cartesian coordinates $u_\mu \in
(-\infty, \infty)$ $(\mu =0,1,2,3)$; 2) Euler coordinates $u, \alpha, \beta, \gamma$;
3) double polar coordinates $\rho_1, \rho_2, \varphi_1, \varphi_2$;
4) canonical spherical coordinates $u, \psi, \theta, \varphi$;
5) four-dimensional spheroidal coordinates $\xi, \eta, \alpha, \gamma$.

Let's start by defining the four-dimensional coordinates we use
(the differential elements of length, volume, and the Laplace operator):

\begin{itemize}
\item
Euler coordinates $u \in [0, \infty)$,
$\alpha \in [0, 2\pi)$, $\beta \in [0, \pi]$, $\gamma \in [0, 4\pi)$
\begin{equation}
u_{0} + iu_{1} = u\cos{\frac{\beta}{2}} e^{i\frac{\alpha -
\gamma}{2}}, \qquad u_{2} + iu_{3} = u\sin{\frac{\beta}{2}}
e^{i\frac{\alpha + \gamma}{2}},
\label{3.8.1}
\end{equation}
\vspace{3mm}
\begin{eqnarray*}
dl_{4}^{2} = du^{2} + \frac{u^{2}}{4}\,\left(d\alpha^{2} + d\beta^{2} + d\gamma^{2}
+ 2\cos\beta\,d\alpha\,d\gamma\right), \qquad dV_{4} = \frac{u^{3}}{8}\,
\sin\beta\,du\,d\beta\,d\alpha\,d\gamma,
\end{eqnarray*}
\vspace{3mm}
\begin{eqnarray*}
\Delta_{4} = \frac{1}{u^{3}}\,\frac{\partial}{\partial u}\,\left(u^{3}\,\frac{\partial}{\partial u}\right)
- \frac{4}{u^{2}}\,{\hat {\bf L}}^2
\end{eqnarray*}
where
\begin{eqnarray}
{\hat {\bf L}}^2 = -\left[\frac{1}{\sin
\beta}\frac{\partial}{\partial \beta} \left(\sin
\beta\frac{\partial}{\partial \beta}\right) +
\frac{1}{\sin^2\beta}\left(\frac{\partial^2}{\partial \alpha^2} -
2\cos\beta\frac{\partial^2}{\partial \alpha \partial \gamma}+
\frac{\partial^2}{\partial \gamma^2}\right)\right].
\label{3.8.2}
\end{eqnarray}
\item
Double polar coordinates $\rho_{1},\rho_{2} \in [0,\infty)$,
$\varphi_{1},\varphi_{2} \in [0,2\pi)$
\begin{equation}
u_{0}+i u_{1}= \rho_{1}\,e^{i\varphi_{1}}, \qquad u_{2}+i u_{3}= \rho_{2}\,
e^{i\varphi_{2}},
\label{3.8.3}
\end{equation}
\vspace{3mm}
\begin{eqnarray*}
dl_{4}^{2} = d\rho_{1}^{2} + d\rho_{2}^{2} + \rho_{1}^{2}\,d\varphi_1^{2}
+ \rho_{2}^{2}\,d\varphi_2^{2}, \\ [5mm]
dV_{4} = \rho_{1}\,\rho_{2}\, d\rho_{1}\,d\rho_{2}\,d\varphi_1\,d\varphi_2,
\end{eqnarray*}
\vspace{3mm}
\begin{eqnarray*}
\Delta_4 = \frac{1}{\rho_{1}} \frac{\partial}{\partial
\rho_{1}}\left(\rho_{1}\frac{\partial} {\partial \rho_{1}}\right) +
\frac{1}{\rho_{2}} \frac{\partial}{\partial
\rho_{2}}\left(\rho_{2}\frac{\partial} {\partial \rho_{2}}\right) +
\frac{1}{\rho_{1}^{2}}\frac{\partial^2}{\partial \varphi_{1}^{2}}+
+\frac{1}{\rho_{2}^{2}}\frac{\partial^2}{\partial \varphi_{2}^{2}}.
\end{eqnarray*}
\item
Canonical spherical coordinates $u \in
[0, \infty)$, $\psi \in [0, \pi]$, $\theta \in [0, \pi]$, $\varphi \in [0,
2\pi)$
\begin{eqnarray}
u_{0}+iu_{1}= u\sin\psi \sin\theta\, e^{i\phi}, \qquad
u_{2}= u\sin\psi \cos\theta, \qquad  u_{3}= u\cos\psi,
\label{3.8.4}
\end{eqnarray}
\begin{eqnarray*}
dl_{4}^{2} = du^{2} + u^{2}\,\left[d\psi^{2} + \sin^{2}\psi\,
\left(d\theta^{2} + \sin^{2}\theta\,d\phi^{2}\right)\right],
\\  [5mm]
dV_{4} = u^{3}\,\sin^{2}\psi\,sin\theta\,du\,d\psi\,d\theta\,d\phi,
\end{eqnarray*}
\vspace{3mm}
\begin{eqnarray*}
\Delta_4 = \frac{1}{u^{3}} \frac{\partial}{\partial
u}\left(u^{3}\frac{\partial} {\partial u}\right) + \frac{1}{u^{2}}
\left[\frac{1}{\sin^2\psi}\frac{\partial}{\partial
\psi}\left(\sin^2\psi\frac{\partial} {\partial \psi}\right)
-\frac{{\hat l}^2}{\sin^2\psi}\right],
\end{eqnarray*}
where
\begin{eqnarray*}
{\hat l}^2 = -\left[\frac{1}{\sin\theta}\frac{\partial}{\partial
\theta}\left(\sin\theta\frac{\partial}{\partial \theta}\right)
+\frac{1}{\sin^2\theta}\frac{\partial^2}{\partial \phi^2}\right].
\end{eqnarray*}
\item
Spheroidal coordinates
\begin{eqnarray}
u_0+iu_1 = \frac{d}{2}\sqrt{(\xi+1)(1+\eta)}\,\,
e^{i\frac{\alpha+\gamma}{2}}, \quad u_2+iu_3 =
\frac{d}{2}\sqrt{(\xi-1)(1-\eta)}\,\,
e^{i\frac{\alpha-\gamma}{2}}.
\label{3.8.5}
\end{eqnarray}
Here $d$ is the spheroidal parameter $(0\leq d < \infty)$.
\begin{eqnarray*}
dl_{4}^{2} &=& \frac{d^{2}}{8}\,\left[(\xi - \eta)\,\left(\frac{d\xi^{2}}{\xi^{2} - 1} +
\frac{d\eta^{2}}{1 - \eta^{2}}\right) + (\xi + \eta)\,
(d\alpha^{2} + d\gamma^{2}) + (\xi\,\eta + 1)\,d\alpha\,d\gamma\right], \\ [7mm]
dV_{4} &=& \frac{d^{4}}{8}\,(\xi - \eta)\,d\xi\,d\eta\,d\alpha\,d\gamma,
\end{eqnarray*}
\vspace{3mm}
\begin{eqnarray*}
\Delta_4 &=& \frac{8}{d^2(\xi - \eta)}
\Biggl\{\frac{\partial}{\partial \xi}\left[(\xi^2-1)
\frac{\partial}{\partial \xi}\right] + \frac{\partial}{\partial
\eta}\left[(1- \eta^2) \frac{\partial}{\partial \eta}\right]-   \\
[7mm]
&-&\frac{1}{2}\left(\frac{1}{\xi + 1} - \frac{1}{1 +
\eta}\right) \left(\frac{\partial}{\partial \alpha} +
\frac{\partial}{\partial \gamma} \right)^2 +
\frac{1}{2}\left(\frac{1}{\xi - 1} + \frac{1}{1 - \eta}\right)
\left(\frac{\partial}{\partial \alpha} - \frac{\partial}{\partial
\gamma} \right)^2\Biggr\}.
\end{eqnarray*}
\end{itemize}

Now we will give explicit expressions for the fundamental bases of a four-dimensional
isotropic oscillator in Cartesian, Euler, double polar and canonical spherical
coordinate systems:
\begin{itemize}
\item
The Cartesian basis
\begin{eqnarray}
\left|\,{\cal C}\,\right\rangle =\left|N_0, N_1, N_2,
N_3\right\rangle = a^{2}\,\overline{H}_{N_0}\left(a u_0
\right)\,\overline{H}_{N_1}\left(a u_1\right)\,
\overline{H}_{N_2}\left(a u_2\right)\,\overline{H}_{N_3}\left(a u_3\right),
\label{3.8.6}
\end{eqnarray}
where $a= (\mu_0\omega/\hbar)^{1/2}$, and $\overline{H}_{n}\left(x\right)$ is
the orthonormalized Hermite polynomial, which is related to the ordinary Hermite
polynomial $H_{n}(x)$ as follows
\begin{eqnarray*}
\overline{H}_{n}\left(x\right) = \frac{e^{-x^{2}/2}}{\pi^{1/4}\sqrt{2^{n}\,n!}}\,
H_{n}\left(x\right).
\end{eqnarray*}
\item
The Euler basis
\begin{eqnarray}
\left|\,{\cal E}\,\right\rangle \equiv \left|N, L, M, M'\right\rangle =
\sqrt{\frac{2L+1}{2\pi^2}}R_{NL}\left(u\right)
D_{M M'}^L(\alpha,\, \beta,\, \gamma),
\label{3.8.7}
\end{eqnarray}
where
\begin{eqnarray}
R_{NL}\left(u\right) =
\sqrt{\frac{2\left(\frac{N}{2}+L+1\right)!}
{\left(\frac{N}{2}-L\right)!}}\,
\frac{a^2(a u)^{2L}}{(2L+1)!}\,e^{-\frac{a^2 u^2}{2}}
F\left(-\frac{N}{2}+L; 2L+2; a^2 u^2\right).
\label{3.8.8}
\end{eqnarray}
The Euler basis of the four-dimensional oscillator (\ref{3.8.7}) is
simultaneously proper for the operators ${\hat {\bf L}}^2$, ${\hat
L}_3=-i\partial/\partial \alpha$ and ${\hat L}_{3'}=-i\partial/\partial
\gamma$, with
\begin{eqnarray}
{\hat {\bf L}}^2\left|N, L, M, M'\right\rangle  &=& L(L+1)\left|N,
L, M, M'\right\rangle,  \nonumber \\ [5mm]
{\hat L}_3 \left|N,
L,M, M'\right\rangle &=& M \left|N, L, M, M'\right\rangle,
\label{3.8.9}
\\ [5mm]
{\hat L}_{3'}\left|N, L, M, M'\right\rangle
&=& M' \left|N, L, M, M'\right\rangle. \nonumber
\end{eqnarray}
In Cartesian coordinates, the square of the moment operator ${\hat {\bf L}}^2$ (\ref{3.8.2}) has the form
\begin{eqnarray}
{\hat {\bf L}}^2 = -\frac{1}{4}\,\sum\limits_{\mu < \nu}\,
\left(u_{\mu}\,\frac{\partial}{\partial u_{\nu}} -
u_{\nu}\,\frac{\partial}{\partial u_{\mu}}\right)^{2}.
\label{3.8.10}
\end{eqnarray}
\item
The double polar basis
\begin{eqnarray}
\left|\,2P\,\right\rangle \equiv \left|N_{\rho_1}, N_{\rho_2}, k_1,
k_2\right\rangle = \frac{1}{2\pi}f_{N_{\rho_1},k_1}(a^2\rho_1^2)\,
f_{N_{\rho_2},k_2}(a^2\rho_2^2)\, e^{ik_1\varphi_1}\,e^{ik_2\varphi_2},
\label{3.8.11}
\end{eqnarray}
where
\begin{eqnarray}
f_{pq}(x) = \frac{a}{|q|!} \sqrt{\frac{2(p+|q|)!}{p!}}
e^{-x/2}x^{|q|/2} F(-p,\, |q|+1,\, x).
\label{3.8.12}
\end{eqnarray}
Quantum numbers $N_{\rho_1}$, $N_{\rho_2}$, $k_1$ and $k_2$ are related to the
principal quantum number $N$ as follows:
\begin{eqnarray}
N = 2N_{\rho_1} + 2N_{\rho_2} + |k_1| + |k_2|.
\label{3.8.13}
\end{eqnarray}
The double polar basis is an eigenvalue of the following operators:
\begin{eqnarray}
{\hat {\cal P}} \left|N_{\rho_1}, N_{\rho_2}, k_1,
k_2\right\rangle &=& (2N_{\rho_1}-2N_{\rho_2}+|k_1|-|k_2|)
\left|N_{\rho_1}, N_{\rho_2}, k_1, k_2\right\rangle,
\nonumber  \\ [5mm]
{\hat L}_3 \left|N_{\rho_1}, N_{\rho_2}, k_1, k_2\right\rangle &=&
\frac{k_1 - k_2}{2} \left|N_{\rho_1}, N_{\rho_2}, k_1,
k_2\right\rangle, \label{3.8.14} \\ [5mm]
{\hat L}_{3'} \left|N_{\rho_1},
N_{\rho_2}, k_1, k_2\right\rangle &=& \frac{k_1 + k_2}{2}
\left|N_{\rho_1}, N_{\rho_2}, k_1, k_2\right\rangle, \nonumber
\end{eqnarray}
where the operator ${\hat {\cal P}}$ is
\begin{eqnarray}
{\hat {\cal P}} = \frac{\hbar}{2\mu_0\omega}\left(
-\frac{\partial^2}{\partial u_0^2} - \frac{\partial^2}{\partial
u_1^2} + \frac{\partial^2}{\partial u_2^2} +
\frac{\partial^2}{\partial u_3^2} \right) + \frac{\mu_0
\omega}{2\hbar}\left(u_0^2+u_1^2-u_2^2-u_3^2\right).
\label{3.8.15}
\end{eqnarray}
\item
The canonical basis
\begin{eqnarray}
\left|\,{\cal K}\,\right\rangle \equiv \left|N, J, l, \overline{m}
\right\rangle = R_{NJ}(u)Y_{Jl\overline{m}}(\psi, \theta, \phi),
\label{3.8.16}
\end{eqnarray}
where
\begin{eqnarray}
R_{NJ}(u) = \frac{a^2(au)^{J}}{(J+1)!}
\sqrt{\frac{2\left(\frac{N+J}{2}+1\right)!}
{\left(\frac{N-J}{2}\right)!}}\, e^{-\frac{a^2 u^2}{2}}
F\left(-\frac{N-J}{2}; J+2; a^2u^2\right)
\label{3.8.17}
\end{eqnarray}
and
\begin{eqnarray}
Y_{Jl\overline{m}}(\psi, \theta, \phi) =
2^l\,l!\sqrt{\frac{(2J+2)(J-l)!}{\pi
(J+l+1)!}}\,\left(\sin\psi\right)^l\,C_{J-l}^{l+1}\left(\cos\psi\right)\,
Y_{l\overline{m}}\left(\theta, \phi\right).
\label{3.8.19}
\end{eqnarray}
Here  $C_{n}^{\lambda}(x)$ is the Gegenbauer polynomial and
$Y_{l\overline{m}}\left(\theta, \phi\right)$ is an ordinary spherical function.

For the canonical basis, the following spectral problem takes place:
\begin{eqnarray}
{\hat{\bf J}}^2\,\left|N, J, l, \overline{m}\right\rangle &=&
J(J+2)\,\left|N, J, l, \overline{m}\right\rangle,
\nonumber \\ [5mm]
{\hat{\bf l}}^2\,\left|N, J, l, \overline{m}\right\rangle &=&
l(l+1)\,\left|N, J, l, \overline{m}\right\rangle, \label{3.8.19} \\ [5mm]
{\hat l}_3\,\left|N, J, l, \overline{m}\right\rangle &=&
\overline{m}\,\left|N, J, l, \overline{m}\right\rangle.
\nonumber
\end{eqnarray}
The ranges of variation of quantum numbers are: $J=0,2,...,N$ or $J=1,3,...,N$
for even and odd $N$, respectively, and $0 \leq l \leq J$, $|\overline{m}| \leq l$.
\end{itemize}

\section{Interbasis expansions in a $4D$ oscillator}
\markboth{CHAPTER 3. THE MIC-KEPLER PROBLEM}{3.9. INTERBASIS EXPANSIONS IN A $4D$ OSCILLATOR}

Now we will consider the problem of interbasis expansions of the four-dimensional isotropic
oscillator. The general number of nontrivial interbasis expansions equals
twelve. By virtue of the unitarity condition the coefficients of "direct" and
"inverse" expansions are obtained from each other by complex conjugation. Thus,
it is sufficient to study the following six expansions:
\begin{eqnarray}
\left|\,{\cal C}\,\right\rangle
&=&
{\sum_{m_1=-N_0-N_1}^{N_0+N_1}}'\,{\sum_{m_2=
-N_2-N_3}^{N_2+N_3}}'\,\left\langle\, 2P\,|\,{\cal C}
\,\right\rangle\, \left|\,2P\, \right\rangle,
\label{3.9.1}
\\[5mm]
\left|\,{\cal C}\,\right\rangle
&=&
\sum_{L=\frac{|M|+|M'|}{2}}^{N/2}\,{\sum_{M=-N_0-N_1}^{N_0+N_1}}'\,
{\sum_{M'=-N_2-N_3}^{N_2+N_3}}'\,\left\langle\, {\cal E} \,|\,{\cal C}
\,\right\rangle\, \left|\,{\cal E}\, \right\rangle,
\label{3.9.2}
\\[5mm]
\left|\,{\cal C}\,\right\rangle
&=&
{\sum_{J=0,1}^{N}}\,'\,\sum_{l=0}^{J}\,\sum_{\overline{m}=
-l}^{l}\,\left\langle\, {\cal K} \,|\,{\cal C}\,\right\rangle\,
\left|\,{\cal K}\, \right\rangle,
\label{3.9.3}
\\[5mm]
\left|\,{\cal K}\,\right\rangle
&=&
\sum_{L=\frac{|M|+|M'|}{2}}^{N/2}\,\sum_{M,M'=-L}^{L}\,\left\langle\,
{\cal E} \,|\,{\cal K}\,\right\rangle\, \left|\,{\cal E}\,
\right\rangle,
\label{3.9.4}
\\[5mm]
\left|\,2P\,\right\rangle
&=&
\sum_{L=\frac{|M|+|M'|}{2}}^{N/2}\,\sum_{M,M'=-L}^{L}\,\left\langle\,
{\cal E} \,|\,2P\,\right\rangle\, \left|\,{\cal E}\,
\right\rangle,
\label{3.9.5}
\\[5mm]
\left|\,{\cal K}\,\right\rangle
&=&
{\sum_{J=0}^{N/2}}\,'\,\sum_{l=0}^{J}\,\sum_{\overline{m}=-l}^{l}\,
\left\langle\,2P\,|\,{\cal K}\,\right\rangle\, \left|\,2P\,\right\rangle.
\label{3.9.6}
\end{eqnarray}
Primes over the summation symbol means that the corresponding summation is only
over those values of indices whose parity coincides with that of a maximum
possible value indicated in the sum.

All six coefficients (\ref{3.9.1}) - (\ref{3.9.6}) are explicitly expressed in
terms of three structural elements: the Wigner $d$ function
$d_{m,m'}^j(\pi/2)$, Clebsch--Gordan coefficients
$C_{a,\alpha;b,\beta}^{c,\gamma}$ and the objects ${\cal
M}_{a,\alpha;b,\beta}^{c,\gamma}$ obtained by an analytic continuation of
ordinary Clebsch--Gordan coefficients $C_{a,\alpha;b,\beta}^{c,\gamma}$ to the
region of one-fourth values of some indices. According to \cite{VAR},
\begin{eqnarray*}
{\cal M}_{a,\alpha;b,\beta}^{c,\gamma} &=&
\frac{\delta_{\gamma,\alpha+\beta}\Delta(a,b,c)}
{\Gamma(a+b-c+1)\Gamma(c-b+\alpha+1)\Gamma(c-a-\beta+1)}\times
\\ [6mm] &\times&
\left[\frac{(2c+1)\Gamma(a+\alpha+1)\Gamma(b-\beta+1)
\Gamma(c+\gamma+1)\Gamma(c-\gamma+1)}{\Gamma(a-\alpha+1)
\Gamma(b+\beta+1)}\right]^{1/2}\times \\ [6mm]
&\times& {_3F}_2
\left\{
\begin{array}{l}
c-a-b, -a + \alpha, -b - \beta \\ [3mm]
c-a-\beta +1, c-b + \alpha +1\\
\end{array}
\biggr| 1 \right\}.
\end{eqnarray*}
In the last formula
\begin{eqnarray*}
\Delta(a,b,c) =
\left[\frac{\Gamma(a+b-c+1)\Gamma(a-b+c+1)\Gamma(d-a+c+1)}
{\Gamma(a+b+c+2)}\right]^{1/2}.
\end{eqnarray*}
Note that the objects ${\cal M}_{a,\alpha;b,\beta}^{c,\gamma}$ appear already in the
problem of interbasis expansions in a three \cite{POGOSYAN6} and
multidimensional isotropic oscillator \cite{POGOSYAN2}. Some problems of the quantum theory
of angular momentum also lead to them \cite{SMOR-SHEL}.

Let us write down the formulae determining the explicit form of the
coefficients of interbasis expansions (\ref{3.9.1}) - (\ref{3.9.6})
\begin{eqnarray}
\left\langle\, 2P\,|\,{\cal C}\,\right\rangle
&=& e^{i\pi\left(N_1+N_3-|k_1|-|k_2|\right)/2}\,
d_{\frac{k_1}{2},\,\frac{N_0-N_1}{2}}^{\frac{N_0+N_1}{2}}
\left(\frac{\pi}{2}\right)\,
d_{\frac{k_2}{2},\,\frac{N_2-N_3}{2}}^{\frac{N_2+N_3}{2}}
\left(\frac{\pi}{2}\right),
\label{3.9.7}
\\[7mm]
\left\langle\, {\cal E} \,|\,{\cal C}\,\right\rangle
&=& e^{i\pi
\left(N+N_1-N_3-2|M|-M'\right)/2}\,
d_{\frac{M+M'}{2},\,\frac{N_0-N_1}{2}}^{\frac{N_0+N_1}{2}}
\left(\frac{\pi}{2}\right)\,
d_{\frac{M-M'}{2},\,\frac{N_2-N_3}{2}}^{\frac{N_2+N_3}{2}}
\left(\frac{\pi}{2}\right)\times
\nonumber \\[3mm]
\label{3.9.8}
\\[1mm]
&\times&\, C^{L, \frac{|M|+|M'|}{2}}_{\frac{N-|M|+|M'|}{4},\,
\frac{N-2N_0-2N_1+|M|+|M'|}{4};
\,\frac{N+|M|-|M'|}{4},\,\frac{2N_0+2N_1-N+|M|+|M'|}{4}}\,,
\nonumber
\\[5mm]
\left\langle\, {\cal K} \,|\,{\cal C}\,\right\rangle\,
&=&
e^{i\pi \left(|M|-N_1\right)/2}\,
d_{\frac{M}{2},\,\frac{N_0-N_1}{2}}^{\frac{N_0+N_1}{2}}
\left(\frac{\pi}{2}\right)\,{\cal M}_{\frac{N+l+1}{4},
\frac{N-2N_3+l+1}{4};\,\frac{N-l-1}{4},\frac{2N_3-N+l-1}{4}}
^{\frac{J}{2},\frac{l}{2}}\times\,
\nonumber \\[3mm]
\label{3.9.9}
\\[7mm]
&\times&\,{\cal M}_{\frac{N-N_4+|M|}{4},\frac{N_0+N_1-N_2+|M|}{4};\,
\frac{N-N_4-|M|-1}{4},
\frac{N_2-N_1-N_0+|M|-1}{4}}^{\frac{2l-1}{4},\,\frac{2|M|-1}{4}\gamma}\,,
\nonumber
\\[7mm]
\left\langle\, {\cal E} \,|\,{\cal K}\,\right\rangle
&=& e^{i\pi
\left(J+l+2|M|-|M'|\right)/2}\,\delta_{L,\frac{J}{2}}\,
\delta_{\overline{m},M+M'}
C_{\frac{J}{2}, M'; \frac{J}{2}, M}^{l,\overline{m}},
\label{3.9.10}
\\[7mm]
\left\langle\, {\cal E} \,|\,2P\,\right\rangle
&=&
(-1)^{N_{\rho_1}+\frac{k_2+|k_2|}{2}}\,\delta_{M,\frac{k_1-k_2}{2}}\,
\delta_{M',\frac{k_1+k_2}{2}}\times\,
\nonumber \\[3mm]
\label{3.9.11}
\\[1mm]
&\times&\, C^{L,
\frac{|k_1|+|k_2|}{2}}_{\frac{N_{\rho_1}+N_{\rho_2}+|k_2|}{2},\,
\frac{N_{\rho_2}-N_{\rho_1}+|k_2|}{2}\,;
\,\frac{N_{\rho_1}+N_{\rho_2}+|k_1|}{2},\,
\frac{N_{\rho_1}-N_{\rho_2}+|k_1|}{2}}\,,
\nonumber
\\[7mm]
\left\langle\, 2P \,|\,{\cal K}\,\right\rangle
&=& e^{i\pi
\left(2N_{\rho_1}+J+l+2|k_1|-|k_2|\right)/2}\,
\delta_{\overline{m}, k_1}\, C_{\frac{J}{2}, \frac{k_1-k_2}{2};
\frac{J}{2}, \frac{k_1+k_2}{2}}^{l,\overline{m}}\times
\nonumber \\[3mm]
\label{3.9.12}
\\[3mm]
&\times& C^{\frac{J}{2},
\frac{|k_1|+|k_2|}{2}}_{\frac{N_{\rho_1}+N_{\rho_2}+|k_2|}{2},\,
\frac{N_{\rho_2}-N_{\rho_1}+|k_2|}{2}\,;
\,\frac{N_{\rho_1}+N_{\rho_2}+|k_1|}{2},\,
\frac{N_{\rho_1}-N_{\rho_2}+|k_1|}{2}}.
\nonumber
\end{eqnarray}

We now describe the methods used in calculating interbasis coefficients. In
calculating the coefficients (\ref{3.9.7}), (\ref{3.9.9}), (\ref{3.9.10}) and
(\ref{3.9.11}) we made use of the known recipes suggested in the fourth chapter
and paper \cite{KIL} which are useful for the isotropic oscillator of arbitrary
dimension. The results (\ref{3.9.9}) и (\ref{3.9.12}) are obtained by the
formulae
\begin{eqnarray*}
\left\langle\,{\cal E}\,|\,{\cal C}\,\right\rangle =
\sum_{\left|\,2P\,\right\rangle}\,\left\langle\, {\cal E}
\,|\,2P\,\right\rangle\,\left\langle\,2P\,|\,{\cal
C}\,\right\rangle, \qquad
\left\langle\,2P\,|\,{\cal K}\,\right\rangle =
\sum_{\left|\,{\cal E}\,\right\rangle}\,\left\langle\, 2P \,|\,{\cal
E}\,\right\rangle\,\left\langle\,{\cal E}\,|\,{\cal
K}\,\right\rangle.
\end{eqnarray*}

\section{Dyon-oscillator correspondence}
\markboth{CHAPTER 3. THE MIC-KEPLER PROBLEM}{3.10. DYON-OSCILLATOR CORRESPONDENCE}

The connection between the KS transformation and the physics of hydrogen-like atoms in external
electric and strong magnetic fields has been noted in \cite{Kibler-Negadi-1,Kibler-Negadi-2,Kibler-Negadi-3,
Kibler-Negadi-4,Kibler4,Chen-1,Chen-2,Chen-3}. On this basis \cite{Kibler4}, the problem of expansions
connecting the bases of the hydrogen atom and a four-dimensional isotropic oscillator was formulated.
General selection rules and bilinear relationships were derived, to which the expansion coefficients
must obey, numerical and analytical calculation programs were compiled on a computer, explicit forms of
some coefficients were obtained for specific (some of the first) values of quantum numbers.
For arbitrary values of quantum numbers, the explicit form of these coefficients was established
by us in \cite{M-19}.

In this section we calculate the expansion coefficients relating to each other the fundamental bases
of the bound states of the MIC-Kepler problem and the four-dimensional isotropic oscillator.

Write down all eight expansions
\begin{eqnarray}
\left|\,n j m s\,\right\rangle &=& \sum\,\left\langle\,N L M
M'\,|\,n j m s\,\right\rangle\, \left|\,N L M M'\, \right\rangle,
\label{3.10.1}
\\ [3mm]
\left|\,n j m s\,\right\rangle &=&
\sum\,\left\langle\,N_{\rho_1} N_{\rho_2} k_1 k_2\,|\,n j m s
\,\right\rangle\, \left|\,N_{\rho_1} N_{\rho_2} k_1 k_2\,
\right\rangle,
\label{3.10.2}
\\ [3mm]
\left|\,n j m s\,\right\rangle &=& \sum\,\left\langle\,N J l
\overline{m}\,|\,n j m s \,\right\rangle\, \left|\,N J l
\overline{m}\, \right\rangle,
\label{3.10.3}
\\ [3mm]
\left|\,n j m s\,\right\rangle &=& \sum\,\left\langle\,N_0 N_1 N_2
N_3 \,|\,n j m s \,\right\rangle\, \left|\,N_0 N_1 N_2 N_3\,
\right\rangle,
\label{3.10.4}
\\ [3mm]
\left|\,n_1 n_2 m s\,\right\rangle &=&
\sum\,\left\langle\,N_{\rho_1} N_{\rho_2} k_1 k_2 \,|\,n_1 n_2 m s
\,\right\rangle\, \left|\,N_{\rho_1} N_{\rho_2} k_1 k_2\,
\right\rangle,
\label{3.10.5}
\\ [3mm]
\left|\,n_1 n_2 m s\,\right\rangle &=& \sum\,\left\langle\,N L M
M' \,|\,n_1 n_2 m s \,\right\rangle\, \left|\,N L M M'\,
\right\rangle,
\label{3.10.6}
\\ [3mm]
\left|\,n_1 n_2 m s\,\right\rangle &=& \sum\,\left\langle\,N J l
\overline{m}\,|\,n_1 n_2 m s \,\right\rangle\, \left|\,N J l
\overline{m}\, \right\rangle,
\label{3.10.7}
\\ [3mm]
\left|\,n_1 n_2 m s\,\right\rangle &=& \sum\,\left\langle\,N_0 N_1
N_2 N_3\,|\,n_1 n_2 m s \,\right\rangle\, \left|\,N_0 N_1 N_2
N_3\, \right\rangle.
\label{3.10.8}
\end{eqnarray}
Here, for convenience, we have carried out the following notations of the
spherical and parabolic wave function of the MIC-Kepler problem
\begin{eqnarray*}
\left|\,n j m s\,\right\rangle = \psi_{n j m}^{(s)}\,\frac{e^{is\gamma}}{4\pi}, \qquad
\left|\,n_1 n_2 m s\,\right\rangle =\psi_{n_1 n_2 m}^{(s)}\,\frac{e^{is\gamma}}{4\pi}.
\end{eqnarray*}
Let us first consider expansion (\ref{3.10.1}). From the formulas (\ref{3.1.4}), (\ref{3.7.3})
and (\ref{3.7.9}) it follows that $r = u^{2}, \theta = \beta$ and $\varphi = \alpha$.
Next, substituting into (\ref{3.10.1}) explicit expressions for the spherical basis of the
MIC--Kepler problem (\ref{3.1.11}) and the Euler basis of the four-dimensional isotropic
oscillator (\ref{3.8.7}), passing in the left-hand side from the spherical coordinates to the Euler ones,
and taking finally into account relation (\ref{3.1.9}), we get that
\begin{eqnarray}
\left\langle\,N L M M'\,|\,n j m s\,\right\rangle =
4n\,\sqrt{r_0}\,\delta_{n,\frac{N}{2}+1}\,\delta_{L,j}\,
\delta_{M,m}\,\delta_{M',s}.
\label{3.10.9}
\end{eqnarray}
In an analogous way we can show that the transition matrix in expansion
(\ref{3.10.5}) is also diagonal and has the form
\begin{eqnarray}
\left\langle\,N_{\rho_1} N_{\rho_2} k_1 k_2 \,|\,n_1 n_2 m s
\,\right\rangle = 4n\, \sqrt{r_0}\, \delta_{N_{\rho_1},
n_1}\,\delta_{N_{\rho_2}, n_2}\,
\delta_{k_1,m-s}\,\delta_{k_2,m+s}.
\label{3.10.10}
\end{eqnarray}
In deriving the last formula we proceeded from that the parabolic and double
polar coordinates were related with each other by: $\mu = 2\rho_1^2$, $\nu =
2\rho_2^2$, $\alpha = \varphi_1 + \varphi_2$, $\gamma = \varphi_1 - \varphi_2$.
The latter can easily be established with the use of the definitions of these
coordinates and the $KS$ transformation (\ref{3.7.9}).

Using now formulae (\ref{3.2.8}), (\ref{3.2.10}), (\ref{3.10.9}), and
(\ref{3.10.5}), and the identities
\begin{eqnarray*}
\left\langle\,N_{\rho_1} N_{\rho_2} k_1 k_2 \,|\,n j m s
\,\right\rangle
&=& \sum_{\left|\,n_1 n_2 m
s\,\right\rangle}\,\left\langle\,N_{\rho_1} N_{\rho_2} k_1 k_2
\,|\,n_1 n_2 m s \,\right\rangle \left\langle\,n_1 n_2 m s \,|\,n
j m s \,\right\rangle,
\\[3mm]
\left\langle\,N L M M' \,|\,n_1 n_2 m s \,\right\rangle
&=&
\sum_{\left|\,n j m s\,\right\rangle}\left\langle\,N L M M' \,|\,n
j m s \,\right\rangle \left\langle\,n j m s \,|\,n_1 n_2 m s
\,\right\rangle
\end{eqnarray*}
we find explicit expressions for the expansion coefficients(\ref{3.10.2}) and
(\ref{3.10.6}):
\begin{eqnarray}
\left\langle\,N_{\rho_1} N_{\rho_2} k_1 k_2 \,|\,n j m s
\,\right\rangle = (-1)^{N_{\rho_1}+\frac{m-s+|m-s|}{2}}\, 4n\,
\sqrt{r_0}\, \delta_{k_1,m-s}\,\delta_{k_2,m+s}\times
\nonumber \\ [1mm]
\label{3.10.11}
\\ [1mm]
\times \, C_{\frac{N_{\rho_1}+N_{\rho_2}+|m+s|}{2},
\frac{N_{\rho_2}-N_{\rho_1}+|m+s|}{2};
\frac{N_{\rho_1}+N_{\rho_2}+|m-s|}{2},
\frac{N_{\rho_1}-N_{\rho_2}+|m-s|}{2}}^
{j,\,m_+}\,,
\nonumber
\end{eqnarray}
\begin{eqnarray}
\left\langle\,N L M M' \,|\,n_1 n_2 m s \,\right\rangle =
(-1)^{n_1+\frac{m-s+|m-s|}{2}}\, 4n\,
\sqrt{r_0}\,\delta_{n,\frac{N}{2}+1}\,\delta_{M,m-s}\,\delta_{M',s}\times
\nonumber \\ [1mm]
\label{3.10.12}
\\ [1mm]
\times \,
C_{\frac{n+m_{-} -1}{2},\frac{n+m_{-} -1}{2}-n_1; \frac{n-m_-
-1}{2},\frac{m_{+} +|m+s|-n+1}{2}+n_1}^ {L, m_+}\,.
\nonumber
\end{eqnarray}
For complete solution of the problem we need to calculate the coefficients
$\left\langle\,N J l \overline{m}\,|\,n j m s \,\right\rangle$,\,\,\,
$\left\langle\,N_0 N_1 N_2 N_3 \,|\,n j m s \,\right\rangle$,\,\,\,
$\left\langle\,N J l \overline{m}\,|\,n_1 n_2 m s \,\right\rangle$\, и\,
$\left\langle\,N_0 N_1 N_2 N_3\,|\,n_1 n_2 m s \,\right\rangle$. The
coefficients can be obtained by the following formulae:
\begin{eqnarray*}
\left\langle\,N J l \overline{m} \,|\,n j m s \,\right\rangle
&=&
\sum\,\left\langle\,N J l \overline{m} \,|\,N L M M'
\,\right\rangle \left\langle\,N L M M' \,|\,n j m s
\,\right\rangle\,,
\\[3mm]
\left\langle\,N_0 N_1 N_2 N_3\,|\,n j m s \,\right\rangle
&=&
\sum\,\left\langle\,N_0 N_1 N_2 N_3 \,|\,N L M M' \,\right\rangle
\left\langle\,N L M M' \,|\,n j m s \,\right\rangle\,,
\\[3mm]
\left\langle\,N J l \overline{m} \,|\,n_1 n_2 m s \,\right\rangle
&=& \sum\,\left\langle\,N J l \overline{m} \,|\,N_{\rho_1}
N_{\rho_2} k_1 k_2\,\right\rangle \left\langle\,N_{\rho_1}
N_{\rho_2} k_1 k_2 \,|\,n_1 n_2 m s \,\right\rangle\,,
\\[3mm]
\left\langle\,N_0 N_1 N_2 N_3\,|\,n_1 n_2 m s \,\right\rangle
&=&
\sum\,\left\langle\,N_0 N_1 N_2 N_3 \,|\,N_{\rho_1} N_{\rho_2} k_1
k_2 \,\right\rangle \left\langle\,N_{\rho_1} N_{\rho_2} k_1 k_2
\,|\, n_1 n_2 m s\,\right\rangle\,.
\end{eqnarray*}
Here
\begin{eqnarray*}
\left\langle\,N J l \overline{m} \,|\,N L M M' \,\right\rangle &\equiv&
\left\langle\,{\cal K}\,|\,{\cal E}\,\right\rangle, \qquad
\left\langle\,N_0 N_1 N_2 N_3 \,|\,N L M M' \,\right\rangle \equiv
\left\langle\,{\cal C}\,|\,{\cal E}\,\right\rangle, \\ [3mm]
\left\langle\,N J l \overline{m} \,|\,N_{\rho_1}
N_{\rho_2} k_1 k_2\,\right\rangle &\equiv& \left\langle\,{\cal
K}\,|\,2P\,\right\rangle, \qquad
\left\langle\,N_0 N_1 N_2 N_3 \,|\,N_{\rho_1}
N_{\rho_2} k_1 k_2 \,\right\rangle \equiv \left\langle\,{\cal
C}\,|\,2P\,\right\rangle.
\end{eqnarray*}
Then using explicit expressions for the coefficients (\ref{3.9.7}),
(\ref{3.9.8}), (\ref{3.9.10}), (\ref{3.9.12}), (\ref{3.10.9}), and
(\ref{3.10.10}), we have
\begin{eqnarray}
\left\langle\,N J l \overline{m} \,|\,n j m s \,\right\rangle =
4n\sqrt{r_0}\,e^{i\pi (2j+l+2|m|-|s|)/2}\,\delta_{n,\frac{N}{2}+1}
C_{j,s;j,m}^{l,m+s},
\label{3.10.13}
\end{eqnarray}
\begin{eqnarray}
\left\langle\,N_0 N_1 N_2 N_3\,|\,n j m s \,\right\rangle =
4n\sqrt{r_0}\,e^{i\pi
(N+N_1-N_3-2|m|-s)/2}\,\delta_{n,\frac{N}{2}+1} \,
d_{\frac{m+s}{2},\,\frac{N_0-N_1}{2}}^{\frac{N_0+N_1}{2}}
\left(\frac{\pi}{2}\right)\times \nonumber \\ [1mm]
\label{3.10.14}
\\ [1mm]
\times\,d_{\frac{m-s}{2},\,\frac{N_2-N_3}{2}}^{\frac{N_2+N_3}{2}}
\left(\frac{\pi}{2}\right)\,
C^{j, \frac{|m|+|s|}{2}}_{\frac{N-|m|+|s|}{4},\,
\frac{N-2N_0-2N_1+|m|+|s|}{4};
\,\frac{N+|m|-|s|}{4},\,\frac{2N_0+2N_1-N+|m|+|s|}{4}}\,,
\nonumber
\end{eqnarray}
\begin{eqnarray}
\left\langle\,N J l \overline{m} \,|\,n_1 n_2 m s \,\right\rangle
= 4n\sqrt{r_0}\,e^{i\pi (2n_1+ J +l +2|m-s|-|m+s|)/2}\,
\delta_{n,\frac{N}{2}+1}\, \delta_{\overline{m},m+s}\times
\nonumber \\ [1mm]
\label{3.10.15}
\\ [1mm]
\times\, C_{\frac{J}{2}, s; \frac{J}{2}, m}^{l, m+s}\,
C^{\frac{J}{2}, m_{+}}_{\frac{n_1+n_2+|m+s|}{2},\,
\frac{n_2-n_1 +|m+s|}{2};
\,\frac{n_1+n_2+|m-s|}{2},\,\frac{n_1-n_2+|m-s|}{2}}\,,
\nonumber
\end{eqnarray}
\begin{eqnarray}
\left\langle\,N_0 N_1 N_2 N_3\,|\,n_1 n_2 m s \,\right\rangle =
4n\sqrt{r_0}\,e^{i\pi (N_1+N_3-|m+s|-|m-s|)/2}\,
\delta_{n,\frac{N}{2}+1}\times \nonumber \\ [1mm]
\label{3.10.16}
\\ [1mm]
\times\,d_{\frac{m+s}{2},\,\frac{N_0-N_1}{2}}^{\frac{N_0+N_1}{2}}
\left(\frac{\pi}{2}\right)\,
d_{\frac{m-s}{2},\,\frac{N_2-N_3}{2}}^{\frac{N_2+N_3}{2}}
\left(\frac{\pi}{2}\right)\,.
\nonumber
\end{eqnarray}
In the end, we note that the results obtained in this section are a generalization of the
correspondence between the Coulomb and oscillator wave functions obtained by us in \cite{M-20}.

\section{Spheroidal basis of the $4D$ oscillator}
\markboth{CHAPTER 3. THE MIC-KEPLER PROBLEM}{3.11. SPHEROIDAL BASIS OF THE $4D$ OSCILLATOR}

According to the definition of the spheroidal coordinates (\ref{3.8.5}),
$u^{2} = d^{2}\,(\xi + \eta)/2$, then, after substitution,
\begin{eqnarray*}
\psi(\xi, \eta, \alpha, \gamma; d) = X(\xi; d)\,Y(\eta; d)\,\frac{e^{iM\alpha}\,e^{iM'\gamma}}{\sqrt{8\pi^{2}}},
\end{eqnarray*}
the variables in the Schr\"{o}dinger equation (\ref{3.7.6}) are separated and we arrive at the equations
\begin{eqnarray}
\frac{d}{d\xi}\left[(\xi^{2} - 1)\frac{dX}{d\xi}\right] +
\left[\frac{\mu_{0}Ed^{2}}{4\hbar^{2}}\,\xi - \frac{a^{4}d^{4}}{16}\,(\xi^{2} - 1)
+ \frac{(M+M')^{2}}{2(\xi + 1)} - \frac{(M-M')^{2}}{2(xi - 1)} - Q\right]X = 0,
 \nonumber \\ [1mm]
\label{3.11.1}
\\ [1mm]
\frac{d}{d\eta}\left[(1 - \eta^{2})\frac{dY}{d\eta}\right] -
\left[\frac{\mu_{0}Ed^{2}}{4\hbar^{2}}\,\eta - \frac{a^{4}d^{4}}{16}\,(1 - \eta^{2})
+ \frac{(M+M')^{2}}{2(1 + \eta)} + \frac{(M-M')^{2}}{2(1 - \eta)} - Q\right]Y = 0.
\nonumber
\end{eqnarray}
Here, Q is the separation constant.

Further, using the method proposed by Miller \cite{MIL1}, i.e. excluding energy $E$
from equations (\ref{3.11.1}), we conclude that the wave function
$\psi \equiv \left| N p M M';d\right\rangle$ is an eigenfunction of the operator $\hat{Q}$, which has the form:
\begin{eqnarray*}
\hat{Q} &=& - \frac{1}{\xi - \eta}\,\left\{\eta\,\frac{\partial}{\partial \xi}\,
\left[(\xi^{2} - 1)\,\frac{\partial}{\partial \xi}\right] + \xi\,\frac{\partial}{\partial \eta}\,
\left[(1 - \eta^{2})\,\frac{\partial}{\partial \eta}\right]\right\} - \\ [3mm]
&-& \frac{\xi + \eta - 1}{2(xi - 1)\,(1 - \eta)}\,
\left(\frac{\partial}{\partial \alpha} - \frac{\partial}{\partial \gamma}\right)^{2} +
\frac{a^{4}d^{4}}{16}\,(\xi \eta + 1).
\end{eqnarray*}
In the last expression, passing to Cartesian coordinates and taking into account
(\ref{3.8.10}) and (\ref{3.8.15}) we obtain that
\begin{eqnarray}
\hat{Q} = {\hat{\bf L}}^{2} + \frac{a^{2}d^{2}}{4}\,{\hat {\cal P}}.
\label{3.11.2}
\end{eqnarray}

Thus, the spheroidal basis of a four-dimensional isotropic oscillator is determined by
solving the following system of equations:
\begin{eqnarray}
{\hat H}\,\left|N p M M';d\right\rangle &=& E
\,\left|N p M M';d\right\rangle, \qquad
{\hat Q}\,\left|N p M, M';d\right\rangle =
Q_p\,\left|N p M M';d\right\rangle, \nonumber \\
\label{3.11.2-1}
\\
{\hat L}_3 \left|N p M M';d\right\rangle &=&
M \left|N p M M';d\right\rangle, \qquad
{\hat L}_{3'}\left|N p M M';d\right\rangle
= M' \left|N p M M';d\right\rangle, \nonumber
\end{eqnarray}
where ${\hat H}$ is the Hamiltonian of the four-dimensional isotropic
oscillator and the subscript $p$ numbers the discrete values ​​of the
spheroidal separation constant $Q$.

The explicit form of the spheroidal integral of motion suggests that the
spheroidal basis of the four-dimensional isotropic oscillator can most
conveniently  be represented as expansion over the Euler and double polar
bases, i.e., in the form
\begin{eqnarray}
\left|\,NpMM';d\,\right\rangle =
\sum_L\,
U_{NpMM'}^{L}(d)\left|\,NLMM'\,\right\rangle,
\label{3.11.3}
\end{eqnarray}
\begin{eqnarray}
\left|\,NpMM';d\,\right\rangle =
\sum_{N_{\rho_1}}
V_{NpMM'}^{N_{\rho_1}}(d)\,
\left|\,N_{\rho_1}, N_{\rho_2}, (M+M')/2,
(M-M')/2\,\right\rangle,
\label{3.11.4}
\end{eqnarray}
where summation over $L$ and over $N_{\rho_1}$ is carried out within the
limits
\begin{eqnarray*}
\frac{|M+M'|+|M-M'|}{2} \leq L \leq \frac{N}{2}, \qquad
0 \leq N_{\rho_1} \leq \frac{N-|M+M'|+|M-M'|}{2}.
\end{eqnarray*}

The coefficients $U_{NpMM'}^{L}(d)$ and $V_{NpMM'}^{N_{\rho_1}}(d)$ are
determined from the following three-term recurrence
relations:
\begin{eqnarray}
A_{L+1} U_{NpMM'}^{L+1} + A_L U_{NpMM'}^{L-1}
&=&
\left\{\frac{4}{a^2 d^2}\left[Q_p-L(L+1)\right]-B_L\right\}
U_{NpMM'}^{L}, \nonumber \\
\label{3.11.5}
\\
C_{N_{\rho_1}+1}V_{NpMM'}^{N_{\rho_1}+1} +
C_{N_{\rho_1}} V_{NpMM'}^{N_{\rho_1}-1}
&=&
\left[Q_p - a^2 d^2 \left(N_{\rho_1} - \frac{N -
|M+M'|}{4}\right) - D_{N_{\rho_1}}\right]V_{NpMM'}^{N_{\rho_1}},
\nonumber
\end{eqnarray}
and the orthonormalization conditions
\begin{eqnarray*}
\sum_{L}\,U_{Np_1ms}^L (d)\,U_{Np_2ms}^{L*} (d) = \delta_{p_1p_2},
\qquad \sum_{N_{\rho_1}}\,{\tilde V}_{Np_1MM'}^{N_{\rho_1}} (d)
\,{\tilde V}_{Np_2MM'}^{N_{\rho_1}*} (d)= \delta_{p_1p_2}.
\end{eqnarray*}
The quantities $A_L$, $B_L$, $C_{N_{\rho_1}}$ and $D_{N_{\rho_1}}$ are given by
the relations
\begin{eqnarray*}
A_{L} &=& - \left\{\frac{[(N+2)^2-4L^2](L^2-M_+^2)
(L^2-M_-^2)}
{L^2(2L-1)(2L+1)}\right\}^{\frac{1}{2}},  \qquad
B_L = \frac{M_+M_-(N+2)}{L(L+1)},
\\ [3mm]
C_{N_{\rho_1}} &=& -\sqrt{\,N_{\rho_1}\,(N_{\rho_2}+1)\,
\left(N_{\rho_1}+\frac{|M+M'|}{2}\right)\,
\left(N_{\rho_2}+\frac{|M-M'|}{2}+1\right)},      \\ [3mm]
D_{N_{\rho_1}} &=&
\frac{N(N+4)}{8}+\frac{1}{8}M_-^2
+\frac{1}{2}\left(N_{\rho_1}-N_{\rho_2} + \frac{|M+M'|}{2}\right)
\left(N_{\rho_2}-N_{\rho_1}+\frac{|M-M'|}{2}\right),
\end{eqnarray*}
where for convenience we introduced the notation $M_{\pm} = (M \pm M')/2$.

It is easy to notice that from formulas (\ref{3.3.1}), (\ref{3.7.9}), (\ref{3.8.5})
it follows that the spheroidal parameters $R$ and $d$ are related by the relation
$R = d^{2}$, and the transformation coefficients
$\left\langle N p M M'; d | n q m s; R\right\rangle$
connecting the spheroidal functions of the MIC-Kepler problem and the four-dimensional
isotropic oscillator, diagonal and have the form:
\begin{eqnarray*}
\left\langle\, NpMM';d\,|\,nqms;R\,\right\rangle = 4n\sqrt{r_0}
\delta_{n, \frac{N}{2}+1}\, \delta_{p,q}\, \delta_{M,m}\,
\delta_{M',s}.
\end{eqnarray*}

Now let’s construct a spheroidal basis of a four-dimensional isotropic oscillator
for the case of $N=2$ and $M=M’=0$. From expansion (\ref{3.11.3}) we have
\begin{eqnarray}
\left|\,2 p 0 0\,\right\rangle = U_{2 p 0 0}^{0}\,\left|\,2 0 0 0\,\right\rangle
+ U_{2 p 0 0}^{1}\,\left|\,2 1 0 0\,\right\rangle.
\label{3.11.6}
\end{eqnarray}
Doing exactly as in paragraph $3.3$, for the MIK-Kepler problem, we obtain
\begin{eqnarray*}
U_{2 p 0 0}^{0} = \frac{1}{\sqrt{2}}\,\left(\frac{Q_{1,2} - 2}{Q_{1,2} - 1}\right)^{1/2},
\qquad
U_{2 p 0 0}^{1} = - \frac{1}{\sqrt{2}}\,\left(\frac{Q_{1,2}}{Q_{1,2} - 1}\right)^{1/2}
\end{eqnarray*}
Further, substituting the explicit expressions $U_{2 p 0 0}^{0}$, $U_{2 p 0 0}^{1}$
in (\ref{3.11.6}) and using the explicit form of the Eulerian wave function (\ref{3.8.7}), we obtain
\begin{eqnarray*}
\left|\,2 p 0 0\,\right\rangle = \frac{a^{2}}{\pi}\,\left(\frac{Q_{1,2} - 2}{Q_{1,2} - 1}\right)^{1/2}\,
\left[1 + \frac{2Q_{1,2}\,(\xi - 1)}{a^{2}d^{2} + 2Q_{1,2}}\right]\,
\left[1 + \frac{2Q_{1,2}\,(1 + \eta)}{a^{2}d^{2} - 2Q_{1,2}}\right]\,e^{-a^{2}d^{2}\,(\xi + \eta)/4}.
\end{eqnarray*}
Here $Q_{1,2} = 1 \pm \sqrt{1 + a^{4}d^{4}/4}$ are the solutions to the quadratic equation
$Q(Q-2)=a^{4}d^{4}/4$, which follows from the equality to zero of the determinant of the
system of equations obtained from the first recurrence relation (\ref{3.11.5}).

\section{Perturbation theory}
\markboth{CHAPTER 3. THE MIC-KEPLER PROBLEM}{3.12. PERTURBATION THEORY}

An infinite set of three-dimensional prolate spheroidal coordinates (\ref{3.3.1}) differ from
each other by the value of the parameter $R$. For arbitrary $R$, ​​prolate spheroidal coordinates
are related to spherical and parabolic coordinates as follows
\begin{eqnarray}
\xi = \frac{r}{R} + \sqrt{1 - \frac{2r}{R}\,\cos\theta + \frac{r^{2}}{R^{2}}}, \qquad
\eta = \frac{r}{R} - \sqrt{1 - \frac{2r}{R}\,\cos\theta + \frac{r^{2}}{R^{2}}},
\label{3.12.1}
\end{eqnarray}
\begin{eqnarray}
\xi = \frac{\mu + \nu}{R} + \sqrt{1 - \frac{\mu - \nu}{R}\,\cos\theta + \frac{(\mu + \nu)^{2}}{R^{2}}},
\nonumber \\
\label{3.12.2}
\\
\eta = \frac{\mu + \nu}{R} - \sqrt{1 - \frac{\mu - \nu}{R}\,\cos\theta + \frac{(\mu + \nu)^{2}}{R^{2}}}.
\nonumber
\end{eqnarray}
Formulas (\ref{3.12.1}) and (\ref{3.12.2}) allow us to investigate the behavior of elongated
spheroidal coordinates at small and large $R$. It is easy to show that when $R \to 0$
\begin{eqnarray*}
\xi \to \frac{2r}{R}, \qquad \eta \to \cos\theta ,
\end{eqnarray*}
and when $R \to \infty$
\begin{eqnarray*}
\xi \to \frac{\nu}{R} + 1, \qquad \eta \to \frac{\mu}{R} - 1,
\end{eqnarray*}
i.e., prolate spheroidal coordinates within the limits of $R \to 0$ and $R \to \infty$
a transform into spherical and parabolic coordinates, respectively.
From the above it follows that within the limits of small and large values ​​of the parameter
$R$, the spheroidal wave functions of the MIC-Kepler problem $\psi^{(s)}_{nqm}(\xi, \eta, \varphi; R)$
should transform into spherical and parabolic bases, respectively.
Indeed, for small R the second term in the operator $\hat{\Lambda}$ (\ref{3.3.5}) is a
perturbation, and the spherical basis is the zeroth approximation.
The equation for the eigenvalues ​​and eigenfunctions of the operator $\hat{\Lambda}$ for
large $R$ after dividing it by $\hbar/R\sqrt{\mu_0}$ is
\begin{eqnarray*}
\left({\hat I}_z + \frac{\hbar}{R\sqrt{\mu_0}}\,{\hat{\bf J}}^2\right)\,
\psi^{(s)}_{nqm}(\xi, \eta, \varphi; R) = \frac{\hbar \Lambda_{q}}{R\sqrt{\mu_0}}\,
\psi^{(s)}_{nqm}(\xi, \eta, \varphi; R).
\end{eqnarray*}
As can be seen, now the perturbation is the term $\hbar {\hat{\bf J}}^2/R\sqrt{\mu_0}$,
and the parabolic basis is the zero approximation.
After what has been said, the calculation of corrections to $\Lambda_{q}$ and
$\psi^{(s)}_{nqm}(\xi, \eta, \varphi; R)$ for small and large R takes on a standard character.
Using the standard formulas of perturbation theory and expressions for matrix elements
${\hat{\bf J}}^2$ and ${\hat I}_z$ (\ref{3.3.10}), we obtain
\begin{eqnarray*}
\Lambda_{q}(R) = j(j+1) + \frac{R}{r_{0}n}\,B_{j} + \frac{R^{2}}{8r_{0}^{2}n^{2}}\,
\left(\frac{1}{j}\,A_{j}^{2} - \frac{1}{j+1}\,A_{j+1}^{2}\right), \\ [3mm]
\psi^{(s)}_{nqm}(\xi, \eta, \varphi; R) = \psi^{(s)}_{njm} +
\frac{R}{2r_{0}n}\,\left(\frac{1}{j}\,A_{j}\,\psi^{(s)}_{njm}
- \frac{1}{j+1}\,A_{j+1}\psi^{(s)}_{n,j+1,m}\right).
\end{eqnarray*}
\begin{eqnarray*}
\Lambda_{q}(R) = \frac{R}{r_{0}n}\,(n_{1} - n_{2} + m_{-})+ D_{n_{1}} +
\frac{2r_{0}n}{R}\,\left(C_{n_{1}}^{2} - C_{n_{1}+1}^{2}\right), \\ [3mm]
\psi^{(s)}_{nqm}(\xi, \eta, \varphi; R) = \psi^{(s)}_{n_{1}-1,n_{2}+1,m} +
\frac{r_{0}n}{2R}\,\left(C_{n_{1}}\,\psi^{(s)}_{n_{1},n_{2},m} -
C_{n_{1}+1}\,\psi^{(s)}_{n_{1}+1,n_{2}-1,m}\right).
\end{eqnarray*}
The corrections for the spheroidal separation constant are obtained in the second order,
and the wave functions in the first order of perturbation theory.
Higher-order corrections are calculated similarly. At $s=0$, the obtained relationships transform
into the corresponding formulas for the hydrogen atom obtained by us in \cite{M-35}.

Four-dimensional spheroidal coordinates $\xi, \eta$ as functions of the parameter $d$
and the Eulerian  coordinates $u, \beta$ have the form:
\begin{eqnarray}
\xi = \frac{u^{2}}{d^{2}} + \sqrt{1 - \frac{2u^{2}}{d^{2}}\,\cos\beta + \frac{u^{4}}{d^{4}}}, \qquad
\eta = \frac{u^{2}}{d^{2}} - \sqrt{1 - \frac{2u^{2}}{d^{2}}\,\cos\beta + \frac{u^{4}}{d^{4}}},
\label{3.12.3}
\end{eqnarray}
and with double polar coordinates are related as follows:
\begin{eqnarray}
\xi = \frac{\rho_{1}^{2}+ \rho_{2}^{2}}{d^{2}} + \sqrt{1 - \frac{2}{d^{2}}\,(\rho_{1}^{2} - \rho_{2}^{2}) +
\frac{1}{d^{4}}\,(\rho_{1}^{2} + \rho_{2}^{2})^{2}},
\nonumber \\
\label{3.12.4}
\\
\eta = \frac{\rho_{1}^{2}+ \rho_{2}^{2}}{d^{2}} - \sqrt{1 - \frac{2}{d^{2}}\,(\rho_{1}^{2} - \rho_{2}^{2}) +
\frac{1}{d^{4}}\,(\rho_{1}^{2} + \rho_{2}^{2})^{2}}.
\nonumber
\end{eqnarray}
From formulas (\ref{3.12.3}) and (\ref{3.12.4}) it follows directly that in the limit $d \to 0$
\begin{eqnarray*}
\xi \to \frac{2u^{2}}{d^{2}}, \qquad \eta \to \cos\beta,
\end{eqnarray*}
and as $d \to \infty$
\begin{eqnarray*}
\xi \to 1 + \frac{2}{d^{2}}\,\rho_{2}^{2}, \qquad \eta \to \frac{2}{d^{2}}\,\rho_{1}^{2} - 1.
\end{eqnarray*}
So, four-dimensional spheroidal coordinates (\ref{3.8.5}) within $d \to 0$
and $d \to \infty$ transform into Eulerian and double polar coordinates, respectively.

Now we calculate the spheroidal corrections to the Eulerian and double polar bases.
For small $d$, the second term in the operator (\ref{3.11.3}) is a perturbation,
and the Eulerian wave functions are a zero approximation. For large $d$ from the equation
\begin{eqnarray*}
\left({\hat{\cal P}} +\frac{4}{a^{2}d^{2}}\,{\hat{\bf L}}^{2}\right)\,
\left|N p M M';d\right\rangle = \frac{4Q_{p}}{a^{2}d^{2}}\,\left|N p M M';d\right\rangle,
\end{eqnarray*}
which is obtained from the second equation (\ref{3.11.2-1}) by dividing it by
$a^{2}d^{2}/2$, it can be seen that the term $4{\hat{\bf L}}^{2}/a^{2}d^{2}$
is a perturbation, and the double polar basis is the zeroth approximation.
Let us present the final results.

The region of small $d$:
\begin{eqnarray*}
Q_{p}(d) = L(L+1) + \frac{a^{2}d^{2}}{4}\,B_{L} + \frac{a^{4}d^{4}}{8}\,
\left(\frac{1}{L}\,A_{L}^{2} - \frac{1}{L+1}\,A_{L+1}^{2}\right), \\ [2mm]
\left|N p M M';d\right\rangle = \left|N L M M'\right\rangle +
\frac{a^{2}d^{2}}{8}\,
\left(\frac{A_{L}}{L}\,\left|N, L-1, M M'\right\rangle - \frac{A_{L+1}}{L+1}\,
\left|N, L+1, M M'\right\rangle \right).
\end{eqnarray*}
The region of large $d$:
\begin{eqnarray*}
Q_{p}(d) &=& a^{2}d^{2}\left(N_{\rho_{1}} - \frac{N-2|m_{1}|}{4}\right) + D_{N_{\rho_{1}}} +
\frac{4}{a^{2}d^{2}}\,\left(C_{N_{\rho_{1}}}^{2} - C_{N_{\rho_{1}}+1}^{2}\right), \\ [2mm]
\left|N p M M';d\right\rangle &=& \left|N_{\rho_{1}} N_{\rho_{2}} m_{1} m_{2}\right\rangle +
\frac{1}{a^{2}d^{2}}\,\Biggl(C_{N_{\rho_{1}}}\,\left|N_{\rho_{1}}-1, N_{\rho_{2}}+1, m_{1} m_{2}\right\rangle - \\ [2mm]
&-& C_{N_{\rho_{1}}+1}^{2}\,\left|N_{\rho_{1}}+1, N_{\rho_{2}}-1, m_{1} m_{2}\right\rangle\Biggr).
\end{eqnarray*}
Here, too, the corrections for spheroidal wave functions are calculated in the first order,
and for the separation constant in the second order of perturbation theory.

\newpage

\chapter{Generalized ring-shaped potentials}
\markboth{CHAPTER 4. GENERALIZED RING-SHAPED POTENTIALS}{}

\section{Ring-shaped functions}
\markboth{CHAPTER 4. GENERALIZED RING-SHAPED POTENTIALS}{4.1. RING-SHAPED FUNCTIONS}

The problem we are considering in this section relates to the quantum mechanics of ring-shaped
potentials \cite{M-21}. In general, such potentials have the form:
\begin{eqnarray}
U(r, \theta) = U(r) + \frac{\lambda}{r^{2}\,\sin\theta}.
\label{4.1.1}
\end{eqnarray}
Here $U(r)$ is an arbitrary centrally symmetric field, $\theta$ is the angle between the $z$
axis and the radius vector of the particle, and $\lambda$ is a nonnegative parameter
characterizing the intensity of the axially symmetric term added to the “bare” potential $U(r)$.
For $U(r) = -\alpha/r$ and $U(r) = \mu_{0}\omega^{2}r^{2}/2$, we respectively speak about a ring-shaped
hydrogen atom and a ring-shaped spatial oscillator. These two models are exactly solvable in both
classical and quantum mechanics \cite{LL-1,EVANS-1,EVANS-2,EVANS-3,FMSUW,MSVW,FSUW}.
For the potentials $U(r) = -\alpha/r$ and $U(r) = \mu_{0}\omega^{2}r^{2}/2$, the variables are
respectively separable in spherical and parabolic coordinates and in spherical and cylindrical
coordinates \cite{HARTMAN-1,HARTMAN-2,HARTMAN-3,Quesne}.
Both the bases corresponding to the indicated coordinates and the matrices of interbasis expansions are
known \cite{Lutsenko-1, Lutsenko-2}. The symmetry and statistical aspects of these models have also been
investigated (see \cite{Gerry,KIB-WIN-1,GRAN2,Zhedanov} and \cite{Lutsenko-3,Lutsenko-4} respectively).
These systems were studied from different standpoints in \cite{Kibler-Negadi-5} - \cite{GROP1}.

The separation of variables in the spherical coordinates $r, \theta, \varphi$ generates a function \\
$Z_{l m}(\theta, \varphi; \delta)$ describing the angular dependence of the wave function for an
arbitrary “bare” potential $U(r)$, i.e., a function playing the same role for a ring-shaped field
as that played by spherical functions for a centrally symmetric field. The parameter $\delta$
is related to $\lambda$ by the formula
\begin{eqnarray}
\delta = \sqrt{m^{2} + 2\Lambda},
\label{4.1.2}
\end{eqnarray}
where $\Lambda = \mu_{0}\,\lambda/\hbar^{2}$, and the quantum numbers $m$ and $l$
assume the values $m = 0,\pm 1, \pm 2, . . . ,\pm l$ and
$l = |m|, |m| + 1, . . .$ . For $\delta = 0$, they have the meaning of the magnetic
and orbital quantum numbers.

Now let's find the explicit form of the ring-shaped function.

Let ${\hat{\bf l}}^{2}$ and $\hat{l}_{z}$ be the respective operators for the square of the orbital
momentum and for its $z$ projection, and let $\hat{M}$ be an operator of the form
\begin{eqnarray}
\hat{M} = {\hat{\bf l}}^{2} + \frac{2\Lambda}{\sin^{2}\theta}.
\label{4.1.3}
\end{eqnarray}
The ring-shaped functions $Z_{l m}(\theta, \varphi; \delta)$ are defined as common eigenfunctions
for $\hat{l}_{z}$ and $\hat{M}$ corresponding to the eigenvalues $m$ and
$(l + \delta)(l + \delta + 1)$. Hence,
\begin{eqnarray}
\hat{l}_{z}\,Z_{l m}(\theta, \varphi; \delta) = m\,Z_{l m}(\theta, \varphi; \delta),
\label{4.1.4}
\end{eqnarray}
\begin{eqnarray}
\hat{M}\,Z_{l m}(\theta, \varphi; \delta) = (l + \delta)(l + \delta + 1)\,
Z_{l m}(\theta, \varphi; \delta).
\label{4.1.5}
\end{eqnarray}
The functions $Z_{l m}(\theta, \varphi; \delta)$ normalized by the condition
\begin{eqnarray}
\int\limits_{0}^{\pi}\,\int\limits_{0}^{2\pi}\,Z_{l' m'}^{*}(\theta, \varphi; \delta)\,
Z_{l m}(\theta, \varphi; \delta)\,\sin\theta\,d\theta\,d\varphi = \delta_{l l'}\,
\delta_{m m'}.
\label{4.1.6}
\end{eqnarray}
are written as
\begin{eqnarray}
Z_{l m}(\theta, \varphi; \delta) = N_{l m}(\delta)\,(\sin\theta)^{|m| + \delta}\,
C_{l - |m|}^{|m| + \delta + 1/2}(\cos\theta)\,e^{im\varphi},
\label{4.1.7}
\end{eqnarray}
where $C_{n}^{\nu})x)$ are the Gegenbauer polynomials and the normalizing factor
$N_{l m}(\delta)$) is given by the formula
\begin{eqnarray*}
N_{l m}(\delta) = (-1)^{(m-|m|)/2}\,2^{|m| + \delta}\,\Gamma\left(|m| + \delta + \frac{1}{2}\right)\,
\left[\frac{(2l + 2\delta + 1)\,(l - |m|)!}{4\pi^{2}\,\Gamma\left(l + |m| + 2\delta + 1\right)}\right]^{1/2}.
\end{eqnarray*}
Here, the chosen phase factor is such that for $\delta = 0$, the functions
$Z_{l m}(\theta, \varphi; \delta)$ pass into spherical functions
with the phase accepted in \cite{VAR},
\begin{eqnarray*}
Y_{l m}(\theta, \varphi) = (-1)^{(m+|m|)/2}\,
\left[\frac{(2l + 1)\,(l - |m|)!}{4\pi\,(l + |m| )!}\right]^{1/2}\,
P_{l}^{|m|}(\cos\theta)\,e^{im\varphi}.
\end{eqnarray*}
What has been said can be easily shown using the formula \cite{BE2}
\begin{eqnarray}
P_{l}^{|m|}(\cos\theta) = \frac{(-2)^{|m|}}{\sqrt{\pi}}\,
\Gamma\left(|m| + \frac{1}{2}\right)\,(\sin\theta)^{|m|}\,
C_{l - |m|}^{|m| + 1/2}(\cos\theta).
\label{4.1.7a}
\end{eqnarray}
It is convenient to determine the numerical values of the function
$Z_{l m}(\theta, \varphi; \delta)$ using formula (\ref{1.9.6})
expressing the Gegenbauer polynomials in terms of the hypergeometric function.

Finally, we give the representation for the ring-shaped function
$Z_{l m}(\theta, \varphi; \delta)$,

\begin{eqnarray}
Z_{l m}(\theta, \varphi; \delta) &=& \frac{(-1)^{(m+|m|)/2}\,e^{i\pi\,\delta}\,e^{im\varphi}}
{2^{l + \delta}\,\Gamma\left(l + \delta + 1\right)}\,
\left[\frac{(2l + 2\delta + 1)\,\Gamma\left(l + |m| + 2\delta + 1\right)}{4\pi\,(l - |m|)!}\right]^{1/2} \times
\nonumber \\ [3mm]
\label{4.1.8} \\ [3mm]
&\times& (\sin\theta)^{-|m| - \delta}\,\frac{d^{l - |m|}}{(d\cos\theta)^{l - |m|}}\,
\left(cos^{2}\theta - 1\right)^{l +  \delta}, \nonumber
\end{eqnarray}
which can be easily derived using the Rodrigues formula for Gegenbauer polynomials \cite{BE2}.

\section{The general expansion}
\markboth{CHAPTER 4. GENERALIZED RING-SHAPED POTENTIALS}{4.2. THE GENERAL EXPANSION}

We now consider the expansion of the ring-shaped function $Z_{l m}(\theta, \varphi; \delta')$ in the
functions $Z_{l m}(\theta, \varphi; \delta)$ relating to different operators $\hat{M}$.
The expansion matrix is diagonal with respect to $m$, and therefore
\begin{eqnarray}
Z_{l m}(\theta, \varphi; \delta) = \sum\limits_{l' = |m|}^{\infty}\,
T_{l l'}^{m}(\delta, \delta')\,Z_{l' m}(\theta, \varphi; \delta').
\label{4.2.1}
\end{eqnarray}
Because the system of functions $Z_{l m}(\theta, \varphi; \delta')$ is orthonormalized,
the expansion matrix is determined by the integral \cite{VILEN1}
\begin{eqnarray}
T_{l l'}^{m}(\delta, \delta') = D_{l l'}^{m}(\delta, \delta')\,\int\limits_{-1}^{1}
\,(1 - x^{2})^{|m| + (\delta + \delta')/2}\,C_{l - |m|}^{|m| + \delta + 1/2}(x)\,
C_{l' - |m|}^{|m| + \delta' + 1/2}(x)\,dx,
\label{4.2.2}
\end{eqnarray}
where $x = \cos\theta$ and
\begin{eqnarray}
&&D_{l l'}^{m}(\delta, \delta') = 2^{2|m| + \delta +\delta'}\,
\Gamma\left(|m| + \delta + 1\right)\,\Gamma\left(|m| + \delta' + 1\right)\times
\nonumber \\ [3mm]
\label{4.2.3}
\\ [3mm]
&\times& \left[\frac{(2l + 2\delta + 1)\,(2l + 2\delta' + 1)\,(l - |m|)!\,(l' - |m|)!}
{4\pi^{2}\,\Gamma\left(l + |m| + 2\delta + 1\right)\,\Gamma\left(l' + |m| + 2\delta' + 1\right)}\right]^{1/2}
\nonumber
\end{eqnarray}
By the symmetry condition $C_{n}^{\nu}(-x) = (-1)^{n}\,C_{n}^{\nu}(x)$, the integral in
(\ref{4.2.2}) is nonzero only if the numbers $l$ and $l'$ have the same parity.
Another symmetry property follows obviously from (\ref{4.2.2}) and (\ref{4.2.3}),
\begin{eqnarray}
T_{l l'}^{m}(\delta, \delta') = T_{l' l}^{m}(\delta', \delta).
\label{4.2.4}
\end{eqnarray}

According to \cite{Rashid}, we have
\begin{eqnarray*}
&&\int\limits_{-1}^{1}\,(1 - x^{2})^{(\alpha + \beta - 3)/2 - p}\,
C_{n}^{\alpha}(x)\,C_{k}^{\beta}(x)\,dx =
{_4F_3}\left\{\matrix{ \frac{\alpha - \beta}{2} + p +1,\,\,\frac{\alpha - \beta}{2} - p,\,\, -\frac{n}{2},\,\, -\frac{n-1}{2}\cr \cr
\frac{k-n}{2} + 1,\,\,-\frac{n+k}{2} - \beta + 1,\,\,\alpha + \frac{1}{2}
\cr}\Biggr|1\right\}\times \\ [7mm]
&\times& \frac{(-1)^{(k-n)/2}}{2^{2\alpha - 1}}\,\frac{\pi \Gamma(n+2\alpha)\,
\Gamma\left(\frac{\alpha + \beta - 1}{2} - p\right)\,\Gamma\left(\frac{\alpha - \beta}{2} - p\right)\,
\Gamma\left(\frac{n + k }{2} + \beta\right)}{n!\,\left(\frac{k-n}{2}\right)!\,
\Gamma\left(\alpha\right)\,\Gamma\left(\beta\right)\,\Gamma\left(\alpha + \beta + \frac{1}{2}\right)\,
\Gamma\left(\frac{n + \alpha - k - \beta}{2} - p\right)\,
\Gamma\left(\frac{n + \alpha + k + \beta}{2} - p\right)}.
\end{eqnarray*}
It is assumed here that $k \geq n$. We have $p = −1$ in the case under consideration, and therefore
\begin{eqnarray}
T_{l l'}^{m}(\delta, \delta'; l'\geq l) &=& \frac{(-1)^{\frac{l'-l}{2}}\,E_{l l'}^{m}(\delta, \delta')}
{\Gamma\left(\frac{l' - l}{2} + 1\right)\,\Gamma\left(\frac{l' - l + \delta - \delta'}{2} + 1\right)}\times \nonumber \\ [5mm]
\label{4.2.5}
\\  [5mm]
&\times& {_4F_3}\left\{\matrix{ -\frac{l - |m|}{2},\,\,-\frac{l - |m| - 1}{2},\,\, \frac{\delta - \delta'}{2},\,\, 1 + \frac{\delta - \delta'}{2}\cr \cr
|m| + \delta + 1,\,\,\frac{l' - l}{2} + 1,\,\,\frac{l + l' - 1}{2} - \delta'
\cr}\Biggr|1\right\}, \nonumber
\end{eqnarray}
where
\begin{eqnarray}
E_{l l'}^{m}(\delta, \delta') &=& \frac{\Gamma\left(\frac{\delta - \delta'}{2} + 1\right)\,
\Gamma\left(\frac{\delta + \delta'}{2} +  |m| + 1\right)\,\Gamma\left(\frac{l + l' + 1}{2} + \delta'\right)}
{2^{\delta - \delta' + 1}\,\Gamma\left(|m| + \delta + 1\right)\,\Gamma\left(\frac{l + l' + \delta + \delta' + 3}{2}\right)}\times
\nonumber \\  [5mm]
\label{4.2.6}
\\ [5mm]
&\times& \left[\frac{(2l + 2\delta + 1)\,(2l + 2\delta' + 1)\,(l' - |m|)!\,\Gamma\left(l + |m| + 2\delta + 1\right)}
{(l - |m|)!\,\Gamma\left(l' + |m| + 2\delta' + 1\right)}\right]^{1/2}.
 \nonumber
\end{eqnarray}
Knowing the formula for $T_{l l'}^{m}(\delta, \delta'; l'\geq l)$, we can use it in c
combination with symmetry property (\ref{4.2.4}) to find $T_{l l'}^{m}(\delta, \delta'; l'\leq l)$ as well.
It follows from expansion (\ref{4.2.1}) that the relation
\begin{eqnarray}
T_{l l'}^{m}(\delta, \delta') = \delta_{l l'}.
\label{4.2.7}
\end{eqnarray}
must hold for $\delta = \delta'$. It can be easily shown that
\begin{eqnarray*}
T_{l l'}^{m}(\delta, \delta) &=&
\left[\frac{(2l + 2\delta + 1)\,(2l' + 2\delta + 1)\,(l' - |m|)!\,\Gamma\left(l + |m| + 2\delta + 1\right)}
{(l - |m|)!\,\Gamma\left(l' + |m| + 2\delta + 1\right)}\right]^{1/2}\times  \\ [5mm]
&\times& \frac{(-1)^{\frac{l'-l}{2}}}{l + l' + 2\delta + 1}
\left[\Gamma\left(\frac{l' - l}{2} + 1\right)\,\Gamma\left(\frac{l - l'}{2} + 1\right)\right]^{-1}.
\end{eqnarray*}
The resulting expression vanishes for $l' \neq l$ because of the factors
$\Gamma\left((l' - l)/2 + 1\right)$ and $\Gamma\left((l - l')/2 + 1\right)$ in
the right-hand side of this relation, and it is equal to unity for $l' = l$.
We thus obtain identity (\ref{4.2.7}).

We now consider expansion (\ref{4.2.1}) for the case $l = |m| = 0$. We have
\begin{eqnarray*}
\frac{\sqrt{\pi}\,\Gamma\left(\frac{\delta' - \delta}{2} \right)}
{\Gamma\left(\delta + \frac{1}{2}\right)\,\Gamma\left(\frac{\delta + \delta'}{2} + 1\right)}\,
\left(\sin\theta\right)^{\delta - \delta'} = \sum\limits_{l'=0}^{\infty}\,
\frac{\left(l' + \delta' + \frac{1}{2}\right)\,\Gamma\left(\frac{l' + 1}{2}\right)\,
\Gamma\left(\frac{l' - \delta + \delta'}{2}\right)\,}
{\Gamma\left(\frac{l' + \delta + \delta' + 3}{2}\right)\,\Gamma\left(\frac{l'}{2} + \delta' + 1\right)}\,
C_{l'}^{\delta' + 1/2}(\cos\theta).
\end{eqnarray*}
whence the relation
\begin{eqnarray*}
\int\limits_{0}^{\pi}\,\left(\sin\theta\right)^{\delta + \delta' + 1}\,
C_{l'}^{\delta' + 1/2}(\cos\theta)\,d\theta =
\frac{\sqrt{\pi}\,\Gamma\left(\frac{\delta + \delta'}{2} + 1\right)
\,\Gamma\left(\frac{l + 1}{2} + \delta'\right)\,
\Gamma\left(\frac{l' - \delta + \delta'}{2}\right)}
{\Gamma\left(\frac{l}{2} + 1\right)\,\Gamma\left(\delta' + \frac{1}{2}\right)\,
\Gamma\left(\frac{l' + \delta + \delta' + 3}{2}\right)\,
\Gamma\left(\frac{\delta' - \delta}{2}\right)}.
\end{eqnarray*}
follows in view of the orthogonality of Gegenbauer polynomials \cite{BE2}. The same result
can be obtained independently from the relation \cite{BE2}
\begin{eqnarray*}
\int\limits_{-1}^{1}\,\left(1 - x\right)^{\alpha}\,\left(1 + x\right)^{\beta}\,
C_{n}^{\nu}(x)\,dx &=&
\frac{2^{\alpha + \beta + 1}\,\Gamma\left(\alpha + 1\right)\,\Gamma\left(\beta + 1\right)\,
\Gamma\left(n + 2\nu\right)}{n!\,\Gamma\left(2\nu\right)\,\Gamma\left(\alpha + \beta + 2\right)}\times \\ [5mm]
&\times& {_3F_2}\left\{\matrix{ -n,\,\,n + 2\nu,\,\, \alpha + 1\cr \cr
\nu + \frac{1}{2},\,\,\alpha + \beta + 2
\cr}\Biggr|1\right\}
\end{eqnarray*}
and the formula \cite{BA}
\begin{eqnarray*}
{_3F_2}\left\{\matrix{ a,\,\,b,\,\, c\cr \cr
\frac{a + b + 1}{2},\,\,2c
\cr}\Biggr|1\right\} = \frac{\sqrt{\pi}\,\Gamma\left(\frac{a + b + 1}{2}\right)\,
\Gamma\left(c - \frac{a + b - 1}{2}\right)\,\Gamma\left(c + \frac{1}{2}\right)}
{\Gamma\left(\frac{a + 1}{2}\right)\,\Gamma\left(\frac{b + 1}{2}\right)\,
\Gamma\left(c - \frac{a - 1}{2}\right)\,\Gamma\left(c - \frac{b - 1}{2}\right)}.
\end{eqnarray*}
We thus obtain another confirmation that our results are correct
(besides for the case $\delta' = \delta'$).

Now we give explicit expressions for the coefficients $T_{l l'}^{m}(\delta, \delta')$
for even and odd values ​​of $l-|m|$.

It is known \cite{BA} that if in the function $_4F_3$ the sum of the upper parameters is
greater than the sum of the lower parameters by exactly one, then the formula is valid
\begin{eqnarray}
{_4F_3}\left\{\matrix{ x,\,\,y,\,\, z,\,\, -n\cr \cr
u,\,\,v,\,\,w
\cr}\Biggr|1\right\} = \frac{(v-z)_{n}\,(w-z)_{n}}{(v)_{n}\,(w)_{n}}\times \nonumber \\ [3mm]
\label{4.2.8}
\\ [3mm]
{_4F_3}\left\{\matrix{ u-x,\,\,u-y,\,\, z,\,\, -n\cr \cr
u,\,\,1-v+z-n,\,\,1-w+z-n
\cr}\Biggr|1\right\}. \nonumber
\end{eqnarray}
In our case, the above condition is satisfied, and therefore formula (\ref{4.2.8})
can be applied. Below are the final results:
\begin{eqnarray*}
T_{l l'}^{m(+)}(\delta, \delta') = \frac{1}{\pi}\,\sin\frac{\pi(\delta - \delta')}{2}\,
K_{l l'}^{m}(\delta, \delta')\,Q_{l l'}^{m(+)}(\delta, \delta'),
\end{eqnarray*}
\vspace{2mm}
\begin{eqnarray*}
T_{l l'}^{m(-)}(\delta, \delta') = \frac{1}{\pi}\,\sin\frac{\pi(\delta' - \delta)}{2}\,
K_{l l'}^{m}(\delta, \delta')\,Q_{l l'}^{m(-)}(\delta, \delta').
\end{eqnarray*}
These formulas use the following notation:
\begin{eqnarray*}
K_{l l'}^{m}(\delta, \delta') = A_{l}^{m}(\delta)\, B_{l'}^{m}(\delta'),
\end{eqnarray*}
\begin{eqnarray*}
A_{l}^{m}(\delta) = \left[\frac{\left(l + \delta + \frac{1}{2}\right)\,
\Gamma\left(\frac{l - |m|}{2} + 1\right)\,\Gamma\left(\frac{l + |m|}{2} + \delta + 1\right)}
{\Gamma\left(\frac{l - |m| + 1}{2}\right)\,\Gamma\left(\frac{l + |m| + 1}{2} + \delta\right)}\right]^{1/2},
\\ [5mm]
B_{l'}^{m}(\delta') = \left[\frac{\left(l' + \delta' + \frac{1}{2}\right)\,
\Gamma\left(\frac{l' - |m| + 1}{2}\right)\,\Gamma\left(\frac{l' + |m| + 1}{2} + \delta'\right)}
{\Gamma\left(\frac{l' - |m| + 1}{2}\right)\,\Gamma\left(\frac{l' + |m|}{2} + \delta' + 1\right)}\right]^{1/2},
\end{eqnarray*}
\begin{eqnarray*}
Q_{l l'}^{m(+)}(\delta, \delta') = \sum\limits_{s=0}^{(l-|m|)/2}\,
\frac{\Gamma\left(\frac{\delta - \delta'}{2} + s + 1\right)\,
\Gamma\left(|m| + \frac{\delta + \delta'}{2} + s + 1\right)}
{s!\,\Gamma\left(|m| + \delta + s + 1\right)}\,M_{l l'}^{m s}(\delta, \delta'), \\ [7mm]
Q_{l l'}^{m(-)}(\delta, \delta') = \sum\limits_{s=0}^{(l-|m|-1)/2}\,
\frac{\Gamma\left(\frac{\delta - \delta'}{2} + s + 1\right)\,
\Gamma\left(|m| + \frac{\delta + \delta'}{2} + s + 1\right)}
{s!\,\Gamma\left(|m| + \delta + s + 1\right)}\,N_{l l'}^{m s}(\delta, \delta'),
\end{eqnarray*}
\begin{eqnarray*}
M_{l l'}^{m s}(\delta, \delta') &=& \frac{\Gamma\left(\frac{l + |m| + 1}{2} + \delta + s\right)\,
\Gamma\left(\frac{l' - |m| - \delta + \delta'}{2} -s\right)}
{\Gamma\left(\frac{l - |m|}{2} - s + 1\right)\,\Gamma\left(\frac{l' + |m| +\delta + \delta' + 3}{2} + s\right)},
\\ [7mm]
N_{l l'}^{m s}(\delta, \delta') &=& \frac{\Gamma\left(\frac{l + |m|}{2} + \delta + s + 1\right)\,
\Gamma\left(\frac{l' - |m| - \delta + \delta' - 1}{2} -s\right)}
{\Gamma\left(\frac{l - |m| + 1}{2} - s\right)\,\Gamma\left(\frac{l' + |m| +\delta + \delta' + 3}{2} + s\right)}.
\end{eqnarray*}
Here $(+)$ and $(-)$ refer to the even and odd $l-|m|$ respectively.

At the end of this section we will present the expansion of ring-shaped functions
in terms of spherical ones. Let us write this expansion in the form
\begin{eqnarray*}
Z_{l m}(\theta, \varphi; \delta) = \sum\limits_{l'=|m|}^{\infty}\,
W_{l l'}^{m}(\delta)\,Y_{l' m}(\theta, \varphi).
\end{eqnarray*}
The matrix $W_{l l'}^{m}(\delta)$ satisfies the completeness condition
\begin{eqnarray*}
\sum\limits_{l'=|m|}^{\infty}\,W_{l l'}^{m}(\delta)\,
W_{l l'}^{m*}(\delta) = 1.
\end{eqnarray*}
It follows that the inverse expansion has the form
\begin{eqnarray*}
Y_{l m}(\theta, \varphi) = \sum\limits_{l'=|m|}^{\infty}\,
W_{l l'}^{m*}(\delta)\,Z_{'l m}(\theta, \varphi; \delta).
\end{eqnarray*}
Referring to the general formulas (\ref{4.2.1}), (\ref{4.2.2}), (\ref{4.2.3}),
(\ref{4.2.5}) and (\ref{4.2.6}), it can be shown that
\begin{eqnarray*}
W_{l l'}^{m}(\delta; l' \geq l) &=& \frac{(-1)^{(l' - l)/2}}{2^{l + \delta}}\,
\frac{\Gamma\left(1 + \frac{\delta}{2}\right)\,\Gamma\left(|m| + \frac{\delta}{2} + 1\right)\,
\Gamma\left(\frac{l + l' +1}{2}\right)}{\Gamma\left(\frac{l' - l}{2} + 1\right)\,
\Gamma\left(\frac{l - l' + \delta}{2} + 1\right)\,\Gamma\left(|m| + \delta + 1\right)\,
\Gamma\left(\frac{l + l' + \delta + 3}{2}\right)}\times \\ [5mm]
&\times& \left[\frac{(2l + 2\delta + 1)\,(2l' + 1)\,(l' - |m|)!\,\Gamma\left(l + |m| + 2\delta + 1\right)}
{(l - |m|)!\,(l' + |m|)!}\right]^{1/2}\times \\ [5mm]
&\times& {_4F_3}\left\{\matrix{ -\frac{l - |m|}{2},\,\,-\frac{l - |m| - 1}{2},\,\, \frac{\delta}{2},\,\, 1 + \frac{\delta}{2}\cr \cr
|m| + \delta + 1,\,\,\frac{l' - l}{2} + 1,\,\,-\frac{l + l' - 1}{2}
\cr}\Biggr|1\right\},
\end{eqnarray*}
\vspace{5mm}
\begin{eqnarray*}
W_{l l'}^{m}(\delta; l' \leq l) &=& \frac{(-1)^{(l - l')/2}}{2^{l - \delta}}\,
\frac{\Gamma\left(1 - \frac{\delta}{2}\right)\,\Gamma\left(|m| + \frac{\delta}{2} + 1\right)\,
\Gamma\left(\frac{l + l' +1}{2}\right)}{|m|!\,\Gamma\left(\frac{l - l'}{2} + 1\right)\,
\Gamma\left(\frac{l' - l - \delta}{2} + 1\right)\,\Gamma\left(\frac{l + l' + \delta + 3}{2}\right)}\times \\ [5mm]
&\times& \left[\frac{(2l' + 2\delta + 1)\,(2l + 1)\,(l - |m|)!\,(l' + |m|)!}
{(l - |m|)!\,\Gamma\left(l + |m| + 2\delta + 1\right)}\right]^{1/2}\times \\ [5mm]
&\times& {_4F_3}\left\{\matrix{ -\frac{l' - |m|}{2},\,\,-\frac{l' - |m| - 1}{2},\,\, -\frac{\delta}{2},\,\, 1 - \frac{\delta}{2}\cr \cr
|m| + 1,\,\,\frac{l - l'}{2} + 1,\,\,-\frac{l + l' - 1}{2} - \delta
\cr}\Biggr|1\right\}.
\end{eqnarray*}

We note that the resulting expansion relates to a kind of general result for ring-shaped models because it
is independent of the form of the centrally symmetric potential $U(r)$ in the expression $U(r, \theta)$.
It follows that it can be used in problems of expanding the bases for ring-shaped models in the bases for
the corresponding centrally symmetric systems.

\section{Relation of $T_{ll'}^{m}(\delta, \delta')$ coefficients with $6j$ -– symbols}
\markboth{CHAPTER 4. GENERALIZED RING-SHAPED POTENTIALS}
{4.3. RELATION OF $T_{ll'}^{m}(\delta)$ COEFFICIENTS WITH $6j$ –- SYMBOLS}

According to (\ref{4.2.6}), the matrices $T_{ll'}^{m}(\delta, \delta')$ can be
expressed in terms of the hypergeometric function $_4F_3$. We show that this allows
relating them to the $6j$-symbols. We recall that the Wigner $6j$-symbols \cite{Wigner}
are defined via transformations between different addition schemes for three angular momenta.
For this, we use the formula \cite{VAR}
\begin{eqnarray*}
&& \left\{\matrix{ a,\,\,b,\,\, c \cr \cr
d,\,\,e,\,\,f \cr}\right\}  =
\frac{(a + b + d + e + 1)!\,\Delta(abc)\,\Delta(cde)\,\Delta(aef)\,\Delta(bdf)}
{(d-c+e)!\, (a+e-f)!\,(b+d-f)!\,(a+c-d+f)!\,(c-b-e+f)!}\times \\ [3mm]
&\times& \frac{(-1)^{a+b+d+e}}{(a+b-c)!}\,
{_4F_3}\left\{\matrix{ -a-b+c,\,\,c-d-e,\,\, -a-e+f,\,\, -b-d+f\cr \cr
-a-b-d-e-1,\,\,-a+c-d+f+1,\,\,-b+c-e+f+1
\cr}\Biggr|1\right\},
\end{eqnarray*}
where
\begin{eqnarray*}
\Delta(abc) = \left[\frac{(a+b-c)!\,(a-b+c)!\,(-a+b+c)!}
{(a+b+c+1)!}\right]^{1/2}.
\end{eqnarray*}
We now compare this formula with expressions (\ref{4.2.6}) and (\ref{4.2.7}) to verify that
\begin{eqnarray}
T_{ll'}^{m}(\delta, \delta') &=& e^{-i\pi (l+l'+1)/2 +\delta'}\,
\left\{\matrix{ a,\,\,b,\,\, c \cr \cr
a-\frac{3}{4},\,\,b+\frac{1}{4},\,\,f \cr}\right\}\times \nonumber \\
\label{4.3.1}
\\
&\times& \left[\frac{\left(l + \delta + \frac{1}{2}\right)\,\left(l + \delta' + \frac{1}{2}\right)\,
\left(\delta - \delta'\right)\,\left(2|m| + \delta + \delta'\right)}
{\left(l - l' + \delta - \delta'\right)\,
\left(l + l' + \delta + \delta' + 1\right)}\right]^{1/2}, \nonumber
\end{eqnarray}
where
\begin{eqnarray}
a &=& \frac{l + |m| + \delta + \delta'}{4}, \qquad
b = \frac{l' - |m| + \delta' - \delta - 2}{4}, \nonumber \\
\label{4.3.2}
\\
c &=& \frac{l' - l + 2|m| + 2\delta' - 2}{4}, \qquad
d = \frac{l + l' + 2\delta - 1}{4}. \nonumber
\end{eqnarray}

We now find a recursive relation for $T_{ll'}^{m}(\delta, \delta')$. For this,
we consider the action of the operator $\hat{M}$ on expansion (\ref{4.2.1}) and use
Eq. (\ref{4.1.5}) to obtain
\begin{eqnarray*}
(l + \delta)\,(l + \delta + 1)\,Z_{l m}(\theta, \varphi; \delta) =
\sum\limits_{l'=|m|}^{\infty}\,T_{ll'}^{m}(\delta, \delta')\,
\left[(l' + \delta)\,(l' + \delta' + 1) + \frac{2(\Lambda - \Lambda')}
{\sin^{2}\theta}\right]\,Z_{l' m}(\theta, \varphi; \delta').
\end{eqnarray*}
We multiply the two sides of this relation by $\sin\theta Z_{l'' m}^{*}(\theta, \varphi; \delta)$
and then integrate over the solid angle $d\Omega = \sin\theta\,d\theta\,d\varphi$.
In view of the orthogonality of the functions
$Z_{l m}(\theta, \varphi; \delta)$, this results in
\begin{eqnarray*}
2(\Lambda - \Lambda')\,,T_{ll''}^{m}(\delta, \delta') =
\sum\limits_{l'=|m|}^{\infty}\,T_{ll'}^{m}(\delta, \delta')\,
\left(l + \delta -l' - \delta'\right)\,\left(l + \delta + l' + \delta' + 1\right)\,
\left(\sin^{2}\theta\right)_{l'' l'},
\end{eqnarray*}
where
\begin{eqnarray*}
\left(\sin^{2}\theta\right)_{l'' l'} = \int\,Z_{l'' m}^{*}(\theta, \varphi; \delta)\,
\sin^{2}\theta\,Z_{l' m}(\theta, \varphi; \delta)\,d\Omega.
\end{eqnarray*}
Clearly,
\begin{eqnarray*}
\left(\sin^{2}\theta\right)_{l'' l'} = \delta_{l'' l'}\, \left(\cos^{2}\theta\right)_{l'' l'}.
\end{eqnarray*}
To calculate the matrix entry $\left(\cos^{2}\theta\right)_{l'' l'}$, we use the recursive relation for
Gegenbauer polynomials \cite{BE2}
\begin{eqnarray*}
2(n + \nu)\,x\,C_{n}^{\nu}(x) = (2\nu + n - 1)\,C_{n-1}^{\nu}(x)
+ (n + 1\,C_{n+1}^{\nu}(x).
\end{eqnarray*}
The ultimate result is
\begin{eqnarray*}
\left(\sin^{2}\theta\right)_{l'' l'} = H_{l' m}(\delta')\,\delta_{l'', l'-2} +
R_{l' m}(\delta')\,\delta_{l'', l'} + H_{l'+2, m}(\delta')\,\delta_{l'', l'+2},
\end{eqnarray*}
where
\begin{eqnarray*}
H_{l' m}(\delta') &=& -\left[\frac{(l'-|m|)\,(l'-|m|-1)\,(l'+|m|+2\delta' -1)\,
(l'+|m|+2\delta')}{(2l' + 2\delta' - 1)^{2}\,(2l' + 2\delta' - 3)\,(2l' + 2\delta' + 1)}\right]^{1/2}, \\ [3mm]
R_{l' m}(\delta') &=& \frac{2(l'+\delta')\,(l'+\delta' + 1) + 2(|m| + \delta' - 1)\,(|m| + \delta' + 1)}
{(2l' + 2\delta' - 1)\,(2l' + 2\delta' + 3)}.
\end{eqnarray*}
Using this result, it can be shown that the matrix $T_{ll'}^{m}(\delta, \delta')$
satisfies the three-term recursive relation
\begin{eqnarray*}
&&\left[2(\Lambda - \Lambda') - \left(l + \delta -l' - \delta'\right)\,
\left(l + \delta + l' + \delta' + 1\right)\,R_{l' m}(\delta')\right]\,
T_{ll'}^{m}(\delta, \delta') = \\ [3mm]
&=& \left(l + \delta -l' - \delta' + 2\right)\,
\left(l + \delta + l' + \delta' - 1\right)\,H_{l' m}(\delta')\,T_{l,l'-2}^{m}(\delta, \delta') + \\ [3mm]
&+& \left(l + \delta -l' - \delta' - 2\right)\,
\left(l + \delta + l' + \delta' + 3\right)\,H_{l'+2, m}(\delta')\,T_{l,l'+2}^{m}(\delta, \delta').
\end{eqnarray*}
Substituting (\ref{4.3.1}) and (\ref{4.3.2}) in this relation, we conclude that
the $6j$-symbols must satisfy the recursive relation
\begin{eqnarray*}
\left\{\matrix{ a,\,\,b,\,\, c \cr \cr
a-\frac{3}{4},\,\,b+\frac{1}{4},\,\,f \cr}\right\}\,F =
\left\{\matrix{ a,\,\,b-\frac{1}{2},\,\, c-\frac{1}{2} \cr \cr
a-\frac{3}{4},\,\,b-\frac{1}{4},\,\,f-\frac{1}{2} \cr}\right\}\,G_{1} +
\left\{\matrix{ a,\,\,b+\frac{1}{2},\,\, c+\frac{1}{2} \cr \cr
a-\frac{3}{4},\,\,b+\frac{3}{4},\,\,f+\frac{1}{2} \cr}\right\}\,G_{2},
\end{eqnarray*}
where
\begin{eqnarray*}
F = \left(a-b+c\right)\,\left(-a+b+f-\frac{1}{4}\right) - \left(a-b-c-1\right)\,
\left(a+b+f+\frac{9}{4}\right)\,L, \\ [3mm]
L = \frac{\left(2b+c+f+7/4\right)\,\left(2b+c+f+11/4\right) -
\left(2a+c-f-3/4\right)\,\left(2a+c-f+5/4\right)}
{2\,\left(2b+c+f+5/4\right)\,\left(2b+c+f+13/4\right)},
\end{eqnarray*}
\begin{eqnarray*}
G_{1} &=& \frac{(a-b-c)\,\left[(b-a+f+1/4)\,(b-a+f+3/4)\right]^{1/2}}
{(2b+c+f+5/4)\,(2b+c+f+9/4)}\times \\ [3mm]
&\times& \left\{(a+b+c+1/2)\left[a^{2} - (b+c+1)^{2}\right]\,
(a+b+f+1/4)\,(a+b+f+5/4)\right\}^{1/2},
\end{eqnarray*}
\begin{eqnarray*}
G_{2} &=& \frac{\left[(a-b-c-1)\,(b-a+f+5/4)\,(b-a+f+7/4)\right]^{1/2}}
{(2b+c+f+9/4)\,(2b+c+f+13/4)}\times \\ [3mm]
&\times& \left\{(a+b+c+3/2)\left[a^{2} - (b+c+2)^{2}\right]\,
(a+b+f+5/4)\,(a+b+f+9/4)\right\}^{1/2}.
\end{eqnarray*}
We did not find this recursive relation for the Wigner $6j$-symbols in
\cite{VAR}, and it is not excluded that it is new.

\section{Free motion in a ring-shaped model}
\markboth{CHAPTER 4. GENERALIZED RING-SHAPED POTENTIALS}
{4.4. FREE MOTION IN A RING--SHAPED MODEL}

According to (\ref{4.1.1}), free motion in a ring-shaped model with the “bare” potential
$U(r) = 0$ is described by the equation \cite{M-22}
\begin{eqnarray}
\left(-\frac{\hbar^{2}}{2\mu_{0}}\,\Delta + \frac{\lambda}{r^{2}\,\sin^{2}\theta}\right)\,
\psi(\bf r) = E\,\psi(\bf r).
\label{4.4.1}
\end{eqnarray}

Let's consider the solution to this equation in spherical and cylindrical coordinates.

The spherical basis in the ring-shaped model, i.e., the solution of Eq. (\ref{4.4.1})
in the spherical coordinates, has the form
\begin{eqnarray}
\psi_{k l m}(r, \theta, \varphi; \delta) = \sqrt{\frac{2\pi k}{r}}\,
J_{l + \delta + 1/2}(kr)\,Z_{l m}(\theta, \varphi; \delta),
\label{4.4.2}
\end{eqnarray}
where $k = \sqrt{2\mu_{0}E}/\hbar$ and the radial wave function
\begin{eqnarray*}
R_{k l}(r; \delta) = \sqrt{\frac{2\pi k}{r}}\,
J_{l + \delta + 1/2}(kr)
\end{eqnarray*}
is normalized by the condition
\begin{eqnarray*}
\int\limits_{0}^{\infty}\,R_{k' l}(r; \delta)\,R_{k l}(r; \delta)\,dr = 2\pi\,\delta(k-k').
\end{eqnarray*}
This function is the solution of the radial equation
\begin{eqnarray*}
\frac{d^{2}\,R}{dr^{2}} + \frac{2}{r}\,\frac{d\,R}{dr}
+ \left[k^{2} - \frac{(l + \delta)\,(l + \delta + 1)}{r^{2}}\right]\,
R(r; \delta) = 0.
\end{eqnarray*}

In the cylindrical coordinates, the factorization
\begin{eqnarray*}
\psi(\rho, \varphi, z; \delta) = R(\rho; \delta)\,e^{ik_{z}z}\,\frac{e^{im\varphi}}{\sqrt{2\pi}}
\end{eqnarray*}
corresponds to the separation of variables scheme, which leads to the radial equation
\begin{eqnarray}
\frac{d^{2}\,R}{d\rho^{2}} + \frac{1}{\rho}\,\frac{d\,R}{d\rho}
+ \left[\omega^{2} - \frac{(|m| + \delta)^{2}}{\rho^{2}}\right]\,
R(\rho; \delta) = 0,
\label{4.4.3}
\end{eqnarray}
where $\omega = \sqrt{k^{2} - k_{z}^{2}}$. The regular solution of
Eq. (\ref{4.4.3}) has the representation
\begin{eqnarray*}
R_{\omega, m}(\rho; \delta) = \sqrt{2\pi \omega}\,
J_{|m|+\delta}(\omega \rho)
\end{eqnarray*}
in terms of the Bessel functions. The cylindrical basis normalized by the condition
\begin{eqnarray*}
\int\,\psi_{\omega' k'_{z} m'}^{*}(\rho, \varphi, z; \delta)\,
\psi_{\omega k_{z} m}(\rho, \varphi, z; \delta)\,dV = 4\pi^{2}\,
\delta(\omega - \omega')\,\delta(k_{z} - k'_{z})\,\delta^{m m'}
\end{eqnarray*}
hence has the form
\begin{eqnarray}
\psi_{k \gamma m}(\rho, \varphi, z; \delta) = \sqrt{k\sin\gamma}\,
J_{|m|+\delta}(k \rho \sin\gamma)\,e^{ikz\cos\gamma}\,e^{im\varphi},
\label{4.4.4}
\end{eqnarray}
where the notation
\begin{eqnarray*}
k_{z} = k\cos\gamma, \qquad \omega = k\sin\gamma, \qquad 0\leq \gamma \leq \pi,
\end{eqnarray*}
is introduced for convenience.

\section{The generalization of the Rayleigh formula}
\markboth{CHAPTER 4. GENERALIZED RING-SHAPED POTENTIALS}
{4.5. THE GENERALIZATION OF THE RAYLEIGH FORMULA}

We consider the expansion of cylindrical basis (\ref{4.4.4}) for free motion in a ring-shaped model
in spherical basis (\ref{4.4.2}). This expansion can be written in the form
\begin{eqnarray}
\psi_{k \gamma m}(\rho, \varphi, z; \delta) = \sum\limits_{l=|m|}^{\infty}\,
W_{k \gamma m}^{l}(\delta)\,\psi_{k l m}(r, \theta, \varphi; \delta),
\label{4.5.1}
\end{eqnarray}
Our objective is to calculate the matrix $W_{k \gamma m}^{l}(\delta)$, for which
the following steps should performed:

a. substitute (\ref{4.4.2}) and (\ref{4.4.4}) in (\ref{4.5.1}),

b. multiply the two sides of (\ref{4.5.1}) by $Z_{l' m}^{*}(\theta, \varphi; \delta)$,
integrate over the solid angle, apply

orthonormality condition (\ref{4.1.6}) for ring-shaped functions, and

c. use the expansion
\begin{eqnarray*}
J_{|m|+\delta}(k \rho \sin\gamma) = \sum\limits_{s=0}^{\infty}\,
\frac{(-1)^{s}\,(k \rho \sin\gamma)^{|m|+\delta + s}}
{2^{|m|+\delta + s}\,s!\,\Gamma(|m|+\delta + s + 1)}, \qquad
e^{ikz\cos\gamma} = \sum\limits_{t=0}^{\infty}\,
\frac{(ikz\cos\gamma)^{t}}{t!},
\end{eqnarray*}

d. to pass from the cylindrical coordinates in the left-hand side of (\ref{4.5.1})
to spherical coordinates. Performing these steps, we obtain the expression
\begin{eqnarray}
W_{k \gamma m}^{l}(\delta)\,\sqrt{\frac{2\pi}{r}}\,J_{l+\delta + 1/2}(kr) &=&
\frac{\sqrt{\sin\gamma}}{\Gamma(|m|+\delta + 1)}\,\sum\limits_{s=0}^{\infty}\,
\frac{(-1)^{s}\,\left(\left(k \rho \sin\gamma\right)/2\right)^{|m|+\delta + s}}
{s!\,\Gamma(|m|+\delta + s + 1)}\times \nonumber \\
\label{4.5.2}
\\
&\times& \sum\limits_{t=0}^{\infty}\,
\frac{(ik\cos\gamma)^{t}}{t!}\,r^{|m|+\delta + 2s + t}\,Q_{s t}^{l m}(\delta),
\nonumber
\end{eqnarray}
where
\begin{eqnarray*}
Q_{s t}^{l m}(\delta) = \int\,\left(\cos\theta\right)^{t}\,
\left(\sin\theta\right)^{2s + |m| + \delta}\,e^{im\varphi}\,
Z_{l m}^{*}(\theta, \varphi; \delta)\,d\Omega.
\end{eqnarray*}
Next, we use the Rodrigues formula for ring-shaped functions (\ref{4.1.8}) and integrate
by parts consecutively to ensure that the integral $Q_{s t}^{l m}(\delta)$ is nonzero only
under the condition $2s + t + |m| - l \geq 0$. Hence, all the terms in the series contain
nonnegative powers of $r$, and we therefore conclude that
\begin{eqnarray}
W_{k \gamma m}^{l}(\delta) &=& \frac{i^{l - |m|}}{\sqrt{\pi k}}\,
\frac{\Gamma\left(l + \delta + 3/2\right)}{\Gamma\left(|m| + \delta + 1\right)}\,
\left(2\cos\gamma\right)^{l - |m|}\,\left(\sin\gamma\right)^{|m| + \delta + 1/2}\times \nonumber \\
\label{4.5.3}
\\
&\times&\,\sum\limits_{s=0}^{\left[\left(l - |m|\right)/2\right]}\,
\frac{\left(\tan\gamma\right)^{2s}}
{4^{s}\,s!\,\left(|m|+\delta + 1\right)_{s}}\,
\frac{Q_{s, l - |m| - 2s}^{l m}(\delta)}{\left(l - |m| - 2s\right)!}
\nonumber
\end{eqnarray}
in the limit as $r \to 0$, where
\begin{eqnarray*}
\left[\frac{l - |m|}{2}\right] =
\Bigg\{
\begin{array}{cc}
\frac{l - |m|}{2} & \rm{for} \quad l - |m| = 2n,\\ [5mm]
\frac{l - |m| - 1}{2} & \rm{for} \quad l - |m| = 2n + 1. \\
\end{array}
\end{eqnarray*}
For $t = l - |m| - 2s$, the integral $Q_{s t}^{l m}(\delta)$ passes into the expression
\begin{eqnarray*}
Q_{s, l - |m| - 2s}^{l m}(\delta) &=& (-1)^{s + \frac{m - |m|}{2}}\,
\frac{2^{l + \delta + 1}\,\Gamma\left(l + \delta + 1\right)}
{\Gamma\left(2l + 2\delta + 2\right)}\times \\ [3mm]
&\times& \left[\pi\,\left(2l + 2\delta + 1\right)\,\left(l - |m|\right)!\,
\Gamma\left(l + |m| + 2\delta + 1\right)\right]^{1/2}.
\end{eqnarray*}
We now substitute this expression in (\ref{4.5.3}) and apply the formula \cite{BE1}
\begin{eqnarray*}
\Gamma\left(2z\right) = \frac{2^{2z-1}}{\sqrt{\pi}}\,
\Gamma\left(z\right)\,\Gamma\left(z + \frac{1}{2}\right),
\end{eqnarray*}
which results in the relation for the matrix $W_{k \gamma m}^{l}(\delta)$,
\begin{eqnarray*}
&&W_{k \gamma m}^{l}(\delta) = (-1)^{\frac{m-|m|}{2}}\,i^{l-|m|}\,
\left[\frac{\pi\,(2l + 2\delta + 1)\,\Gamma\left(l + |m| + 2\delta + 1\right)
\,\sin\gamma}{k\,(l - |m|)!}\right]^{1/2}\times \\ [3mm]
&\times& \frac{\left(\cos\gamma\right)^{l-|m|}\,\left(\sin\gamma\right)^{|m|+\delta}}
{2^{|m|+\delta}\,\Gamma\left(|m| + \delta + 1\right)}\,
_{2}F_{1}\left(-\frac{l - |m|}{2}, -\frac{l - |m| - 1}{2}; |m| + \delta + 1; -\tan^{2}\gamma\right).
\end{eqnarray*}
Furthermore, using the formulas (see \cite{BE1} and \cite{BE2})
\begin{eqnarray*}
_{2}F_{1}\left(\alpha, \beta; \gamma; z\right) &=& \frac{\Gamma\left(\gamma\right)\,
\Gamma\left(\beta - \alpha\right)}{\Gamma\left(\beta\right)\,\Gamma\left(\gamma - \alpha\right)}\,
(1-z)^{-\alpha}\,_{2}F_{1}\left(\alpha, \gamma - \beta; \alpha + 1 -\beta; \frac{1}{1-z}\right) + \\ [3mm]
&+& \frac{\Gamma\left(\gamma\right)\,\Gamma\left(\alpha - \beta\right)}
{\Gamma\left(\alpha\right)\,\Gamma\left(\gamma - \beta\right)}\,
(1-z)^{-\beta}\,_{2}F_{1}\left(\beta, \gamma - \alpha; \beta + 1 -\alpha; \frac{1}{1-z}\right),
\end{eqnarray*}
\begin{eqnarray*}
C_{2n}^{\lambda} = (-1)^{n}\,\frac{(\lambda)_{n}}{n!}\,_{2}F_{1}\left(-n, n + \lambda; \frac{1}{2}; x^{2}\right),
\\ [3mm]
C_{2n+1}^{\lambda} = (-1)^{n}\,\frac{(\lambda)_{n+1}}{n!}\,2x\,_{2}F_{1}\left(-n, n + \lambda + 1; \frac{3}{2}; x^{2}\right),
\end{eqnarray*}
we obtain an expression for the matrix $W_{k \gamma m}^{l}(\delta)$ in terms of Gegenbauer polynomials,
\begin{eqnarray}
W_{k \gamma m}^{l}(\delta) &=& (-1)^{\frac{m-|m|}{2}}\,i^{l-|m|}\,
\Gamma\left(|m| + \delta + \frac{1}{2}\right)\,
\left[\frac{(2l + 2\delta + 1)\,(l - |m|)!}{2k\,
\Gamma\left(l + |m| + 2\delta + 1\right)}\right]^{1/2}\times
\nonumber \\
\label{4.5.4}
\\
&\times&\,\left(\sin\gamma\right)^{|m|+\delta + 1/2}\,
C_{l-|m|}^{|m|+\delta + 1/2}(\cos\gamma),
\nonumber
\end{eqnarray}
which holds for both even and odd values of $l - |m|$.

Formulas (\ref{4.5.1}) and (\ref{4.5.4}) completely determine the expansion of the cylindrical
basis for free motion in a ring-shaped model in the spherical basis.

For $m = \delta = \gamma = 0$, it can be easily verified that expansion (\ref{4.5.1})
becomes the well-known Rayleigh expansion of a plane wave in spherical waves,
\begin{eqnarray*}
e^{ikz} = \sum\limits_{l=0}^{\infty}\,\frac{i}{k}\,\sqrt{\pi\,(2l + 1)}\,
\psi_{k l 0}(r, \theta, \varphi), \qquad
\psi_{k l m}(r, \theta, \varphi) = \sqrt{\frac{2\pi k}{r}}\,J_{l + 1/2}(k r )\,Y_{l m}(\theta, \varphi).
\end{eqnarray*}

Expression (\ref{4.5.1}) can be brought to a form well known in the theory of special functions.
Substituting explicit expressions (\ref{4.4.2}), (\ref{4.4.4}), and (\ref{4.5.4}) for the cylindrical and spherical
bases and for the matrix $W_{k \gamma m}^{l}(\delta)$ and passing to the spherical coordinates, we obtain
\begin{eqnarray}
&&\left(\sin\theta\,\sin\gamma\right)^{-|m|-\delta}\,
J_{|m|+\delta}\left(kr\,\sin\theta\,\sin\gamma\right)\,
e^{ikr\,\cos\theta\,\cos\gamma} =
\frac{4^{|m|+\delta}}{\sqrt{\pi\,kr}}
\left[\Gamma\left(|m| + \delta + \frac{1}{2}\right)\right]^{2}\times \nonumber \\
\label{4.5.5}
\\
&\times&\sum\limits_{l=|m|}^{\infty}\,\frac{i^{l-|m|}\,\Gamma\left(l + \delta + \frac{1}{2}\right)\,
(l - |m|)!}{\Gamma\left(l + |m| + 2\delta + 1\right)}\,
J_{l+\delta + 1/2}\left(kr\right)\,C_{l-|m|}^{|m|+\delta + 1/2}(\cos\theta)\,
C_{l-|m|}^{|m|+\delta + 1/2}(\cos\gamma). \nonumber
\end{eqnarray}

We introduce the new notation
\begin{eqnarray*}
y = kr, \qquad \lambda = |m| + \delta + \frac{1}{2}, \qquad n = l - |m|
\end{eqnarray*}
and rewrite formula (\ref{4.5.5} as
\begin{eqnarray}
&&\left(\sin\theta\,\sin\gamma\right)^{1/2-\lambda}\,
J_{\lambda +1/2}\left(y\,\sin\theta\,\sin\gamma\right)\,
e^{iy\,\cos\theta\,\cos\gamma} = \nonumber \\
\label{4.5.6}
\\
&=& \frac{2^{2\lambda - 1/2}}{\sqrt{\pi y}}\,\left[\Gamma\left(2\lambda\right)\right]^{2}\,
\sum\limits_{n=0}^{\infty}\,\frac{i^{n}\,n!\,\left(\lambda + n\right)}
{\Gamma\left(2\lambda +n\right)}\,
J_{\lambda+ n}\left(y\right)\,C_{\lambda}^{n}(\cos\theta)\,
C_{n}^{\lambda}(\cos\gamma). \nonumber
\end{eqnarray}
Next, taking the relations
\begin{eqnarray*}
J_{\lambda}(z) = \sum\limits_{n=0}^{\infty}\,\frac{(-1)^{n}\,(z/2)^{2n + \lambda}}
{n!\,\Gamma\left(n + \lambda + 1\right)}, \qquad C_{n}^{\lambda}1) =
\frac{(\lambda)_{n}}{n!}
\end{eqnarray*}
into account, we conclude that for $\gamma = 0$, (\ref{4.5.6}) becomes the formula
\begin{eqnarray}
e^{ixy} = \Gamma\left(\lambda\right)\,\left(\frac{y}{2}\right)^{-\lambda}\,
\sum\limits_{n=0}^{\infty}\,i^{n}\,\left(\lambda + n\right)\,
J_{\lambda+ n}\left(y\right)\,C_{\lambda}^{n}(x), \qquad x = \cos\theta.
\label{4.5.7}
\end{eqnarray}
Both formulas (\ref{4.5.6} and (\ref{4.5.7}) are given in \cite{BE2}.

First, we have verified the correctness of our result (\ref{4.5.4}); second, we have established
the relation between the free ring-shaped model and a specific division of the theory of special functions.

\section{The generalized Kepler-Coulomb system}
\markboth{CHAPTER 4. GENERALIZED RING-SHAPED POTENTIALS}
{4.6. THE GENERALIZED KEPLER-COULOMB SYSTEM}

By the generalized Kepler-Coulomb system we mean the system that corresponds to the three-dimensional
axially symmetric potential \cite{M-23}:
\begin{eqnarray}
U(x, y, z) = \frac{\alpha}{\sqrt{x^{2} + y^{2} + z^{2}}} +
\frac{z}{\sqrt{x^{2} + y^{2} + z^{2}}}\,\frac{\beta}{x^{2} + y^{2}} +
\frac{\gamma}{x^{2} + y^{2}},
\label{4.6.1}
\end{eqnarray}
(in Cartesian coordinates), where $\alpha, \beta$ and $\gamma$ are constants such that
$\alpha < 0$ and $\gamma \geq |\beta|$. If $\beta = \gamma = 0$, we have the ordinary
(spherically symmetric) Kepler-Coulomb potential.In the case where $\beta = 0$ and
$\gamma \neq 0$, the potential (\ref{4.6.1}) reduces to the Hartmann potential that
has been used for describing axially symmetric systems like ring-shaped molecules
\cite{HARTMAN-1, HARTMAN-2, HARTMAN-3} and investigated from different points of view in the
\cite{Kibler-Negadi-5} - \cite{Kibler5}. The generalized Kepler-Coulomb system corresponding
to $\beta \neq 0$p and $\gamma \neq 0$ was worked out in \cite{Guha,Carpio-1,DRAGAN,Grosche,Kibler5,Zhedanov}.
In particular, the (quantum mechanical) discrete spectrum for the generalized Kepler-Coulomb
system is well known \cite{Guha,Carpio-1,DRAGAN}, even for the so-called $(q, p)$-analog of this system
\cite{DRAGAN}. Furthermore, a path integral treatment of the potential (\ref{4.6.1}) was given in
\cite{Carpio-1,Grosche}. Recently, the dynamical symmetry of the generalized Kepler-Coulomb system
was studied in \cite{DRAGAN,Kibler5,Zhedanov}, the classical motion of a particle moving in the potential
(\ref{4.6.1}) was considered in \cite{Kibler5}, and the coefficients connecting the parabolic and spherical
bases were identified in \cite{Zhedanov} as Clebsch-Gordan coefficients of the pseudounitary group $SU(1,1)$.

First, consider the spherical basis of the generalized Kepler-Coulomb system.

Now, assuming that $\alpha = -e^{2}$ and introducing non-negative constants $\lambda_{i}\, (i=1,2)$,
so that $\beta = \lambda_{2} - \lambda_{1}$ and $\gamma  = \lambda_{1} +\lambda_{2}$, then the potential
(\ref{4.6.1}) in spherical coordinates can be rewritten as the P\"{o}schl-Teller potential \cite{Flugge}:
\begin{eqnarray}
U(r, \theta) = -\frac{e^{2}}{r} +
\frac{\lambda_{1}}{2r^{2}\,\cos^{2}\theta/2} +
\frac{\lambda_{2}}{2r^{2}\,\sin^{2}\theta/2}.
\label{4.6.2}
\end{eqnarray}
The Schr\"{o}dinger equation for the potential (\ref{4.6.2}) may be solved by searching
for a wave function in the form
\begin{eqnarray*}
\psi(r, \theta, \varphi) = R(r)\,Z(\theta, \varphi).
\end{eqnarray*}
This leads to the two coupled differential equations
\begin{eqnarray}
\frac{1}{\sin\theta}\,\frac{\partial}{\partial \theta}\,\left(\sin\theta\,
\frac{\partial Z}{\partial \theta}\right) + \frac{1}{\sin^{2}\theta}\,
\frac{\partial^{2}\,Z}{\partial \varphi^{2}} - \left(\frac{\Lambda_{1}}{\cos^{2}\theta}
+ \frac{\Lambda_{2}}{\sin^{2}\theta}\right)\,Z = -AZ, \nonumber \\
\label{4.6.3}
\end{eqnarray}
\begin{eqnarray}
\frac{1}{r^{2}}\,\frac{d}{dr}\left(r^{2}\,\frac{dR}{dr}\right) - \frac{A}{r^{2}}\,R
+ \frac{2\mu_{0}}{\hbar^{2}}\,\left(E + \frac{e^{2}}{r}\right)\,R = 0,
\label{4.6.4}
\end{eqnarray}
where $A$ is a separation constant in spherical coordinates, and
$\Lambda_{i} = \mu_{0}\lambda_{i}/\hbar^{2}$.

Normalized by condition
\begin{eqnarray}
\int\limits_{0}^{\pi}\,\int\limits_{0}^{2\pi}\,
Z_{l' m'}^{*}(\theta, \varphi; \delta_{1}, \delta_{2})\,
Z_{l m}(\theta, \varphi; \delta_{1}, \delta_{2})\,\sin\theta\,
d\theta\,d\varphi = \delta_{l l'}, \delta_{m m'}
\label{4.6.5}
\end{eqnarray}
the solution of Eq. (\ref{4.6.3}) has the form
\begin{eqnarray}
Z_{l m}(\theta, \varphi; \delta_{1}, \delta_{2}) =
N_{l m}(\delta_{1}, \delta_{2})\left(\cos\frac{\theta}{2}\right)^{|m| + \delta_{1}}
\left(\sin\frac{\theta}{2}\right)^{|m| + \delta_{2}}
P_{l - |m|}^{\left(|m| + \delta_{2}, |m| + \delta_{1}\right)}(\cos\theta)\e^{im\varphi},
\label{4.6.6}
\end{eqnarray}
where $l \in \textbf{N}, m \in \textbf{Z}, \delta_{i} = \sqrt{m^{2} + 4\Lambda_{i}}$
(for $i =1,2$) and $P_{n}^{\alpha, \beta}(x)$ denotes a Jacobi polynomial.
Furthermore, the separation constant $A$ is quantized as
\begin{eqnarray}
A = \left(l + \frac{\delta_{1} + \delta_{2}}{2}\right)\,
\left(l + \frac{\delta_{1} + \delta_{2}}{2} + 1\right).
\label{4.6.7}
\end{eqnarray}
The normalization constant $N_{l m}(\delta_{1}, \delta_{2})$ in (\ref{4.6.6})
is given (up to a phase factor) by
\begin{eqnarray}
N_{l m}(\delta_{1}, \delta_{2}) = (-1)^{\frac{m-|m|}{2}}\,
\sqrt{\frac{\left(2l + \delta_{1} + \delta_{2} + 1\right)\,(l-|m|)!\,
\Gamma\left(l + |m| + \delta_{1} + \delta_{2} + 1\right)}{4\pi\,
\Gamma\left(l + \delta_{1} + 1\right)\,\Gamma\left(l + \delta_{2} + 1\right)}}.
\label{4.6.8}
\end{eqnarray}

Using the formula \cite{BE2}
\begin{eqnarray}
\left(\lambda + \frac{1}{2}\right)_{n}\,C_{n}^{\lambda} = \left(2\lambda\right)_{n}\,
P_{n}^{(\lambda - \frac{1}{2}, \lambda - \frac{1}{2})}(x)
\label{4.6.9}
\end{eqnarray}
it is easy to show that when $\delta_{1} = \delta_{2} = \delta$
it transforms into a ring-shaped function (\ref{4.1.7}).

Let us go now to the radial equation. The introduction of (\ref{4.6.7}) into
(\ref{4.6.4}) leads to
\begin{eqnarray}
\frac{1}{r^{2}}\frac{d}{dr}\left(r^{2}\frac{dR}{dr}\right) - \frac{1}{r^{2}}
\left(l + \frac{\delta_{1} + \delta_{2}}{2}\right)
\left(l + \frac{\delta_{1} + \delta_{2}}{2} + 1\right)R
+ \frac{2\mu_{0}}{\hbar^{2}}\,\left(E + \frac{e^{2}}{r}\right)R = 0,
\label{4.6.10}
\end{eqnarray}
which is reminiscent of the radial equation for the hydrogen atom except that the orbital
quantum number $l$ is replaced here by $l + (\delta_{1} + \delta_{2})/2$.
The solution of Eq. (\ref{4.6.10}) for the discrete spectrum is
\begin{eqnarray}
R_{n l}\left(r; \delta_{1}, \delta_{2}\right) = C_{n l}\left(r; \delta_{1}, \delta_{2}\right)
\left(2\varepsilon r\right)^{l + \frac{\delta_{1} + \delta_{2}}{2}} e^{-\varepsilon r}
F\left(-n + l +1; 2l + \delta_{1} + \delta_{2} + 2; 2\varepsilon r\right),
\label{4.6.11}
\end{eqnarray}
where $n \in \textbf{N} - \left\{0\right\}$. In Eq. (\ref{4.6.11}),
the normalization factor $C_{n l}\left(r; \delta_{1}, \delta_{2}\right)$ reads
\begin{eqnarray}
C_{n l}\left(r; \delta_{1}, \delta_{2}\right) = \frac{2\varepsilon^{2}\sqrt{r_{0}}}
{\Gamma\left(2l + \delta_{1} + \delta_{2} + 2\right)}\,
\sqrt{\frac{\Gamma\left(n + l + \delta_{1} + \delta_{2} + 1\right)}{(n-l-1)!}},
\label{4.6.12}
\end{eqnarray}
and the parameter $\varepsilon$ is defined by
\begin{eqnarray*}
\varepsilon = \frac{1}{r_{0}\left(n + \frac{\delta_{1} + \delta_{2}}{2}\right)}.
\end{eqnarray*}
The eigenvalues $E$ are then given by
\begin{eqnarray}
E_{n} = - \frac{\mu_{0}e^{4}}{2\hbar^{2}\,\left(n + \frac{\delta_{1} + \delta_{2}}{2}\right)^{2}},
\qquad n = 1,2,3...,
\label{4.6.13}
\end{eqnarray}
in agreement with \cite{Carpio-1} (see also \cite{Guha,DRAGAN}).
Equations (\ref{4.6.11})-(\ref{4.6.13}) can be specialized to the cases
$\delta_{1} = \delta_{2} = \delta$ (with $\delta \neq 0$ or $\delta = 0$).
In the limiting case $\delta = 0$, we recover the familiar results for hydrogenlike atoms.

Note also that the spherical wave function of the generalized Kepler-Coulomb system
\begin{eqnarray}
\psi_{n l m}\left(r, \theta, \varphi; \delta_{1}, \delta_{2}\right) =
R_{n l}\left(r; \delta_{1}, \delta_{2}\right)\,
Z_{l m}\left(\theta, \varphi; \delta_{1}, \delta_{2}\right)
\label{4.6.14}
\end{eqnarray}
is an eigenfunction of the commuting operators $\hat{M}, \hat{l}_{z}$ and the following spectral problem takes place
\begin{eqnarray}
\hat{M}\psi_{n l m}\left(r, \theta, \varphi; \delta_{1}, \delta_{2}\right) =
\left(l + \frac{\delta_{1} + \delta_{2}}{2}\right)\,
\left(l + \frac{\delta_{1} + \delta_{2}}{2} + 1\right)\,
\psi_{n l m}\left(r, \theta, \varphi; \delta_{1}, \delta_{2}\right),
\label{4.6.15}
\end{eqnarray}
where
\begin{eqnarray}
\hat{M} = \hat{l}^{2} + \frac{2\Lambda_{1}}{1 + \cos\theta} +
\frac{2\Lambda_{2}}{1 - \cos\theta}.
\label{4.6.16}
\end{eqnarray}
The operator $\hat{M}$ in Cartesian coordinates has the form
\begin{eqnarray}
\hat{M} = -r^{2}\Delta + x_{i}x_{j}\frac{\partial^{2}}{\partial x_{i} \partial x_{j}}
+ 2x_{i}\frac{\partial}{\partial x_{i}} + \frac{2\Lambda_{1}r}{r + z} +
\frac{2\Lambda_{2}r}{r - z}.
\label{4.6.17}
\end{eqnarray}

Now let us consider the parabolic basis of the generalized Kepler-Coulomb system.

In the parabolic coordinates (\ref{3.1.13}) the potential U (\ref{4.6.2}) reads
\begin{eqnarray*}
U(\mu, \nu) = -\frac{2e^{2}}{\mu + \nu} + \frac{2\lambda_{1}}{\mu (\mu + \nu)} +
\frac{2\lambda_{2}}{\nu (\mu + \nu)}.
\end{eqnarray*}
By looking for a solution of the Schrodinger equation for this potential in the form
\begin{eqnarray*}
\psi(\mu, \nu, \varphi) = \phi_{1}(\mu)\,\phi_{2}(\nu)\,\frac{e^{im\varphi}}{\sqrt{2\pi}},
\end{eqnarray*}
we obtain the two coupled equations
\begin{eqnarray}
\frac{d}{d\mu}\,\left(\mu\,\frac{d\phi_{1}}{d\mu}\right) + \left[\frac{\mu_{0}\,E}{2\hbar^{2}}\mu
- \frac{(|m| + \delta_{1})^{2}}{4\mu} + \frac{\sqrt{\mu_{0}}}{2\hbar}
\Omega\left(\delta_{1}, \delta_{2}\right) + \frac{1}{r_{0}}\right]\,\phi_{1} = 0,
\label{4.6.18}
\end{eqnarray}
\begin{eqnarray}
\frac{d}{d\nu}\,\left(\nu\,\frac{d\phi_{2}}{d\nu}\right) + \left[\frac{\mu_{0}\,E}{2\hbar^{2}}\nu
- \frac{(|m| + \delta_{2})^{2}}{4\nu} - \frac{\sqrt{\mu_{0}}}{2\hbar}
\Omega\left(\delta_{1}, \delta_{2}\right) + \frac{1}{r_{0}}\right]\,\phi_{1} = 0,
\label{4.6.19}
\end{eqnarray}
where $\Omega$ is the separation constant in parabolic coordinates.
By solving (\ref{4.6.18}) and (\ref{4.6.19}), we get the normalized
wave function
\begin{eqnarray}
\psi_{n_{1} n_{2} m}(\mu, \nu, \varphi) = \varepsilon^{2}\,\sqrt{\frac{r_{0}}{\pi}}
\phi_{n_{1} m}(\mu; \delta_{1})\,\phi_{n_{2} m}(\nu;  \delta_{2})\,e^{im\varphi},
\label{4.6.20}
\end{eqnarray}
where
\begin{eqnarray}
\phi_{n_{i} m}(t_{i}; \delta_{i}) = \sqrt{\frac{\Gamma\left(n_{i}+ |m| + \delta_{i} + 1\right)}
{\left(n_{i}\right)!}}\,\frac{e^{-\varepsilon t_{i}}\left(\varepsilon t_{i}\right)^{(|m| + \delta_{i})/2}}
{\Gamma\left(|m| + \delta_{i} + 1\right)}\,F\left(-n_{i}; |m| + \delta_{i} + 1; \varepsilon t_{i}\right),
\label{4.6.21}
\end{eqnarray}
with $i = 1,2$ ($t_{1} \equiv \mu$,  and $t_{2} \equiv \nu$). In Eq. (\ref{4.6.21}), we have
\begin{eqnarray}
n_{1} = -\frac{|m| + \delta_{1} + 1}{2} + \frac{\sqrt{\mu_{0}}}{2\hbar \varepsilon}\Omega
+ \frac{1}{2r_{0} \varepsilon}, \qquad
n_{2} = -\frac{|m| + \delta_{2} + 1}{2} - \frac{\sqrt{\mu_{0}}}{2\hbar \varepsilon}\Omega
+ \frac{1}{2r_{0} \varepsilon}.
\label{4.6.22}
\end{eqnarray}
Here, again, $m \in \textbf{Z}$ in order to ensure the univaluedness of $\psi_{n_{1} n_{2} m}$.
In addition, $n_{i} \in \textbf{N}$ (for $i = 1,2$) in order that $\psi_{n_{1} n_{2} m}$ be in
$L^{2}(\textbf{R}^{3})$. Therefore, the quantized values of the energy $E$ are given by
(\ref{4.6.13}) where now the quantum number $n$ is $n = n_{1} + n_{2} + |m| + 1$,
a number that parallels the principal quantum number of the hydrogen atom in parabolic coordinates.

Following Kibler and Campigotto \cite{Kibler5}, we can obtain a further integral of motion besides
$E$. In parabolic coordinates, this integral corresponds to the Hermitean operator
\begin{eqnarray}
\hat{\Omega} = \frac{\hbar}{\sqrt{\mu_{0}}}\,\Biggl\{\frac{2}{\mu + \nu}\,
\left[\mu\frac{\partial}{\partial \nu}\left(\nu \frac{\partial}{\partial \nu}\right) -
\nu\frac{\partial}{\partial \mu}\left(\mu \frac{\partial}{\partial \mu}\right)\right]
+ \frac{\mu - \nu}{2\mu \nu}\,\frac{\partial^{2}}{\partial \varphi^{2}}+ \nonumber \\
\label{4.6.23}
\\
+ \frac{2\Lambda_{1}\nu}{\mu (\mu + \nu)} - \frac{2\Lambda_{2}\mu}{\nu (\mu + \nu)}
+ \frac{\mu - \nu}{r_{0}(\mu + \nu)}\Biggr\}.
\nonumber
\end{eqnarray}
The eigenvalues of $\hat{\Omega}$ are
\begin{eqnarray}
\Omega\left(\delta_{1}, \delta_{2}\right) = \frac{\hbar}{r_{0}\sqrt{\mu_{0}}}\,
\frac{n_{1} - n_{2} + \frac{\delta_{1} - \delta_{2}}{2}}{n + \frac{\delta_{1} + \delta_{2}}{2}},
\label{4.6.24}
\end{eqnarray}
i.e. we have that
\begin{eqnarray}
\hat{\Omega}\,\psi_{n_{1} n_{2} m}\left(\mu, \nu, \varphi; \delta_{1}, \delta_{2}\right) =
\frac{\hbar}{r_{0}\sqrt{\mu_{0}}}\,
\frac{n_{1} - n_{2} + \frac{\delta_{1} - \delta_{2}}{2}}{n + \frac{\delta_{1} + \delta_{2}}{2}}\,
\psi_{n_{1} n_{2} m}\left(\mu, \nu, \varphi; \delta_{1}, \delta_{2}\right).
\label{4.6.25}
\end{eqnarray}
In Cartesian coordinates, the operator $\hat{\Omega}$ can be rewritten as
\begin{eqnarray}
\hat{\Omega} = \frac{\hbar}{\sqrt{\mu_{0}}}\,\Biggl[z\left(\frac{\partial^{2}}{\partial x^{2}} +
\frac{\partial^{2}}{\partial y^{2}}\right) - x\frac{\partial^{2}}{\partial x \partial z}
- y\frac{\partial^{2}}{\partial y \partial z} - \frac{\partial}{\partial z} + \nonumber \\
\label{4.6.26}
\\
+ \Lambda_{1}\,\frac{r - z}{r(r + z)} - \Lambda_{2}\,\frac{r + z}{r(r - z)}
+ \frac{z}{r_{0}r}\Biggr],
\nonumber
\end{eqnarray}
so that it immediately follows that $\hat{\Omega}$ is connected to the $z$-component $A_{z}$ of the
Laplace - Runge - Lenz- Pauli vector \cite{Laplace,Runge,Lenz,Pauli} via
\begin{eqnarray}
\hat{\Omega} = A_{z} + \Lambda_{1}\,\frac{r - z}{r(r + z)} - \Lambda_{2}\,\frac{r + z}{r(r - z)}
\label{4.6.27}
\end{eqnarray}
and coincides with $A_{z}$ when $\beta = \gamma = 0$.

\section{Interbasis expansions and the spheroidal basis}
\markboth{CHAPTER 4. GENERALIZED RING-SHAPED POTENTIALS}
{4.7. INTERBASIS EXPANSIONS AND SPHEROIDAL BASIS}

Before moving on to finding the spheroidal basis of the generalized Kepler-Coulomb system,
let us establish the connection between the spherical and parabolic bases of this system.

For a given subspace corresponding to a fixed value of $E_{n}$, we can write the parabolic wave
function (\ref{4.6.20}) in terms of the spherical wave functions (\ref{4.6.14}) as
\begin{eqnarray}
\psi_{n_{1} n_{2} m}\left(\mu, \nu, \varphi; \delta_{1}, \delta_{2}\right) =
\sum\limits_{l=|m|}^{n-1}\,W_{n_{1} n_{2} m}^{l}\left(\delta_{1}, \delta_{2}\right)\,
\psi_{n l m}\left(r, \theta, \varphi; \delta_{1}, \delta_{2}\right).
\label{4.7.1}
\end{eqnarray}
By virtue of Eq. (\ref{3.2.3}), the left-hand side of (\ref{4.7.1}) can be rewritten in
spherical coordinates. Then, by substituting $\theta = 0$ in the so-obtained equation and
by taking into account that (\ref{1.7.4}), we get an equation that depends only on the
variable $r$. Now using the condition of orthogonality of the radial wave functions
(\ref{4.6.11}) with respect to the orbital quantum number $l$
\begin{eqnarray*}
\int\limits_{0}^{\infty}\,R_{n l'}\left(r; \delta_{1}, \delta_{2}\right)\,
R_{n l}\left(r; \delta_{1}, \delta_{2}\right) = \frac{2\varepsilon^{3}\,r_{0}}
{2l + \delta_{1} + \delta_{2} + 1}\,\delta_{l l'}
\end{eqnarray*}
we arrive at the relationship
\begin{eqnarray}
W_{n_{1} n_{2} m}^{l}\left(\delta_{1}, \delta_{2}\right) = (-1)^{\frac{m-|m|}{2}}
\frac{\sqrt{\left(2l + \delta_{1} + \delta_{2} + 1\right)
\left(l-|m|\right)!\Gamma\left(l + \delta_{1} + 1\right)}}
{\Gamma\left(|m| + \delta_{1} + 1\right)\Gamma\left(2l + \delta_{1} + \delta_{2} + 2\right)}
E_{n_{1} n_{2}}^{l m}K_{n n_{1}}^{l m},
\label{4.7.2}
\end{eqnarray}
where
\begin{eqnarray}
E_{n_{1} n_{2}}^{l m} = \left[\frac{\Gamma\left(n_{1} + |m| + \delta_{1} + 1\right)\,
\Gamma\left(n_{2} + |m| + \delta_{2} + 1\right)\,\Gamma\left(n + l + \delta_{1} + \delta_{2} + 1\right)}
{\left(n_{1}\right)!\,\left(n_{2}\right)!\,\Gamma\left(l + \delta_{2} + 1\right)\,
\Gamma\left(l + |m| + \delta_{1} + \delta_{2} + 1\right)}\right]^{1/2},
\label{4.7.3}
\end{eqnarray}
\begin{eqnarray*}
K_{n n_{1}}^{l m} = \int\limits_{0}^{\infty}\,x^{l+|m|+\delta_{1} + \delta_{2}}
e^{-x}F\left(-n_{1}; |m| + \delta_{1} + 1; x\right)\,
F\left(-n+l+1; 2l + \delta_{1} + \delta_{2} + 1; x\right)dx,
\end{eqnarray*}
and $x = 2\varepsilon r$. To calculate the integral $K_{n n_{1}}^{l m}$, it is sufficient
to write the confluent hypergeometric function $F\left(-n_{1}; |m| + \delta_{1} + 1; x\right)$
as a series, to integrate according to (\ref{1.6.5}) and to use the formula
(\ref{1.6.6}) for the summation of the Gauss hypergeometric function
$_{2}F_{1}(a, b, c; 1)$. We thus obtain
\begin{eqnarray}
K_{n n_{1}}^{l m} &=& \frac{\left(n - |m| - 1\right)!\,
\Gamma\left(2l + \delta_{1} + \delta_{2} + 2\right)\,
\Gamma\left(l + |m| + \delta_{1} + \delta_{2} + 1\right)}
{\left(l-|m|\right)!\,\Gamma\left(n + l + \delta_{1} + \delta_{2} + 1\right)}\times
\nonumber \\
\label{4.7.4}
\\
&\times&
{_3F_2}\left\{\matrix{ -n_{1},\,\,-l + |m|,\,\, l + |m| + \delta_{1} + \delta_{2} + 1\cr \cr
|m| + \delta_{1} + 1,\,\,-n + |m| + 1
\cr}\Biggr|1\right\}. \nonumber
\end{eqnarray}
The introduction of (\ref{4.7.4}) into (\ref{4.7.2}) and owing to (\ref{4.7.3}), we end up with
\begin{eqnarray}
&&W_{n_{1} n_{2} m}^{l}\left(\delta_{1}, \delta_{2}\right) = (-1)^{\frac{m-|m|}{2}}
\frac{\left(n - |m| - 1\right)!}{\Gamma\left(|m| + \delta_{1} + 1\right)}
\frac{\sqrt{\left(2l + \delta_{1} + \delta_{2} + 2\right)\,\Gamma\left(l + \delta_{1} + 1\right)}}
{\Gamma\left(l + \delta_{2} + 1\right)\,\Gamma\left(n + l + \delta_{1} + \delta_{2} + 1\right)}\times \nonumber \\ [3mm]
&\times& \left[\frac{\Gamma\left(l + |m| + \delta_{1} + \delta_{2} + 1\right)
\Gamma\left(n_{1} + |m| + \delta_{1} + 1\right)\Gamma\left(n_{2} + |m| + \delta_{2} + 1\right)}
{\left(n_{1}\right)!\,\left(n_{2}\right)!\left(l - |m|\right)!\,\left(n - l - 1\right)!}\right]^{1/2}\times
\label{4.7.5}
\\
&\times&
{_3F_2}\left\{\matrix{ -n_{1},\,\,-l + |m|,\,\, l + |m| + \delta_{1} + \delta_{2} + 1\cr \cr
|m| + \delta_{1} + 1,\,\,-n + |m| + 1
\cr}\Biggr|1\right\}. \nonumber
\end{eqnarray}
which constitutes a closed-form expression for the interbasis coefficients.

By comparing Eqs. (\ref{1.7.12}) and (\ref{4.7.5}), we finally obtain
\begin{eqnarray}
W_{n_{1} n_{2} m}^{l}\left(\delta_{1}, \delta_{2}\right) = (-1)^{n_{1} + \frac{m-|m|}{2}}\,
C_{\frac{n+\delta_{2}-1}{2}, \frac{|m|+n_{2}-n_{1}+\delta_{2}}{2}; \frac{n+\delta_{1}-1}{2},
\frac{|m|+n_{1}-n_{2}+\delta_{1}}{2}}^{l+\frac{\delta_{1} + \delta_{2}}{2}, |m|+\frac{\delta_{1} + \delta_{2}}{2}}.
\label{4.7.6}
\end{eqnarray}
Equation (\ref{4.7.6}) proves that the coefficients for the expansion of the parabolic basis in terms of
the spherical basis are nothing but the analytic continuation, for real values of their arguments,
of the $SU(2)$ Clebsch-Gordan coefficients.

The inverse of Eq. (\ref{4.7.1}), namely,
\begin{eqnarray}
\psi_{n l m}\left(r, \theta, \varphi; \delta_{1}, \delta_{2}\right)  =
\sum\limits_{n_{1}=0}^{n-|m|-1}\,\tilde{W}_{n l m}^{n_{1}}\left(\delta_{1}, \delta_{2}\right)\,
\psi_{n_{1} n_{2} m}\left(\mu, \nu, \varphi; \delta_{1}, \delta_{2}\right),
\label{4.7.7}
\end{eqnarray}
is an immediate consequence of the orthonormality property of the $SU(2)$ Clebsch-Gordan
coefficients. The expansion coefficients in (\ref{4.7.7}) are thus given by
\begin{eqnarray}
\tilde{W}_{n l m}^{n_{1}}\left(\delta_{1}, \delta_{2}\right) = (-1)^{n_{1} + \frac{m-|m|}{2}}\,
C_{\frac{n+\delta_{2}-1}{2}, \frac{n+\delta_{2}-1}{2} - n_{1}; \frac{n+\delta_{1}-1}{2},
n_{1} + |m| - \frac{n+\delta_{1}-1}{2}}^{l+\frac{\delta_{1} + \delta_{2}}{2}, |m|+\frac{\delta_{1} + \delta_{2}}{2}}.
\label{4.7.8}
\end{eqnarray}
and may be expressed in terms of the $_{3}F_{2}$ function through (\ref{1.7.10}) or (\ref{1.7.12}).

It should be mentioned that (\ref{4.7.6}) and (\ref{4.7.8}) generalize the well-known
result, corresponding to $\delta_{1} = \delta_{2} = 0$, for the interbasis expansion between parabolic and
spherical bases obtained in \cite{PARK,TARTER,M-1,ARUT1} in the case of the hydrogen atom.
Furthermore, by taking $\delta_{1} = \delta_{2} \neq 0$ in (\ref{4.7.6}) and (\ref{4.7.8}),
we recover our former result \cite{Lutsenko-1} for the Hartmann system.

Now, using the interbasis expansions we obtained in this section, we will construct
a spheroidal basis of the generalized Kepler-Coulomb system.

In the system of prolate spheroidal coordinates (\ref{3.3.1}), the potential (\ref{4.6.2}) can be written as
\begin{eqnarray}
U(\xi, \eta) = -\frac{2e^{2}}{R(\xi + \eta)} + \frac{4}{R^{2}(\xi + \eta)}\,
\left[\frac{\lambda_{1}}{(\xi + 1)\,(1 + \eta)} +
\frac{\lambda_{2}}{(\xi - 1)\,(1 - \eta)}\right].
\label{4.7.9}
\end{eqnarray}
The Schrodinger equation for the potential (\ref{4.7.9}) is separable in prolate spheroidal coordinates.
As a point of fact, by looking for a solution of this equation in the form
\begin{eqnarray}
\psi(\xi, \eta, \varphi) = \Phi_{1}(\xi)\,\Phi_{2}(\eta)\,e^{im\varphi}, \qquad
m \in \textbf{Z},
\label{4.7.10}
\end{eqnarray}
we obtain the two ordinary differential equations
\begin{eqnarray}
\left[\frac{d}{d \xi}\left(\xi^{2} - 1\right)\frac{d}{d \xi} +
\frac{\left(|m| + \delta_{1}\right)^{2}}{2(\xi + 1)} -
\frac{\left(|m| + \delta_{2}\right)^{2}}{2(\xi - 1)} +
\frac{ER^{2}}{2r_{0}e^{2}}\left(\xi^{2} - 1\right) + \frac{R}{r_{0}}\xi\right]
\Phi_{1} = \Lambda\Phi_{1},
\label{4.7.11}
\end{eqnarray}
\begin{eqnarray}
\left[\frac{d}{d \eta}\left(1 - \eta^{2}\right)\frac{d}{d \eta} -
\frac{\left(|m| + \delta_{1}\right)^{2}}{2(1 + \eta)} -
\frac{\left(|m| + \delta_{2}\right)^{2}}{2(1 - \eta)} +
\frac{ER^{2}}{2r_{0}e^{2}}\left(1 - \eta^{2}\right) - \frac{R}{r_{0}}\eta\right]
\Phi_{2} = -\Lambda\Phi_{2},
\label{4.7.12}
\end{eqnarray}
where $\Lambda(R)$ is a separation constant in prolate spheroidal coordinates. By eliminating
the energy $E$ from Eqs. (\ref{4.7.11}) and (\ref{4.7.12}), we produce the operator
\begin{eqnarray}
\hat{\Lambda} &=& \frac{1}{\xi^{2} - \eta^{2}}\left[\left(1 - \eta^{2}\right)
\frac{\partial}{\partial \xi}\left(\xi^{2} - 1\right)\frac{\partial}{\partial \xi} -
\left(\xi^{2} - 1\right)\frac{\partial}{\partial \eta}\left(1 - \eta^{2}\right)\frac{\partial}{\partial \eta}\right]
- \frac{\xi^{2} + \eta^{2} - 2}{\left(\xi^{2} - 1\right)\left(1 - \eta^{2}\right)}\frac{\partial^{2}}{\partial \eta^{2}} +
\nonumber \\
\label{4.7.13}
\\
&+& \frac{2\mu_{0}}{\hbar^{2}}\left[\lambda_{1}\frac{\left(\xi + \eta\right)^{2}
+ (\xi - 1)\,(1 - \eta)}{\left(\xi + \eta\right)(\xi + 1)\,(1 + \eta)}
\lambda_{2}\frac{\left(\xi + \eta\right)^{2}
- (\xi + 1)\,(1 + \eta)}{\left(\xi + \eta\right)(\xi - 1)\,(1 - \eta)}\right]
+ \frac{R}{r_{0}}\frac{\xi \eta + 1}{\xi + \eta}, \nonumber
\end{eqnarray}
the eigenvalues of which are $\Lambda(R)$) and the eigenfunctions of which are
$\psi(\xi, \eta, \varphi)$. The significance of the (self-adjoint) operator
$\hat{\Lambda}$ can be found by switching to Cartesian coordinates.
A long calculation gives
\begin{eqnarray}
\hat{\Lambda} = \hat{M} + \frac{\sqrt{\mu_{0}R}}{\hbar}\,\hat{\Omega},
\label{4.7.14}
\end{eqnarray}
where $\hat{M}$ and $\hat{\Omega}$ are the constants of motion
(\ref{4.6.16}), (\ref{4.6.26}).

The prolate spheroidal wave functions $\Phi_{1}$ and $\Phi_{2}$ could be obtained by solving
(\ref{4.7.11}) and (\ref{4.7.12}). However, it is more economical to proceed in the following
way that shall give us, at the same time, the global wave function (\ref{4.7.10}), i .e.,
$\psi(\xi, \eta, \varphi; R, \delta_{1}, \delta_{2})$, and the interbasis expansion coefficients.

From what has preceded, we have three sets of commuting operators, viz., ${\hat{H}, \hat{l}_{z}, \hat{M}}$ ,
${\hat{H}, \hat{l}_{z}, \hat{\Omega}}$ , and ${\hat{H}, \hat{l}_{z}, \hat{\Lambda}}$ corresponding to the
spherical, parabolic, and prolate spheroidal coordinates, respectively. (The operators $\hat{l}_{z}$ and
$\hat{H}$ are the $z$-component of the angular momentum and the Hamiltonian for the generalized
Kepler-Coulomb system, respectively.) In particular, we have Eqs. (\ref{4.6.15}), (\ref{4.6.26})
for the spherical, parabolic, and
\begin{eqnarray}
\hat{\Lambda}\,\psi_{n q m}(\xi, \eta, \varphi; R, \delta_{1}, \delta_{2}) =
\Lambda_{q}(R, \delta_{1}, \delta_{2})\,\psi_{n q m}(\xi, \eta, \varphi; R, \delta_{1}, \delta_{2})
\label{4.7.15}
\end{eqnarray}
for the prolate spheroidal bases, respectively. In Eq. (\ref{4.7.15}), the index
$q$ labels the eigenvalues of the operator $\hat{\Lambda}$ and varies in the
range $0 \leq q \leq n - |m| -1$. We are now ready to deal with the interbasis expansions
\begin{eqnarray}
\psi_{n q m}(\xi, \eta, \varphi; R, \delta_{1}, \delta_{2}) = \sum\limits_{l=|m|}^{n-1}\,
U_{n q m}^{l}(R, \delta_{1}, \delta_{2})\,
\psi_{n l m}(r, \theta, \varphi; \delta_{1}, \delta_{2}),
\label{4.7.16}
\end{eqnarray}
\begin{eqnarray}
\psi_{n q m}(\xi, \eta, \varphi; R, \delta_{1}, \delta_{2}) = \sum\limits_{n_{1}=0}^{n- |m| - 1}\,
V_{n q m}^{n_{1}}(R, \delta_{1}, \delta_{2})\,
\psi_{n_{1} n_{2} m}(\mu, \nu, \varphi; \delta_{1}, \delta_{2}),
\label{4.7.17}
\end{eqnarray}
for the prolate spheroidal basis in terms of the spherical and parabolic bases.
[Equation (\ref{4.7.16}) was first considered by Coulson and Joseph \cite{COJO}
in the particular case $\delta_{1} = \delta_{2} = 0$.]

First, we consider Eq. (\ref{4.7.16}). Let the operator $\hat{\Lambda}$ act on
both sides of (\ref{4.7.16}). Then, by using Eqs. (\ref{4.7.14}), (\ref{4.6.15}), and
(\ref{4.7.15}) as well as the orthonormality property of the spherical basis, we
find that
\begin{eqnarray}
\left[\Lambda_{q}(R, \delta_{1}, \delta_{2}) - \left(l + \frac{\delta_{1} + \delta_{2}}{2}\right)
\left(l + \frac{\delta_{1} + \delta_{2}}{2} + 1\right)\right]\,U_{n q m}^{l} =
\frac{\mu_{0}R}{\hbar}\,\sum\limits_{l'=|m|}^{n-1}\,U_{n q m}^{l'}
\left(\hat{\Omega}\right)_{ll'},
\label{4.7.18}
\end{eqnarray}
where
\begin{eqnarray}
\left(\hat{\Omega}\right)_{ll'} = \int\limits_{0}^{\infty}\,\int\limits_{0}^{\pi}\,
\int\limits_{0}^{2\pi}\,\psi_{n l m}^{*}(r, \theta, \varphi; \delta_{1}, \delta_{2})\,
\hat{\Omega}\,\psi_{n l' m}(r, \theta, \varphi; \delta_{1}, \delta_{2})\,r^{2}\,
\sin\theta\,dr\,d\theta\,d\varphi.
\label{4.7.19}
\end{eqnarray}
The calculation of the matrix element $\left(\hat{\Omega}\right)_{ll'}$ can be done
by expanding the spherical wave functions in (\ref{4.7.19}) in terms of parabolic
wave functions [see Eq. (\ref{4.7.7})] and by making use of the eigenvalue equation for
$\hat{\Omega}$ [see Eq. (\ref{4.6.25})]. This leads to
\begin{eqnarray}
\left(\hat{\Omega}\right)_{ll'} = \frac{2e^{2}\sqrt{\mu_{0}}}{\hbar\,
\left(2n + \delta_{1} + \delta_{2}\right)}\,\sum\limits_{n_{1}=0}^{n- |m| - 1}\,
\left(2n_{1} - n + |m| + \frac{\delta_{1} - \delta_{2}}{2} + 1\right)\,
\tilde{W}_{n l m}^{n_{1}}\,\tilde{W}_{n l' m}^{n_{1}}.
\label{4.7.20}
\end{eqnarray}
Then, by using Eq. (\ref{4.7.7}) together with the recursion relation (\ref{1.8.10})
and the orthonormality condition (\ref{1.8.11}), we find that  $\left(\hat{\Omega}\right)_{ll'}$
is given by
\begin{eqnarray}
\left(\hat{\Omega}\right)_{ll'} = \frac{e^{2}\sqrt{\mu_{0}}}{\hbar}
\left[\frac{\left(|m| + \frac{\delta_{1} + \delta_{2}}{2}\right)\,\left(\delta_{1} - \delta_{2}\right)}
{2\left(l + \frac{\delta_{1} + \delta_{2}}{2}\right)\left(l + \frac{\delta_{1} + \delta_{2}}{2} + 1\right)}\,
\delta_{l',l}
-\frac{A_{n m}^{l+1}\,\delta_{l',l+1} + A_{n m}^{l}\,\delta_{l',l-1}}
{\left(n + \frac{\delta_{1} + \delta_{2}}{2}\right)}\right],
\label{4.7.21}
\end{eqnarray}
where
\begin{eqnarray}
A_{n m}^{l} = \left[\frac{4(l-|m|)(l+|m|+\delta_{1} + \delta_{2})
(l+\delta_{1})(l+\delta_{2})(n-l)(n+l+\delta_{1} + \delta_{2})}
{(2l+\delta_{1} + \delta_{2})^{2}(2l+\delta_{1} + \delta_{2}-1)
(2l+\delta_{1} + \delta_{2}+1)}\right]^{1/2}.
\label{4.7.22}
\end{eqnarray}
Now by introducing (\ref{4.7.22}) into (\ref{4.7.18}), we get the following
three-term recursion relation for the coefficient $U_{n q m}^{l}$:
\begin{eqnarray}
&&\Biggl[\left(l+\frac{\delta_{1} + \delta_{2}}{2}\right)
\left(l+\frac{\delta_{1} + \delta_{2}}{2}+1\right) -
\frac{R(2|m|+\delta_{1} + \delta_{2})(\delta_{1} - \delta_{2})}
{r_{0}(2l+\delta_{1} + \delta_{2})(2l+\delta_{1} + \delta_{2}+2)} -\nonumber
\\
\label{4.7.23}
\\
&-& \Lambda_{q}(R, \delta_{1}, \delta_{2})\Biggr]
U_{n q m}^{l}(R,\delta_{1}, \delta_{2}) =
\frac{R}{r_{0}(2n + \delta_{1} + \delta_{2})}
\left[A_{n m}^{l+1}\,U_{n q m}^{l+1}(R,\delta_{1}, \delta_{2})
+A_{n m}^{l}\,U_{n q m}^{l-1}(R,\delta_{1}, \delta_{2})\right].\nonumber
\end{eqnarray}
The recursion relation (\ref{4.7.23}) provides us with a system of $n - |m|$ linear
homogeneous equations that can be solved by taking into account the normalization condition
\begin{eqnarray*}
\sum\limits_{l=|m|}^{n-1}\,\left|U_{n q m}^{l}(R,\delta_{1}, \delta_{2})\right|^{2} = 1.
\end{eqnarray*}
The eigenvalues $\Lambda_{q}(R, \delta_{1}, \delta_{2})$ of the operator
$\hat{\Lambda}$ then follow from the vanishing of the determinant
for the latter system.

Second, let us concentrate on the expansion (\ref{4.7.17}) of the prolate spheroidal
basis in terms of the parabolic basis. By employing a technique similar to the one
used for deriving Eq. (\ref{4.7.18}), we get
\begin{eqnarray}
\Biggl[\Lambda_{q}(R, \delta_{1}, \delta_{2}) - \frac{R}{r_{0}}\,
\frac{2n_{1} - 2n_{2} + \delta_{1} - \delta_{2}}{2n + \delta_{1} + \delta_{2}}\Biggr]\,
V_{n q m}^{n_{1}} = \sum\limits_{n'_{1}=0}^{n-|m|-1}\,V_{n q m}^{n'_{1}}\,
\left(\hat{M}\right)_{n_{1} n'_{1}},
\label{4.7.24}
\end{eqnarray}
where
\begin{eqnarray}
\left(\hat{M}\right)_{n_{1} n'_{1}} =\int\limits_{0}^{\infty}\,
\int\limits_{0}^{\infty}\,\int\limits_{0}^{2\pi}\,
\psi_{n_{1} n_{2} m}^{*}(\mu, \nu, \varphi; \delta_{1}, \delta_{2})\,
\hat{M}\,\psi_{n'_{1} n'_{2} m}(\mu, \nu, \varphi; \delta_{1}, \delta_{2})\,
\frac{\mu + \nu}{4}\,d\mu\,d\nu\,d\varphi.
\label{4.7.25}
\end{eqnarray}
The matrix elements $\left(\hat{M}\right)_{n_{1} n'_{1}}$ can be calculated in the same
way as $\left(\hat{\Omega}\right)_{ll'}$ except that now we must use the relation
(\ref{1.8.15}) and the orthonormality condition (\ref{1.8.16})
instead of Eqs. (\ref{1.8.10}) and (\ref{1.8.10}). This yields the matrix element
\begin{eqnarray}
&&\left(\hat{M}\right)_{n_{1} n'_{1}} = -\left[n_{1}(n - n_{1} - |m|)(n_{1} + |m| + \delta_{1})
(n - n_{1} + \delta_{2})\right]^{1/2}\,\delta_{n'_{1},n_{1}-1}-  \nonumber \\ [3mm]
&-&\left[(n_{1} + 1)(n_{1} + |m| + \delta_{1} + 1)
(n - n_{1} - |m| - 1)(n - n_{1} + \delta_{2} - 1)\right]^{1/2}\,\delta_{n'_{1},n_{1}+1} +
\label{4.7.26} \\ [3mm]
&+&\left[(n_{1} + 1)(n - n_{1} - |m| - 1) + (n_{1} + |m| + \delta_{1})(n - n_{1} + \delta_{2}) +
\frac{1}{4}(\delta_{1} - \delta_{2})(\delta_{1} - \delta_{2} - 2)\right]\,\delta_{n'_{1},n_{1}}.
\nonumber
\end{eqnarray}
Finally, the introduction (\ref{4.7.26}) into (\ref{4.7.24}) leads to the three-term
recursion relation
\begin{eqnarray}
&&\Biggl[(n_{1} + 1)(n - n_{1} - |m| - 1) + (n_{1} + |m| + \delta_{1})(n - n_{1} + \delta_{2}) +
\frac{1}{4}(\delta_{1} - \delta_{2})(\delta_{1} - \delta_{2} - 2) + \nonumber \\ [3mm]
&+& \frac{R}{r_{0}}\,
\frac{2n_{1} - 2n_{2} + \delta_{1} - \delta_{2}}{2n + \delta_{1} + \delta_{2}} -
\Lambda_{q}(R, \delta_{1}, \delta_{2})\Biggr]\,V_{n q m}^{n_{1}} = \nonumber \\
\label{4.7.27}
\\
&=& \left[n_{1}(n - n_{1} - |m|)(n_{1} + |m| + \delta_{1})
(n - n_{1} + \delta_{2})\right]^{1/2}\,V_{n q m}^{n_{1}-1}-  \nonumber \\ [3mm]
&-&\left[(n_{1} + 1)(n_{1} + |m| + \delta_{1} + 1)
(n - n_{1} - |m| - 1)(n - n_{1} + \delta_{2} - 1)\right]^{1/2}\,
V_{n q m}^{n_{1}+1},  \nonumber
\end{eqnarray}
for the expansion coefficients $V_{n q m}^{n_{1}}$. This relation can be iterated by taking account
of the normalization condition
\begin{eqnarray*}
\sum\limits_{n_{1}=0}^{n-|m|-1}\,\left|V_{n q m}^{n_{1}}(R,\delta_{1}, \delta_{2})\right|^{2} = 1.
\end{eqnarray*}
Here, again, the eigenvalues $\Lambda_{q}(R, \delta_{1}, \delta_{2})$
may be obtained by solving a system of $n - |m|$ linear
homogeneous equations.

In the case $\delta_{1} = \delta_{2} = 0$, from Eqs. (\ref{4.7.23}) and
(\ref{4.7.27}), we obtain three-term recursion relations for the coefficients
of interbasis expansions of the prolate spheroidal basis in spherical and
parabolic bases for the hydrogen atom; these coefficients were calculated in
\cite{COJO,M-1,M-35}.

Finally, it should be noted that the following four limits:
\begin{eqnarray*}
\lim\limits_{R \to \infty}\,V_{n q m}^{n_{1}}(R,\delta_{1}, \delta_{2})
= \delta_{n_{1} q}, \qquad
\lim\limits_{R \to 0}\,V_{n q m}^{n_{1}}(R,\delta_{1}, \delta_{2})
= \tilde{W}_{n q m}^{n_{1}}(\delta_{1}, \delta_{2}), \\ [4mm]
\lim\limits_{R \to 0}\,U_{n q m}^{l}(R,\delta_{1}, \delta_{2})
= \delta_{l q}, \qquad
\lim\limits_{R \to \infty}\,U_{n q m}^{l}(R,\delta_{1}, \delta_{2})
= W_{n q m}^{l}(\delta_{1}, \delta_{2}),
\end{eqnarray*}
furnish a useful means for checking the calculations presented in this Section.

\section{The Generalized oscillator system. Spherical basis}
\markboth{CHAPTER 4. GENERALIZED RING-SHAPED POTENTIALS}
{4.8. THE GENERALIZED OSCILLATOR SYSTEM. SPHERICAL BASIS}

In the following sections of this chapter we will study the quantum-mechanical
motion of a particle in a three-dimensional axially symmetric field with potential
\cite{M-24,M-25}
\begin{eqnarray}
U = \frac{\Omega^{2}}{2}\,\left(x^{2} + y^{2} + z^{2}\right)+
\frac{P}{2z^{2}} + \frac{Q}{2\,\left(x^{2} + y^{2}\right)},
\label{4.8.1}
\end{eqnarray}
where $\Omega, P$, and $Q$ are constants with $\Omega > 0$, $P > -1/4$,
and $Q \geq 0$. The Schr\"{o}dinger and Hamilton-Jacobi equations for this generalized
oscillator potential are separable in spherical, cylindrical, and spheroidal
(prolate and oblate) coordinates. In the case when $P = 0$ we get the
well-known ring-shape oscillator potential which was investigated in many articles
\cite{Quesne,Lutsenko-2,KIB-WIN-2,GRAN3,KIB-WIN-3}. If $P = Q = 0$, we have
the ordinary isotropic harmonic oscillator in three dimensions.

Before proceeding to the study of the generalized oscillator system, we note
that for convenience, we will take Planck's constant $\hbar$ and the mass of the
particle $\mu_{0}$ equal to one, and the letter $s$ will denote the fraction $1/2$.

The Schr\"{o}dinger equation in spherical coordinates for the potential (\ref{4.8.1}), i.e.,
\begin{eqnarray*}
U = \frac{\Omega^{2}}{2}\,r^{2} + \frac{P}{2r^{2}\,\cos^{2}\theta} +
\frac{Q}{2\,r^{2}\,\sin^{2}\theta},
\end{eqnarray*}
may be solved by seeking a wave function $\psi$ of the form
\begin{eqnarray}
\psi(r, \theta, \varphi) = R(r)\,\Theta(\theta)\,\frac{e^{im\varphi}}{2\pi},
\label{4.8.2}
\end{eqnarray}
with $m \in \textbf{Z}$. This amounts to finding the eigenfunctions of the set
$\left\{\hat{H}, \hat{L}_{z}, \hat{M}\right\}$ of commuting operators,
where the constant of motion $\hat{M}$ reads
\begin{eqnarray}
\hat{M} = \hat{L}^{2} + \frac{P}{\cos^{2}\theta} + \frac{Q}{\sin^{2}\theta}
\label{4.8.3}
\end{eqnarray}
($\hat{L}^{2}$ is the square of the angular momentum and $\hat{L}_{z}$
its $z$-component). We are thus left with the system of coupled differential equations:
\begin{eqnarray}
\left(\hat{M} - A\right)\,\Theta(\theta) = 0,
\label{4.8.4}
\end{eqnarray}
\begin{eqnarray}
\left[\frac{1}{r^{2}}\,\frac{d}{d r}\,\left(r^{2}\,\frac{d}{d r}\right) + 2E -
\Omega^{2}\,r^{2} - \frac{A}{r^{2}}\right]\,R(r) = 0,
\label{4.8.5}
\end{eqnarray}
where $A$ is a (spherical) separation constant.

Let us consider the angular equation (\ref{4.8.4}). By putting
$\Theta(\theta) = f(\theta)/\sin\theta$, we can rewrite Eq. (\ref{4.8.4})
in the P\"{o}schl-Teller form:
\begin{eqnarray}
\left(\frac{d^{2}}{d \theta^{2}} + A + \frac{1}{4} - \frac{b^{2} - \frac{1}{4}}{\cos^{2}\theta}
- \frac{c^{2} - \frac{1}{4}}{\sin^{2}\theta}\right)\,f(\theta) = 0, \quad
b = \sqrt{1 + \frac{1}{4}}, \quad c = \sqrt{m^{2} + Q}.
\label{4.8.6}
\end{eqnarray}
In the case where $b > s$, the angular potential is repulsive for $\theta = \pi/2$.
In this case, the $\theta$ domain is separated in two regions: $\theta \in ]0, \pi/2[$
and $\theta \in ]\pi/2, \pi[$ and the "motion" takes place in one or another region.
Furthermore, in this case Eq. (\ref{4.8.5}) corresponds to a genuine P\"{o}schl-Teller potential.
In the case where $0 < b < s$, we can call the angular potential an attractive
P\"{o}schl-Teller potential. When $b = 0$, i.e., $P = 0$, we get the well-known
ring-shape oscillator potential \cite{Quesne,Lutsenko-2,KIB-WIN-2,KIB-WIN-3,GRAN3}.
The solution $\Theta(\theta) \equiv \Theta_{q}(\theta; c, \pm b)$ of Eq. (\ref{4.8.4}) (for both
$0 < b < s$ and $b > s$), with the conditions $\Theta(0) = \Theta(\pi/2) = 0$ is easily found
to be (cf., \cite{EVANS-3,Flugge})
\begin{eqnarray}
\Theta_{q}(\theta; c, \pm b) = N_{q}(c, \pm b)\,\left(\sin\theta\right)^{c}\,
\left(\cos\theta\right)^{s \pm b}\,P_{q}^{(c, \pm b}\left(\cos2\theta\right),
\label{4.8.7}
\end{eqnarray}
with $q \in \textbf{N}$, where $P_{n}^{\alpha, \beta}(x)$ denotes a Jacobi polynomial.
Then, the constant $A$ is quantized as
\begin{eqnarray}
A_{q}(c, \pm b) = \left(2q + c \pm b + s\right)\,
\left(2q + c \pm b + 3s\right).
\label{4.8.8}
\end{eqnarray}
The normalization constant $N_{q}(c, \pm b)$ in (\ref{4.8.7}) is
given (up to a phase factor) by
\begin{eqnarray}
\int\limits_{0}^{\pi/2}\,\Theta_{q'}(\theta; c, \pm b)\,
\Theta_{q}(\theta; c, \pm b)\,\sin\theta\,d\theta = \frac{1}{2}\,\delta_{q' q}.
\label{4.8.8a}
\end{eqnarray}
This leads to
\begin{eqnarray}
N_{q}(c, \pm b) = \sqrt{\frac{\left(2q + c \pm b + 1\right)\,q!\,
\Gamma\left(q + c \pm b + 1\right)}{\Gamma\left(q + c + 1\right)\,
\Gamma\left(q \pm b + 1\right)}}.
\label{4.8.9}
\end{eqnarray}
Note that only the positive sign in front of $b$ has to be taken with $b > s$
while both the positive and negative signs have to be considered for $0 < b < s$.

Let us go to the radial equation (\ref{4.8.5}). The introduction of (\ref{4.8.8}) into
(\ref{4.8.5}) yields an equation that is very reminiscent of the radial equation for the
three-dimensional isotropic oscillator except that the orbital quantum number
$l$ is replaced by $2q + c \pm b + s$. The solution $R(r) \equiv R_{n_{r} q}(r; c, \pm b)$ of
the obtained equation, in terms of Laguerre polynomials $L_{n}^{\alpha}$, is
\begin{eqnarray}
R_{n_{r} q}(r; c, \pm b) = N_{n_{r} q}(c, \pm b)\,
\left(\sqrt{\Omega} r\right)^{2q + c \pm b + s}\,
e^{-s\Omega r^{2}}\,L_{n_{r}}^{2q + c \pm b + 1}
\left(\Omega r^{2}\right),
\label{4.8.10}
\end{eqnarray}
with $n_{r} \in \textbf{N}$. In Eq. (\ref{4.8.10}), the radial wave functions
$R_{n_{r} q}$ satisfy the orthogonality relation
\begin{eqnarray*}
\int\limits_{0}^{\infty}\,R_{n'_{r} q}(r; c, \pm b))\,
R_{n_{r} q}(r; c, \pm b)\,r^{2}\,dr = \delta_{n'_{r} n_{r}}
\end{eqnarray*}
so that the normalization factor $N{n_{r} q}(c, \pm b)$ is
\begin{eqnarray}
N_{n_{r} q}(c, \pm b) = \sqrt{\frac{2\Omega^{3s}\,\left(n_{r}\right)!}
{\Gamma\left(n_{r} + 2q + c \pm b + 2\right)}}.
\label{4.8.11}
\end{eqnarray}

The normalized total wave function $\psi(r, \theta, \varphi) \equiv
\psi_{n_{r} q m}(r, \theta, \varphi; c, \pm b)$ is then given by Eqs.
(\ref{4.8.2}), (\ref{4.8.7}), (\ref{4.8.9}), (\ref{4.8.10}), and (\ref{4.8.11}).
The energies $E$ corresponding to $n_{r} + q$ fixed are
\begin{eqnarray}
E_{n}(c, \pm b) = \Omega\,\left(2n + c \pm b + 2\right),
\label{4.8.12}
\end{eqnarray}
with $n = n_{r} + q$. Equation (\ref{4.8.12}) shows that, for each quantum number $n$,
we have two levels (for $+b$ and $-b$) in the $0 < b < s$ region and one level (for $+b$)
in the $b > s$ region. Note that this spectrum was obtained through a path integral
approach in \cite{Carpio-2, Carpio-3, Carpio-4} for the $b > s$ case and in \cite{GROP1}
for the general case (see also Refs. \cite{Zhedanov,Kibler6}).

In the $0 < b < s$ region, for the limiting situation where $b = s^{-}$, i.e., $P = 0^{-}$,
we have for the separation constant $A$:
\begin{eqnarray}
A_{q}(c, + s) = \left(2q + c + 1\right)\,\left(2q + c + 2\right), \quad
A_{q}(c, - s) = \left(2q + c\right)\,\left(2q + c + 1\right).
\label{4.8.13}
\end{eqnarray}
Then, by using the connecting formulas \cite{BE2}
\begin{eqnarray*}
C_{2n + 1}^{\lambda}(x) = \frac{(\lambda)_{n + 1}}{(s)_{n + 1}}\,x\,
P_{n}^{(\lambda - s. +s)}\left(2x^{2} - 1\right), \qquad
C_{2n}^{\lambda}(x) = \frac{(\lambda)_{n}}{(s)_{n}}\,
P_{n}^{(\lambda - s. -s)}\left(2x^{2} - 1\right),
\end{eqnarray*}
between the Jacobi polynomial $P_{n}^{(\alpha, \beta)}$ and the
Gegenbauer polynomial $C_{n}^{\lambda}$, we have the following
odd and even angular solutions (with respect to
$\cos\theta \to - cos\theta$):
\begin{eqnarray}
\Theta_{q}(\theta; c, +s) = \sqrt{\frac{(4q + c + 3)\,(2q + 1)!}{2\pi\,\Gamma(q + c + 2)}}\,
2^{c}\,\Gamma(c + s)\left(\sin\theta\right)^{c}\,C_{2q + 1}^{c + s}(\cos\theta,),
\label{4.8.14}
\end{eqnarray}
\begin{eqnarray}
\Theta_{q}(\theta; c, -s) = \sqrt{\frac{(4q + c + 1)\,(2q)!}{2\pi\,\Gamma(q + c + 1)}}\,
2^{c}\,\Gamma(c + s)\left(\sin\theta\right)^{c}\,C_{2q}^{c + s}(\cos\theta,).
\label{4.8.15}
\end{eqnarray}
Let us introduce (a new orbital quantum number)
$l$ and (a new principal quantum number) $N$ through
\begin{eqnarray}
l - |m| &=& \left\{\matrix{ 2q + 1,\,\,\rm{for\,\, the\,\,\, "+"\, sign}
,\,\, \cr \cr
2q,\,\,\rm{for\,\, the\,\,\, "-"\, sign}
\cr}\right\},  \nonumber \\
\label{4.8.15a}
\\
N - |m| &=& \left\{\matrix{ 2n + 1,\,\,\rm{for\,\, the\,\,\, "+" sign}
,\,\, \cr \cr
2n,\,\,\rm{for\,\, the\,\,\, "-"\, sign}
\cr}\right\}. \nonumber
\end{eqnarray}
Note that $N = 2n_{r} + l$ both for the $"+"$ and $"-"$ signs.
Then, the separation constant [Eq. (\ref{4.8.13})] and
the energy [Eq. (\ref{4.8.12})] can be expressed as
\begin{eqnarray}
A_{q}(c, \pm s) \equiv  A_{l}(\delta) = \left(l + \delta\right)\,\left(l + \delta + 1\right), \,\,
E_{n}(c, \pm b) \equiv E_{N}(\delta) = \Omega\,\left(N + \delta + 3s\right),
\label{4.8.16}
\end{eqnarray}
respectively, where
\begin{eqnarray*}
\delta = \sqrt{Q + m^{2}} - |m|.
\end{eqnarray*}
Thus, the two parts of the energy spectrum for the signs $"\pm"$ correspond now to odd
(for $"+"$) and even (for $"-"$) values of $N-|m|$. In terms of $N, l$, and $\delta$,
the functions $R_{n_{r} q}(r; c, \pm b) \equiv R_{N l}(r; \delta)$ [cf. Eq. (\ref{4.8.10})]
and $\Theta_{q}(\theta; c, \pm s) \equiv \Theta_{l}(\theta; \delta)$ [cf. Eqs. (\ref{4.8.14})
and (\ref{4.8.15})] can be rewritten as
\begin{eqnarray}
R_{n_{r} q}(r; c, \pm b) = \sqrt{\frac{2\Omega^{3s}\,\left(\frac{N-l}{2}\right)!}
{\Gamma\left(\frac{N+l}{2} + \delta + 3s\right)}}\,
\left(\sqrt{\Omega} r\right)^{l + \delta}\,
e^{-s\Omega r^{2}}\,L_{\frac{N-l}{2}}^{l + \delta + s}
\left(\Omega r^{2}\right),
\label{4.8.17}
\end{eqnarray}
\begin{eqnarray}
\Theta_{l m}(\theta; \delta) =\Gamma(|m|+\delta + s)
\sqrt{\frac{(2l + 2\delta + 1)(l - |m|)!}{2\pi\,\Gamma(l + |m| + 2\delta + 1)}}
\left(2\sin\theta\right)^{|m|+\delta}C_{l - |m|}^{|m|+\delta + s}(\cos\theta).
\label{4.8.18}
\end{eqnarray}
Equations (\ref{4.8.17}) and (\ref{4.8.18}) compare with the corresponding
results for the ring-shape oscillator in \cite{Quesne,Lutsenko-2}.
Note that (\ref{4.8.18}) was given in terms of Legendre functions in
Refs. \cite{Quesne} and \cite{Lutsenko-2}.

In the $b > 0$ region, for the limiting situation where $b = s^{+}$,
i.e., $P = 0^{+}$, we have only odd solutions. In other words when $P \to 0^{+}$,
the eigenvalues and eigenfunctions of the generalized oscillator
do not restrict to the eigenvalues and eigenfunctions,
respectively, of the ring-shape oscillator. This fact may be explained in the
following manner. To make $P = 0$ in the wave function
$\psi_{n_{r} q m}(r, \theta, \varphi; c, \pm b)$ amounts to changing the
Hamiltonian into a Hamiltonian corresponding to $P = 0$ and to introducing
an unpenetrable barrier. (Another way to describe this phenomenon is to
say that for very small $P$, the potential $U$ is infinite in the
$\theta = \pi/2$ plan and equal to the ring-shape potential only for $P = 0$).
This phenomenon is known as the Klauder phenomenon \cite{Klauder}.

A further limit can be obtained in the case when $\delta = 0$, i.e.,
$Q = 0$. It is enough to use the connecting formula (\ref{4.1.7a})
between the Gegenbauer polynomial $C_{n}^{\lambda}$ and the
Legendre polynomial $P_{l}^{|m|}$. In fact, for $Q = 0$, Eq. (\ref{4.8.18})
can be reduced to
\begin{eqnarray*}
\Theta_{l m}(\theta; 0) = (-1)^{|m|}\,
\sqrt{\frac{(2l + 1)\,(l - |m|)!}{2\,(l + |m|)!}}\,P_{l}^{|m|}(\cos\theta),
\end{eqnarray*}
so that $\Theta_{l m}(\theta; 0)e^{im\varphi}/\sqrt{2\pi}$ coincides with the usual
spherical harmonic $Y_{lm}(\theta, \varphi)$ (up to a phase factor, e.g., see \cite{VAR}.
The wave functions$\Theta_{l m}(\theta; c, \pm b)e^{im\varphi}/\sqrt{2\pi}$
may thus be considered as a generalization of the spherical harmonics.

\section{Cylindrical basis and connection with the spherical}
\markboth{CHAPTER 4. GENERALIZED RING-SHAPED POTENTIALS}
{4.9. CYLINDRICAL BASIS AND CONNECTION WITH THE SPHERICAL}

In the cylindrical coordinates $(\rho, \varphi, z)$ the potential
$U$ (\ref{4.8.1}) reads
\begin{eqnarray*}
U = \frac{\Omega^{2}}{2}\,(\rho^{2} + z^{2}) + \frac{P}{2z^{2}} + \frac{Q}{2\rho^{2}}.
\end{eqnarray*}
The Schr\"{o}dinger equation, with this potential, admits a solution
$\psi$ of the form
\begin{eqnarray}
\psi(\rho, \varphi, z) = R(\rho)\,Z(z)\,\frac{e^{im\varphi}}{\sqrt{2\pi}},
\label{4.9.1}
\end{eqnarray}
where $z \in \textbf{Z}$. In other words, we look for the eigenfunctions of the set
$\hat{H}H, \hat{L}_{z}, \hat{N}$ of commuting operators, where the constant of
motion $N\hat{N}$ is
\begin{eqnarray}
\hat{N} = \hat{D}_{z z} + \frac{P}{z^{2}},
\label{4.9.2}
\end{eqnarray}
$\hat{D}_{z z}$ being the $zz$ component of
\begin{eqnarray*}
\hat{D}_{x_{i} x_{j}} = - \partial_{x_{i} x_{j}} + \Omega^{2}\,x_{i} x_{j},
\end{eqnarray*}
the so-called Demkov tensor \cite{DEMKOV1} for the isotropic
harmonic oscillator in $\textbf{R}^{3}$. It is sufficient to solve
the two coupled equations
\begin{eqnarray}
\left(2E_{z} - \hat{N}\right)\,Z(z) = \left(\frac{d^{2}}{dz^{2}} - \Omega^{2}z^{2}
- \frac{P}{z^{2}} + 2E_{z}\right)\,Z(z) = 0,
\label{4.9.3}
\end{eqnarray}
\begin{eqnarray}
\left[\frac{1}{\rho}\,\frac{d}{d \rho}\,\left(\rho\,\frac{d}{d \rho}\right) +
2E_{\rho} - \Omega^{2}\rho^{2} - \frac{Q + m^{2}}{\rho^{2}}\right]\,R(\rho) = 0,
\label{4.9.4}
\end{eqnarray}
where the two cylindrical separation constants $E_{\rho}$
and $E_{z}$ obey  $E_{\rho} + E_{z} = E$. The solutions
$\psi(\rho, \varphi, z) \equiv \psi_{n_{\rho} p m}(\rho, \varphi, z; c, \pm b)$
of Eqs. (\ref{4.9.3}) and (\ref{4.9.4}) lead to the normalized wave function
\begin{eqnarray}
\psi_{n_{\rho} p m}(\rho, \varphi, z; c, \pm b) = R_{n_{\rho}}(\rho; c)\,Z_{p}(z; \pm b)\,
\frac{e^{im\varphi}}{\sqrt{2\pi}},
\label{4.9.5}
\end{eqnarray}
where
\begin{eqnarray}
R_{n_{\rho}}(\rho; c) = \sqrt{\frac{2\Omega\,\left(n_{\rho}\right)!}
{\Gamma\left(n_{\rho} + c + 1\right)}}\,e^{-s\Omega\,\rho^{2}}\,
\left(\sqrt{\Omega}\,\rho\right)^{c}\,L_{n_{\rho}}^{c}\left(\Omega\,\rho^{2}\right)
\label{4.9.6}
\end{eqnarray}
and
\begin{eqnarray}
Z_{p}(z; \pm b) = (-1)^{p}\,\sqrt{\frac{2\Omega^{s}\,p!}
{\Gamma\left(p \pm b + 1\right)}}\,e^{-s\Omega\,z^{2}}\,
\left(\sqrt{\Omega}\,z\right)^{s \pm b}\,
L_{n_{p}}^{\pm b}\left(\Omega\,z^{2}\right),
\label{4.9.7}
\end{eqnarray}
with $n_{\rho} \in \textbf{N}$ and $p \in \textbf{N}$. The normalization of the
wave function (\ref{4.9.5}) is ensured by
\begin{eqnarray*}
\int\limits_{0}^{\infty}\,R_{n'_{\rho}}(\rho; c)\,R_{n_{\rho}}(\rho; c)\,
\rho\,d\rho = \delta_{n'_{\rho} n_{\rho}}, \qquad
\int\limits_{0}^{\infty}\,Z_{p'}(z; \pm b)\,Z_{p}(z; \pm b)\,dz = \frac{1}{2}\,\delta_{p' p}.
\end{eqnarray*}
Furthermore, the constants $E_{z}$ and $E_{\rho}$ in Eqs. (\ref{4.9.3})
and (\ref{4.9.3}) become
\begin{eqnarray}
E_{z}(p; \pm b) = \Omega\,(2p \pm b + 1), \qquad
E_{\rho}\left(n_{\rho}, c\right) = \Omega\,\left(2n_{\rho} + c + 1\right).
\label{4.9.8}
\end{eqnarray}
Therefore, the quantized values of the energy $E$ are given by (\ref{4.8.12})
where now the quantum number $n$ is $n = n_{\rho} + p$. As in the second section,
the sign in front of $b$ in Eqs. (\ref{4.9.5})-(\ref{4.9.7}) may be only positive
when $b > s$. When $0 < b < s$ , both the signs $"+"$ and $"-"$ are admissible.

In the $0 < b < s$ region, in the limiting case where $b = s^{-}$, due to the
connecting formulas \cite{BE2}
\begin{eqnarray*}
{\it{H}}_{2n + 1}(x) = (-1)^{n}\,2^{2n+1}\,n!\,x\,L_{n}^{+s}(x^{2}), \qquad
 {\it{H}}_{2n}(x) = (-1)^{n}\,2^{2n}\,n!\,L_{n}^{-s}(x^{2}),
\end{eqnarray*}
between the odd ${\it{H}}_{2n + 1}$ and even ${\it{H}}_{2n}$ Hermite
polynomials and the Laguerre polynomials $L_{n}^{\pm s}$,
we immediately have
\begin{eqnarray*}
Z_{p}(z; +s) = \left(\frac{\Omega}{\pi}\right)^{1/4}\,\frac{e^{-s\Omega\,z^{2}}}
{\sqrt{2^{2p+1}\,(2p + 1)!}}\,{\it{H}}_{2p + 1}(\sqrt{\Omega}\,z), \\ [4mm]
Z_{p}(z; -s) = \left(\frac{\Omega}{\pi}\right)^{1/4}\,\frac{e^{-s\Omega\,z^{2}}}
{\sqrt{2^{2p}\,(2p)!}}\,{\it{H}}_{2p}(\sqrt{\Omega}\,z).
\end{eqnarray*}
Introducing (a new quantum number) $n_{3}$ such that $n_{3} = 2p + 1$
for the $"+"$ sign and $n_{3} = 2p$ for the $"-"$ sign, we obtain
\begin{eqnarray*}
Z_{p}(z; \pm s) = \left(\frac{\Omega}{\pi}\right)^{1/4}\,\frac{e^{-s\Omega\,z^{2}}}
{\sqrt{2^{n_{3}}\,(n_{3})!}}\,{\it{H}}_{n_{3}}(\sqrt{\Omega}\,z).
\end{eqnarray*}
The energy is then given by (]\ref{4.8.16}) where $N = 2n_{\rho} + n_{3} + |m|$.
Note that the spectrum in the case $b = s^{-}$, which corresponds to the ring-shape
oscillator system, was obtained in Refs. \cite{Quesne,Carpio-3,Carpio-4,Lutsenko-2}.

In the $b > s$ region, in the limiting situation where $b = s^{+}$, we get only
the odd solution of the ring-shape oscillator system.

Now let us consider the expansion of the cylindrical basis (\ref{4.9.5}) in terms of the spherical basis.

According to first principles, any cylindrical wave function (\ref{4.9.1}) corresponding to a given value
of $E$ can be developed in terms of the spherical wave functions (\ref{4.8.2}) associated
to the eigenvalue $E$ (see also Ref. \cite{Zhedanov}). Thus, we have
\begin{eqnarray}
\psi_{n_{\rho} p m}(\rho, \varphi, z; c, \pm b) = \sum\limits_{q=0}^{n}\,W_{n p}^{q}(c, \pm b)\,
\psi_{n_{r} q m}(r, \theta, \varphi; c, \pm b),
\label{4.9.9}
\end{eqnarray}
where $n_{\rho} + p = n_{r} + q = n$. In Eq. (\ref{4.9.9}), it is understood
that the wave functions in the left- and right-hand sides are written in spherical
coordinates $(r, \theta, \varphi)$ owing to $\rho = r\sin\theta$ and
$z = r\cos\theta$. The dependence on $e^{im\varphi}$ can be eliminated in both
sides of Eq. (\ref{4.9.9}). Furthermore, by using the formula $L_{n}^{\alpha}(x) \sim (-1)^{n}x^{n}/n!$,
valid for $x$ arbitrarily large, (\ref{4.9.9}) yields an equation that depends only on
the variable $\theta$. Thus, by using the orthonormality relation (\ref{4.8.8a}),
for the quantum numbers $q$, we can derive the following expression for the interbasis
expansion coefficients:
\begin{eqnarray}
W_{n p}^{q}(c, \pm b) =(-1)^{q-p}\,B_{n p}^{q}(c, \pm b)\,E_{n p}^{q}(c, \pm b),
\label{4.9.10}
\end{eqnarray}
where
\begin{eqnarray}
B_{n p}^{q}(c, \pm b) = \sqrt{\frac{(2q + c \pm b + 1)(n - q)!q!\Gamma(q + c \pm b + 1)
\Gamma(n + q + c \pm b + 2)}{(n - p)!p!\Gamma(q + c + 1)\Gamma(q \pm b + 1)
\Gamma(n - p + c + 1)\Gamma(p \pm b + 1)}}, \nonumber \\
\label{4.9.11}
\\
E_{n p}^{q}(c, \pm b) = 2\int\limits_{0}^{\pi/2}\left(\sin\theta\right)^{2n-2p+2c}
\left(\cos\theta\right)^{2p+1 \pm 2b}P_{q}^{(c, \pm b)}\left(\cos2\theta\right)\,
\sin\theta\,d\theta. \nonumber
\end{eqnarray}
By making the change of variable $x = \cos2\theta$ and by using the Rodrigues formula for the Jacobi
polynomial (\ref{2.10.4}), Eqs. (\ref{4.9.10}) and (\ref{4.9.11}) lead to the integral expression
\begin{eqnarray}
W_{n p}^{q}(c, \pm b) &=& \sqrt{\frac{(2q + c \pm b + 1)(n - q)!\Gamma(q + c \pm b + 1)
\Gamma(n + q + c \pm b + 2)}{p!q!(n - p)!\Gamma(q + c + 1)\Gamma(q \pm b + 1)
\Gamma(n - p + c + 1)\Gamma(p \pm b + 1)}}\times \nonumber \\
\label{4.9.12}
\\
&\times& \frac{(-1)^{p}}{2^{n + q + c \pm b + 1}}\,\int\limits_{-1}^{1}\left(1 - x\right)^{n-p}
\left(1 + x\right)^{p}\,\frac{d^{q}}{d x^{q}}\left[\left(1 - x\right)^{q+c}
\left(1 + x\right)^{q \pm b}\right]\,dx    \nonumber
\end{eqnarray}
for the coefficient $W_{n p}^{q}(c, \pm b)$. Equation (\ref{4.9.12}) can be compared with the integral
representation \cite{VAR}
\begin{eqnarray*}
C_{a \alpha; b \beta}^{c \gamma} &=& \delta_{\alpha+\beta, \gamma}\sqrt{\frac{(2c + 1)(J + 1)!(J - 2c)!
(c + \gamma)!}{(J - 2a)!(J - 2b)!(a - \alpha)!(a + \alpha)!(b - \beta)!(b + \beta)!\Gamma(c - \gamma)!}}\times  \\ [4mm]
&\times& \frac{(-1)^{a - c + \beta}}{2^{J + 1}}\,\int\limits_{-1}^{1}\left(1 - x\right)^{a - \alpha}
\left(1 + x\right)^{b - \beta}\,\frac{d^{c - \gamma}}{d x^{c - \gamma}}\left[\left(1 - x\right)^{J - 2a}
\left(1 + x\right)^{J - 2b}\right]\,dx    \nonumber
\end{eqnarray*}
for the Clebsch-Gordan coefficients of the compact Lie group $SU(2)$. This yields
\begin{eqnarray}
W_{n p}^{q}(c, \pm b) = (-1)^{n - q}\,C_{\frac{n \pm b}{2}, p - \frac{n \mp b}{2};
\frac{n + c}{2}, \frac{n + c}{2} - p}^{q + \frac{c \pm b}{2}, \frac{c \pm b}{2}}.
\label{4.9.13}
\end{eqnarray}
Since the quantum numbers in Eq. (\ref{4.9.13}) are not necessarily integers or half
of odd integers, the coefficients for the expansion of the cylindrical basis in terms
of the spherical basis may be considered as analytical continuation, for real values of
their arguments, of the $SU(2)$. Clebsch-Gordan coefficients. The inverse of
Eq. (\ref{4.9.9}), namely
\begin{eqnarray}
\psi_{n_{r} q m}(r, \theta, \varphi; c, \pm b) =
\sum\limits_{p=0}^{n}\,\tilde{W}_{n q}^{p}(c, \pm b)\,
\psi_{n_{\rho} p m}(\rho, \varphi, z; c, \pm b)
\label{4.9.14}
\end{eqnarray}
follows from the orthonormality property of the $SU(2)$ Clebsch-Gordan coefficients.
The expansion coefficients in (\ref{4.9.14}) are thus
\begin{eqnarray*}
\tilde{W}_{n q}^{p}(c, \pm b) = W_{n p}^{q}(c, \pm b).
\end{eqnarray*}
Note that in order to compute the coefficients $W_{n p}^{q}(c, \pm b)$ through
(\ref{4.9.13}), we can use the $_{3}F_{2}(a, b, c; d, e; 1)$ representation
\cite{VAR} of the $SU(2)$ Clebsch-Gordan coefficients.

We close this section with some considerations concerning the limiting cases
$(P  = 0, Q \neq 0)$ and $(P  = 0, Q = 0)$. It is to be observed that the pas-
sage from $(P \neq 0, Q \neq 0)$. to $(P  = 0, Q \neq 0)$ needs some caution.
Indeed for $b = s^{-}$, Eq. (\ref{4.9.13}) can be rewritten in terms of the
quantum numbers $N, l$, and $n_{3}$ as
\begin{eqnarray}
W_{n p}^{q}(c, \pm s) = (-1)^{\frac{N-l}{2}}\,
C_{\frac{N-|m|-s \pm s}{4}, \frac{2n_{3}-N+|m|-s \pm s}{4}; \frac{N+|m|-s \mp s}{4}
+\frac{\delta}{2}, \frac{N+|m|-2n_{3}+s \pm s}{4}+\frac{\delta}{2}}^{\frac{2l-1}{4}+\frac{\delta}{2},
\frac{|m| \pm s +delta}{2}}.
\label{4.9.15}
\end{eqnarray}
By using the ordinary symmetry property \cite{VAR}
\begin{eqnarray*}
C_{a \alpha; b \beta}^{c \gamma} = (-1)^{a+b-c}C_{a, -\alpha; b, -\beta}^{c, -\gamma}
\end{eqnarray*}
and the Regge symmetry \cite{VAR}
\begin{eqnarray*}
C_{a \alpha; b \beta}^{c \gamma} = C_{\frac{a+b+\gamma}{2}, \frac{a-b+\alpha - \beta}{2};
\frac{a+b-\gamma}{2}, \frac{a-b-\alpha + \beta}{2}}^{c, a-b}
\end{eqnarray*}
in Eq. (\ref{4.9.15}) with the sign $"+"$q and by using the
ordinary symmetry property \cite{VAR}
\begin{eqnarray*}
C_{a \alpha; b \beta}^{c \gamma} = (-1)^{a+b-c}\,C_{b, \beta;
a, \alpha}^{c \gamma}
\end{eqnarray*}
in Eq. (\ref{4.9.15}) with the sign $"-"$, we get
\begin{eqnarray}
W_{n p}^{q}(c, \pm s) \equiv W_{N m n_{3}}^{l}(\delta) =
C_{\frac{N+|m|}{4} +\frac{\delta}{2}, \frac{N + |m| - 2n_{3}}{4}+\frac{\delta}{2};
\frac{N - |m|- 1}{4}, \frac{2n_{3} - N + |m| - 1}{4}}^{\frac{2l-1}{4}+\frac{\delta}{2},
\frac{2|m| - 1}{4} + \frac{\delta}{2}}.
\label{4.9.16}
\end{eqnarray}
As a conclusion, when $b = s^{-}$ we have an expansion
of the type \cite{Lutsenko-2}
\begin{eqnarray}
\psi_{N m n_{3}}(\rho, \varphi, z; \delta) = \sum\limits_{l}\,
W_{N m n_{3}}^{l}(\delta)\,
\psi_{N l m}(r, \theta, \varphi; \delta),
\label{4.9.17}
\end{eqnarray}
where the summation on $l$ goes, by steps of 2, from $|m|$ or $|m|+1$ to $N$ according to
whether as $N - |m|$ is even or odd (because $N - l$ is always even).
Equations (\ref{4.9.16}) and (\ref{4.9.17}) were obtained in Ref. \cite{Lutsenko-2}
for the ring-shape oscillator system. Finally,the case $P = Q = 0$ can be easily
deduced from (\ref{4.9.16}) and (\ref{4.9.17}) by taking $\delta = 0$:
we thus recover the result obtained in Refs. \cite{POGOSYAN2, POGOSYAN6} for the
isotropic harmonic oscillator in three dimensions. Note that in the case $P = Q = 0$,
the expansion coefficients in Eq. (\ref{4.9.17}) become Clebsch-Gordan
coefficients for the noncompact Lie group $SU(1, 1)$ (cf. Ref. \cite{KPS}).

\section{Spheroidal bases}
\markboth{CHAPTER 4. GENERALIZED RING-SHAPED POTENTIALS}
{4.10. SPHEROIDAL BASES}
\subsection{Prolate spheroidal basis}

The prolate spheroidal coordinates $(\xi, \eta, \varphi)$ are
defined via
\begin{eqnarray}
x = \frac{R}{2}\sqrt{(\xi^{2}-1)(1-\eta^{2})}\cos\varphi, \quad
y = \frac{R}{2}\sqrt{(\xi^{2}-1)(1-\eta^{2})}\sin\varphi, \quad
z = \frac{R}{2}\xi \eta,
\label{4.10.1}
\end{eqnarray}
(with $1 \leq \xi \leq \infty, -1 \leq \eta \leq 1, 0 \leq \varphi < 2\pi)$,
where $R$ is the interfocus distance.

The different definitions of prolate spheroidal coordinates (\ref{3.3.1}) and (\ref{4.10.1}) are due to the
fact that the variables in the Schr\"{o}dinger equation for oscillator-like potentials are separated
only in the case when the attractive center is located in the geometric center of the ellipse,
and in none of its focuses, as for Coulomb-like systems.

As is well known \cite{KOMPOSLA}, in the limits where $R \to 0$ and $R \to \infty$, the prolate
spheroidal coordinates reduce to the spherical coordinates and the cylindrical coordinates, respectively.
In prolate spheroidal coordinates, the potential $U$ (\ref{4.8.1}) reads
\begin{eqnarray}
U = \frac{\Omega^{2}R^{2}}{8}\,\left(\xi^{2} + \eta^{2} - 1\right) +
\frac{2}{R^{2}}\,\left[\frac{P}{\xi^{2} \eta^{2}} +
\frac{Q}{\left(\xi^{2}-1\right)\left(1-\eta^{2}\right)}\right].
\label{4.10.2}
\end{eqnarray}
By looking for a solution $\psi$ of the Schr\"{o}dinger equation, with the
potential (\ref{4.10.2}), in the form
\begin{eqnarray}
\psi(\xi, \eta, \varphi) = \psi_{1}(\xi)\,\psi_{2}(\eta)\,\frac{e^{im\varphi}}{\sqrt{2\pi}},
\label{4.10.3}
\end{eqnarray}
with $m \in \textbf{Z}$, we obtain the two ordinary differential
equations
\begin{eqnarray}
\left[\frac{d}{d \xi}\left(\xi^{2}-1\right)\frac{d}{d \xi} - \frac{Q+m^{2}}{\xi^{2}-1} +
\frac{ER^{2}}{2}\xi^{2} - \frac{\Omega^{2}R^{4}}{16}\xi^{2}\left(\xi^{2}-1\right) +
\frac{P}{\xi^{2}}\right]\psi_{1} = \lambda \psi_{1},
\label{4.10.4}
\end{eqnarray}
\begin{eqnarray}
\left[\frac{d}{d \eta}\left(1 - \eta^{2}\right)\frac{d}{d \eta} - \frac{Q+m^{2}}{1 - \eta^{2}} -
\frac{ER^{2}}{2}\eta^{2} - \frac{\Omega^{2}R^{4}}{16}\eta^{2}\left(1 - \eta^{2}\right) -
\frac{P}{\eta^{2}}\right]\psi_{2} = -\lambda \psi_{2},
\label{4.10.5}
\end{eqnarray}
where $\lambda(R)$ is a separation constant in prolate spheroidal coordinates.
The combination of Eqs. (\ref{4.10.4}) and (\ref{4.10.5}), leads to the operator
\begin{eqnarray*}
\hat{\Lambda} = - \frac{1}{\xi^{2} - \eta^{2}}
\left[\eta^{2}\frac{\partial}{\partial \xi}\left(\xi^{2}-1\right)\frac{\partial}{\partial \xi}  +
\xi^{2}\frac{\partial}{\partial \eta}\left(1 - \eta^{2}\right)\frac{\partial}{\partial \eta}\right] + \\ [4mm]
+ \frac{\xi^{2} + \eta^{2} - 1}{\left(\xi^{2}-1\right)\left(1-\eta^{2}\right)}
\left(Q - \frac{\partial^{2}}{\partial \varphi^{2}}\right) +
\frac{\Omega^{2}R^{4}}{16}\xi^{2}\eta^{2} + P\frac{\xi^{2} + \eta^{2}}{\xi^{2} \eta^{2}},
\end{eqnarray*}
after eliminating the energy $E$. The eigenvalues of the operator $\hat{\Lambda}$
are $\lambda(R)$ while its eigenfunctions are given by (\ref{4.10.3}.
The significance of the (self-adjoint) operator $\hat{\Lambda}$ is to be found
in the connecting formula
\begin{eqnarray}
\hat{\Lambda} = \hat{M} + \frac{R^{2}}{4}\,\hat{N},
\label{4.10.6}
\end{eqnarray}
where $\hat{M}$ and $\hat{N}$ are the constants of motion (\ref{4.8.4})
and (\ref{4.9.2}). The operator $\hat{\Lambda}$ is of pivotal importance for the derivation of the
interbasis expansion coefficients from the spherical basis or the cylindrical
basis to the prolate spheroidal basis. In particular, it allows us to derive the latter coefficients
without knowing the wave functions in prolate spheroidal basis. Therefore, we shall not derive the prolate
spheroidal wave functions $\psi_{1}$ and $\psi_{2}$ which could be obtained by solving
Eqs. (\ref{4.10.4}) and (\ref{4.10.5}). It is more economical to proceed in the following way
that presents the advantage of giving, at the same time, the global wave function
$\psi(\xi, \eta, \varphi) \equiv \psi(\xi, \eta, \varphi; c, \pm b)$ and the interbasis expansion
coefficients.

The three constants of motion $\hat{M}, \hat{N}$, and $\hat{\Lambda}$, which occur in
Eq. (\ref{4.10.6}, can be seen to satisfy the following eigenequations
\begin{eqnarray}
\hat{M}\,\psi_{n_{r} q m}(r, \theta, \varphi; c, \pm b) = A_{q}(c, \pm b)\,
\psi_{n_{r} q m}(r, \theta, \varphi; c, \pm b),
\label{4.10.7}
\end{eqnarray}
\begin{eqnarray}
\hat{N}\,\psi_{n_{\rho} p m}(\rho, \varphi, z,; c, \pm b) = 2E_{z}(c, \pm b)\,
\psi_{n_{\rho} p m}(\rho, \varphi, z,; c, \pm b),
\label{4.10.8}
\end{eqnarray}
and
\begin{eqnarray}
\hat{\Lambda}\,\psi_{n k m}(\xi, \eta, \varphi; c, \pm b) = \lambda_{k}(R)\,
\psi_{n k m}(\xi, \eta, \varphi; c, \pm b).
\label{4.10.9}
\end{eqnarray}
for the spherical, cylindrical, and prolate spheroidal bases, respectively.
[In Eq. (\ref{4.10.9}, the index $k$ labels the eigenvalues of the operator
$\hat{\Lambda}$ and varies in the range $0 \leq k \leq n$]. The spherical,
cylindrical, and prolate spheroidal bases are indeed eigenbases for
the three sets of commuting operators $\left\{\hat{H}, \hat{L}_{z}, \hat{M}\right\}$,
$\left\{\hat{H}, \hat{L}_{z}, \hat{N}\right\}$, and $\left\{\hat{H}, \hat{L}_{z}, \hat{\Lambda}\right\}$
respectively. We are now in a position to deal with the interbasis expansions
\begin{eqnarray}
\psi_{n k m}(\xi, \eta, \varphi; c, \pm b) = \sum\limits_{p=0}^{n}\,
U_{n k}^{p}(R; c, \pm b)\,\psi_{n_{\rho} p m}(\rho, \varphi, z,; c, \pm b),
\label{4.10.10}
\end{eqnarray}
\begin{eqnarray}
\psi_{n k m}(\xi, \eta, \varphi; c, \pm b) = \sum\limits_{q=0}^{n}\,
T_{n k}^{q}(R; c, \pm b)\,\psi_{n_{r} q m}(r, \theta, \varphi; c, \pm b)
\label{4.10.11}
\end{eqnarray}
for the prolate spheroidal basis in terms of the cylindrical and spherical bases.

First, we consider Eq. (\ref{4.10.10}). Let the operator $\hat{\Lambda}$
act on both sides of (\ref{4.10.10}). Then, by using Eqs. (\ref{4.10.6}),
(\ref{4.10.8}), and (\ref{4.10.9}), along with the orthonormality property
of the cylindrical basis, we find that
\begin{eqnarray}
\frac{1}{2}\,\left[\lambda_{k}(R) - \frac{R^{2}}{2}\,E_{z}(c, \pm b)\right]\,
U_{n k}^{p}(R; c, \pm b) = \sum\limits_{p'=0}^{n}\,U_{n k}^{p'}(R; c, \pm b)\,
M_{p p'}^{(\pm)},
\label{4.10.12}
\end{eqnarray}
where
\begin{eqnarray}
M_{p p'}^{(\pm)} = \int\limits_{0}^{\infty}\,\int\limits_{0}^{2\pi}\,\int\limits_{0}^{\infty}\,
\psi_{n_{\rho} p m}^{*}(\rho, \varphi, z; c, \pm b)\,\hat{M}\,
\psi_{n_{\rho} p' m}(\rho, \varphi, z; c, \pm b)\,\rho\,d\rho\,d\varphi\,dz.
\label{4.10.13}
\end{eqnarray}
The calculation of the matrix element $M_{p p'}^{(\pm)}$ can be done by expanding the
cylindrical wave functions in (\ref{4.10.13}) in terms of spherical wave functions
[see Eq. (\ref{4.9.9})] and by making use of the eigenvalue equation for $\hat{M}$
[see Eq. (\ref{4.10.7})]. This leads to
\begin{eqnarray}
M_{p p'}^{(\pm)} = \frac{1}{2}\,\sum\limits_{q=0}^{n}\,A_{q}(c, \pm b)\,
W_{n p}^{q}(c, \pm b)\,W_{n p'}^{q}(c, \pm b).
\label{4.10.14}
\end{eqnarray}
To calculate the sum in Eq. (\ref{4.10.14}), we need some recursion relation
for the coefficient $W_{n p}^{q}(c, \pm b)$ involving $p - 1, p$, and $p + 1$.
Owing to Eq. (\ref{4.9.13}), this amounts to using the recursion relation
(\ref{1.8.15}) and the orthonormality condition (\ref{1.8.16}) for the
Clebsch-Gordan coefficients, we find that $M_{p p'}^{(\pm)}$ is given by
\begin{eqnarray}
M_{p p'}^{(\pm)} &=& \left[s(c \mp b + s)(c \mp b + 3s) + 2(p+1)(n-p) +
2(p \pm b)(n + c - p + 1)\right]\,\delta_{p', p}+ \nonumber \\ [3mm]
&+& 2\,\left[p(p \pm b)(n - p + 1)(n + c - p + 1)\right]^{s}\delta_{p', p-1}+
\label{4.10.15}
\\  [3mm]
&+& 2\,\left[(p + 1)(p \pm b + 1)(n - p)(n + c - p)\right]^{s}\,\delta_{p', p+1}. \nonumber
\end{eqnarray}
Now by introducing (\ref{4.10.15}) into (\ref{4.10.12}), we get the
following three-term recursion relation:
\begin{eqnarray}
&&\Bigl[(p+1)(n-p) + \frac{1}{4}(c \mp b + s)(c \mp b + 3s)
+(p \pm b)(n + c - p + 1) + \frac{R^{2}}{8}E_{z}(c; \pm b) - \nonumber \\ [3mm]
&-&  \frac{1}{4}\lambda_{k}(R)\Bigr]\,U_{n k}^{p} +
\left[p(p \pm b)(n - p + 1)(n + c - p + 1)\right]^{s}\,U_{n k}^{p-1} +
\label{4.10.16}
\\ [3mm]
&+&\left[(p + 1)(p \pm b + 1)(n - p)(n + c - p)\right]^{s}\,U_{n k}^{p+1} = 0. \nonumber
\end{eqnarray}
for the expansion coefficients $U_{n k}^{p} \equiv U_{n k}^{p}(R; c, \pm b)$.
The recursion relation (\ref{4.10.16}) provides us with a
system of $n + 1$ linear homogeneous equations which can be solved by taking into
account the normalization condition
\begin{eqnarray*}
\sum\limits_{p=0}^{n}\,\left|U_{n k}^{p}(R; c, \pm b)\right|^{2} = 1.
\end{eqnarray*}
The eigenvalues $\lambda_{k}(R)$ of the operator $\hat{\Lambda}$ then
follow from the vanishing of the determinant for the latter system.

Second, let us concentrate on the expansion (\ref{4.10.11}) of the prolate
spheroidal basis in terms of the spherical basis. By employing a technique similar
to the one used for deriving Eq. (\ref{4.10.11}), we get
\begin{eqnarray}
\frac{1}{2}\,\left[\lambda_{k}(R) - A_{q}(c, \pm b)\right]\,
T_{n k}^{q}(R; c, \pm b) = \frac{R^{2}}{2}\sum\limits_{q'=0}^{n}\,T_{n k}^{q'}(R; c, \pm b)\,
N_{q q'}^{(\pm)},
\label{4.10.17}
\end{eqnarray}
\begin{eqnarray*}
N_{q q'}^{(\pm)} = \int\limits_{0}^{\infty}\,\int\limits_{0}^{\pi/2}\,\int\limits_{0}^{2\pi}\,
\psi_{n_{r} q m}^{*}(r, \theta, \varphi; c, \pm b)\,\hat{N}\,
\psi_{n_{r} q' m}(r, \theta, \varphi; c, \pm b)\,r^{2}\,\sin\theta\,dr\,d\theta\,d\varphi.
\end{eqnarray*}
The matrix elements $N_{q q'}^{(\pm)}$ can be calculated in the same way as
$M_{p p'}^{(\pm)}$ except that we must use the relation (\ref{1.8.10})
and the orthonormality condition (\ref{1.8.11}), instead of Eqs. (\ref{1.8.15})
and (\ref{1.8.16}). This produces the matrix element
\begin{eqnarray}
N_{q q'}^{(\pm)} &=& E_{n}(R; c, \pm b)\,\frac{2q(q+1) + (c \pm b)(2q \pm b + 1)}
{(2q + c \pm b)(2q + c \pm b + 2)}\,\delta_{q' q} - \nonumber \\
\label{4.10.18}
\\
&-& 2\Omega\,\left[A_{n}^{q+1}(c, \pm b)\,\delta_{q', q + 1} +
A_{n}^{q-1}(c, \pm b)\,\delta_{q', q - 1}\right], \nonumber
\end{eqnarray}
where
\begin{eqnarray*}
A_{n}^{q}(c, \pm b) = \left[\frac{q(n-q+1)(q+c\pm b)(q \pm b)(q+c)(n+q+c \pm b +1)}
{(2q+c\pm b)^{2}(2q+c\pm b -1)(2q+c\pm b +1)}\right]^{s}.
\end{eqnarray*}
Finally, the introduction of (\ref{4.10.18}) into (\ref{4.10.17}) leads to
the three-term recursion relation
\begin{eqnarray}
&&\left[\lambda_{k}(R) - A_{q}(c, \pm b) - \frac{R^{2}}{2}\,E_{n}(c, \pm b)\,
\frac{2q(q+1) + (c \pm b)(2q \pm b + 1)}
{(2q + c \pm b)(2q + c \pm b + 2)}\right]\,T_{n k}^{q}+ \nonumber \\
\label{4.10.19}
\\
&+& \Omega\,R^{2}\left[A_{n}^{q+1}(c, \pm b)\,T_{n k}^{q+1} +
A_{n}^{q}(c, \pm b)\,T_{n k}^{q-1}\right]=0  \nonumber
\end{eqnarray}
for the expansion coefficients $T_{n k}^{q} \equiv T_{n k}^{q}(R; c, \pm b)$.
This relation can be iterated by taking into account the normalization condition
\begin{eqnarray*}
\sum\limits_{q=0}^{n}\,\left|T_{n k}^{q}(R; c, \pm b)\right|^{2} = 1.
\end{eqnarray*}
Here again, the eigenvalues $\lambda_{k}(R)$ may be obtained from the vanishing of
the determinant of a system of nq1 linear homogeneous equations.

\subsection{Limiting cases}

Putting $b = s^{-}$, i.e., $P = 0$, in the matrix element
(\ref{4.10.18}) with $Q \neq 0$ and by using (\ref{4.8.15a}), we have
\begin{eqnarray*}
N_{q q'}^{(\pm)} = E_{n}(\delta)\,\frac{2A_{l}(\delta) - 2(|m|+ \delta)^{2}- 1}
{(2l + 2\delta - 1)(2l + 2\delta + 3)}\,\delta_{l' l} -
2\Omega\,\left[A_{N}^{l+2}(\delta)\,\delta_{l', l + 2} +
A_{N}^{l}(\delta)\,\delta_{l', l - 2}\right],
\end{eqnarray*}
where
\begin{eqnarray*}
A_{l}(\delta) = \left[\frac{l_{-}(l_{-}-1)(l_{+}+2\delta)(l_{+}+2\delta -1)
(N-l+2)(N+l+2\delta +1)}{4(2l+2\delta -1)^{2}(2l+2\delta -3)(2l+2\delta +1)}\right]^{s}
\end{eqnarray*}
(with $l_{\pm} = l \pm |m|$) and finally we get the following
three-term recursion relation
\begin{eqnarray*}
\left[\lambda_{k}(R) - A_{l}(\delta) - \frac{R^{2}}{2}
E_{n}(\delta)\,\frac{2A_{l}(\delta) - 2(|m|+ \delta)^{2}- 1}
{(2l + 2\delta - 1)(2l + 2\delta + 3)}\right]\,T_{N k}^{l}(R; \delta) + \\ [3mm]
+ \Omega\,R^{2}\,\left[A_{N}^{l+2}(\delta)\,T_{N k}^{l+2}(R; \delta) +
A_{N}^{l}(\delta)\,T_{N k}^{l-2}(R; \delta)\right] = 0
\end{eqnarray*}
for $T_{N k}^{l}(R; \delta) \equiv T_{N k}^{l}(R; c, \pm s)$. By analogy it is easy
to obtain a three-term recursion relation for the interbasis expansion coefficients
$U_{N k}^{n_{3}}(R; \delta) \equiv U_{N k}^{n_{3}}(R; c, \pm s)$. We get
\begin{eqnarray*}
\left[(2n_{3} + 1)(N-n_{3}+\delta + 1) + (|m|+ \delta)^{2} - 1 +
\frac{\Omega R^{2}}{4}\,(2n_{3} + 1) - \lambda_{k}(R)\right]\,
U_{N k}^{n_{3}}(R; \delta) + \\ [3mm]
+ \left[(n_{3} + 1)(n_{3} + 2)(N-|m|-n_{3})(N+|m|-n_{3}+2\delta)\right]^{s}\,
U_{N k}^{n_{3}+2}(R; \delta) + \\ [3mm]
+ \left[n_{3}(n_{3} - 1)(N-|m|-n_{3}+2)(N+|m|-n_{3}+2\delta + 2)\right]^{s}\,
U_{N k}^{n_{3}-2}(R; \delta) = 0.
\end{eqnarray*}
Consequently, when $b = s^{-}$ we have the expansions
[cf. Eqs. (\ref{4.10.10}) and (\ref{4.10.10})]
\begin{eqnarray*}
\psi_{N k m} = \sum\limits_{n_{3}}\,
U_{N k}^{n_{3}}(R; \delta)\,\psi_{N n_{3} m},
\end{eqnarray*}
\begin{eqnarray*}
\psi_{N k m} = \sum\limits_{l}\,
T_{N k}^{l}(R; \delta)\,\psi_{N l m},
\label{4.10.11}
\end{eqnarray*}
for the ring-shape oscillator. The summations on $l$ and $n_{3}$ go, by steps of 2,
from $|m|$ or $|m|+1$ to $N$ and from $0$ or $1$ to $N - |m|$ according to whether as
$N - |m|$ is even or odd (because $N - l$ and $N - |m|- n_{3}$ are always even).

The next limiting case $\delta = 0$, i.e., $Q = 0$, is trivial and the corresponding
results for the isotropic harmonic oscillator agree with the ones obtained
in Ref. \cite{MPSTA8,MPSTA9,MPSTA10}.

Finally, it should be noted that the following two limits
\begin{eqnarray*}
\lim\limits_{R \to 0}\,U_{n k}^{p}(R; c, \pm b) = \tilde{W}_{n k}^{p}(c, \pm b),
\qquad
\lim\limits_{R \to \infty}\,T_{n k}^{q}(R; c, \pm b) = W_{n k}^{q}(c, \pm b)
\end{eqnarray*}
furnish a useful means for checking the calculations
presented in the 4.9 and 4.10 sections.

\subsection{Oblate spheroidal basis}

The oblate spheroidal coordinates $(\bar{\xi}, \bar{\eta}, \varphi)$ are
defined by
\begin{eqnarray*}
x = \frac{\bar{R}}{2}\,\sqrt{\left(\bar{\xi}^{2} + 1\right)\left(1 - \bar{\eta}^{2}\right)}\,
\cos\varphi, \qquad
y = \frac{\bar{R}}{2}\,\sqrt{\left(\bar{\xi}^{2} + 1\right)\left(1 - \bar{\eta}^{2}\right)}\,
\sin\varphi, \qquad
z = \frac{\bar{R}}{2}\,\bar{\xi}\,\bar{\eta},
\end{eqnarray*}
(with $0\leq \bar{\xi} \leq \infty,\, -1 \leq \bar{\eta} \leq 1$, and
$0 \leq \varphi < 2\pi$, where $\bar{R}$ is the interfocus distance in the oblate
spheroidal coordinate system. As in the prolate system, in the limits $\bar{R} \to 0$ and
$\bar{R} \to \infty$, the oblate spheroidal coordinates give the spherical and
cylindrical coordinates, respectively \cite{KOMPOSLA,MPSTA8, MPSTA9, MPSTA10}.

The potential $U$ (\ref{4.8.1}), the Schr\"{o}dinger equation, the oblate spheroidal
constant of motion $\hat{\bar{\Lambda}}$, and the interbasis expansion coefficients for
the oblate spheroidal coordinates can be obtained from the corresponding expressions
for the prolate spheroidal coordinates by means of the trick: $\xi \to i\bar{\xi}$
and $R \to -i\bar{R}$.

\section{Spheroidal corrections to the spherical and cylindrical bases}
\markboth{CHAPTER 4. GENERALIZED RING-SHAPED POTENTIALS}
{4.11. SPHEROIDAL CORRECTIONS}

As we have already mentioned, the spheroidal system of coordinates is one of the most general
one-parameter systems of coordinates which contains spherical and cylindrical coordinates as some
limiting cases. Accordingly, the prolate spheroidal basis of the generalized oscillator as
$\bar{R} \to 0$ and $\bar{R} \to \infty$, degenerates into the spherical and cylindrical
bases that can be treated as zeroth-order approximations in some perturbation series.
The three-term recursion relations for the expansion coefficients of the prolate spheroidal
basis in the cylindrical and spherical bases, which have been obtained in the fifth section,
may serve as a basis for constructing an algebraic perturbation theory, respectively, at large
$(R \gg 1)$ and small $(R \ll 1)$ values of the interfocus distance $R$. Thus it is possible to
derive prolate spheroidal corrections for the spherical and cylindrical bases.

\subsection{The case $(R \ll 1)$}

Let us rewrite the three-term recursion relation (\ref{4.10.19}) in the following form:
\begin{eqnarray}
&&\left[\lambda_{k}(R) - A_{q}(c, \pm b) - \Omega\,R^{2}\,B_{n}^{q}(c, \pm b)\right]
\,T_{n k}^{q}+ \nonumber \\
\label{4.11.1}
\\
&+& \Omega\,R^{2}\left[A_{n}^{q+1}(c, \pm b)\,T_{n k}^{q+1} +
A_{n}^{q}(c, \pm b)\,T_{n k}^{q-1}\right]=0,  \nonumber
\end{eqnarray}
where
\begin{eqnarray*}
B_{n}^{q}(c, \pm b) = (2n + c \pm b +2)\,\frac{2q(q+1) + (c \pm b)(2q \pm b + 1)}
{2(2q + c \pm b)(2q + c \pm b + 2)}.
\end{eqnarray*}
The zeroth-order approximation for the separation constant $\lambda_{k}(R)$ and the
coefficients $T_{n k}^{q}(c, \pm b)$ can immediately be derived from the recursion
relation (\ref{4.11.1}). Indeed, from Eq. (\ref{4.11.1}), we obtain
\begin{eqnarray*}
\lim\limits_{R \to 0}\,\lambda_{k}(R) = A_{k}(c, \pm b), \qquad
\lim\limits_{R \to 0}\,T_{n k}^{q}(R; c, \pm b) = \delta_{k q},
\end{eqnarray*}
so that, for the wave function, we have
\begin{eqnarray*}
\lim\limits_{R \to 0}\,\psi_{n k m}(\xi, \eta, \varphi; R, c, \pm b) =
\psi_{n_{r} k m}(r, \theta, \varphi; c, \pm b).
\end{eqnarray*}
As is seen from these limiting relations, the quantum number $k$, labeling the spheroidal separation
constant and being (according to the oscillation theorem \cite{KOMPOSLA}) the number of zeros of
the prolate angular spheroidal function $\psi_{2}(\eta)$ in the interval $-1 \leq \eta \leq 1$,
turns into a spherical quantum number determining the number of zeros of the angular function
(\ref{4.8.7}). It is clear that this fact is a consequence of the independence of the number of zeros
of the wave function on $R(r)$.

In order to calculate higher order corrections, we represent the interbasis coefficients
$T_{n k}^{q}(R; c, \pm b)$  and the spheroidal separation constant $\lambda_{k}(R)$
as expansions in powers of $\Omega R^{2}$:
\begin{eqnarray}
T_{n k}^{q} = \delta_{k q} + \sum\limits_{j=1}^{\infty}\,T_{n k}^{(j)}\,
\left(\Omega R^{2}\right)^{j},
\label{4.11.2}
\end{eqnarray}
\begin{eqnarray}
\lambda_{k}(R) = A_{k}(c, \pm b) + \sum\limits_{j=1}^{\infty}\,\lambda_{k}^{(j)}\,
\left(\Omega R^{2}\right)^{j}.
\label{4.11.3}
\end{eqnarray}
Substituting Eqs. (\ref{4.11.2}) and (\ref{4.11.3}) into the three-term
recursion relation (\ref{4.11.1}) and equating the coefficients
with the same power of $R$, we arrive at the equation for the coefficients
$T_{n k}^{(j)}$ and $\lambda_{k}^{(j)}$
\begin{eqnarray}
&&4(k - q)(k + q + c \pm b + 1)T_{k q}^{(j)} = - A_{n}^{q+1}(c, \pm b)\,T_{k, q+1}^{(j-1)}
+ \nonumber \\
\label{4.11.4}
\\
&+& B_{n}^{q}(c, \pm b)\,T_{k, q}^{(j-1)} - A_{n}^{q}(c, \pm b)\,T_{k, q-1}^{(j-1)}
- \sum\limits_{t=0}^{j-1}\,\lambda_{k}^{(j-t)}\,T_{k q}^{(t)}. \nonumber
\end{eqnarray}
Equation (\ref{4.11.4}) with the initial condition $T_{k q}^{(0)} = \delta_{k q}$
and the condition  $T_{q q}^{(j)} = \delta_{j 0}$ arising in the standard
perturbation theory \cite{LL} allow us to derive a formula expressing
$\lambda_{k}^{(j)}$ for $j \geq 1$ through the coefficients
$T_{k k}^{(j-1)}$ and $T_{k, k \pm 1}^{(j-1)}$:
\begin{eqnarray}
\lambda_{k}^{(j)} = - A_{n}^{k+1}(c, \pm b)\,T_{k, k+1}^{(j-1)}
+ B_{n}^{k}(c, \pm b)\,T_{k, k}^{(j-1)} - A_{n}^{k}(c, \pm b)\,T_{k, k-1}^{(j-1)}.
\label{4.11.5}
\end{eqnarray}
This gives a possibility to determine, step by step, the coefficients $\lambda_{k}^{(j)}$
and $T_{k q}^{(j)}$ in Eqs. (\ref{4.11.2}) and (\ref{4.11.3}). As an example, let us write
down the first- and second-order corrections in (\ref{4.11.3}) for $\lambda_{k}(R)$ and the
first-order correction in (\ref{4.11.2}) for $T_{n k}^{q}(c, \pm b)$. It follows from
Eq. (\ref{4.11.4}) that
\begin{eqnarray*}
\lambda_{k}^{(1)} = B_{n}^{k}(c, \pm b), \qquad
\lambda_{k}^{(2)} = - A_{n}^{k+1}(c, \pm b)\,T_{k, k+1}^{(1)}
- A_{n}^{k}(c, \pm b)\,T_{k, k-1}^{(1)}
\end{eqnarray*}
and Eq. (\ref{4.11.4}) for $j = 1$ results in
\begin{eqnarray}
T_{k, q}^{(1)} = - \frac{A_{n}^{k}(c, \pm b)}{4(2k + c \pm b)}\,\delta_{q, k-1}
+ \frac{A_{n}^{k+1}(c, \pm b)}{4(2k + c \pm b + 2)}\,\delta_{q, k+1}.
\label{4.11.6}
\end{eqnarray}
Thus, for the spheroidal separation constant, with an accuracy up to the term
$\left(\Omega R^{2}\right)^{2}$, we get
\begin{eqnarray*}
\lambda_{k}(R) = A_{k}(c, \pm b) + \Omega R^{2}\,B_{n}^{k}(c, \pm b) +
\frac{\Omega^{2} R^{4}}{4}\,\left\{\frac{\left[A_{n}^{k}(c, \pm b)\right]^{2}}
{2k + c \pm b} - \frac{\left[A_{n}^{k+1}(c, \pm b)\right]^{2}}
{2k + c \pm b + 2}\right\}.
\end{eqnarray*}
Introducing (\ref{4.11.6}) into (\ref{4.11.2}) and then using \ref{4.10.11}) for
the expansion of the prolate spheroidal basis over the spherical one, we get the
following approximate formula:
\begin{eqnarray*}
&&\psi_{n k m}(\xi, \eta, \varphi; R, c, \pm b) = \psi_{n k m}(r, \theta, \varphi; c, \pm b) -
\frac{\Omega R^{2}}{4}\times \\ [4mm]
&\times& \left[\frac{A_{n}^{k}(c, \pm b)}
{2k + c \pm b}\,\psi_{n, k-1, m}(r, \theta, \varphi; c, \pm b) -
\frac{A_{n}^{k+1}(c, \pm b)}{2k + c \pm b + 2}\,\psi_{n, k+1, m}(r, \theta, \varphi; c, \pm b)\right].
\end{eqnarray*}

\subsection{The case $(R \gg 1)$}

Now let us consider the case $(R \gg 1)$. The three-term recursion relation
(\ref{4.10.16}) can be written as
\begin{eqnarray}
&&\left[D_{n}^{p}(c, \pm b) + \frac{R^{2}}{8}\,E_{z}(c, \pm b) - \frac{1}{4}\,\lambda_{k}(R)\right]\,
U_{n k}^{p} + \nonumber \\
\label{4.11.7}
\\
&+& \left[C_{n}^{p+1}(c, \pm b)\,U_{n k}^{p+1} + C_{n}^{p(c, \pm b)}
\,U_{n k}^{p-1}\right] = 0, \nonumber
\end{eqnarray}
where
\begin{eqnarray*}
C_{n}^{p}(c, \pm b) = \left[p(p \pm b)(n-p+1)(n + c - p + 1)\right]^{s},
\end{eqnarray*}
\begin{eqnarray*}
D_{n}^{p}(c, \pm b) = (p+1)(n-p) + (p \pm b)(n + c - p + 1) +
\frac{1}{4}(c \mp b + s)(c \mp b + 3s).
\end{eqnarray*}
It follows from Eq. (\ref{4.11.7}) that
\begin{eqnarray*}
\lim\limits_{R \to \infty}\,\frac{\lambda_{k}(R)}{R^{2}} =
\frac{1}{2}\,E_{z}(k, \pm b), \qquad
\lim\limits_{R \to \infty}\,U_{n k}^{p}(R; c, \pm b) = \delta_{k p}.
\end{eqnarray*}
For $(R \gg 1)$, the interbasis expansion coefficients and the spheroidal
separation constant are developed in negative powers of $\Omega R^{2}$:
\begin{eqnarray}
U_{n k}^{p}(R; c, \pm b) = \delta_{k p} + \sum\limits_{j=1}^{\infty}\,
U_{k p}^{(j)}\,\left(\Omega R^{2}\right)^{-j}
\label{4.11.8}
\end{eqnarray}
\begin{eqnarray}
\frac{\lambda_{k}(R)}{R^{2}} = \frac{1}{2\Omega}\,E_{z}(k, \pm b)
+ \sum\limits_{j=1}^{\infty}\,\lambda_{k}^{(j)}\,
\left(\Omega R^{2}\right)^{-j}.
\label{4.11.9}
\end{eqnarray}
Substituting Eqs. (\ref{4.11.8}) and (\ref{4.11.9}) into Eq. (\ref{4.11.7}), we get
\begin{eqnarray}
\frac{1}{4}(p - k)\,U_{k p}^{(j)} + C_{n}^{p+1}(c, \pm b)\,U_{k, p+1}^{(j-1)}
+ D_{n}^{p}(c, \pm b)\,U_{k p}^{(j-1)} + \nonumber \\
\label{4.11.10}
\\
+ C_{n}^{p}(c, \pm b)\,U_{k, p-1}^{(j-1)} - \frac{1}{4}\,
\sum\limits_{t=1}^{j-1}\,\lambda_{k}^{(j-t)}\,
U_{k p}^{(t)} = 0. \nonumber
\end{eqnarray}
Using the conditions $U_{k p}^{(0)} = \delta_{k p}$ and $U_{p p}^{(j)} = \delta_{j 0}$,
one easily obtains
\begin{eqnarray}
\frac{1}{4}\lambda_{k}^{(j)} = C_{n}^{p+1}(c, \pm b)\,U_{k, p+1}^{(j-1)}
+ D_{n}^{p}(c, \pm b)\,U_{k p}^{(j-1)} +
+ C_{n}^{p}(c, \pm b)\,U_{k, p-1}^{(j-1)}.
\label{4.11.11}
\end{eqnarray}
Equations (\ref{4.11.10}) and (\ref{4.11.11}) completely solve the problem
of determining the expansion coefficients $\lambda_{k}^{(j)}$
and $U_{k p}^{(j)}$. For instance, we have the approximate formulas
\begin{eqnarray*}
\frac{\lambda_{k}(R)}{\Omega R^{2}} =
\frac{1}{2\Omega}\,E_{z}(k, \pm b) + \frac{4}{\Omega R^{2}}D_{n}^{k}(c, \pm b)
+ \frac{16}{\left(\Omega R^{2}\right)^{2}}\,\left\{\left[C_{n}^{k}(c, \pm b)\right]^{2}
- \left[C_{n}^{k+1}(c, \pm b)\right]^{2}\right\},
\end{eqnarray*}
\begin{eqnarray*}
&&\psi_{n k m}(\xi, \eta, \varphi; R, c, \pm b) = \psi_{n k m}(\rho, \varphi, z; c, \pm b) +
\frac{4}{\Omega R^{2}}\times \\ [4mm]
&\times&\left[C_{n}^{k}(c, \pm b)\,\psi_{n, k-1, m}(\rho, \varphi, z; c, \pm b)
- C_{n}^{k+1}(c, \pm b)\psi_{n, k+1, m}(\rho, \varphi, z; c, \pm b)\right].
\end{eqnarray*}

Higher order corrections are calculated in a similar way.

\newpage

\chapter{The generalized MIC-Kepler problem}
\markboth{CHAPTER 5. THE GENERALIZED MIC-KEPLER PROBLEM}{}

A minimally superintegrable system described by a Hamiltonian
\begin{eqnarray}
\hat{\mathcal{H}} = \frac{1}{2\mu_{0}}\,\left(-i\hbar\,\nabla - \frac{e}{c}\,{\bf A}^{(\pm)}\right)^{2}
+ \frac{\hbar^{2}s^{2}}{2\mu_{0}r^{2}} - \frac{e^{2}}{r} + \frac{\lambda_{1}}{r(r+z)}
+ \frac{\lambda_{2}}{r(r-z)},
\label{5.0.1}
\end{eqnarray}
where $\lambda_{1}$ and $\lambda_{2}$ are nonnegative constants, later on will be called the
generalized MIC-Kepler (GMICK) problem \cite{M-26}. The Hamiltonian (\ref{5.1.1}) for $s = 0$
and $\lambda_{i}$ $(i = 1, 2)$ reduces to the Hamiltonian
\begin{eqnarray*}
\hat{H} = -\frac{\hbar^{2}}{2\mu_{0}}\,\triangle  - \frac{e^{2}}{r} + \frac{\lambda_{1}}{r(r+z)}
+ \frac{\lambda_{2}}{r(r-z)}
\end{eqnarray*}
of the generalized Kepler–Coulomb system, which we studied in the previous chapter.
In the following paragraphs we will show that the variables in the Schr\"{o}dinger equation for
the generalized Kepler-MIC problem, as well as for the Kepler-MIC problem and the generalized
Kepler-Coulomb system, are separated into spherical, parabolic, and prolate spheroidal coordinates.
We also show that the generalized MIK-Kepler problem is dual to the four-dimensional double singular
oscillator and that the duality transformation is a generalized version of the KS-transformation \cite{M-27}.

We also note that the generalized MIK-Kepler problem and its dual four-dimensional system of a
double singular oscillator were considered from different points of view in
\cite{Nuri,Giri-1,Giri-2,Giri-3,MARC-1,MARC-2,Hoque-1,Hoque-2,Salazar-1,Salazar-2,SHMAVON,Lavrenov}.

\section{Spherical basis of the GMICK problem}
\markboth{CHAPTER 5. THE GENERALIZED MIC-KEPLER PROBLEM}{5.1. SPHERICAL BASIS OF THE GMICK PROBLEM}

The Schr\"{o}dinger equation with Hamiltonian (\ref{5.1.1}) in spherical coordinates $(r, \theta, \varphi)$
may be solved by seeking a wave function $\psi$ of the form
\begin{eqnarray}
\psi\left(r,\theta, \varphi\right)=R(r)\,Z(\theta, \varphi).
\label{5.1.1}
\end{eqnarray}
After substitution the expression (\ref{5.1.1}) the variables in the Schr\"{o}dinger equation
are separated and we arrive at the at the following system of coupled differential equations:
\begin{eqnarray}
&& \frac{1}{\sin\theta}\frac{\partial}{\partial\theta}\left(\sin\theta \frac{\partial Z}{\partial\theta}\right)+
\frac{1}{4\cos^{2}\frac{\theta}{2}}\left(\frac{\partial^{2}}{\partial \varphi^{2}}-4\Lambda_{1}\right)Z +
\nonumber \\
\label{5.1.2}
\\
&+& \frac{1}{4\sin^{2}\frac{\theta}{2}}\left[\left(\frac{\partial}{\partial \varphi} + 2is\right)^{2}-4\Lambda_{2}\right]Z=
-\it{A}Z,
\nonumber
\end{eqnarray}
\begin{eqnarray}
\frac{1}{r^{2}}\frac{d}{dr}\left(r^{2}\frac{dR}{dr}\right)-\frac{\it{A}}{r^{2}}R +
\frac{2\mu_{0}}{\hbar^{2}}\left(E+\frac{e^{2}}{r}\right)R=0,
\label{5.1.3}
\end{eqnarray}
where $\Lambda_{i} =\mu_{0}\lambda_{i}/\hbar^{2}$, and $\it{A}$ is the separation constant in spherical coordinates.

The solution of Eq. (\ref{5.1.2}) normalized by the condition
\begin{eqnarray}
\int\limits_{0}^{\pi} \sin\theta d\theta \int\limits_{0}^{2\pi}
Z^{(s)}_{jm}\left(\theta, \varphi;\delta_{1}^{(s)},\delta_{2}^{(s)}\right)
Z^{(s)*}_{j'm'}\left(\theta, \varphi;\delta_{1}^{(s)},\delta_{2}^{(s)}\right)d\varphi = \delta_{jj'} \delta_{mm'},
\label{5.1.4}
\end{eqnarray}
has the form
\begin{eqnarray}
Z^{(s)}_{jm}\left(\theta, \varphi;\delta_{1}^{(s)},\delta_{2}^{(s)}\right)
= N_{jm}^{(s)}\left(\delta_{1}^{(s)},\delta_{2}^{(s)}\right)\left(\cos\frac{\theta}{2}\right)^{m_{1}}\,
\left(\sin\frac{\theta}{2}\right)^{m_{2}} P^{(m_{2},m_{1})}_{j-m_{+}}\left(\cos\theta\right)
\,e^{i(m-s)\varphi},
\label{5.1.5}
\end{eqnarray}
where $m_{1}=\left|m-s\right|+\delta_{1}^{(s)}=\sqrt{\left(m-s\right)^{2}+4\Lambda_{1}},\,
m_{2}=\left|m+s\right|+\delta_{2}^{(s)}=\sqrt{\left(m+s\right)^{2}+4\Lambda_{2}}$,
$P^{(a,b)}_{n}(x)$ denotes a Jacobi polinomial, and $m_{\pm}$ is defined by the formula (\ref{3.1.10}).
The quantum numbers $m$ and $j$ run through values: $m=-j,-j+1,\ldots,j-1,j$, and
$j=m_{+}, m_{+}+1,\ldots$. The quantum numbers $j, m$ characterize the total momentum of
the system and its projection on the axis $z$. For the (half)integer $s$ $j, m$ (half)integers.

Furthermore, the separation constant $\it {A}$ is quantized as
\begin{eqnarray}
\it{A}=\left(j+\frac{\delta_{1}^{(s)}+\delta_{2}^{(s)}}{2}\right)\left(j+\frac{\delta_{1}^{(s)}+\delta_{2}^{(s)}}{2}+1\right).
\label{5.1.6}
\end{eqnarray}
The normalization constant $N_{jm}^{(s)}\left(\delta_{1}^{(s)},\delta_{2}^{(s)}\right)$ in (\ref{5.1.5})
is given (up to a phase factor) by
\begin{eqnarray}
N_{jm}^{(s)}\left(\delta_{1},\delta_{2}\right)=
\sqrt{\frac{(2j+\delta_{1}^{(s)}+\delta_{2}^{(s)}+1)(j-m_{+})!
\Gamma(j+m_{+}+\delta_{1}^{(s)}+\delta_{2}^{(s)}+1)}
{4\pi \Gamma(j - m_{-}+\delta_{1}^{(s)}+1) \Gamma(j + m_{-}+\delta_{2}^{(s)}+1)}}.
\label{5.1.7}
\end{eqnarray}

The angular wave function $Z^{(s)}_{jm}$ [see Eq. (\ref{5.1.5})] is called a Tamm ring-shaped monopole
harmonics by analogy with the term "monopole harmonics" that were studied by Tamm \cite{TAMM}.
The Tamm ring-shaped monopole harmonics generalize the functions studied by Hartmann in \cite{HARTMAN-3}
in the case $s=0$, $\delta_{1}^{(s)} = \delta_{2}^{(s)}$. Due to the connecting formula \cite{BE2},
\begin{eqnarray*}
\left(\lambda+\frac{1}{2}\right)_{n}C^{\lambda}_{n}(x)=\left(2\lambda\right)_{n} P^{(\lambda-1/2, \lambda-1/2)}_{n}(x),
\end{eqnarray*}
between the Jacobi polynomial $P_{n}^{(a,b)}$ and the Gegenbauer polynomial $C_{n}^{\lambda}$, the case
$s=0$, $\delta_{1}^{(s)} = \delta_{2}^{(s)}=\delta$ yields
\begin{eqnarray}
Z^{(0)}_{jm}\left(\theta, \varphi;\delta,\delta\right)
&=& 2^{|m|+\delta}\Gamma\left(|m|+\delta +\frac{1}{2}\right)
\sqrt{\frac{(2j+2\delta +1)(j-|m|)!}{4\pi^{2}\Gamma(j+|m|+2\delta +1)}}\times  \nonumber
\\
\label{5.1.8}
\\
&\times&\left(\sin\theta\right)^{|m|+\delta}C^{|m|+\delta +1/2}_{j-|m|}(\cos\theta)
e^{im\varphi}, \nonumber
\end{eqnarray}
the result already obtained in \cite{M-22,M-23}. In the case $\delta = 0$, i.e.,
$\Lambda_{1} = \Lambda_{2} = 0$, using the formula relating the Gegenbauer and Legandre polynomials \cite{BE2},
\begin{eqnarray*}
P^{|m|}_{j}(\cos\theta)=\frac{(-2)^{|m|}}{\sqrt{\pi}}\Gamma\left(|m|+\frac{1}{2}\right)
\left(\sin\theta\right)^{|m|}C^{|m|+1/2}_{j-|m|}(\cos\theta),
\end{eqnarray*}
we can reduce relation (\ref{5.1.8}) to the form
\begin{eqnarray*}
Z^{(0)}_{jm}\left(\theta, \varphi;0,0\right)= (-1)^{|m|}\sqrt{\frac{(2j+1)(j-|m|)!}{4\pi (j+|m|)!}}
P^{|m|}_{j}(\cos\theta) e^{im\varphi},
\end{eqnarray*}
which up to a phase factor coincides with the usual spherical function $Y_{l m}(\theta, \varphi)$.

Let us go now to radial equation (\ref{5.1.3}). The introduction of (\ref{5.1.6}) into the (\ref{5.1.3})
leads to
\begin{eqnarray}
\frac{1}{r^{2}}\frac{d}{dr}\left(r^{2}\frac{dR}{dr}\right) &-& \frac{1}{r^{2}}\,
\left(j+\frac{\delta_{1}^{(s)}+\delta_{2}^{(s)}}{2}\right)\left(j+\frac{\delta_{1}^{(s)}+\delta_{2}^{(s)}}{2}+1\right)\,R +
\nonumber \\
\label{5.1.9}
\\
&+& \frac{2\mu_{0}}{\hbar^{2}}\left(E+\frac{e^{2}}{r}\right)R=0,
\nonumber
\end{eqnarray}
which is reminiscent of the radial equation for the hydrogen atom except that the orbital quantum
number $l$ is replaced here by $j + \left(\delta_{1}^{(s)}+\delta_{2}^{(s)}\right)/2$. Normalized by the condition
\begin{eqnarray}
\int\limits_{0}^{\infty} r^{2} R^{(s)}_{nj}\left(r; \delta_{1}^{(s)}, \delta_{2}^{(s)}\right)
R^{(s)*}_{n'j}\left(r; \delta_{1}^{(s)}, \delta_{2}^{(s)}\right) dr =\delta_{n n'}
\label{5.1.10}
\end{eqnarray}
the solution of Eq. (\ref{5.1.9}) for the discrete spectrum is
\begin{eqnarray}
R^{(s)}_{nj}\left(r; \delta_{1}^{(s)}, \delta_{2}^{(s)}\right)&=& C^{(s)}_{nj}\left(\delta_{1}^{(s)}, \delta_{2}^{(s)}\right)
\left(2\varepsilon r\right)^{j+\frac{\delta_{1}^{(s)}+ \delta_{2}^{(s)}}{2}}
e^{-\varepsilon r} \times \nonumber \\
\label{5.1.11} \\
&\times& F\left(-n + j + 1; 2j+\delta_{1}^{(s)}+ \delta_{2}^{(s)}+2; 2\varepsilon r\right),
\nonumber
\end{eqnarray}
where $n = |s|+1, |s|+2, \ldots$. In (\ref{5.1.11}), the normalization factor
$C^{(s)}_{nj}\left(\delta_{1}^{(s)}, \delta_{2}^{(s)}\right)$ reads
\begin{eqnarray}
C^{(s)}_{nj}\left(\delta_{1}^{(s)}, \delta_{2}^{(s)}\right) = \frac{2\varepsilon^{2}\sqrt{r_{0}}}
{\Gamma\left(2j+\delta_{1}^{(s)}+ \delta_{2}^{(s)}+2\right)}\,
\sqrt{\frac{\Gamma\left(n + j + \delta_{1}^{(s)} + \delta_{2}^{(s)} + 1\right)}{\left(n - j -1\right)!}}
\label{5.1.12}
\end{eqnarray}
and the parameter $\varepsilon$  is defined by
\begin{eqnarray}
\varepsilon = \sqrt{-\frac{2\mu_{0} E}{\hbar^{2}}} = \frac{\mu_{0} e^{2}}{\hbar^{2}\,
\left(n + \frac{\delta_{1}^{(s)}+ \delta_{2}^{(s)}}{2}\right)}.
\label{5.1.13}
\end{eqnarray}
The eigenvalues $E$ are then given by and the parameter $\varepsilon$  is defined by
\begin{eqnarray}
E \equiv E_{n}^{(s)}= - \frac{\mu_{0} e ^{4}}{2\hbar^{2}\left(n + \frac{\delta_{1}^{(s)}+ \delta_{2}^{(s)}}{2}\right)^{2}}.
\label{5.1.14}
\end{eqnarray}

Note also that the spherical wave function of the generalized MIK-Kepler system
\begin{eqnarray}
\psi_{n j m}^{(s)} \equiv \psi_{n j m}^{(s)}\left(r, \theta, \varphi; \delta_{1}^{(s)}, \delta_{2}^{(s)}\right) =
R_{n j m}^{(s)}\left(r; \delta_{1}^{(s)}, \delta_{2}^{(s)}\right)\,
Z^{(s)}_{jm}\left(\theta, \varphi;\delta_{1}^{(s)},\delta_{2}^{(s)}\right)
\label{5.1.15}
\end{eqnarray}
is an eigenfunction of the system of commuting operators $\left\{\hat{H}, \hat{M}^{(s)}, \hat{J}_{z}\right\}$
and the following spectral problems take place:
\begin{eqnarray}
\hat{M}^{(s)}\,\psi_{n j m}^{(s)} =
\left(j+\frac{\delta_{1}^{(s)}+\delta_{2}^{(s)}}{2}\right)\left(j+\frac{\delta_{1}^{(s)}+\delta_{2}^{(s)}}{2}+1\right)\,
\psi_{n j m}^{(s)},
\label{5.1.16}
\end{eqnarray}
\begin{eqnarray}
\hat{J}_{z}\,\psi_{n j m}^{(s)}\left(r, \theta, \varphi; \delta_{1}^{(s)}, \delta_{2}^{(s)}\right) =
\left(s - i \frac{\partial}{\partial \varphi}\right) =
m\,\psi_{n j m}^{(s)}\left(r, \theta, \varphi; \delta_{1}^{(s)}, \delta_{2}^{(s)}\right),
\label{5.1.17}
\end{eqnarray}
where
\begin{eqnarray}
\hat{M}^{(s)} = \hat{J}^{2} + \frac{\Lambda_{1}}{1 + \cos\theta} +
\frac{\Lambda_{2}}{1 - \cos\theta}.
\label{5.1.18}
\end{eqnarray}
Now, taking into account the explicit form of the operator $ \hat{J}^{2}$ in Cartesian coordinates
(\ref{3.3.3}), for the operator $\hat{M}^{(s)}$ we obtain the expression
\begin{eqnarray}
\hat{M}^{(s)} = x_{i}x_{j}\frac{\partial^{2}}{\partial x_{i} \partial x_{j}} - r^{2}\Delta +
2x_{i}\frac{\partial}{\partial x_{i}} + \frac{2isr}{r-z}\left(y\frac{\partial}{\partial x} -
x\frac{\partial}{\partial y} - is\right) + \frac{2\Lambda_{1}r}{r + z} + \frac{2\Lambda_{2}r}{r - z},
\label{5.1.19}
\end{eqnarray}
which we will need when constructing the prolate spheroidal basis of the generalized MIC-Kepler problem.

\section{Parabolic basis of the GMICK problem}
\markboth{CHAPTER 5. THE GENERALIZED MIC-KEPLER PROBLEM}{5.2. PARABOLIC BASIS OF THE GMICK PROBLEM}

In the parabolic coordinates (\ref{3.1.13}) the substitution
\begin{eqnarray*}
\psi\left(\mu, \nu, \varphi\right) = \Phi_{1}\left(\mu\right)\,\Phi_{2}\left(\nu\right)
\,\frac{e^{i(m - s)\varphi}}{\sqrt{2\pi}}
\end{eqnarray*}
separates the variables in the Schr\"{o}dinger equation and we arrive at the following system of equations:
\begin{eqnarray}
\frac{d}{d\mu}\left(\mu\frac{d\Phi_{1}}{d\mu}\right)+\left[\frac{\mu_{0}E}{2\hbar^{2}}\mu - \frac{m_{1}^{2}}{4\mu}
+\frac{\sqrt{\mu_{0}}}{2\hbar}\Omega^{(s)}\left(\delta_{1}^{(s)}, \delta_{2}^{(s)}\right) + \frac{1}{2r_{0}}\right]\Phi_{1}=0,
\label{5.2.1}
\end{eqnarray}
\begin{eqnarray}
\frac{d}{d\nu}\left(\nu\frac{d\Phi_{2}}{d\nu}\right)+\left[\frac{\mu_{0}E}{2\hbar^{2}}\nu - \frac{m_{2}^{2}}{4\nu}
-\frac{\sqrt{\mu_{0}}}{2\hbar}\Omega^{(s)}\left(\delta_{1}^{(s)}, \delta_{2}^{(s)}\right) + \frac{1}{2r_{0}}\right]\Phi_{2}=0,
\label{5.2.2}
\end{eqnarray}
where $\Omega^{(s)}\left(\delta_{1}^{(s)}, \delta_{2}^{(s)}\right)$ is the parabolic separation constant, and $r_{0}$ is the Bohr radius.

For $s = 0$ these equations coincide with equations (\ref{4.6.18}) and (\ref{4.6.19}) for the generalized Kepler-Coulomb
system and therefore we can immediately write that
\begin{eqnarray}
\psi_{n_{1} n_{2} m}^{(s)}\left(\mu, \nu, \varphi; \delta_{1}^{(s)}, \delta_{2}^{(s)}\right) = \varepsilon^{2}\,\sqrt{\frac{r_{0}}{\pi}}\,
\Phi_{n_{1} m_{1}}\left(\mu; \delta_{1}^{(s)}\right)\,\Phi_{n_{2} m_{2}}\left(\nu; \delta_{2}^{(s)}\right)
\,e^{i(m - s)\varphi},
\label{5.2.3}
\end{eqnarray}
where
\begin{eqnarray}
\Phi_{n_{i} m_{i}}\left(t_{i}; \delta_{i}^{(s)}\right) = \sqrt{\frac{\Gamma\left(n_{i} + m_{i} +1\right)}{\left(n_{i}\right)!}}\,
\frac{e^{-\varepsilon t_{i}/2}\,\left(\varepsilon t_{i}\right)^{m_{i}/2}}{\Gamma\left(m_{i} +1\right)}\,
F\left(-n_{i}; m_{i} +1; \varepsilon t_{i}\right).
\label{5.2.4}
\end{eqnarray}
Here $n_{1}$ and $n_{2}$ are non-negative integers
\begin{eqnarray}
n_{1} = - \frac{|m-s| + \delta_{1}^{(s)} + 1}{2} + \frac{\sqrt{\mu_{0}}}{2\varepsilon \hbar}
\Omega^{(s)}\left(\delta_{1}^{(s)}, \delta_{2}^{(s)}\right) + \frac{1}{2r_{0}\varepsilon},
\nonumber \\
\label{5.2.5}
\\
n_{2} = - \frac{|m+s| + \delta_{2}^{(s)} + 1}{2} - \frac{\sqrt{\mu_{0}}}{2\varepsilon \hbar}
\Omega^{(s)}\left(\delta_{1}^{(s)}, \delta_{2}^{(s)}\right) + \frac{1}{2r_{0}\varepsilon}.
\nonumber
\end{eqnarray}
From the last relations, taking into account (\ref{5.1.14}), we get that the parabolic quantum numbers
$n_{1}$ and $n_{2}$ are connected with the principal quantum number $n$ as follows:
\begin{eqnarray}
n = n_{1} + n_{2} + m_{+} + 1.
\label{5.2.6}
\end{eqnarray}
Excluding the energy $E$ from Eqs. (\ref{5.2.1}) and (\ref{5.2.2}), we obtain the additional integral
of motion,
\begin{eqnarray*}
\hat{\Omega}^{(s)}&=&\frac{\hbar}{\sqrt{\mu_{0}}}\Biggl\{\frac{2}{\mu + \nu}
\left[\mu\frac{\partial}{\partial \nu}\left(\nu\frac{\partial}{\partial \nu}\right)
-\nu\frac{\partial}{\partial \mu}\left(\mu\frac{\partial}{\partial \mu}\right)\right]+
\frac{\mu - \nu}{2\mu \nu}\frac{\partial^{2}}{\partial \varphi^{2}} -
\frac{2is\mu}{\nu \left(\mu + \nu\right)}\frac{\partial}{\partial \varphi}- \\
\\
&-& \frac{2s^{2}\mu}{\nu \left(\mu + \nu\right)} + \frac{2\Lambda_{1}\nu}{\mu \left(\mu + \nu\right)}-
\frac{2\Lambda_{2}\mu}{\nu \left(\mu + \nu\right)}+\frac{\mu - \nu}{r_{0}\left(\mu + \nu\right)}\Biggr\}
\end{eqnarray*}
with the eigenvalues
\begin{eqnarray}
\Omega^{(s)}\left(\delta_{1}^{(s)}, \delta_{2}^{(s)}\right)=
\frac{\varepsilon \hbar}{\sqrt{\mu_{0}}}\left(n_{1} - n_{2} - m_{-} + \frac{\delta_{1}^{(s)} - \delta_{2}^{(s)}}{2}\right)
\label{5.2.7}
\end{eqnarray}
and eigenfunctions $\psi_{n_{1} n_{2} m}^{(s)}\left(\mu, \nu, \varphi; \delta_{1}^{(s)}, \delta_{2}^{(s)}\right)$.

In Cartesian coordinates, the operator $\hat{\Omega}^{(s)}$ can be rewritten as
\begin{eqnarray}
\hat{\Omega}^{(s)} &=& \frac{\hbar}{\sqrt{\mu_{0}}}\,\Biggl[z\left(\frac{\partial^{2}}{\partial x^{2}} +
\frac{\partial^{2}}{\partial y^{2}}\right) - x\frac{\partial^{2}}{\partial x \partial z}
- y\frac{\partial^{2}}{\partial y \partial z} - is\frac{r+z}{r\left(r-z\right)}\left(y\frac{\partial}{\partial x}
-x\frac{\partial}{\partial y} - is\right) -  \nonumber \\
\label{5.2.8}
\\
&-& \frac{\partial}{\partial z} + \frac{z}{r_{0}r} + \Lambda_{1}\,\frac{r-z}{r\left(r+z\right)} -
\Lambda_{2}\,\frac{r+z}{r\left(r-z\right)}\Biggr],
\nonumber
\end{eqnarray}
so that it immediately follows that $\hat{\Omega}^{(s)}$ is connected to the $z$ - component $\hat{I}_{z}$
of the analog of the Runge-Lenz vector (\ref{3.1.3}) via
\begin{eqnarray}
\hat{\Omega}^{(s)} = \hat{I}_{z} + \frac{\hbar}{\sqrt{\mu_{0}}}\,
\left[\Lambda_{1}\,\frac{r-z}{r\left(r+z\right)} -
\Lambda_{2}\,\frac{r+z}{r\left(r-z\right)}\right]
\label{5.2.9}
\end{eqnarray}
and coincides with $\hat{I}_{z}$ when $\Lambda_{1} = \Lambda_{2} = 0$.

Thus we have solved the spectral problem in parabolic coordinates
\begin{eqnarray}
\hat{\mathcal{H}}\psi = E\psi, \qquad
\hat{\Omega}^{(s)} \psi = \Omega^{(s)} \psi, \qquad
\hat{J}_{z}\psi = m\psi,
\label{5.2.10}
\end{eqnarray}
where $\hat{\mathcal{H}}, \hat{\Omega}^{(s)}$ and $\hat{J}_{z}$ are defined by the expressions
(\ref{5.1.1}), (\ref{5.2.9}) and (\ref{3.1.3}).

It is mentioned that all the formulas obtained in this Chapter for $s=0$ yield the corresponding
formulas for the generalized Kepler - Coulomb system.

\section{Interbasis expansions "parabola-sphere"}
\markboth{CHAPTER 5. THE GENERALIZED MIC-KEPLER PROBLEM}{5.3. INTERBASIS EXPANSIONS "PARABOLA-SPHERE"}

First, we note that for a fixed value of the principal number $n$, the radial wave function of the generalized MIK-Kepler
system (\ref{5.1.11}), like the radial wave functions of all systems considered by us in the previous chapters,
is orthogonal with respect to the orbital quantum number $j$ and this condition has the form:
\begin{eqnarray}
I_{j j'} = \int\limits_{0}^{\infty}\,R^{(s)}_{nj'}\left(r; \delta_{1}^{(s)}, \delta_{2}^{(s)}\right)\,
R^{(s)}_{nj}\left(r; \delta_{1}^{(s)}, \delta_{2}^{(s)}\right)dr =
\frac{2\varepsilon^{3}r_{0}}{2j + \delta_{1}^{(s)} + \delta_{2}^{(s)} +1}\,\delta_{j j'}.
\label{5.3.1}
\end{eqnarray}
Indeed, substituting into (\ref{5.3.1}) the explicit expressions for the radial wave functions according
to (\ref{5.1.11}), writing the confluent hypergeometric function in
$R^{(s)}_{nj}\left(r; \delta_{1}^{(s)}, \delta_{2}^{(s)}\right)$ as a finite sum, integrating according to
formula (\ref{1.6.5}) and taking into account (\ref{1.6.6}), we obtain
\begin{eqnarray*}
I_{j j'} &=& \frac{\Gamma\left(j + j' + \delta_{1}^{(s)} + \delta_{2}^{(s)} + 1\right)}
{\Gamma\left(2j + \delta_{1}^{(s)} + \delta_{2}^{(s)} +2\right)}\Biggl[
\frac{\Gamma\left(n + j + \delta_{1}^{(s)} + \delta_{2}^{(s)} +1\right)}{\left(n - j - 1\right)!
\left(n - j' - 1\right)!\Gamma\left(n + j' + \delta_{1}^{(s)} + \delta_{2}^{(s)} +1\right)}\Biggr]^{1/2}\times \\ [3mm]
&\times& 2\varepsilon^{3}r_{0}\,\sum\limits_{p=0}^{n-j-1}\,
\frac{\left(-n + j + 1\right)_{p}\left(j + j' + \delta_{1}^{(s)} + \delta_{2}^{(s)} +1\right)_{p}}
{p!\left(2j + \delta_{1}^{(s)} + \delta_{2}^{(s)} + 2\right)_{p}}\,\frac{\Gamma\left(n - j - p\right)}
{\Gamma\left(j' - j - p + 1\right)}.
\end{eqnarray*}
Further, using formula (\ref{1.6.8}), the sum over $p$ can be represented through the hypergeometric function
$_{2}F_{1}$ of the unit argument. Now, using formula (\ref{1.6.6}) again, we obtain
\begin{eqnarray}
I_{j j'} &=& \frac{2\varepsilon^{3}r_{0}}{j + j' + \delta_{1}^{(s)} + \delta_{2}^{(s)} +1}\,
\Biggl[\frac{\left(n - j - 1\right)!\Gamma\left(n + j + \delta_{1}^{(s)} + \delta_{2}^{(s)} +1\right)}
{\left(n - j' - 1\right)!\Gamma\left(n + j' + \delta_{1}^{(s)} + \delta_{2}^{(s)} +1\right)}\Biggr]^{1/2}
\times \nonumber
\\
\label{5.3.2}
\\
&\times& \frac{1}{\Gamma\left(j - j' + 1\right)\Gamma\left(j' - j + 1\right)}. \nonumber
\end{eqnarray}
Equation (\ref{5.3.1}) then easily follows from (\ref{5.3.2}) since
$\left[\Gamma\left(j - j' + 1\right)\Gamma\left(j' - j + 1\right)\right]^{-1} = \delta_{j j'}$.

Now we turn to the problem of interbasis expansions for the generalized MIC-Kepler system.

For fixed values ​​of the system energy $E_{n}^{(s)}$, the parabolic bound states (\ref{5.2.4}) as a
coherent quantum mixture the spherical bound states (\ref{5.1.15}):
\begin{eqnarray}
\psi_{n_{1} n_{2} m}^{(s)}\left(\mu, \nu, \varphi; \delta_{1}^{(s)}, \delta_{2}^{(s)}\right) =
\sum\limits_{j=m_{+}}^{n-1}\,W_{n_{1} n_{2} m s}^{j}\left(\delta_{1}^{(s)}, \delta_{2}^{(s)}\right)\,
\psi_{n j m}^{(s)}\left(r, \theta, \varphi; \delta_{1}^{(s)}, \delta_{2}^{(s)}\right).
\label{5.3.3}
\end{eqnarray}
To find the expansion coefficient $W_{n_{1} n_{2} m s}^{j}\left(\delta_{1}^{(s)}, \delta_{2}^{(s)}\right)$
of the parabolic basis in terms of the spherical basis of the generalized MIC-Kepler system, we proceed
in a similar manner to the calculation of the expansion coefficient "parabola - sphere" for the MIC-Kepler
problem (see section 3.2), i.e.:

1. In the left-hand of the expansion (\ref{5.3.3}) we pass from parabolic coordinates to spherical ones
according to Eqs. (\ref{3.2.3});

2. Then, by substituting  $\theta = 0$ into the so-obtained equation and by taking into account the formula
(\ref{1.7.4}), we get an expression that depends only on the variable $r$;

3.Further, using the condition of orthogonality of the radial wave function on the orbital momentum (\ref{5.3.1}), we obtain
\begin{eqnarray}
W_{n_{1} n_{2} m s}^{j}\left(\delta_{1}^{(s)}, \delta_{2}^{(s)}\right) =
\frac{\sqrt{\left(2j + \delta_{1}^{(s)} + \delta_{2}^{(s)} + 1\right)\left(j - m_{+}\right)!
}}
{\Gamma\left(m_{1} + 1\right)\Gamma\left(2j + \delta_{1}^{(s)} + \delta_{2}^{(s)} + 2\right)}\,
E_{n_{1} n_{2}}^{j m s}\,K_{j m s}^{n n_{1}},
\label{5.3.4}
\end{eqnarray}
where
\begin{eqnarray}
E_{n_{1} n_{2}}^{j m s} &=& \sqrt{\frac{\Gamma\left(j - m_{-}  + \delta_{1}^{(s)} + 1\right)}
{\left(n_{1}\right)!\left(n_{2}\right)!}}\times \nonumber
\\
\label{5.3.5}
\\
&\times& \Biggl[\frac{\Gamma\left(n_{1} + m_{1} + 1\right)
\Gamma\left(n_{2} + m_{2} + 1\right)\Gamma\left(n + j + \delta_{1}^{(s)} + \delta_{2}^{(s)} + 1\right)}
{\left(n - j - 1\right)!
\Gamma\left(j + m_{-}  + \delta_{2}^{(s)} + 1\right)
\Gamma\left(j + m_{+}  + \delta_{1}^{(s)} + \delta_{2}^{(s)} + 1\right)}\Biggr]^{1/2},
\nonumber
\end{eqnarray}
and
\begin{eqnarray*}
K_{j m s}^{n n_{1}} = \int\limits_{0}^{\infty}\,e^{-x}x^{j + m_{+} +\delta_{1}^{(s)} + \delta_{2}^{(s)}}
F\left(-n_{1}; m_{1} + 1; x\right)F\left(-n + j + 1; 2j + \delta_{1}^{(s)} + \delta_{2}^{(s)} +2 ; x\right)dx,
\end{eqnarray*}
where $x=2\varepsilon r$. To calculate the integral $K_{j m s}^{n n_{1}}$, it is sufficient to write the confluent
hypergeometric function $F\left(-n_{1}; m_{1} + 1; x\right)$ as a series, integrate according to (\ref{1.6.5}),
and use formula (\ref{1.6.6} for the summation of the hypergeometric function $_{2}F_{1}$.We thus obtain
\begin{eqnarray}
K_{j m s}^{n n_{1}} &=& \frac{\left(n - m_{+} - 1\right)!\Gamma\left(2j + \delta_{1}^{(s)} + \delta_{2}^{(s)} + 2\right)
\Gamma\left(j + m_{+}  + \delta_{1}^{(s)} + \delta_{2}^{(s)} + 1\right)}{\left(j - m_{+}\right)!
\Gamma\left(n + j + \delta_{1}^{(s)} + \delta_{2}^{(s)} + 1\right)} \times \nonumber
\\
\label{5.3.6}
\\
&\times& {_3F_2}\left\{\matrix{ -n_{1},\,\,-j + m_{+},\,\, j + m_{+}  + \delta_{1}^{(s)} + \delta_{2}^{(s)} + 1\cr \cr
m_{1} + 1,\,\, -n + m_{+} + 1 \cr}\Biggr|1\right\}.
\nonumber
\end{eqnarray}
The introduction of (\ref{5.3.5}) and (\ref{5.3.6}) into (\ref{5.3.4}) gives
\begin{eqnarray}
&&W_{n_{1} n_{2} m s}^{j}\left(\delta_{1}^{(s)}, \delta_{2}^{(s)}\right) =  \sqrt{\frac{\left(2j + \delta_{1}^{(s)} + \delta_{2}^{(s)} + 1\right)
\Gamma\left(n_{1} + m_{1} + 1\right)\Gamma\left(n_{2} + m_{2} + 1\right)}{\left(n_{1}\right)!\left(n_{2}\right)!\left(n - j - 1\right)!
\left(j - m_{+}\right)!\Gamma\left(j + m_{-}  + \delta_{2}^{(s)} + 1\right)}}\times \nonumber \\ [3mm]
&\times& \frac{\left(n - m_{+} - 1\right)!}{\Gamma\left(m_{1} + 1\right)}
\sqrt{\frac{\Gamma\left(j - m_{-}  + \delta_{1}^{(s)} + 1\right)\Gamma\left(j + m_{+} + \delta_{1}^{(s)} + \delta_{2}^{(s)} + 1\right)}
{\Gamma\left(n + j + \delta_{1}^{(s)} + \delta_{2}^{(s)} + 1\right)}}\times
\label{5.3.7} \\ [3mm]
&\times&
{_3F_2}\left\{\matrix{ -n_{1},\,\,-j + m_{+},\,\, j + m_{+}  + \delta_{1}^{(s)} + \delta_{2}^{(s)} + 1\cr \cr
m_{1} + 1,\,\, -n + m_{+} + 1 \cr}\Biggr|1\right\}.
\nonumber
\end{eqnarray}
Finally, comparing (\ref{1.7.12}) and (\ref{5.3.7}), we obtain the representation
\begin{eqnarray}
W_{n_{1} n_{2} m s}^{j}\left(\delta_{1}^{(s)}, \delta_{2}^{(s)}\right) =
(-1)^{n_1} C_{\frac{n_1 + n_2 + m_{2}}{2},\frac{n_2 - n_1 + m_{2}}{2};
\frac{n_1 + n_2 + m_{1}}{2},\frac{n_1 - n_2 + m_{1}}{2}}^
{j + \frac{\delta_{1}^{(s)} + \delta_{2}^{(s)}}{2}, \,\frac{m_{1} + m_{2}}{2}}.
\label{5.3.8}
\end{eqnarray}
Equation (\ref{5.3.8}) proves that the coefficients for the expansion of the parabolic basis in terms of the spherical
basis are nothing but the analytical continuation, for real values of their arguments, of the $SU(2)$ Clebsch--Gordan coefficients.

The inverse of Eq. (\ref{5.3.3}), namely
\begin{eqnarray}
\psi_{n j m}^{(s)}\left(r, \theta, \varphi; \delta_{1}^{(s)}, \delta_{2}^{(s)}\right) =
\sum\limits_{n_{1}=0}^{n-m_{+}-1}\,\tilde{W}_{n_{1} n_{2} m s}^{j}\left(\delta_{1}^{(s)}, \delta_{2}^{(s)}\right)\,
\psi_{n_{1} n_{2} m}^{(s)}\left(\mu, \nu, \varphi; \delta_{1}^{(s)}, \delta_{2}^{(s)}\right),
\label{5.3.9}
\end{eqnarray}
is an immediate consequence of the orthonormality property of the $SU(2)$ Clebsch–Gordan coefficients.
The expansion coefficients in (\ref{5.3.9}) are thus given by
\begin{eqnarray}
{\tilde W}_{n j ms}^{n_1}\left(\delta_{1}^{(s)}, \delta_{2}^{(s)}\right) = (-1)^{n_1}
C_{\frac{n + m_{-} + \delta_{2}^{(s)} - 1}{2},\frac{n + m_{-} + \delta_{2}^{(s)} - 1}{2} - n_{1};
\frac{n - m_{-} + \delta_{1}^{(s)} - 1}{2}, n_{1} - \frac{n - m_{+} - |m-s| + \delta_{1}^{(s)} - 1}{2}}
^{j + \frac{\delta_{1}^{(s)} + \delta_{2}^{(s)}}{2},\,
\frac{m_{1} + m_{2}}{2}}
\label{5.3.10}
\end{eqnarray}
and may be expressed in terms of the $_{3}F_{2}$ function through (\ref{1.7.10}) or (\ref{1.7.12}).

\section{Prolate spheroidal basis of the GMICK problem}
\markboth{CHAPTER 5. THE GENERALIZED MIC-KEPLER PROBLEM}{5.4. PROLATE SPHEROIDAL BASIS OF THE GMICK PROBLEM}

After the substitution
\begin{eqnarray*}
\psi(\xi, \eta, \varphi) = \Phi_{1}(\xi)\,\Phi_{2}(\eta)\,\frac{e^{i(m - s)}\varphi}{\sqrt{2\pi}},
\end{eqnarray*}
the variables in the Schr\"{o}dinger equation for the generalized MIC-Kepler problem are separated
\begin{eqnarray}
\left[\frac{d}{d\xi}(\xi^2-1)\frac{d}{d\xi} +
\frac{m_{1}^2}{2(\xi+1)}- \frac{m_{2}^2}{2(\xi -1)} +
\frac{ER^2} {2r_{0}e^{2}}(\xi^2-1) + \frac{R}{r_0}\xi
-\Lambda^{(s)}\right]\Phi_{1}(\xi) = 0,
\label{5.4.1}
\end{eqnarray}
\begin{eqnarray}
\left[\frac{d}{d\eta}(1-\eta^2)\frac{d}{d\eta} -
\frac{m_{1}^2}{2(1+\eta)}- \frac{m_{2}^2}{2(1-\eta)} +
\frac{ER^2} {2r_{0}e^{2}}(1-\eta^2) - \frac{R}{r_0}\eta +
\Lambda^{(s)}\right]\Phi_{2}(\eta) = 0,
\label{5.4.2}
\end{eqnarray}
where $\Lambda^{(s)}(R)$ is a separation constant in prolate spheroidal coordinates.
By eliminating the energy $E$ from Eqs. (\ref{5.4.1}) and (\ref{5.4.2}), we produce the operator
\begin{eqnarray}
&&\hat \Lambda^{(s)} =
\frac{1}{\xi^2-\eta^2}\left[(1-\eta^2)\frac{\partial}{\partial
\xi} (\xi^2-1)\frac{\partial}{\partial \xi}- (\xi^2-1)
\frac{\partial}{\partial \eta}(1-\eta^2)\frac{\partial}{\partial
\eta}\right] + \frac{R}{r_0}\frac{\xi\eta+1}{\xi+\eta} - \nonumber \\ [3mm]
&-&\frac{\xi^2+\eta^2-2}{(\xi^2-1)(1-\eta^2)}
\frac{\partial^2}{\partial \varphi^2} +
\frac{2s}{(\xi - 1)(1-\eta)}
\left(\xi+\eta-1-\frac{\xi\eta+1}{\xi+\eta}
\right)\left(s - i\frac{\partial}{\partial \varphi}\right) +
\label{5.4.3} \\ [3mm]
&+& 2\Lambda_{1}\frac{(\xi + \eta)^2 + (\xi - 1)(1 - \eta)}{(\xi + \eta)(\xi + 1)(1 + \eta)}
+ 2\Lambda_{2}\frac{(\xi + \eta)^2 - (\xi + 1)(1 + \eta)}{(\xi + \eta)(\xi - 1)(1 - \eta)},
\nonumber
\end{eqnarray}
the eigenvalues of which are $\Lambda^{(s)}(R)$ and the eigenfunctions
of which are $\psi(\xi, \eta, \varphi)$. The significance of the
self-adjoint operator $\hat \Lambda^{(s)}$ can be found by switching to
Cartesian coordinates. Passing to Cartesian coordinates in (\ref{5.4.3})
and taking (\ref{5.1.19}) and (\ref{5.2.8}) into account, we obtain
\begin{eqnarray}
\hat{\Lambda}^{(s)} = \hat{M}^{(s)} + \frac{\sqrt{R\mu_{0}}}{\hbar}\hat{\Omega}^{(s)}.
\label{5.4.4}
\end{eqnarray}
Therefore,
\begin{eqnarray}
\hat{\Lambda}^{(s)}\,\psi_{nqm}^{(s)}\left(\xi, \eta, \varphi; R, \delta_{1}^{(s)}, \delta_{2}^{(s)}\right) =
\Lambda^{(s)}(R)\,\psi_{nqm}^{(s)}\left(\xi, \eta, \varphi; R, \delta_{1}^{(s)}, \delta_{2}^{(s)}\right),
\label{5.4.5}
\end{eqnarray}
where index $q$ labels the eigenvalues of the operator $\hat{\Lambda}^{(s)}$
and varies in the range $0 \leq q \leq n - m_{+} - 1$.

We are now ready to deal with the interbasis expansions
\begin{eqnarray}
\psi_{nqm}^{(s)}\left(\xi,\eta,\varphi; R, \delta_{1}^{(s)}, \delta_{2}^{(s)}\right)
&=& \sum_{j= m_+}^{n-1}
V_{nqms}^j \left(R\right)
\psi_{nqm}^{(s)}\left(r,\theta,\varphi; \delta_{1}^{(s)}, \delta_{2}^{(s)}\right),
\label{5.4.6}
\end{eqnarray}
\begin{eqnarray}
\psi_{nqm}^{(s)}\left(\xi,\eta,\varphi;R;\delta_{1}^{(s)}, \delta_{2}^{(s)}\right)
&=& \sum_{n_1=0}^{n-m_+ -1}
U_{nqms}^{n_1} \left(R\right)\,
\psi_{n_1n_2m}^{(s)}\left(\mu,\nu,\varphi; \delta_{1}^{(s)}, \delta_{2}^{(s)}\right)
\label{5.4.7}
\end{eqnarray}
for the prolate spheroidal basis in terms of the spherical and parabolic bases.

First, we consider Eq. (\ref{5.4.6}). Let the operator $\hat{\Lambda}^{(s)}$ act
on both sides of (\ref{5.4.6}). Then, by using Eqs. (\ref{5.4.5}), (\ref{5.4.6}),
and (\ref{5.1.16}) as well as the orthonormality property of the spherical basis, we find that
\begin{eqnarray}
\left[\Lambda_{q}^{(s)}(R)-\left(j + \frac{\delta_{1}^{(s)} + \delta_{2}^{(s)}}{2}\right)
\left(j + \frac{\delta_{1}^{(s)} + \delta_{2}^{(s)}}{2} + 1\right)\right]V_{nqms}^{j}(R)
= \nonumber \\
\label{5.4.8}
\\
= \frac{R\sqrt{\mu_0}}{\hbar}
\sum_{j'}\,V_{nqms}^{j'}(R) \left(\hat{\Omega}^{(s)}\right)_{j j'},
\nonumber
\end{eqnarray}
where
\begin{eqnarray}
\left(\hat{\Omega}^{(s)}\right)_{j j'} = \int \psi_{nj
m}^{(s)*}(r,\theta,\varphi)\hat{\Omega}^{(s)} \psi_{nj'm}^{(s)}
(r,\theta,\varphi) dV.
\label{5.4.9}
\end{eqnarray}
The calculation of the matrix element $\left(\hat{\Omega}^{(s)}\right)_{j j'}$ can
be done by expanding the basis in (\ref{5.4.9}) in terms of parabolic wave functions
[see Eq. (\ref{3.2.9})] and by making use of the eigenvalue equation for
$\hat{\Omega}^{(s)}$ [see the second equation of the formula (\ref{5.2.10})].
This leads to
\begin{eqnarray*}
\left(\hat{\Omega}^{(s)}\right)_{j j'} = \frac{2\hbar}{r_{0}\sqrt{\mu_{0}}}
\frac{1}{2n + \delta_{1}^{(s)} + \delta_{2}^{(s)}}\sum\limits_{n_{1} = 0}^{n - m_{+} - 1}
\left(2n_{1} - n + |m - s| + \frac{\delta_{1}^{(s)} - \delta_{2}^{(s)}}{2} + 1\right)
{\tilde W}_{n j ms}^{n_1}{\tilde W}_{n j' ms}^{n_1}.
\end{eqnarray*}
Then, by using Eq. (\ref{5.3.10}) together with the recursion relation (\ref{1.8.10})
and and the orthonormality condition (\ref{1.8.11}) we find that $\left(\hat{\Omega}^{(s)}\right)_{j j'}$
is given by
\begin{eqnarray}
\left(\hat{\Omega}^{(s)}\right)_{j j'} &=& \frac{2\hbar}{r_{0}\sqrt{\mu_{0}}}
\Biggl[\frac{\left(m_{1} + m_{2}\right)\left(m_{1} - m_{2}\right)}
{\left(2j + \delta_{1}^{(s)} + \delta_{2}^{(s)}\right)\left(2j + \delta_{1}^{(s)} + \delta_{2}^{(s)} + 2\right)}
\delta_{j' j} + \nonumber \\
\label{5.4.10}
\\
&+& \frac{4}{2n + \delta_{1}^{(s)} + \delta_{2}^{(s)}}\left(A_{n m}^{j + 1}\delta_{j', j + 1} +
A_{n m}^{j}\delta_{j', j - 1}\right)
\Biggr], \nonumber
\end{eqnarray}
where
\begin{eqnarray*}
A_{n m}^{j} &=& - \frac{\sqrt{\left(n - j\right)\left(n + j + \delta_{1}^{(s)} + \delta_{2}^{(s)}\right)}}
{\left(2j + \delta_{1}^{(s)} + \delta_{2}^{(s)}\right)} \times \nonumber \\ [4mm]
&\times& \Biggl[\frac{\left(j - m_{+}\right)\left(j + m_{+} + \delta_{1}^{(s)} + \delta_{2}^{(s)}\right)
\left(j - m_{-} + \delta_{1}^{(s)}\right)\left(j + m_{-} + \delta_{2}^{(s)}\right)}{
\left(2j + \delta_{1}^{(s)} + \delta_{2}^{(s)} - 1\right)\left(2j + \delta_{1}^{(s)} + \delta_{2}^{(s)} + 1\right)}
\Biggr]^{1/2}.
\end{eqnarray*}

Now, by introducing (\ref{5.4.10}) into (\ref{5.4.8}), we get the
following three-term recursion relation for the coefficient $V_{nqms}^j(R)$:
\begin{eqnarray}
\Biggl[\Lambda_{q}^{(s)}(R)-\left(j + \frac{\delta_{1}^{(s)} + \delta_{2}^{(s)}}{2}\right)
\left(j + \frac{\delta_{1}^{(s)} + \delta_{2}^{(s)}}{2} + 1\right) - \nonumber \\ [3mm]
- \frac{R\left(m_{1} + m_{2}\right)\left(m_{1} - m_{2}\right)}
{r_{0}\left(2j + \delta_{1}^{(s)} + \delta_{2}^{(s)}\right)\left(2j + \delta_{1}^{(s)} + \delta_{2}^{(s)} + 2\right)}
\Biggr]V_{nqms}^{j}(R) - \\ [3mm]
\label{5.4.11}
- \frac{4R}{r_{0}\left(2n + \delta_{1}^{(s)} + \delta_{2}^{(s)}\right)}
\Biggl[A_{n m}^{j + 1}V_{nqms}^{j + 1}(R) + A_{n m}^{j}V_{nqms}^{j - 1}(R)\Biggr] = 0.
\nonumber
\end{eqnarray}
The recursion relation (\ref{5.4.11}) provides us with a system of $n − m_{+}$ linear homogeneous equations
which can be solved by taking into account the normalization condition
\begin{eqnarray*}
\sum\limits_{j = m_{+}}^{n - 1}\left|V_{nqms}^{j}(R)\right|^{2} = 1.
\end{eqnarray*}
The eigenvalues $\Lambda_{q}^{(s)}(R)$ of the operator $\hat{\Lambda}^{(s)}$ then follow
from the vanishing of the determinant for the latter system.

Second, let us concentrate on the expansion (\ref{5.4.7}) of the prolate spheroidal basis in terms
of the parabolic basis. By employing a technique similar to the one used for deriving Eq. (\ref{5.4.8}), we get
\begin{eqnarray}
\left[\Lambda_{q}^{(s)}(R) - \frac{R}{r_{0}}\frac{2n_1 - 2n_2 + m_{1} - m_{2}}{2n + \delta_{1}^{(s)} + \delta_{2}^{(s)}}
\right]U_{nqms}^{n_1}(R) =
\sum_{n_1'}\,U_{nqms}^{n_1'}(R)
\left({\hat M}^{(s)}\right)_{n_1 n_1'},
\label{5.4.12}
\end{eqnarray}
where
\begin{eqnarray*}
\left({\hat M}^{(s)}\right)_{n_1 n_1'} =
\int \psi_{n_1n_2m}^{(s)*}\left(\mu,\nu,\varphi; \delta_{1}^{(s)}, \delta_{2}^{(s)}\right)
{\hat M}^{(s)} \psi_{n_1' n_2' m}^{(s)}\left(\mu,\nu,\varphi; \delta_{1}^{(s)}, \delta_{2}^{(s)}\right)dV.
\end{eqnarray*}
The matrix elements $\left({\hat M}^{(s)}\right)_{n_1 n_1'}$ can be calculated in the
same way as $\left(\hat{\Omega}^{(s)}\right)_{j j'}$ except that now we must use the
relation (\ref{1.8.15}) and the orthonormality condition (\ref{1.8.16}) permits deriving
the formula for the matrix element $\left({\hat M}^{(s)}\right)_{n_1 n_1'}$:
\begin{eqnarray}
\left({\hat M}^{(s)}\right)_{n_1 n_1'} = \left[\left(n + \frac{\delta_{1}^{(s)} + \delta_{2}^{(s)}}{2}\right)
\left(n + \frac{\delta_{1}^{(s)} + \delta_{2}^{(s)}}{2} - 1\right) - n_{1}\left(n_{1} + m_{1}\right)
- n_{2}\left(n_{2} + m_{2}\right)\right]\delta_{n'_1 n_1} - \nonumber \\
\label{5.4.13}
\\
- \sqrt{n_{2}\left(n_{1} + 1\right)\left(n_{1} + m_{1} + 1\right)\left(n_{2} + m_{2}\right)}\delta_{n'_1, n_1 + 1}
- \sqrt{n_{1}\left(n_{2} + 1\right)\left(n_{1} + m_{1}\right)\left(n_{2} + m_{2} + 1\right)}\delta_{n'_1, n_1 - 1}.
\nonumber
\end{eqnarray}
Finally, the introduction of (\ref{5.4.13}) into (\ref{5.4.12}) leads to the three-term recursion relation
\begin{eqnarray}
&&\Biggl[\left(n + \frac{\delta_{1}^{(s)} + \delta_{2}^{(s)}}{2}\right)
\left(n + \frac{\delta_{1}^{(s)} + \delta_{2}^{(s)}}{2} - 1\right) - n_{1}\left(n_{1} + m_{1}\right)
- n_{2}\left(n_{2} + m_{2}\right) +  \nonumber \\ [3mm]
&+& \frac{R}{r_{0}}\frac{2n_1 - 2n_2 + m_{1} - m_{2}}{2n + \delta_{1}^{(s)} + \delta_{2}^{(s)}}
- \Lambda_{q}^{(s)}(R)\Biggr]U_{nqms}^{n_1}(R) = \nonumber \\
\label{5.4.14}
\\
&=& \sqrt{n_{2}\left(n_{1} + 1\right)\left(n_{1} + m_{1} + 1\right)\left(n_{2} + m_{2}\right)}U_{nqms}^{n_1 + 1}(R) +
\nonumber \\ [3mm]
&+&  \sqrt{n_{1}\left(n_{2} + 1\right)\left(n_{1} + m_{1}\right)\left(n_{2} + m_{2} + 1\right)}U_{nqms}^{n_1 - 1}(R)
\nonumber
\end{eqnarray}
for the expansion coefficients $U_{nqms}^{n_1}(R)$. This relation can be iterated by taking
account of the normalization condition
\begin{eqnarray*}
\sum\limits_{n_{1} = 0}^{n -  m_{+} - 1}\left|U_{nqms}^{n_1}(R)\right|^{2} = 1.
\end{eqnarray*}
Here again, the eigenvalues $\Lambda_{q}^{(s)}(R)$ may be obtained by solving a system of
$n - m_{+}$ linear homogeneous equations.

It should be mentioned that formulas (\ref{5.3.8}) and (\ref{5.3.10}) and three-term recursion
relations (\ref{5.4.11}) and (\ref{5.4.14}) generalize the analogous results for the following systems:

$\bullet$ hydrogen atom \cite{PARK, TARTER, COJO, ARUT1, M-1, M-4a}, when $s = \delta_{1}^{(s)} = \delta_{2}^{(s)} = 0$;

$\bullet$ generalized Kepler–Coulomb system, when $s = 0, \delta_{1}^{(s)} \neq \delta_{2}^{(s)} \neq 0$;

$\bullet$ Hartmann system \cite{Lutsenko-1}, when $s = 0, \delta_{1}^{(s)} = \delta_{2}^{(s)} \neq 0$;

$\bullet$ MIC-Kepler system, when $s \neq 0, \delta_{1}^{(s)} = \delta_{2}^{(s)} = 0$.

Finally, it should be noted that the following four limits
\begin{eqnarray*}
\lim\limits_{R \to 0}\,V_{nqms}^j \left(R\right) = \delta_{jq}, \qquad
\lim\limits_{R \to \infty}\,V_{nqms}^j \left(R\right) = W_{n_{1} n_{2} m s}^{j}\left(\delta_{1}^{(s)}, \delta_{2}^{(s)}\right), \\ [3mm]
\lim\limits_{R \to 0}\, U_{nqms}^{n_1}(R) = \tilde{W}_{n_{1} n_{2} m s}^{j}\left(\delta_{1}^{(s)}, \delta_{2}^{(s)}\right), \qquad
\lim\limits_{R \to \infty}\,U_{nqms}^{n_1}(R) = \delta_{n_1 q}
\end{eqnarray*}
furnish a useful means for checking the calculations presented in Sections 5.3 and 5.4.

\section{Interbasis expansions in the continuous spectrum}
\markboth{CHAPTER 5. THE GENERALIZED MIC-KEPLER PROBLEM}{5.5. INTERBASIS EXPANSIONS IN THE CONTINUOUS SPECTRUM}

The spherical wave function of the generalized MIC-Kepler problem in the continuous spectrum
can be written in the following form
\begin{eqnarray}
\psi_{n j m}^{(s)} \equiv \psi_{n j m}^{(s)}\left(r, \theta, \varphi; \delta_{1}^{(s)}, \delta_{2}^{(s)}\right) =
R_{n j m}^{(s)}\left(r; \delta_{1}^{(s)}, \delta_{2}^{(s)}\right)\,
Z^{(s)}_{jm}\left(\theta, \varphi;\delta_{1}^{(s)},\delta_{2}^{(s)}\right),
\end{eqnarray}
where the explicit form of the Tamm ring-shaped monopole harmonic
$Z^{(s)}_{jm}\left(\theta, \varphi;\delta_{1}^{(s)},\delta_{2}^{(s)}\right)$
is given by the formula (\ref{5.1.5}). Normalized by the condition
\begin{eqnarray*}
\int\limits_{0}^{\infty} r^{2} R^{(s)}_{kj}\left(r; \delta_{1}^{(s)}, \delta_{2}^{(s)}\right)
R^{(s)*}_{k'j}\left(r; \delta_{1}^{(s)}, \delta_{2}^{(s)}\right) dr
=2\pi\delta\left(k-k'\right)
\end{eqnarray*}
the radial wave function of the generalized MIC-Kepler system has the form
\begin{eqnarray}
R^{(s)}_{kj}\left(r; \delta_{1}^{(s)}, \delta_{2}^{(s)}\right)&=&C^{(s)}_{kj}\left(\delta_{1}^{(s)}, \delta_{2}^{(s)}\right)
\frac{\left(2ikr\right)^{j+\frac{\delta_{1}^{(s)}+ \delta_{2}^{(s)}}{2}}}{\Gamma\left(2j+\delta_{1}^{(s)}+ \delta_{2}^{(s)}+2\right)}
e^{-ikr} \times \nonumber \\
\label{5.5.2}
\\
&\times&
F\left(j+\frac{\delta_{1}^{(s)}+ \delta_{2}^{(s)}}{2}+1+\frac{i}{kr_{0}}; 2j+\delta_{1}^{(s)}+ \delta_{2}^{(s)}+2; 2ikr\right),
\nonumber
\end{eqnarray}
where $k = \sqrt{2\mu_{0} E}/\hbar$, and the normalization constant $C^{(s)}_{kj}\left(\delta_{1}^{(s)}, \delta_{2}^{(s)}\right)$ is equal to
\begin{eqnarray}
 C^{(s)}_{kj}\left(\delta_{1}^{(s)}, \delta_{2}^{(s)}\right)=2k e^{\pi /2kr_{0}}
\left|\Gamma\left(j+\frac{\delta_{1}^{(s)} + \delta_{2}^{(s)}}{2}+1-\frac{i}{kr_{0}}\right)\right|.
\label{5.5.3}
\end{eqnarray}

The parabolic basis normalized by the condition
\begin{eqnarray*}
\int \psi^{(s)}_{k \Omega^{(s)} m}\left(\mu, \nu, \varphi; \delta_{1}^{(s)}, \delta_{2}^{(s)}\right)
\psi^{(s)*}_{k' \Omega^{(s)'} m'}\left(\mu, \nu, \varphi; \delta_{1}^{(s)}, \delta_{2}^{(s)}\right) dV =
4\pi \delta \left(k-k'\right) \delta \left(\Omega^{(s)}-\Omega^{(s)'}\right) \delta_{mm'}
\end{eqnarray*}
has the form
\begin{eqnarray}
\psi^{(s)}_{k \Omega^{(s)} m}\left(\mu, \nu, \varphi; \delta_{1}^{(s)}, \delta_{2}^{(s)}\right)
= C^{(s)}_{k \Omega^{(s)} m}\left(\delta_{1}^{(s)}, \delta_{2}^{(s)}\right)\Phi^{(s)}_{k \Omega^{(s)} m_{1}}(\mu)
\Phi^{(s)}_{k, -\Omega^{(s)}, m_{2}}(\nu) \frac{e^{i(m-s)\varphi}}{\sqrt{2\pi}},
\label{5.5.4}
\end{eqnarray}
where
\begin{eqnarray}
\Phi^{(s)}_{k \Omega^{(s)} q}(x)= e^{-ikx/2}\frac{(ikx)^{q/2}}{\Gamma(q+1)}
F\left(\frac{q+1}{2} + \frac{i}{2kr_{0}}+i\frac{\sqrt{\mu_{0}}}{2k\hbar}\Omega^{(s)}; q+1; ikx\right),
\label{5.5.5}
\end{eqnarray}
and the normalization constant is equal to
\begin{eqnarray}
C^{(s)}_{k \Omega^{(s)} m}\left(\delta_{1}^{(s)}, \delta_{2}^{(s)}\right) =
\sqrt{\frac{k\sqrt{\mu_{0}}}{2\pi \hbar}}e^{\pi /2kr_{0}} \times \nonumber \\
\label{5.5.6} \\
\times \left|\Gamma\left(\frac{m_{1}+1}{2} - \frac{i}{2kr_{0}} - i\frac{\sqrt{\mu_{0}}}{2k\hbar}\Omega^{(s)}\right)
\Gamma\left(\frac{m_{2}+1}{2}-\frac{i}{2kr_{0}}+i\frac{\sqrt{\mu_{0}}}{2k\hbar}\Omega^{(s)}\right)\right|.
\nonumber
\end{eqnarray}
We note that to calculate normalization factors (\ref{5.5.3}) and (\ref{5.5.3}), we used the asymptotic
representation of the confluent hypergeometric function (\ref{1.13.3}).

Now let's proceed to the problem of interbasis expansions.
First, let us consider the expansion of a parabolic basis in terms of a spherical one.
We write the sought expansion at fixed values of the energy in the form
\begin{eqnarray}
\psi^{(s)}_{k \Omega^{(s)} m}\left(\mu, \nu, \varphi; \delta_{1}^{(s)}, \delta_{2}^{(s)}\right)=
\sum_{j=m_{+}}^{\infty} W^{j}_{k \Omega^{(s)} m s}\left(\delta_{1}^{(s)}, \delta_{2}^{(s)}\right)
\psi^{(s)}_{k j m}\left(r, \theta, \varphi; \delta_{1}^{(s)}, \delta_{2}^{(s)}\right).
\label{5.5.7}
\end{eqnarray}
Multiplying both sides of expansion (\ref{5.5.7}) by
$\sin\theta Z^{(s)}_{jm}\left(\theta, \varphi;\delta_{1}^{(s)},\delta_{2}^{(s)}\right)$, we integrate
over the angle $\theta$ and use orthonormalization condition (\ref{5.1.4}).
We next represent the confluent hypergeometric function contained in the parabolic basis as the
series (\ref{2.10.2}), and pass from parabolic coordinates to to spherical coordinates using the
relation (\ref{3.2.3}). Instead of (\ref{5.5.7}), we then obtain
\begin{eqnarray}
&&W^{j}_{k\beta m s}\left(\delta_{1}^{(s)}, \delta_{2}^{(s)}\right)
F\left(j+\frac{\delta_{1}^{(s)}+ \delta_{2}^{(s)}}{2}+1+i\frac{\alpha}{k}; 2j+\delta_{1}^{(s)}+ \delta_{2}^{(s)}+2; 2ikr\right)=
\frac{N^{(s)}_{jm}\left(\delta_{1}, \delta_{2}\right)}{2^{j+\frac{\delta_{1}+ \delta_{2}}{2}}}
\times \nonumber \\
\label{5.5.8} \\
&\times& \frac{\Gamma\left(2j+\delta_{1}^{(s)}+ \delta_{2}^{(s)}+2\right)}{\Gamma\left(m_{1}+1\right)\Gamma\left(m_{2}+1\right)}
\frac{C^{(s)}_{k\beta m}\left(\delta_{1}^{(s)}, \delta_{2}^{(s)}\right)}{C^{(s)}_{k j}\left(\delta_{1}^{(s)}, \delta_{2}^{(s)}\right)}
\sum_{p=0}^{\infty}\sum_{t=0}^{\infty} \frac{(u)_{p}(v)_{t}}{p! t!}\frac{(ikr)^{p+t-j+m_{+}}}{\left(m_{1}+1\right)_{p}
\left(m_{2}+1\right)_{t}}Q^{pt}_{jms}, \nonumber
\end{eqnarray}
where
\begin{eqnarray*}
u = \frac{m_{1}+1}{2}+\frac{i}{2kr_{0}} + i\frac{\sqrt{\mu_{0}}}{2k\hbar}\Omega^{(s)}, \qquad
v = \frac{m_{2}+1}{2}+\frac{i}{2kr_{0}} - i\frac{\sqrt{\mu_{0}}}{2k\hbar}\Omega^{(s)},
\end{eqnarray*}
\begin{eqnarray*}
Q^{pt}_{jms}=\int\limits_{0}^{\pi} \sin\theta (1+ \cos\theta)^{p+m_{1}} (1- \cos\theta)^{t+m_{2}}
P^{(m_{1}, m_{2}}_{j-m_{+}}(\cos\theta)d\theta.
\end{eqnarray*}
Further, using the Rodrigues formula for Jacobi polynomials (\ref{2.10.4}), and successively integrating by
parts $j - m_{+}$ times, we see that the integral is nonzero only under the condition $p + t -j + m_{+} \geq 0$,
and hence all terms of series (\ref{5.5.8}) contain $r$ to a nonnegative power; therefore,
in the limit $r \to 0$,
\begin{eqnarray}
&&W^{j}_{k \Omega^{(s)} m s}\left(\delta_{1}^{(s)}, \delta_{2}^{(s)}\right)
= \frac{N^{(s)}_{jm}\left(\delta_{1}^{(s)}, \delta_{2}^{(s)}\right)}{2^{j+\frac{\delta_{1}^{(s)}+ \delta_{2}^{(s)}}{2}}}
\frac{\Gamma\left(2j+\delta_{1}^{(s)}+ \delta_{2}^{(s)}+2\right)}{\Gamma\left(m_{1}+1\right)\Gamma\left(m_{2}+1\right)}
\times \nonumber \\
\label{5.5.9}
\\
&\times&
\frac{C^{(s)}_{k \Omega^{(s)} m}\left(\delta_{1}^{(s)}, \delta_{2}^{(s)}\right)}{C^{(s)}_{k j}\left(\delta_{1^{(s)}}, \delta_{2}^{(s)}\right)}
\sum_{p=0}^{j-m_{+}} \frac{(u)_{p}(v)_{j-m_{+}-p}}{p! (j-m_{+}-p)!}\frac{Q^{p,j-m_{+}-p}_{jms}}{\left(m_{1}+1\right)_{p}
\left(m_{2}+1\right)_{j-m_{+}-p}}. \nonumber
\end{eqnarray}
For $t = j - m_{+} - p$, the integral $Q^{pt}_{jms}$ becomes a closed expression
\begin{eqnarray*}
Q^{p,j-m_{+}-p}_{jms} =(-1)^{j-m_{+}-p} 2^{j+m_{+}\delta_{1}^{(s)}+ \delta_{2}^{(s)}+1}
\frac{\Gamma\left(j-m_{-}+\delta_{1}^{(s)}+1\right) \Gamma\left(j+m_{-}+\delta_{2}^{(s)}+1\right)}
{\Gamma\left(2j+\delta_{1}^{(s)}+ \delta_{2}^{(s)}+2\right)}.
\end{eqnarray*}
Substituting the last expression in formula (\ref{5.5.9}) and using the auxiliary equalities
\begin{eqnarray*}
(v)_{j-m_{+}-p} = \frac{(-1)^{p}(v)_{j-m_{+}}}{(1-j+m_{+}-v)_{p}}, \qquad
(j - m_{+} - p)! = \frac{(-1)^{p} (j-m_{+})!}{(-j+m_{+})_{p}},
\end{eqnarray*}
\begin{eqnarray*}
\left(m_{2}+1\right)_{j-m_{+}-p} = \frac{(-1)^{p}\Gamma\left(j+m_{-}+\delta_{2}^{(s)}+1\right)}
{\Gamma\left(m_{2}+1\right) \left(-j-m_{-}-\delta_{2}\right)_{p}},
\end{eqnarray*}
as well as the explicit form of the normalization factor $N_{jm}^{(s)}\left(\delta_{1}^{(s)}, \delta_{2}^{(s)}\right)$
in (\ref{5.1.7}), we obtain the expansion coefficients $W^{j}_{k\beta m s}\left(\delta_{1}^{(s)}, \delta_{2}^{(s)}\right)$
in the form
\begin{eqnarray*}
&&W^{j}_{k \Omega^{(s)} m s}\left(\delta_{1}^{(s)}, \delta_{2}^{(s)}\right) =
\frac{(-1)^{j-m_{+}}(v)_{j-m_{+}}}{\Gamma\left(m_{1}+1\right)}
\frac{C^{(s)}_{k \Omega^{(s)} m}\left(\delta_{1}^{(s)}, \delta_{2}^{(s)}\right)}
{C^{(s)}_{k j}\left(\delta_{1}^{(s)}, \delta_{2}^{(s)}\right)} \times \nonumber \\ [4mm]
&\times& \left[\frac{2(2j+\delta_{1}^{(s)}+ \delta_{2}^{(s)}+1)\Gamma\left(j+m_{+}+\delta_{1}^{(s)}+ \delta_{2}^{(s)}+1\right)
\Gamma\left(j-m_{-}+\delta_{1}^{(s)}+1\right)}{(j-m_{+})! \Gamma\left(j+m_{-}+\delta_{2}^{(s)}+1\right)}\right]^{1/2} \times
\\ [4mm]
&\times&
{_3}F_{2}\left\{
\begin{array}{ccc|c}
-j+m_{+},&-j-m_{-}-\delta_{2}^{(s)},&u &\\ [1mm]
&&&1\\ 
m_{1}+1,&
1-j+m_{+}-v &&
\end{array}
\right\}.
\end{eqnarray*}
Further, using the formula (\ref{1.7.11}), and the explicit expressions for the normalization constants
$C^{(s)}_{k j}\left(\delta_{1}^{(s)}, \delta_{2}^{(s)}\right)$ and
$C^{(s)}_{k \Omega^{(s)} m}\left(\delta_{1}^{(s)}, \delta_{2}^{(s)}\right)$, we finally obtain
\begin{eqnarray}
&&W^{j}_{k \Omega^{(s)} m s}\left(\delta_{1}^{(s)}, \delta_{2}^{(s)}\right) =
\sqrt{\frac{\mu_{0}^{1/2}}{\pi k \hbar}}\frac{(-1)^{j-m_{+}}e^{-i\delta_{j}}}{\Gamma\left(m_{1}+1\right)} \times \nonumber \\ [3mm]
&\times& \frac{\left|\Gamma\left(\frac{m_{1}+1}{2} - \frac{i}{2kr_{0}} - i\frac{\sqrt{\mu_{0}}}{2k\hbar}\Omega^{(s)}\right)
\Gamma\left(\frac{m_{2}+1}{2}-\frac{i}{2kr_{0}}+i\frac{\sqrt{\mu_{0}}}{2k\hbar}\Omega^{(s)}\right)\right|}
{\Gamma\left(m_{+}+\frac{\delta_{1}^{(s)}+ \delta_{2}^{(s)}}{2}+1+\frac{i}{2kr_{0}}\right)}
\times \nonumber \\
\label{5.5.10}
\\
&\times&
\left[\frac{(2j+\delta_{1}^{(s)}+ \delta_{2}^{(s)}+1)\Gamma\left(j+m_{+}+\delta_{1}^{(s)}+ \delta_{2}^{(s)}+1\right)
\Gamma\left(j-m_{-}+\delta_{1}^{(s)}+1\right)}{4(j-m_{+})! \Gamma\left(j+m_{-}+\delta_{2}^{(s)}+1\right)}\right]^{1/2}
\times  \nonumber \\ [3mm]
&\times&
_{3}F_{2}\left\{
\begin{array}{ccc|c}
-j+m_{+},&j+m_{+}+\delta_{1}^{(s)}+\delta_{2}^{(s)}+1,
&\frac{m_{1}+1}{2}-\frac{i}{2kr_{0}}-i\frac{\sqrt{\mu_{0}}}{2k\hbar}\Omega^{(s)} &\\ [3mm]
&&&1\\ 
m_{1}+1,&
m_{+}+\frac{\delta_{1}^{(s)} + \delta_{2}^{(s)}}{2} + 1 + \frac{i}{2kr_{0}} &&
\end{array}
\right\}, \nonumber
\end{eqnarray}
where
\begin{eqnarray*}
\delta_{j} = \arg \Gamma\left(j + \frac{\delta_{1}^{(s)}+ \delta_{2}^{(s)}}{2}+1 - \frac{i}{kr_{0}}\right).
\end{eqnarray*}
Using formula (\ref{1.7.11}), we see that coefficients (\ref{5.5.10}) are real.

Now we consider the expansion of the spherical basis of the generalized MIC–Kepler system with respect
to the parabolic basis. We note that in generally speaking, the parabolic separation constant
$\Omega^{(s)}$ can take both real and complex values. Therefore, it is not clear which region of integration over
$\Omega^{(s)}$ ensures the orthogonality of the coefficients $W^{j}_{k \Omega^{(s)} m s}\left(\delta_{1}^{(s)}, \delta_{2}^{(s)}\right)$
given by (\ref{5.5.10}) with respect to the quantum number $j$.

We prove an orthogonality property important for us:
\begin{eqnarray}
Q_{jj'} = \int\limits_{-\infty}^{\infty} W^{j}_{k \Omega^{(s)} m s}\left(\delta_{1}^{(s)}, \delta_{2}^{(s)}\right)
W^{j'*}_{k \Omega^{(s)} m s}\left(\delta_{1}^{(s)}, \delta_{2}^{(s)}\right)d\Omega^{(s)} = \delta_{j j'}.
\label{5.5.11}
\end{eqnarray}
We substitute expression \ref{5.5.10}) for $W^{j}_{k \Omega^{(s)} m s}\left(\delta_{1}^{(s)}, \delta_{2}^{(s)}\right)$
in the integral \ref{5.5.11}), write the generalized hypergeometric
function $_{3}F_{2}$ as a polynomial, and change the variable as
$z = i\frac{\sqrt{\mu_{0}}}{2k\hbar}\Omega^{(s)}$. We obtain
\begin{eqnarray*}
&&Q_{jj'} = \left(-1\right)^{j+j'-2m_{+}}\frac{e^{i(\delta_{j'}^{(s)}-\delta_{j}^{(s)})}}{\left[\Gamma\left(m_{1}+1\right)\right]^{2}}
\frac{\sqrt{\left(2j+\delta_{1}^{(s)}+ \delta_{2}^{(s)}+1\right)\left(2j'+\delta_{1}^{(s)}+ \delta_{2}^{(s)}+1\right)}}
{\left|\Gamma\left(m_{+}+\frac{\delta_{1}^{(s)}+ \delta_{2}^{(s)}}{2}+1+\frac{i}{2kr_{0}}\right)\right|^{2}}\times \\
&\times& \sqrt{\Gamma\left(j+m_{+}+\delta_{1}^{(s)}+ \delta_{2}^{(s)}+1\right)
\Gamma\left(j-m_{-}+\delta_{1}^{(s)}+1\right)} \times
\\
&\times& \left[\frac{\Gamma\left(j'+m_{+}+\delta_{1}^{(s)}+ \delta_{2}^{(s)}+1\right)
\Gamma\left(j'-m_{-}+\delta_{1}^{(s)}+1\right)}{\left(j-m_{+}\right)! \Gamma\left(j+m_{-}+\delta_{2}^{(s)}+1\right)
\left(j'-m_{+}\right)! \Gamma\left(j'+m_{-}+\delta_{2}^{(s)}+1\right)}\right]^{\frac{1}{2}}\times \\
\\
&\times& \sum_{p=0}^{j-m_{+}}\frac{\left(j-m_{+}\right)_{p}\left(j+m_{+}+\delta_{1}^{(s)}+ \delta_{2}^{(s)}+1\right)_{p}}
{p! \left(m_{+}+\frac{\delta_{1}^{(s)}+ \delta_{2}^{(s)}}{2}+1+\frac{i}{2kr_{0}}\right)_{p}}
\times
\\
&\times&
\sum_{t=0}^{j'-m_{+}}\frac{\left(j'-m_{+}\right)_{t}\left(j'+m_{+}+\delta_{1}^{(s)}+ \delta_{2}^{(s)}+1\right)_{t}}
{t! \left(m_{+}+\frac{\delta_{1}^{(s)}+ \delta_{2}}{2}+1+\frac{i}{2kr_{0}}\right)_{t}}B_{pt},
\end{eqnarray*}
where
\begin{eqnarray*}
B_{pt} &=& \frac{1}{2\pi i}\int\limits_{-i\infty}^{i\infty} \Gamma\left(\frac{m_{1}+1}{2}+p+\frac{i}{2kr_{0}}+z\right)
\Gamma\left(\frac{m_{2}+1}{2}-i\frac{i}{2kr_{0}}+z\right) \times \\
&\times&
\Gamma\left(\frac{m_{1}+1}{2}+t-\frac{i}{2kr_{0}}-z\right)
\Gamma\left(\frac{m_{2}+1}{2}+\frac{i}{2kr_{0}}-z\right)dz.
\end{eqnarray*}
By the Barne's lemma (\ref{2.10.8}) if the poles of $\Gamma(\gamma - s)\Gamma(\delta - s)$ lie to the right of
the path of integration and the poles of $\Gamma(\alpha + s)\Gamma(\beta + s)$
lie to the left, and none of the poles in the first set coincides with any of the poles in the second set. In our
case, the conditions of the lemma are satisfied, and we hence have
\begin{eqnarray*}
&&Q_{jj'} = \left(-1\right)^{j+j'-2m_{+}} e^{i(\delta_{j'}-\delta_{j})} \Gamma\left(m_{2}+1\right)
\frac{\sqrt{\left(2j+\delta_{1}^{(s)}+ \delta_{2}^{(s)}+1\right)\left(2j'+\delta_{1}^{(s)}+ \delta_{2}^{(s)}+1\right)}}
{\Gamma\left(m_{1}+1\right)\Gamma\left(2m_{+}+\delta_{1}^{(s)}+ \delta_{2}^{(s)}+2\right)}\times \\ [3mm]
&\times& \left[\frac{\Gamma\left(j+m_{+}+\delta_{1}^{(s)}+ \delta_{2}^{(s)}+1\right)
\Gamma\left(j-m_{-}+\delta_{1}^{(s)}+1\right)\Gamma\left(j'+m_{+}+\delta_{1}^{(s)}+ \delta_{2}^{(s)}+1\right)}
{\left(j-m_{+}\right)! \Gamma\left(j+m_{-}+\delta_{2}^{(s)}+1\right)
\left(j'-m_{+}\right)! \Gamma\left(j'+m_{-}+\delta_{2}^{(s)}+1\right)}\right]^{\frac{1}{2}}\times \\ [3mm]
&\times& \sqrt{\Gamma\left(j'-m_{-}+\delta_{1}^{(s)}+1\right)}
\sum_{p=0}^{j-m_{+}}\frac{\left(j-m_{+}\right)_{p}\left(j+m_{+}+\delta_{1}^{(s)}+ \delta_{2}^{(s)}+1\right)_{p}}
{p! \left(2m_{+}+\delta_{1}^{(s)}+ \delta_{2}^{(s)}+2\right)_{p}}\times \\ [3mm]
&\times& {_{3}F_{2}}\left\{
\begin{array}{ccc|c}
-j'+m_{+},&j'+m_{+}+\delta_{1}^{(s)}+\delta_{2}^{(s)}+1,
&m_{1}+p+1 &\\ [2mm]
&&&1\\ 
m_{1}+1,&
2m_{+}+\delta_{1}^{(s)}+ \delta_{2}^{(s)}+p+2 &&
\end{array}
\right\}.
\end{eqnarray*}
Now, using the Saalschutz theorem (\ref{2.10.9}) twice, we finally obtain
\begin{eqnarray*}
&&Q_{jj'} = \frac{\left(-1\right)^{2(j+j')e^{i(\delta_{j'}-\delta_{j})}}}
{\Gamma\left(j-j'+1\right)\Gamma\left(j'-j+1\right)}
\frac{\sqrt{\left(2j+\delta_{1}^{(s)}+ \delta_{2}^{(s)}+1\right)\left(2j'+\delta_{1}^{(s)}+ \delta_{2}^{(s)}+1\right)}}
{j+j'+\delta_{1}^{(s)}+\delta_{2}^{(s)}+1} \times \\
\\
&\times& \left[\frac{\left(j'-m_{+}\right)!\Gamma\left(j+m_{+}+\delta_{1}^{(s)}+ \delta_{2}^{(s)}+1\right)
\Gamma\left(j-m_{-}+\delta_{1}^{(s)}+1\right)\Gamma\left(j'+m_{-}+\delta_{2}^{(s)}+1\right)}
{\left(j-m_{+}\right)! \Gamma\left(j'+m_{+}+\delta_{1}^{(s)}+ \delta_{2}^{(s)}+1\right)
\Gamma\left(j'-m_{-}+\delta_{1}^{(s)}+1\right)\Gamma\left(j+m_{-}+\delta_{2}^{(s)}+1\right)}
\right]^{\frac{1}{2}}.
\end{eqnarray*}
Because the numbers $j$ and $j'$ are integers or half-integers simultaneously, we see that the last relation is
zero for $j \neq j'$ due to the product of gamma functions of $(j - j' + 1)$ and $(j' - j + 1)$, and is equal to
$1$ for $j = j'$, i.e.,
\begin{eqnarray}
\int\limits_{-\infty}^{\infty} W^{j}_{k \Omega^{(s)} m s}\left(\delta_{1}^{(s)}, \delta_{2}^{(s)}\right)
W^{j'*}_{k \Omega^{(s)} m s}\left(\delta_{1}^{(s)}, \delta_{2}^{(s)}\right)d\Omega^{(s)} = \delta_{j j'}.
\label{5.5.12}
\end{eqnarray}
It then follows from (\ref{5.5.12}) and (\ref{5.5.7}) that the spherical basis of the generalized MIC–Kepler
problem can be expanded with respect to the parabolic basis as
\begin{eqnarray}
\psi^{(s)}_{k j m}\left(r, \theta, \varphi; \delta_{1}^{(s)}, \delta_{2}^{(s)}\right) =
\int\limits_{-\infty}^{\infty} W^{j}_{k \Omega^{(s)} m s}\left(\delta_{1}^{(s)}, \delta_{2}^{(s)}\right)
\psi^{(s)}_{k \Omega^{(s)} m}\left(\mu, \nu, \varphi; \delta_{1}^{(s)}, \delta_{2}^{(s)}\right)d\Omega^{(s)},
\label{5.5.13}
\end{eqnarray}
where the integrals are taken over the real axis.

The results in this Section at $\lambda_{1} = \lambda_{2} = 0$ coincide with the formulas obtained in
the Section 3.6; at $s = 0$ and$\lambda_{1} = \lambda_{2} = 0$,
they coincide up to a phase factor with the formulas given in our monograph \cite{MONO}.

\section{Scattering problem}
\markboth{CHAPTER 5. THE GENERALIZED MIC-KEPLER PROBLEM}{5.6. SCATTERING PROBLEM}

Because the generalized MIC–Kepler system is a Coulomb-like system, the wave function must be independent
of the azimuthal angle $\varphi$. Substituting $m = s$ in Eqs. (\ref{5.2.1}) and (\ref{5.2.2}) for a
continuous spectrum and accordingly setting  $m_{1} = \delta_{1}^{(s)} = 2\sqrt{\Lambda_{1}}$ and
$m_{2} = 2|s| + \delta_{2}^{(s)} = 2\sqrt{s^{2} + \Lambda_{1}}$, we obtain
\begin{eqnarray*}
\frac{d}{d\mu}\left(\mu\frac{d\Phi_{1}}{d\mu}\right)+\left[\frac{k^{2}}{4}\mu - \frac{{\delta_{1}^{(s)}}^{2}}{4\mu}
+\frac{\sqrt{\mu_{0}}}{2\hbar}\Omega^{(s)}\left(\delta_{1}^{(s)}, \delta_{2}^{(s)}\right) + \frac{1}{2r_{0}}\right]\Phi_{1}=0,
\end{eqnarray*}
\begin{eqnarray*}
\frac{d}{d\nu}\left(\nu\frac{d\Phi_{2}}{d\nu}\right)+\left[\frac{k^{2}}{4}\nu - \frac{\left(2|s| + \delta_{2}^{(s)}\right)^{2}}{4\nu}
-\frac{\sqrt{\mu_{0}}}{2\hbar}\Omega^{(s)}\left(\delta_{1}^{(s)}, \delta_{2}^{(s)}\right) + \frac{1}{2r_{0}}\right]\Phi_{2}=0.
\end{eqnarray*}
If we assume that the parabolic separation constant is
\begin{eqnarray*}
\Omega^{(s)}\left(\delta_{1}^{(s)}, \delta_{2}^{(s)}\right) = -\frac{\hbar}{\sqrt{\mu_{0}}}
\left[\frac{1}{r_{0}} + ik\left(\delta_{1}^{(s)} + 1\right)\right],
\end{eqnarray*}
then by formulas (\ref{5.5.4}) and (\ref{5.5.5}), the solution of the Schr\"{o}dinger equation has the form
\begin{eqnarray*}
\psi^{(s)}_{k}\left(\delta_{1}^{(s)}, \delta_{2}^{(s)}\right) &=&C^{(s)}_{k}\left(\delta_{1}^{(s)}, \delta_{2}^{(s)}\right)
e^{ik(\mu - \nu)/2}(ik\mu)^{\delta_{1}^{(s)}/2} (ik\nu)^{|s|+\delta_{2}^{(s)}/2}\times  \\ [3mm]
&\times& F\left(|s|+\frac{\delta_{2}^{(s)}-\delta_{1}^{(s)}}{2}+\frac{i}{kr_{0}}; 2|s|+\delta_{2}^{(s)}+1; ik\nu\right),
\end{eqnarray*}
where $C^{(s)}_{k}\left(\delta_{1}, \delta_{2}\right)$ is the normalization constant. For large $\eta$,
taking only the first two terms of representation (\ref{1.13.3}) for the confluent hypergeometric function, we obtain
\begin{eqnarray*}
&&F\left(|s|+\frac{\delta_{2}^{(s)}-\delta_{1}^{(s)}}{2}+\frac{i}{kr_{0}}; 2|s|+\delta_{2}^{(s)}+1; ik\nu\right) \approx
e^{-\pi/2kr_{0}}e^{i\pi \left(|s| + \left(\delta_{2}^{(s)} - \delta_{1}^{(s)}\right)/2\right)} \times \\ [3mm]
&\times& \frac{\Gamma\left(2|s|+\delta_{2}^{(s)}+1\right)}{\Gamma\left(|s|+\frac{\delta_{1}^{(s)}+\delta_{2}^{(s)}}{2}+1-\frac{i}{kr_{0}}\right)}
\left(ik\nu\right)^{-|s|-\frac{\delta_{2}^{(s)}-\delta_{1}^{(s)}}{2}}\Biggl\{\Biggl[1-\frac{i}{k\nu}\left(|s|+
\frac{\delta_{2}^{(s)}-\delta_{1}^{(s)}}{2}+\frac{i}{kr_{0}}\right)\times \\ [3mm]
&\times& \left(|s|+\frac{\delta_{1}^{(s)}+\delta_{2}^{(s)}}{2}-\frac{i}{kr_{0}}\right)\Biggr]
e^{-\frac{i}{kr_{0}} \ln k\nu} + e^{-i\pi\left(|s|+\frac{\delta_{2}^{(s)}-\delta_{1}^{(s)}}{2}\right)}
\left(|s|+\frac{\delta_{2}^{(s)}-\delta_{1}^{(s)}}{2}+\frac{i}{kr_{0}}\right)
\times \\ [3mm]
&\times& \frac{\Gamma\left(|s|+\frac{\delta_{1}^{(s)}+\delta_{2}^{(s)}}{2}+1-\frac{i}{kr_{0}}\right)}
{\Gamma\left(|s|+\frac{\delta_{2}^{(s)}-\delta_{1}^{(s)}}{2}+1+\frac{i}{kr_{0}}\right)}
\frac{e^{ik\nu}}{(ik\nu)^{\delta_{1}^{(s)}+1}}e^{\frac{i}{kr_{0}} \ln k\nu}\Biggr\}.
\end{eqnarray*}
Choosing the normalization constant $C^{(s)}_{k}\left(\delta_{1}^{(s)}, \delta_{2}^{(s)}\right)$ in the form
\begin{eqnarray*}
C^{(s)}_{k}\left(\delta_{1}, \delta_{2}\right)=
e^{\pi/2kr_{0}}e^{-i\pi\left(|s|+\frac{\delta_{2}^{(s)}-\delta_{1}^{(s)}}{2}\right)}
\frac{\Gamma\left(|s|+\frac{\delta_{1}^{(s)}+\delta_{2}^{(s)}}{2}+1-\frac{i}{kr_{0}}\right)}
{\Gamma\left(2|s|+\delta_{2}^{(s)} + 1\right)}
\end{eqnarray*}
and passing to spherical coordinates, by formulas (\ref{3.2.3}), we obtain
\begin{eqnarray*}
\psi^{(s)}_{k}\left(\delta_{1}^{(s)}, \delta_{2}^{(s)}\right) &=&
\left(kr\sin\theta\right)^{\delta_{1}^{(s)}} \Biggl[1-\frac{i}{2kr\sin^{2}\theta/2}
\left(|s|+\frac{\delta_{2}^{(s)}-\delta_{1}^{(s)}}{2}+\frac{i}{kr_{0}}\right) \times  \\ [3mm]
&\times& \left(|s|+\frac{\delta_{1}^{(s)}+\delta_{2}^{(s)}}{2}-\frac{i}{kr_{0}}\right)\Biggr]
\exp\left[ikz-\frac{i}{kr_{0}} \ln \left(2kr\sin^{2}\theta/2\right)\right] + \\ [3mm]
&+& \frac{f\left(\theta;\delta_{1}^{(s)}, \delta_{2}^{(s)}\right)}{r} \exp\left(ikr+\frac{i}{kr_{0}} \ln 2kr\right),
\end{eqnarray*}
where
\begin{eqnarray*}
f\left(\theta;\delta_{1}^{(s)}, \delta_{2}^{(s)}\right) =
&=& e^{-i\pi\left(|s|+\delta_{2}^{(s)}/2\right)}
\frac{\left[1-ikr_{0}\left(|s|+\frac{\delta_{2}-\delta_{1}}{2}\right)\right]
\left(\cos\theta/2\right)^{\delta_{1}^{(s)}}}
{2k^{2}r_{0}\left(\sin\theta/2\right)^{\delta_{1}^{(s)}+2}}
\times \\ [3mm]
&\times&
\frac{\Gamma\left(|s|+\frac{\delta_{1}^{(s)}+\delta_{2}^{(s)}}{2}+1-\frac{i}{kr_{0}}\right)}
{\Gamma\left(|s|+\frac{\delta_{2}^{(s)}-\delta_{1}^{(s)}}{2}+1+\frac{i}{kr_{0}}\right)}
\exp\left(\frac{2i}{kr_{0}} \ln \sin\theta/2\right).
\end{eqnarray*}

For $\delta_{1}^{(s)} = 0$ (i.e., $\lambda_{1} = 0$), as in the case of Rutherford scattering, it follows that for negative
$z$ and large values of $r$, the wave function has the form of a plane wave,
\begin{eqnarray}
\psi^{(s)}_{k}\left(0, \delta_{2}^{(s)}\right) &=&
\left\{1-\frac{i}{2kr\sin^{2}\theta/2}
\left[\left(|s|+\frac{\delta_{2}^{(s)}}{2}\right)^{2}+\frac{1}{k^{2}r_{0}^{2}}\right]\right\}
\times \nonumber \\
\label{5.6.1}
\\
&\times& \exp\left[ikz-i\frac{1}{kr_{0}} \ln \left(2kr\sin^{2}\theta/2\right)\right]
+\frac{f\left(\theta;0, \delta_{2}^{(s)}\right)}{r} \exp\left(ikr+\frac{i}{kr_{0}} \ln 2kr\right), \nonumber
\end{eqnarray}
where the scattering amplitude $f\left(\theta;0, \delta_{2}^{(s)}\right)$ is given by
\begin{eqnarray}
f\left(\theta;0, \delta_{2}^{(s)}\right) &=& e^{-i\pi\left(|s|+\delta_{2}/2\right)}
\frac{1-ikr_{0}\left(|s|+ \delta_{2}/2\right)}
{2k^{2}r_{0}\left(\sin\theta/2\right)^{2}}
\times \nonumber \\
\label{5.6.2}
\\
&\times&
\frac{\Gamma\left(|s|+\frac{\delta_{2}^{(s)}}{2}+1-\frac{i}{kr_{0}}\right)}
{\Gamma\left(|s|+\frac{\delta_{2}}{2}+1 + \frac{i}{kr_{0}}\right)}
\exp\left(\frac{2i}{kr_{0}} \ln \sin\theta/2\right). \nonumber
\end{eqnarray}
For the scattering cross section $d\sigma = \left|f\left(\theta;0, \delta_{2}^{(s)}\right)\right|^{2}d\Omega$
(where $d\Omega$ is a solid angle element), we obtain the formula
\begin{eqnarray}
d\sigma = \frac{1+k^{2}r_{0}^{2}\left(|s|+\delta_{2}^{(s)}/2\right)^{2}}
{4k^{4}r_{0}^{2}\left(\sin\theta/2\right)^{4}}d\Omega.
\label{5.6.3}
\end{eqnarray}

Thus, formulas (\ref{5.6.1}) - (\ref{5.6.3}) describe the scattering problem in the system with the Hamiltonian
\begin{eqnarray*}
\hat{\mathcal{H}} = \frac{1}{2\mu_{0}}\,\left(-i\hbar\,\nabla - \frac{e}{c}\,{\bf A}^{(\pm)}\right)^{2}
+ \frac{\hbar^{2}s^{2}}{2\mu_{0}r^{2}} - \frac{e^{2}}{r} + \frac{\lambda_{2}}{r(r-z)}.
\end{eqnarray*}

Further, substituting $\delta_{2}^{(s)} = 0$ (i.e., $\lambda_{2} = 0$) in formulas
(\ref{5.6.1}) - (\ref{5.6.3}), we obtain the relations
\begin{eqnarray*}
\psi^{(s)}_{k}\left(0, 0\right) &=&
\left[1-\frac{i\left(1+k^{2}r_{0}^{2}s^{2}\right)}{2k^{3}r_{0}^{2}r\sin^{2}\theta/2}\right]
\exp\left[ikz-\frac{i}{kr_{0}} \ln \left(2kr\sin^{2}\theta/2\right)\right] +\\
[3mm]
&+&\frac{f(\theta;0, 0)}{r} \exp\left(ikr+\frac{i}{kr_{0}} \ln 2kr\right),
\end{eqnarray*}
\begin{eqnarray*}
f(\theta;0, 0) &=& e^{-i\pi|s|}
\frac{1-ikr_{0}|s|}
{2k^{2}r_{0}\left(\sin\theta/2\right)^{2}}
\frac{\Gamma\left(|s|+1-\frac{i}{kr_{0}}\right)}
{\Gamma\left(|s|+1+\frac{i}{kr_{0}}\right)}
\exp\left(\frac{2i}{kr_{0}} \ln \sin\theta/2\right),
\end{eqnarray*}
\begin{eqnarray*}
d\sigma = \frac{1 + k^{2}r_{0}^{2}s^{2}}
{4k^{4}r_{0}^{2}\left(\sin\theta/2\right)^{4}}d\Omega,
\end{eqnarray*}
which describe the scattering of charged particles in the field of a Dirac dyon (see Section 3.5) and, as is easy to see,
pass into the well-known formulas of the Rutherford scattering for $s = 0$.

\section{Some questions of the theory of KS-transformation}
\markboth{CHAPTER 5. THE GENERALIZED MIC-KEPLER PROBLEM}{5.7. SOME QUESTIONS OF THE THEORY OF KS-TRANSFORMATION}

In Section 3.7 we showed that the MIC-Kepler problem is dual to the four-dimensional isotropic oscillator and that
the duality transformation is a generalized version of the KS--transformation (\ref{3.7.9}) supplemented by the
ansatz $\psi^{s}(\textbf{r}) \rightarrow \psi(\textbf{r}, \gamma)$, $s \rightarrow -i\partial/\partial \gamma$.
Now, before finding out which system the generalized MIC-Kepler problem corresponds to, let’s consider some properties
of the KS--transformation.

The first three equations of relation (\ref{3.7.9}) are a nonbijective quadratic transformation ${\rm I\!R}^4 \to {\rm I \!R}^3$,
which corresponds to each point $\left(u_{0}, u_{1}, u_{2}, u_{3}\right)$ of the four-dimensional space ${\rm I\!R}^4$ with a
point $\left(x, y, z\right)$ of the three-dimensional space ${\rm I \!R}^3$ and can be written in the form \cite{KS}:
\begin{eqnarray*}
X = H(u; 4)\,U,
\end{eqnarray*}
where $X$ and $U$ denote the vectors
\begin{eqnarray*}
X = \left(\begin{array}{c}
x \\y \\z \\0
\end{array}
\right), \qquad U = \left(\begin{array}{c}
u_0 \\u_1 \\u_2 \\u_3
\end{array}
\right),
\end{eqnarray*}
and the matrix $H(u;4)$ has the form
\begin{eqnarray}
H(u;4) =
\left(
\begin{array}{cccc}
u_2& -u_3 &u_0 &-u_1\\
u_3 &u_2 &u_1 &u_0\\
u_0 &u_1 &-u_2 &-u_3\\
u_1 &-u_0 &-u_3 &u_2\\
\end{array}
\right).
\label{5.7.1}
\end{eqnarray}
It is easy to verify that the matrix $H(u;4)$ satisfies the following condition:
\begin{eqnarray}
H_{\mu \lambda}(u;4)\,H_{\lambda \nu }^{T}(u;4) = u^{2}\,\delta_{\mu \nu},
\label{5.7.2}
\end{eqnarray}
which guarantees the observance of the Euler identity. Indeed, from the relation $X^{T}X = U^{T}H^{T}(u;4)H(u;4)U$,
taking into account (\ref{5.7.1}), we have that
\begin{eqnarray*}
r^{2} = x^{2} + y^{2} + z^{2} = \left(u_{0}^{2} + u_{1}^{2} + u_{2}^{2} + u_{3}^{2}\right)^{2} = u^{4}.
\end{eqnarray*}
For differential elements we have
\begin{eqnarray}
\left(\begin{array}{c}
dx \\dy \\dz \\0
\end{array}
\right) = 2
\left(
\begin{array}{cccc}
u_2& -u_3 &u_0 &-u_1\\
u_3 &u_2 &u_1 &u_0\\
u_0 &u_1 &-u_2 &-u_3\\
u_1 &-u_0 &-u_3 &u_2\\
\end{array}
\right)\left(\begin{array}{c}
du_0 \\du_1 \\du_2 \\du_3
\end{array}
\right).
\label{5.7.3}
\end{eqnarray}
From (\ref{5.7.2}) and (\ref{5.7.3}) it follows that
\begin{eqnarray*}
dx^{2} + dy^{2} + dz^{2} = 4u^{2}\left(u_{0}^{2} + u_{1}^{2} + u_{2}^{2} + u_{3}^{2}\right).
\end{eqnarray*}
For partial derivatives, the following relation holds:
\begin{eqnarray}
\left(\begin{array}{c}
\partial /\partial x \\\partial /\partial y \\\partial /\partial z \\ \frac{1}{2r}\hat{X}
\end{array}
\right) = \frac{1}{2r}
\left(
\begin{array}{cccc}
u_2& -u_3 &u_0 &-u_1\\
u_3 &u_2 &u_1 &u_0\\
u_0 &u_1 &-u_2 &-u_3\\
u_1 &-u_0 &-u_3 &u_2\\
\end{array}
\right)\left(\begin{array}{c}
\partial /\partial u_0 \\\partial /\partial u_1 \\\partial /\partial u_2 \\\partial /\partial u_3
\end{array}
\right),
\label{5.7.4}
\end{eqnarray}
where
\begin{eqnarray}
\hat{X} = u_1\frac{\partial}{\partial u_0} - u_0\frac{\partial}{\partial u_1} -
u_3\frac{\partial}{\partial u_2} + u_2\frac{\partial}{\partial u_3}.
\label{5.7.5}
\end{eqnarray}
From (\ref{5.7.4}) we obtain the following relationship between the Laplace operator $\triangle_{3}$ in
${\rm I\!R}^3$ and the Laplace operator $\triangle_{4}$ in ${\rm I\!R}^4$
\begin{eqnarray}
\triangle_{3} = \frac{1}{4u^{2}}\triangle_{4} - \frac{1}{4u^{}}\hat{X}.
\label{5.7.6}
\end{eqnarray}

Now let's establish a connection between the elements of volumes in the spaces $\triangle_{4}$ and
$\triangle_{3}$. This is most convenient to do in Eulerian coordinates (\ref{3.8.1}). It follows directly
from the definition of Eulerian coordinates (\ref{3.8.1}) that
\begin{eqnarray}
\gamma = \frac{i}{2}\,\ln\,\frac{\left(u_{0} - iu_{1}\right)\,\left(u_{2} + iu_{3}\right)}
{\left(u_{0} + iu_{1}\right)\,\left(u_{2} - iu_{3}\right)}.
\label{5.7.7}
\end{eqnarray}
Based on the last relation, we obtain that the operator $\hat{X}$ (\ref{5.7.5}) in Euler coordinates takes a simple form:
\begin{eqnarray}
\hat{X} = -2\,\frac{\partial}{\partial \gamma}.
\label{5.7.8}
\end{eqnarray}
Now, using the KS - transformation (\ref{3.7.9}), it is easy to establish that the Euler coordinates
$u \in [0, \infty ), \beta \in [0, \pi]$, and $\alpha \in [0, 2\pi)$ are related to the three-dimensional spherical coordinates
$r, \theta$, and $\varphi$ as follows:
\begin{eqnarray}
r = u^{2}, \qquad \theta = \beta, \qquad \varphi = \alpha.
\label{5.7.9}
\end{eqnarray}
Then the differential element of the four-dimensional volume (\ref{3.8.1}) can be written in the form
\begin{eqnarray*}
dV_{4} = \frac{u^{3}}{8}\,
\sin\beta\,du\,d\beta\,d\alpha\,d\gamma = \frac{1}{16}r\,\sin\theta\,dr\,d\theta\,d\varphi\,d\gamma.
\end{eqnarray*}
Further, comparing the last relation with the formula for a three-dimensional volume in spherical
coordinates $dV_{3} = r^{2}\,\sin\theta\,dr\,d\theta\,d\varphi$, we establish that
\begin{eqnarray}
dV_{4} = \frac{1}{16r}\, dV_{3}\,d\gamma.
\label{5.7.10}
\end{eqnarray}
Formula (\ref{5.7.10}) also allows us to establish a connection between the volume integrals in
${\rm I\!R}^4$ and ${\rm I\!R}^3$. Integration over $\gamma$ from $0$ to $4\pi$ leads to the relation
\begin{eqnarray}
\int_{{\rm I\!R}^4} \dots d V_{4} = \frac{\pi}{4}\, \int_{{\rm I\!R}^3} \dots d V_{3}.
\label{5.7.11}
\end{eqnarray}

Now we write the four-dimensional Laplace operator $\triangle_{4}$ in coordinates $(x, y, z, \gamma)$.
To do this, we will use the explicit form of the operator $\triangle_{4}$ in Euler coordinates (\ref{3.8.1}).
Taking into account (\ref{5.7.9}) it is easy to establish that
\begin{eqnarray}
\frac{1}{u^{3}}\,\frac{\partial}{\partial u}\,\left(u^{3}\frac{\partial}{\partial u}\right) =
4u^{2}\,\left[\frac{1}{r^{2}}\,\frac{\partial}{\partial r}\left(r^{2}\frac{\partial}{\partial r}\right)\right],
\label{5.7.12}
\end{eqnarray}
\begin{eqnarray}
{\hat {\bf L}}^2 = {\hat {\bf l}}^2 + \frac{1}{\sin^{2}\theta}\left(2\cos\theta\,\frac{\partial^{2}}{\partial \varphi \partial \gamma}
- \frac{\partial^{2}}{\partial \gamma^{2}}\right),
\label{5.7.13}
\end{eqnarray}
where
\begin{eqnarray*}
{\hat {\bf l}}^2 = - \left[\frac{1}{\sin\theta}\,\frac{\partial}{\partial \theta}
\left(\sin\theta \frac{\partial}{\partial \theta}\right) + \frac{1}{\sin^{2}\theta}
\frac{\partial^{2}}{\partial \varphi^{2}}\right]
\end{eqnarray*}
the square of the orbital momentum operator. Thus, for the four-dimensional Laplace operator we obtain the expression:
\begin{eqnarray}
\triangle_{4} = 4u^{2}\,\left[\triangle_{3} -  \frac{1}{\sin^{2}\theta}\left(2\cos\theta\,\frac{\partial^{2}}
{\partial \varphi \partial \gamma} - \frac{\partial^{2}}{\partial \gamma^{2}}\right)\right].
\label{5.7.14}
\end{eqnarray}
Further, taking into account that
\begin{eqnarray*}
\frac{\partial}{\partial \varphi} = - y \frac{\partial}{\partial x} +
x\frac{\partial}{\partial y}
\end{eqnarray*}
equation (\ref{5.7.14}) can be written as
\begin{eqnarray}
\frac{1}{4u^{2}}\triangle_{4} = \left[\frac{\partial}{\partial x} + \frac{yz}{r\left(r^{2} - z^{2}\right)}
\frac{\partial}{\partial \gamma}\right]^{2} +
\left[\frac{\partial}{\partial y} - \frac{xz}{r\left(r^{2} - z^{2}\right)}
\frac{\partial}{\partial \gamma}\right]^{2} + \frac{\partial^{2}}{\partial z^{2}} +
\frac{1}{r^{2}}\frac{\partial^{2}}{\partial \gamma^{2}}.
\label{5.7.15}
\end{eqnarray}
Thus, we have found an explicit expression for the Laplace operator $\triangle_{4}$ in coordinates
$(x, y, z, \gamma)$, which are the coordinates of the four-dimensional space ${\rm I\!R}^4$ represented as a
direct product of the three-dimensional space ${\rm I\!R}^3$ and the one-dimensional sphere $S^{1}$,
i.e. ${\rm I\!R}^3 \bigotimes S^{1}$.

\section{GMICK and $4D$ double singular oscillator}
\markboth{CHAPTER 5. THE GENERALIZED MIC-KEPLER PROBLEM}{5.8. GMICK AND $4D$ DOUBLE SINGULAR OSCILLATOR}

Now let's find out which four-dimensional system corresponds to the generalized MIC-Kepler problem.
We will proceed as follows. First, let us write the Schr\"{o}dinger equation for the generalized
MIC-Kepler problem (\ref{5.0.1}) in Cartesian coordinates
\begin{eqnarray}
\Biggl\{\left[\frac{\partial}{\partial x} - \frac{isy}{r(r-z)}\right]^{2} +
\left[\frac{\partial}{\partial y} + \frac{isx}{r(r-z)}\right]^{2} + \frac{\partial^{2}}{\partial z^{2}} -
\frac{s^{2}}{r^{2}} +  \nonumber \\
\label{5.8.1}
\\
+ \frac{2\mu_{0}}{\hbar^{2}}\left[E + \frac{e^{2}}{r} - \frac{\lambda_{1}}{r(r+z)}
- \frac{\lambda_{2}}{r(r-z)}\right]\Biggr\}\psi^{(s)}({\textbf{r}}) = 0. \nonumber
\end{eqnarray}
Then we rewrite equation (\ref{5.8.1}) in the following form
\begin{eqnarray}
\Biggl\{ \triangle_{3} - \frac{2is}{r(r-z)}\left(y\frac{\partial}{\partial x} - x\frac{\partial}{\partial y}\right)
- \frac{2s^{2}}{r(r-z)} +  \nonumber \\
\label{5.8.2}
\\
+ \frac{2\mu_{0}}{\hbar^{2}}\left[E + \frac{e^{2}}{r} - \frac{\lambda_{1}}{r(r+z)}
- \frac{\lambda_{2}}{r(r-z)}\right]\Biggr\}\psi^{(s)}({\textbf{r}}) = 0. \nonumber
\end{eqnarray}
Next, in equation (\ref{5.8.2}) we making the replacement
\begin{eqnarray}
\psi^{(s)}({\textbf{r}}) \rightarrow \psi({\textbf{r}}, \gamma) = \psi^{(s)}({\textbf{r}})\,
\exp\left(is\arctan\frac{y}{x}\right),
\label{5.8.3}
\end{eqnarray}
and replacing $s\rightarrow -i\partial/\partial \gamma$, we obtain
\begin{eqnarray}
\Biggl\{\left[\frac{\partial}{\partial x} + \frac{yz}{r\left(r^{2} - z^{2}\right)}
\frac{\partial}{\partial \gamma}\right]^{2} +
\left[\frac{\partial}{\partial y} - \frac{xz}{r\left(r^{2} - z^{2}\right)}
\frac{\partial}{\partial \gamma}\right]^{2} + \frac{\partial^{2}}{\partial z^{2}} +
\frac{1}{r^{2}}\frac{\partial^{2}}{\partial \gamma^{2}} + \nonumber \\
\label{5.8.4}
\\
+ \frac{2\mu_{0}}{\hbar^{2}}\left[E + \frac{e^{2}}{r} - \frac{\lambda_{1}}{r(r+z)}
- \frac{\lambda_{2}}{r(r-z)}\right]\Biggr\}\psi({\textbf{r}}, \gamma) = 0. \nonumber
\end{eqnarray}
Now, taking into account formula (\ref{5.7.15}) and relations
\begin{eqnarray*}
r = u^{2}, \qquad r + z = 2\left(u_{0}^{2} + u_{1}^{2}\right), \qquad r - z = 2\left(u_{2}^{2} + u_{3}^{2}\right)
\end{eqnarray*}
equation (\ref{5.8.4}) can be written as
\begin{eqnarray*}
\left\{\frac{1}{4u^{2}}\triangle_{4} +
\frac{2\mu_{0}}{\hbar^{2}}\left[E + \frac{e^{2}}{u^{2}} - \frac{\lambda_{1}}{2u^{2}\left(u_{0}^{2} + u_{1}^{2}\right)}
- \frac{\lambda_{2}}{2u^{2}\left(u_{2}^{2} + u_{3}^{2}\right)}\right]\right\}\psi({\textbf{r}}, \gamma) = 0.
\end{eqnarray*}
Further, multiplying the last equation by $4u^{2}$ and denoting
\begin{eqnarray}
\epsilon = 4e^{2}, \qquad E= - \frac{\mu_{0} \omega^{2}}{8}, \qquad c_{i} = 2\lambda_{i},
\label{5.8.5}
\end{eqnarray}
where $i = 1,2$, finally we get
\begin{eqnarray}
\left[\triangle_{4} +
\frac{2\mu_{0}}{\hbar^{2}}\left(\epsilon - \frac{\mu_{0} \omega^{2}u^{2}}{2} - \frac{c_{1}}{u_{0}^{2} + u_{1}^{2}}
- \frac{c_{2}}{u_{2}^{2} + u_{3}^{2}}\right)\right]\psi({\textbf{u}}) = 0.
\label{5.8.6}
\end{eqnarray}
The four-dimensional system, which is described by the Schr\"{o}dinger equation (\ref{5.8.6}) and dual to
the generalized MIC-Kepler problem, will be called a double singular oscillator in what follows.

Thus, in this case too, the generalized version of the KS - transformation (\ref{3.7.9}) together with
the transformation $s\rightarrow -i\partial/\partial \gamma$ represents a duality transformation.

\section{Bases of $4D$ double singular oscillator}
\markboth{CHAPTER 5. THE GENERALIZED MIC-KEPLER PROBLEM}{5.9. BASES OF $4D$ DOUBLE SINGULAR OSCILLATOR}

The variables in the Schr\"{o}dinger equation for the four-dimensional double singular oscillator are
separated in Euler (\ref{3.8.1}), double polar (\ref{3.8.3}) and spheroidal (\ref{3.8.5}) coordinates.
The following spectral problems take place in these coordinate systems.

1. The Euler basis of the $4D$ double singular oscillator
$\psi_{N L M M'}\left(u, \alpha, \beta, \gamma; \Delta_{1}, \Delta_{2}\right)$
is simultaneously an eigenfunction of the integrals of motion
$\left\{\hat{H}, \hat{\mathcal{A}}, \hat{L}_{3}, \hat{L}_{3}^{'}\right\}$:
\begin{eqnarray}
\hat{H}\psi_{N L M M'} = \hbar \omega \left(N + \Delta_{1} + \Delta_{2} \right)\psi_{N L M M'},
\label{5.9.1}
\end{eqnarray}
\begin{eqnarray}
\hat{\mathcal{A}}\psi_{N L M M'} = \left(L + \frac{\Delta_{1} + \Delta_{2}}{2}\right)
\left(L + \frac{\Delta_{1} + \Delta_{2}}{2} + 1\right)\psi_{N L M M'},
\label{5.9.2}
\end{eqnarray}
\begin{eqnarray}
\hat{L}_{3}\psi_{N L M M'} = -i\frac{\partial \psi_{N L M M'}}{\partial \alpha} = M\psi_{N L M M'},
\label{5.9.3}
\end{eqnarray}
\begin{eqnarray}
\hat{L}_{3}^{'}\psi_{N L M M'} = -i\frac{\partial \psi_{N L M M'}}{\partial \gamma} = M'\psi_{N L M M'}.
\label{5.9.4}
\end{eqnarray}
Here
\begin{eqnarray}
\hat{\mathcal{A}} = \hat{L}^{2} + \frac{c_{1}u^{2}}{2\left(u_{0}^{2} + u_{1}^{2}\right)}
+ \frac{c_{2}u^{2}}{2\left(u_{2}^{2} + u_{3}^{2}\right)},
\label{5.9.5}
\end{eqnarray}
and the explicit expression of the operator $\hat{L}^{2}$ is given in formula (\ref{3.8.2}). In Cartesian
coordinates, the operator $\hat{\mathcal{A}}$ has the form
\begin{eqnarray}
\hat{\mathcal{A}} = -\frac{1}{4}\left(u^{2}\triangle_{4} - u_{i}u_{j}\,\frac{\partial^{2}}{\partial u_{i} \partial u_{j}}
- 3u_{i}\,\frac{\partial}{\partial u_{i}}\right)
 + \frac{c_{1}u^{2}}{2\left(u_{0}^{2} + u_{1}^{2}\right)}
+ \frac{c_{2}u^{2}}{2\left(u_{2}^{2} + u_{3}^{2}\right)},
\label{5.9.6}
\end{eqnarray}
Quantum numbers $N, L, M, M’$ take the values: $N=0,1,2\dots, M,M’ = -L,-L+1,\dots…,L-1, L$ and
\begin{eqnarray*}
L = \frac{\left|M + M'\right|}{2}, \frac{\left|M + M'\right|}{2} + 1, \dots , \frac{N}{2}.
\end{eqnarray*}

The wave function of a four-dimensional double singular oscillator in Euler coordinates has the form:
\begin{eqnarray}
\psi_{N L M M'}\left(u, \alpha, \beta, \gamma; \Delta_{1}, \Delta_{2}\right) = R_{N L}\left(u; \Delta_{1}, \Delta_{2}\right)\,
Z_{L M M'}\left(\alpha, \beta, \gamma; \Delta_{1}, \Delta_{2}\right),
\label{5.9.7}
\end{eqnarray}
where
\begin{eqnarray*}
R_{N L}\left(u; \Delta_{1}, \Delta_{2}\right) = C_{N L}e^{-a^{2}u^{2}/2}\left(au\right)^{2L + \Delta_{1} + \Delta_{2}}\,
F\left(-\frac{N}{2} + L; 2L + \Delta_{1} + \Delta_{2} + 2; a^{2}u^{2}\right),
\end{eqnarray*}
\begin{eqnarray*}
Z_{L M M'}\left(\alpha, \beta, \gamma; \Delta_{1}, \Delta_{2}\right) = N_{L M M'}\left(\cos\frac{\beta}{2}\right)^{M_{1}}\,
\left(\sin\frac{\beta}{2}\right)^{M_{2}}\,P_{L - M_{+}}^{\left(M_{2}, M_{1}\right)}\left(\cos\beta\right)\,
e^{iM\alpha}\,e^{iM'\gamma},
\end{eqnarray*}
and
\begin{eqnarray*}
C_{N L} = \frac{4a^{2}}{\Gamma\left(2L + \Delta_{1} + \Delta_{2} + 2\right)}\,
\sqrt{\frac{\Gamma\left(\frac{N}{2} + L + \Delta_{1} + \Delta_{2} + 2\right)}
{\left(\frac{N}{2} - L\right)!}},
\end{eqnarray*}
\begin{eqnarray*}
C_{N L} = \frac{1}{4\i}\,
\sqrt{\frac{\left(2L + \Delta_{1} + \Delta_{2} + 2\right)\left(L - M_{+}\right)!
\Gamma\left(L + M_{+} + \Delta_{1} + \Delta_{2} + 1\right)}{\Gamma\left(L + M_{-} + \Delta_{1} + 1\right)
\Gamma\left(L - M_{-} + \Delta_{2} + 1\right)}},
\end{eqnarray*}
\begin{eqnarray*}
a = \sqrt{\frac{\mu_{0} \omega}{\hbar}}, \qquad M_{1} = \left|M - M'\right| + \Delta_{1} =
\sqrt{\left(M - M'\right)^{2} + 2\mu_{0}c_{1}/\hbar^{2}}, \\ [3mm]
M_{2} = \left|M + M'\right| + \Delta_{2} =
\sqrt{\left(M + M'\right)^{2} + 2\mu_{0}c_{2}/\hbar^{2}}, \qquad M_{\pm} \frac{\left|M + M'\right| \pm \left|M - M'\right|}{2}.
\end{eqnarray*}

2. The wave function of a four-dimensional double singular oscillator in double polar coordinates (\ref{3.8.3})
$\psi_{N\rho_{1} N\rho_{2} k_{1} k_{2}}\left(\rho_{1}, \rho_{2}, \varphi_{1}, \varphi_{2}; \Delta_{1}, \Delta_{2}\right)$
is a solution of the following spectral problems:
\begin{eqnarray}
\hat{H}\psi_{N\rho_{1} N\rho_{2} k_{1} k_{2}} = \hbar \omega \left(2N\rho_{1} + 2N\rho_{2} + |k_{1}| + |k_{2}| +
\Delta_{1} + \Delta_{2} + 2\right)\psi_{N\rho_{1} N\rho_{2} k_{1} k_{2}},
\label{5.9.8}
\end{eqnarray}
\begin{eqnarray}
\hat{\Pi}\psi_{N\rho_{1} N\rho_{2} k_{1} k_{2}} = \left(2N\rho_{1} - 2N\rho_{2} + |k_{1}| - |k_{2}| +
\Delta_{1} - \Delta_{2}\right)\psi_{N\rho_{1} N\rho_{2} k_{1} k_{2}},
\label{5.9.9}
\end{eqnarray}
\begin{eqnarray}
\hat{L}_{0 1}\psi_{N\rho_{1} N\rho_{2} k_{1} k_{2}} = -i\frac{\partial \psi_{N\rho_{1} N\rho_{2} k_{1} k_{2}}}
{\partial \varphi_{1}} = k_{1}\psi_{N\rho_{1} N\rho_{2} k_{1} k_{2}},
\label{5.9.10}
\end{eqnarray}
\begin{eqnarray}
\hat{L}_{2 3}\psi_{N\rho_{1} N\rho_{2} k_{1} k_{2}} = -i\frac{\partial \psi_{N\rho_{1} N\rho_{2} k_{1} k_{2}}}
{\partial \varphi_{2}} = k_{2}\psi_{N\rho_{1} N\rho_{2} k_{1} k_{2}},
\label{5.9.11}
\end{eqnarray}
where the operators of the integrals of motion $\hat{\Pi}$, and $\hat{L}_{i j}$l in Cartesian coordinates have the form
\begin{eqnarray}
\hat{\Pi} &=& \frac{\hbar}{2\mu_{0}\omega}\left(-\frac{\partial^{2}}{\partial u_{0}^{2}} - \frac{\partial^{2}}{\partial u_{1}^{2}}
+ \frac{\partial^{2}}{\partial u_{2}^{2}} + \frac{\partial^{2}}{\partial u_{3}^{2}}\right) + \frac{\mu_{0}\omega}{2\hbar}
\left(u_{0}^{2} + u_{1}^{2} - u_{2}^{2} - u_{3}^{2}\right) + \nonumber \\
\label{5.9.12}
\\
&+& \frac{1}{\hbar \omega}\left(\frac{c_{1}}{u_{0}^{2} + u_{1}^{2}}
- \frac{c_{2}}{u_{2}^{2} + u_{3}^{2}}\right), \nonumber
\end{eqnarray}
\begin{eqnarray}
\hat{L}_{i j} = i\left(-u_{i}\frac{\partial}{\partial u_{j}} + u_{j}\frac{\partial}{\partial u_{i}}\right).
\label{5.9.13}
\end{eqnarray}
Azimuthal quantum numbers $k_{i}$ change in the regions $|k_{i}| = 0, 1, \dots N\rho_{i}$, where $i=1, 2$.
The non-negative integers $N\rho_{1}$ and $N\rho_{2}$ are related to the principal quantum number $N$ by the relation:
\begin{eqnarray}
N = 2N\rho_{1} + 2N\rho_{2} + |k_{1}| + |k_{2}|.
\label{5.9.14}
\end{eqnarray}

In double polar coordinates, the wave function of a four-dimensional double singular oscillator has the form:
\begin{eqnarray}
\psi_{N\rho_{1} N\rho_{2} k_{1} k_{2}}
= \frac{1}{2\pi}\Phi_{N\rho_{1} k_{1}}\left(a^{2}\rho_{1}^{2}; \Delta_{1}\right)
\Phi_{N\rho_{2} k_{2}}\left(a^{2}\rho_{2}^{2}; \Delta_{2}\right)e^{ik_{1}\varphi_{1}}e^{ik_{2}\varphi_{2}},
\label{5.9.15}
\end{eqnarray}
where
\begin{eqnarray*}
\Phi_{p_{i} k_{i}}\left(x_{i}; \Delta_{i}\right) = \sqrt{\frac{2\Gamma\left(p_{i} + |k_{i}| + \Delta_{i} + 1\right)}
{\left(p_{i}\right)!}}\frac{ae^{-x/2}x^{|k_{i}| +  \Delta_{i}}}{\Gamma\left(|k_{i}| + \Delta_{i} + 1\right)}
F\left(-p_{i}; |k_{i}| + \Delta_{i} + 1; x\right).
\end{eqnarray*}

3. The spheroidal basis $\psi_{N q M M'}\left(\xi, \eta, \alpha, \gamma; \Delta_{1}, \Delta_{2}\right)$ is an eigenfunction
of the following system of commuting operators $\left\{\hat{H}, \hat{X}, \hat{L}_{3}, \hat{L}_{3}^{'}\right\}$:
\begin{eqnarray}
\hat{X}\psi_{N q M M'} = \left(\hat{\mathcal{A}} + \frac{a^{2}d^{2}}{4}\hat{\Pi}\right)\psi_{N q M M'} = X_{q}\psi_{N q M M'},
\label{5.9.16}
\end{eqnarray}
\begin{eqnarray}
\hat{L}_{3}\psi_{N q M M'} = M\psi_{N q M M'}, \qquad \hat{L}_{3}^{'}\psi_{N q M M'} = M'\psi_{N q M M'}.
\label{5.9.17}
\end{eqnarray}
Here the quantum number $q$ numbers the discrete values ​​of the spheroidal separation constant $X$.

It is easy to see that the wave functions of the four-dimensional double singular oscillator (\ref{5.9.7}) and
(\ref{5.9.15}), up to a constant factor, can be obtained from the spherical (\ref{5.1.15}) and parabolic (\ref{5.2.3})
bases of the generalized MIC-Kepler problem, respectively, using the following substitutions of quantum numbers:

$\bullet$ For the Euler basis
\begin{eqnarray*}
n \rightarrow \frac{N}{2} + 1, \qquad j \rightarrow L, \qquad m \rightarrow M, \qquad s \rightarrow M';
\end{eqnarray*}

$\bullet$ For the double polar basis
\begin{eqnarray*}
n_{1} \rightarrow N\rho_{1}, \qquad n_{2} \rightarrow N\rho_{2},
\qquad m \rightarrow \frac{k_{1} + k_{2}}{2}, \qquad s \rightarrow \frac{k_{1} - k_{2}}{2}.
\end{eqnarray*}

Then, from relations (\ref{5.4.11}) and (\ref{5.4.14}), we obtain three-term recurrence relations that
govern the coefficients of the expansions of the spheroidal basis of the four-dimensional double singular
oscillator in the Euler and double polar bases:
\begin{eqnarray}
&&A_{N M M'}^{L + 1}V_{NqMM'}^{L + 1}(d) + A_{N M M'}^{L}V_{NqMM'}^{L - 1}(d) =
\Biggl\{\frac{4}{a^{2}d^{2}}\Biggl[X_{q}\left(d\right) - \nonumber \\
\label{5.9.18}
\\
&-& \left(L + \frac{\Delta_{1} + \Delta_{2}}{2}\right)
\left(L + \frac{\Delta_{1} + \Delta_{2}}{2} + 1\right)\Biggr] -
\frac{\left(N + \Delta_{1} + \Delta_{2} + 2\right)\left(M_{1}^{2}-M_{2}^{2}\right)}
{\left(2L + \Delta_{1} + \Delta_{2}\right)\left(2L + \Delta_{1} + \Delta_{2} + 2\right)}
\Biggr\}V_{NqMM'}^{L}(d), \nonumber
\end{eqnarray}

\begin{eqnarray}
&&\Biggl[\frac{1}{4}\left(N + \Delta_{1} + \Delta_{2}\right)
\left(N + \Delta_{1} + \Delta_{2} + 2\right) - N_{\rho_{1}}\left(N_{\rho_{1}} + |k_{1}| + \Delta_{1}\right)
- N_{\rho_{2}}\left(N_{\rho_{2}} + |k_{2}| + \Delta_{2}\right) +  \nonumber \\ [3mm]
&+& \frac{a^{2}d^{2}}{4}\left(2N_{\rho_{1}} - 2N_{\rho_{}} + |k_{1}| - |k_{2}| + \Delta_{1} - \Delta_{2}\right)
- X_{q}\left(d\right)\Biggr]U_{nqms}^{n_1}(R) = \nonumber \\
\label{5.9.19}
\\
&=& \sqrt{N_{\rho_{2}}\left(N_{\rho_{1}} + 1\right)\left(N_{\rho_{1}} + |k_{1}| + \Delta_{1} + 1\right)
\left(N_{\rho_{2}} + |k_{2}| + \Delta_{2}\right)}U_{nqms}^{N_{\rho_{1}} + 1}(d) +
\nonumber \\ [3mm]
&+&  \sqrt{N_{\rho_{1}}\left(N_{\rho_{2}} + 1\right)\left(N_{\rho_{1}} + |k_{1}| + \Delta_{1}\right)
\left(N_{\rho_{2}} + |k_{2}| + \Delta_{2} + 1\right)}U_{nqms}^{N_{\rho_{1}} - 1}(d).
\nonumber
\end{eqnarray}

Finally, we note that the problem of finding wave functions and coefficients of interbasis expansions
for a four-dimensional double singular oscillator was solved in \cite{Mara}.

\newpage
\pagestyle{headings}

\end{document}